%% file: Dermer_Saas_Fee_arXiv.tex
\pdfoutput=1
\documentclass[vecphys]{svmult}

\usepackage{makeidx}         
\usepackage{graphicx}        
\usepackage{multicol}        
\usepackage{multicol}        
\usepackage[bottom]{footmisc}

\newcommand{\psim}{\lower.5ex\hbox{$\; \buildrel \propto \over\sim \;$}}
\newcommand{\lesssim}{\lower.5ex\hbox{$\; \buildrel < \over\sim \;$}}
\newcommand{\gtrsim}{\lower.5ex\hbox{$\; \buildrel > \over\sim \;$}}

\newcommand{\e}{\epsilon}
\newcommand{\g}{\gamma}
\newcommand{\h}{\hat\gamma}

\newcommand{\gp}{\gamma^\prime}

\newcommand{\zpr}{z^\prime}
\newcommand{\sT}{\sigma_{\rm T}}

\newcommand{\dD}{\delta_{\rm D}}

\newcommand{\zp}{z^\prime}

\newcommand{\iq}{\lower.5ex\hbox{$\; \buildrel \rightarrow \over {a\ll 1} \;$}}
\newcommand{\mq}{\lower.5ex\hbox{$\; \buildrel \rightarrow \over {{\hat \gamma}= 5/3} \;$}}

\newcommand{\tp}{t^\prime}
\newcommand{\ep}{\epsilon^\prime}

\newcommand{\apj}{{\it Astrophysical~Journal}}
\newcommand{\apjl}{{\it Astrophysical~Journal~Letters}}
\newcommand{\apjs}{{\it Astrophysical~Journal~Supplements}}
\newcommand{\aap}{{\it Astronomy~\&~Astrophysics}}
\newcommand{\app}{{\it Astroparticle~Physics}}
\newcommand{\mnras}{{\it Monthly~Notices~of~the~Royal~Astronomical~Society}}
\newcommand{\araa}{{\it Annual~Review~of~Astronomy~and~Astrophysics}}
\newcommand{\jcap}{{\it Journal~of~Cosmology~and~Astroparticle~Physics}}
\newcommand{\aj}{{\it Astronomical~Journal}}
\newcommand{\pasp}{{\it Proceedings~of~the Astronomical Society of the Pacific}}

\newcommand{\nat}{{\it Nature}}
\newcommand{\sci}{{\it Science}}
\newcommand{\prd}{{\it Physical~Review~D}}
\newcommand{\prl}{{\it Physical~Review~Letters}}
\newcommand{\lsim}{\lower.5ex\hbox{$\; \buildrel < \over\sim \;$}}
\newcommand{\gsim}{\lower.5ex\hbox{$\; \buildrel > \over\sim \;$}}

\def\fluxthres{\hat f_{\bar \e}}
\def\fluxeps{f_{_{\rm \epsilon}}}

\def\Estarg{{\cal E}_{*\gamma}}
\def\Estargo{{\cal E}_{*\gamma 0}}
\def\Swift{\emph{Swift}}

\def\beq{\begin{equation}}
\def\eeq{\end{equation}}

\def\ba{\begin{array}}
\def\ea{\end{array}}

\makeindex             

\begin{document}

\setcounter{page}{1}

\pagenumbering{roman}

\title*{Sources of GeV Photons and the Fermi Results}
\author{Charles D.\ Dermer}
\institute{United States Naval Research Laboratory, Code 7653 \\
4555 Overlook Ave., Washington, DC 20375-5352 USA\\
\texttt{charles.dermer@nrl.navy.mil}}
%
%
\maketitle


\begin{tabbing}
\=~~~~~~\=~~~~~~~\=~~~~~~\=~~~~~~~~~~~~~~~~~~~~~~~~~~~~~~~~~~~~~~~~~~~~~~~~~~~~~~~~~~~~~~~~~~~~~~~~~ \=  \\

\> Preface\>\>\>\> iii\\
\> 1. GeV Instrumentation and the GeV Sky\>\>\>\> \\
\>~~~~~~~~~~~~~~~~~with the Fermi Gamma-ray Space Telescope\>\>\>\> 1\\

\>\>1.1 Historical introduction\>\>\> 1 \\

\>\>\>Brief history of GeV astronomy\>\>  \\
\>\>\>The EGRET experiment on CGRO\>\> \\
\>\>\>Point source sensitivity of EGRET\>\>   \\

\>\>1.2 Fermi Gamma-ray Space Telescope\>\>\> 6 \\

\>\>\>GLAST becomes Fermi\>\>  \\
\>\>\>LAT instrument description\>\>  \\
\>\>\>LAT instrument response\>\>  \\

\>\> 1.3 Energy, flux, and luminosity\>\>\>   11 \\
 
\>\>\> Variability information \>\> \\
\>\>\>Extragalactic background light (EBL)\>\>  \\
\>\>\> Limits to the extreme universe\>\>   \\

\> 2. Fermi Gamma-ray Source Catalogs and Fermi Pulsars\>\>\>\> 14\\

\>\>2.1 First Fermi catalog of gamma-ray sources: 1FGL\>\>\>  15\\
\>\>2.2 Second Fermi catalog of gamma ray sources: 2FGL\>\>\> 17 \\
\>\>2.3 Fermi pulsars\>\>\> 18 \\

\>\>\> EGRET pulsars\>\>   \\
\>\>\> Elementary pulsar physics\>\>   \\
\>\>\> Properties of Fermi pulsars\>\>   \\
\>\>\> Millisecond pulsars and globular clusters\>\>   \\
\>\>\> Pulsar wind nebulae\>\>    \\
\>\>\> Crab nebula and flares\>\>    \\
\>\>\> Pulsar physics with Fermi\>\>   \\

\> 3. Fermi AGN Catalogs\>\>\>\> 27\\

\>\> 3.1 LAT Bright AGN Sample (LBAS) and \\
\>~~~~~~~~~~~~~~~~~	the First LAT AGN Catalog (1LAC)\>\>\>\> 27 \\
\>\>3.2 Classification of radio-emitting AGNs and unification\>\>\> 29 \\
\>\>3.3 Properties of Fermi AGNs\>\>\>  30\\

\>\>\>3C 454.3 and FSRQs\>\>  \\
\>\>\> PKS 2155-304, Mrk 501, and BL Lac objects\>\>  \\

\>\> 3.4 Second LAT AGN Catalog (2LAC)\>\>\> 35 \\

\>4. Relativistic Jet Physics\>\>\>\> 36 \\

\>\>4.1 GeV spectral break in LSP blazars\>\>\> 36 \\

\>\>\>$\gamma$ rays from external Compton scattering\>\>  \\
\>\>\>Compton emissivity and electron energy-loss rate\>\>  \\

\>\>4.2 Leptonic jet models\>\>\> 41 \\

\>\>\>Jet Doppler factor\>\>  \\
\>\>\>Variability time scale\>\>  \\
\>\>\>Equipartition field and jet power\>\>  \\

\>\>4.3 Hadronic jet models\>\>\>  45\\

\>\>\>Adiabatic losses, and photopair \>\>  \\
\>~~~~~~~~~~~~~~~~~ and photopion production on the CMBR\>\>\>\> \\
\>\>\>Photopion production efficiency\>\>  \\
\>\>\>Proton models\>\>  \\
\>\>\>\> GRBs\>  \\
 \> \>\>\> Blazars\>  \\

\>\>4.4 Cascade halos and the intergalactic magnetic field (IGMF)\>\>\> 53 \\

\>\>\>Cascade radiation from EBL attenuation of TeV photons\>\>  \\
\>\>\>Derivation of the cascade spectrum\>\>\ \\

\>5. $\gamma$ Rays from Cosmic Rays in the Galaxy \>\>\>\>  62 \\

\>\>5.1 $\gamma$ rays from Solar system objects \>\>\> 63 \\
\>\>5.2 GeV photons from cosmic rays \>\>\>  65 \\ 

\>\>\>Background modeling\>\>  \\
\>\>\>Diffuse Galactic $\g$ rays from cosmic rays\>\>  \\
\>\>\>Other Fermi LAT cosmic ray results\>\>  \\ 

\>\>5.3 Fermi bubbles\>\>\>  71\\
\>\>5.4 $\gamma$-ray supernova remnants\>\>\>  71\\ 
\>\>5.5 Nonrelativistic shock acceleration of electrons\>\>\> 76 \\

\>6. $\gamma$ Rays from Star-Forming Galaxies and Clusters of Galaxies,\>\>\>\> \\
\>~~~~~~~~~~~~~~~~~and the Diffuse $\gamma$-Ray Background \>\>\>\> 80\\

\>\>6.1 $\gamma$ rays from star-forming galaxies\>\>\> 80 \\ 
\>\>6.2 $\gamma$ rays from clusters of galaxies\>\>\>  82\\ 
\>\>6.3 Extragalactic $\gamma$-ray background and populations\>\>\> 83 \\

\>7.  Microquasars, Radio Galaxies, and the EBL \>\>\>\> 86\\

\>\>7.1 $\gamma$-ray binaries\>\>\>  86\\ 
\>\>7.2 Misaligned blazars and radio galaxies\>\>\> 89\\ 
\>\>7.3 EBL\>\>\>  92\\

\>8. Fermi Observations of Gamma Ray Bursts\>\>\>\> 95\\

\>\> 8.1 Fermi LAT observations of GRBs\>\>\>  95\\

\>\>\>Fluence-fluence diagram\>\>  \\
\>\>\>Fermi LAT GRB phenomenology\>\>  \\

\>\>8.2 GRB luminosity function\>\>\>  98\\

\>\>\>Luminosity density from the luminosity function\>\>  \\
\>\>\>Luminosity function in Le \& Dermer model\>\>  \\
\>\>\> Local luminosity density of GRBs\>\>  \\

\>\>8.3 Closure relations \>\>\>  108\\

\>9. Fermi Acceleration and Ultra-High Energy Cosmic Rays\>\>\>\> 109\\

\>\>9.1 Maximum synchrotron photon energy\>\>\> 109 \\
\>\>9.2 $L$-$\Gamma$ diagram\>\>\>  110\\
\>\>9.3 Luminosity density of $\gamma$-ray jet sources\>\>\> 111 \\
\>\>9.4 Origin of UHECRs\>\>\>  112\\
\>\>9.5 Black holes, jets, and the extreme universe\>\>\>  113 \\

\> Acknowledgments\>\>\>\>   119 \\

\> References\>\>\>\>    120\\

\end{tabbing}

The Fermi Gamma-ray Space Telescope, now in its fourth year after launch, continues to make important discoveries and establish new results in all directions: pulsar astronomy, cosmic-ray physics, AGN and black-hole astrophysics,  galactic astronomy, gamma-ray bursts (GRBs), limits on dark matter and Lorentz invariance violations, $\gamma$-ray astronomy of the Sun, moon, and Earth, etc. In this chapter, I survey results at medium energy $\gamma$ rays, from some tens of MeV (at energies above nuclear de-excitation $\gamma$-ray lines) to $\approx 100$ GeV where the ground-based  $\gamma$-ray Cherenkov detector arrays become more sensitive. As shorthand, Fermi and medium-energy $\gamma$-ray astronomy is referred to here as ``GeV astronomy," and ground-based Cherenkov $\gamma$-ray astronomy $\gtrsim 100$ GeV as ``VHE astronomy."

The Fermi results already provided considerably more material than could be presented in the nine lectures that I gave on this subject at the Saas-Fee school on ``Astrophysics at Very High Energies," held 15-20 March 2010 in Les Diablerets, Switzerland. Happily, though, Professor Lars Bergstr\"om  gave a brilliant series of lectures that covered dark matter, so the absence here of extended discussion on dark matter and new physics in GeV astronomy reflects Prof.\ Bergstr\"om's better capabilities to address this subject. My lectures and this book chapter are therefore restricted to astrophysical and astroparticle sources of GeV radiation rather than to $\gamma$ rays with origins in exotic particle physics and dark matter.

Even while the school was in progress, news appeared of a new type of $\gamma$-ray emitter of GeV photons that was identified with the symbiotic binary Nova V407 Cygni.
This extraordinary system reveals an explosive shock evolving 
on a timescale of days to weeks, rather than the hundreds of years for supernova remnants (SNRs). 
I showed the 2010 March 18 ATEL \cite{che10} announcing the Fermi result in my last lecture, capturing the real-time recognition of a new type of galactic $\gamma$-ray source triggered by a thermonuclear explosion on a white dwarf 
fed by its binary red giant's wind. The discovery has been published by the Fermi Collaboration \cite{fer10v407cyg}, and is already triggering a new line of research strongly tied to MeV line astronomy and white-dwarf physics (cf.\ citations to \cite{fer10v407cyg}). 
So these lectures and this write-up capture an early glimpse of the state of knowledge of astrophysical sources of $\gtrsim 100$ MeV and GeV radiation obtained with the Large Area Telescope (LAT) on Fermi at $\approx 2$ -- 3 years into the mission, weighted by the extragalactic interests of the author.

Alongside the LAT on Fermi is the Gamma-ray Burst Monitor (GBM), sensitive to GRBs and bright transients in the $10$ keV -- 30 MeV range. This review can only briefly mention  important GRB results made with the GBM---by itself and  with the LAT---and related GRB science employing Swift, INTEGRAL, and other detectors. Indeed, multiwavelength science is value-added science, and  the possibilities to uncover the underlying physics of the powerful compact systems that are at the heart of high-energy astronomy are multiplied by radio/microwave/sub-mm/IR/optical/UV/X-ray/MeV/TeV/neutrino/gravitational-wave data correlated with the GeV window, now observed with unprecedented clarity due to the LAT on Fermi. 

The GeV field is in full discovery mode, not only due to Fermi but also thanks to AGILE, an EGRET-like sentinel of bright $\gamma$-ray transients, and to ground-based VHE observatories, particularly HESS, VERITAS, and MAGIC.\footnote{AGILE: Astro-rivelatore Gamma a Immagini LEggero (Gamma-ray Light Imaging Detector);
HESS: High Energy Stereoscopic System, in Namibia. VERITAS: Very Energetic Radiation Imaging Telescope Array System, in Arizona. MAGIC: Major Atmospheric Gamma-ray Imaging Cherenkov Telescope in La Palma, Canary Islands} The $\nu F_\nu$ spectral energy distributions (SEDs) based on simultaneous and overlapping data sets are providing valuable information about Galactic sources, blazars and radio galaxies, and starburst and normal galaxies. GeV astronomy with Fermi is still in the midst of an active phase as Fermi accumulates data and increasing time makes faint sources visible and detection of rare cosmic transients more likely. 

This chapter will be divided into sections that follow the course of 
lectures delivered at the Saas-Fee course. Though now somewhat out-of-date, 
these lectures can be found on my website in ppt format.\footnote{
heseweb.nrl.navy.mil/gamma/$\sim$dermer/default.htm}
 The topics of the lectures and the corresponding sections of this chapter are:  
\begin{enumerate}
\item GeV instrumentation and the GeV sky with the Fermi Gamma-ray Space Telescope
\item First Fermi Catalog of Gamma Ray Sources and the Fermi Pulsar Catalog
\item First Fermi AGN Catalog 
\item Relativistic jet physics and blazars
\item $\gamma$ rays from cosmic rays in the Galaxy
\item $\gamma$ rays from star-forming galaxies and clusters of galaxies, and the diffuse extragalactic $\gamma$-ray background 
\item Microquasars, radio galaxies, and the extragalactic background light 
\item Fermi observations of GRBs
\item Fermi acceleration, ultra-high energy cosmic rays, and Fermi at 2 
\end{enumerate}

Besides a discussion of the results of the Fermi Gamma-ray Telescope, I also include here some
 high-energy astrophysical theory essential for analysis of $\gamma$-ray data that
builds on the research presented in my book with Govind Menon, Ref.\ \cite{dm09}: 
``High Energy Radiation from Black Holes: Gamma Rays, Cosmic Rays, and Neutrinos,"
published by
Princeton University Press in 2009. The book itself is focused on theory 
rather than observation, and develops the hypothesis that {\it the most 
energetic and powerful radiations in nature are made 
by particles energized through Fermi acceleration processes in 	  
black-hole jets powered by rotation}. 

It is not possible, even at this early stage of the Fermi mission, to 
adequately summarize all the results from Fermi. 
But together with the accompanying lectures and book, this article
 provides a broad overview of some recent astrophysical advances 
in the Fermi era of $\gamma$-ray astronomy. 


\setcounter{page}{1}

\pagenumbering{arabic}

\section{GeV Instrumentation and the GeV Sky with the Fermi Gamma-ray Space Telescope}
\label{sec_1}

\subsection{Historical introduction}

The year 2010 represents 
a highpoint in high-energy astronomy. 
Astronomical observatories at multiwavelength electromagnetic, neutrino, 
cosmic-ray, and gravitational wavebands are operating and collecting data. 
X-ray astronomy has sensitive 
pointing X-ray telescopes, Chandra and XMM Newton,
deployed in space to observe catalogued sources. 
Broadband X-ray and soft $\gamma$-ray observatories like Swift, INTEGRAL and 
Suzaku are available to measure the X/$\gamma$ spectra of compact objects. 
The Rossi X-ray Timing Explorer (RXTE) continues to operate and provide a monitor 
of the brightest X-ray sources in the sky. 
Already mentioned were AGILE and the ground-based $\gamma$-ray air Cherenkov 
telescopes. The Cherenkov Telescope Array (CTA) consortium \cite{cta10} is 
planning to start building as 
early as 2014 with array completion as early as 2018. Lacking at the moment, however, is an operating 
all-sky water-based Cherenkov telescope successor 
to MILAGRO in the 1 -- 100 TeV range. 
This gap will soon be filled 
by the High Altitude Water Cherenkov (HAWC) experiment \cite{san09} on 
the Sierra Negra mountain near Puebla, Mexico, which 
uses 300 tanks rather than a single pond as utilized by MILAGRO.

The Pierre Auger Observatory, located in the Mendoza province of Argentina and 
covering an area the size of Rhode Island, had its third data release in 2010 \cite{pao10},
giving the spectrum and composition of ultra-high energy cosmic rays (UHECRs) with 
energies  $E \gtrsim 10^{18}$ eV. 
The IceCube Neutrino Observatory, most sensitive to astrophysical neutrinos with energies of 100 TeV -- 100 PeV
($\approx 10^{14}$ -- $10^{18}$ eV), has deployed all 86 of its strings in the latest austral 
summer (December, 2010), and has developed the
DeepCore subarray that is sensitive to lower energy, $\approx 10$ GeV - 100 GeV, neutrinos. 
LIGO, the Laser Interferometry Gravitational-wave
Observatory, hoping to detect gravitational radiation from coalescing compact objects,
is operating at design sensitivity. Development of an order-of-magnitude more sensitive
Advanced LIGO has been approved, with completion expected for 2017 or thereafter. The NASA 
Laser Interferometry Space Antenna (LISA) is supported in the recent Astro-2010 study,
though ESA is developing a separate space-based gravitational wave facility. 
Likewise, Constellation-X has evolved into IXO/ATHENA.\footnote{ESA's Cosmic Vision
International X-ray Observatory/Advanced Telescope for High ENergy Astrophysics, sensitive to photons
with energies 0.1 -- 40 keV} The large-area X-ray 
timing mission RXTE will soon be ended, with the ESA Large Observatory for X-ray Timing (LOFT) mission taking its place. 
Here in the US, a focusing hard X-ray telescope, NuSTAR (Nuclear Spectroscopic Telescope Array), 
in the 5 - 80 keV range, and the Gravity and Extreme Magnetism Small Explorer, GEMS, a NASA mission
to study X-ray polarization of astrophysical sources in the 2 -- 10 keV range, will soon be launched.
 
\subsubsection{Brief history of GeV astronomy}

The progress of GeV astronomy in the range from 
$\approx 10$ MeV to $\approx 100$ GeV
followed a period of remarkable advances 
starting over 40 years ago that culminated with the launch of Fermi. 
Prior to the Fermi-LAT, the most important detectors and some of their 
achievements in the development of medium-energy $\gamma$-ray astronomy are the following:
\begin{itemize}
\item 1967--1968, OSO-3, the Third Orbiting Solar Observatory, carried a Cherenkov counter experiment 
 sensitive to $>50$ MeV $\g$ rays, of which 621 were detected \cite{kra72}. It discovered one source, the 
extended $\g$-ray emission of the Milky Way.
\item 1972--1973, SAS-2, the Small Astronomy Satellite-2 \cite{fic75}, a spark chamber experiment sensitive
to $\gamma$ rays with energies between $\approx 30$ MeV and 200 MeV (and an integral flux above 200 MeV). 
It detected $\approx 8000$ celestial $\g$-rays, making the first $\gamma$-ray identifications 
of the Crab and Vela pulsars, Geminga ($\g 195+5$, then unidentified), and the Cygnus region, and an association with 
Cygnus X-3 was suggested.  A north-south asymmetry in the Galactic $\gamma$-ray plane emission was noted
and attributed to the massive stars in the Gould belt. An isotropic $\gamma$-ray background radiation was
first reported from analysis of SAS-2 data \cite{fkh73,tf82}. 
\item 1975--1982, COS-B, the Cosmic ray Satellite (option B), a magnetic-core, wire-matrix spark chamber 
sensitive to $\g$ rays with energies from $\approx 30$ MeV -- $\approx 5$ GeV), with an effective area 
of $\approx 50$ cm$^2$ at 400 MeV \cite{swa81}. Its orbit resulted in a large and variable background of 
charged particles. During its lifetime, it detected $\approx 200,000$ $\g$-rays, with the COS-B Caravane Collaboration announcing the discovery 
of 25 sources, most along the Galactic plane. These 
included 2CG 135+01, now identified with LSI +61$^\circ$ 303, and 
the first extragalactic source of $\gtrsim 100$ MeV $\g$ rays, 3C 273 \cite{her77}.
\item 1991--2000, EGRET, the Energetic Gamma Ray Experiment Telescope on the {\it Compton Gamma Ray Observatory}, 
was a spark chamber experiment with large, $\gtrsim 1200$ cm$^2$ effective area between 
200 MeV and 1 GeV, accompanied by excellent background rejection \cite{tho93}.  During its 9 year mission lifetime, it collected $>2\times 10^6$ $\g$ rays and 
discovered that blazars and GRBs are luminous $\gamma$-ray sources. Because of the importance of this experiment in the development of GeV astronomy, we describe EGRET in more detail below. 
\item 2007--, AGILE, 
with a Gamma Ray Imaging Detector (GRID) and small calorimeter giving 
$\gamma$-ray sensitivity  from 30 MeV to 50 GeV, roughly comparable to EGRET. 
Along with the GRID is the
accompanying 18 -- 60 keV super-AGILE hard X-ray survey instrument \cite{fer10}. 
The first AGILE GRID catalog consists of 47 $> 100$ MeV $\gamma$-ray sources with significance $>4\sigma$ 
from data taken between July 2007 and June 2008 \cite{pit09}. The brightest sources in the catalog, e.g., Vela, Crab and, during periods of outburst, 3C 454.3, exceed integral photon fluxes of $2\times 10^{-6}$ ph$(> 100$ MeV)/cm$^{2}$-s, while the weakest are fainter than $50\times 10^{-8}$ ph$(> 100$ MeV)/cm$^{2}$-s.
\end{itemize}

\subsubsection{The EGRET experiment on {\it CGRO}}

The {\it Compton Gamma Ray Observatory}, or {\it CGRO}, was a pioneering 
$\gamma$-ray space observatory (5 April 1991 -- 4 June 2000) consisting of 4 instruments, 
OSSE, the Oriented Space Scintillator Experiment (sensitive from $\approx 50$ keV -- 10 MeV); BATSE, the Burst And Transient Source Experiment 
 ($\approx 20$ keV -- few MeV); COMPTEL, the Compton Telescope
  ($\approx 800$ keV -- 30 MeV); and EGRET, the Energetic 
Gamma Ray Experiment Telescope  ($\approx 30$ MeV -- 10 GeV). 
EGRET's detection method utilized a gas-filled spark chamber that tracked the $\g$ ray after it converted
to an electron-positron pair by pair-production interactions with nuclei
in the thin Ta foils in the gas-filled spark chamber. 
Directional information was obtained by time-of-flight coincidence with a scintillator 
array in the lower spark chamber assembly. Charged particles were vetoed by an anticoincidence 
shield and total $\gamma$-ray energy was measured with EGRET's total absorption scintillator counter, 
the TASC, consisting of crystal scintillators and photomultipliers \cite{kur97}.


\begin{figure}[tbp] 
  \centering
  \includegraphics[width=4in]{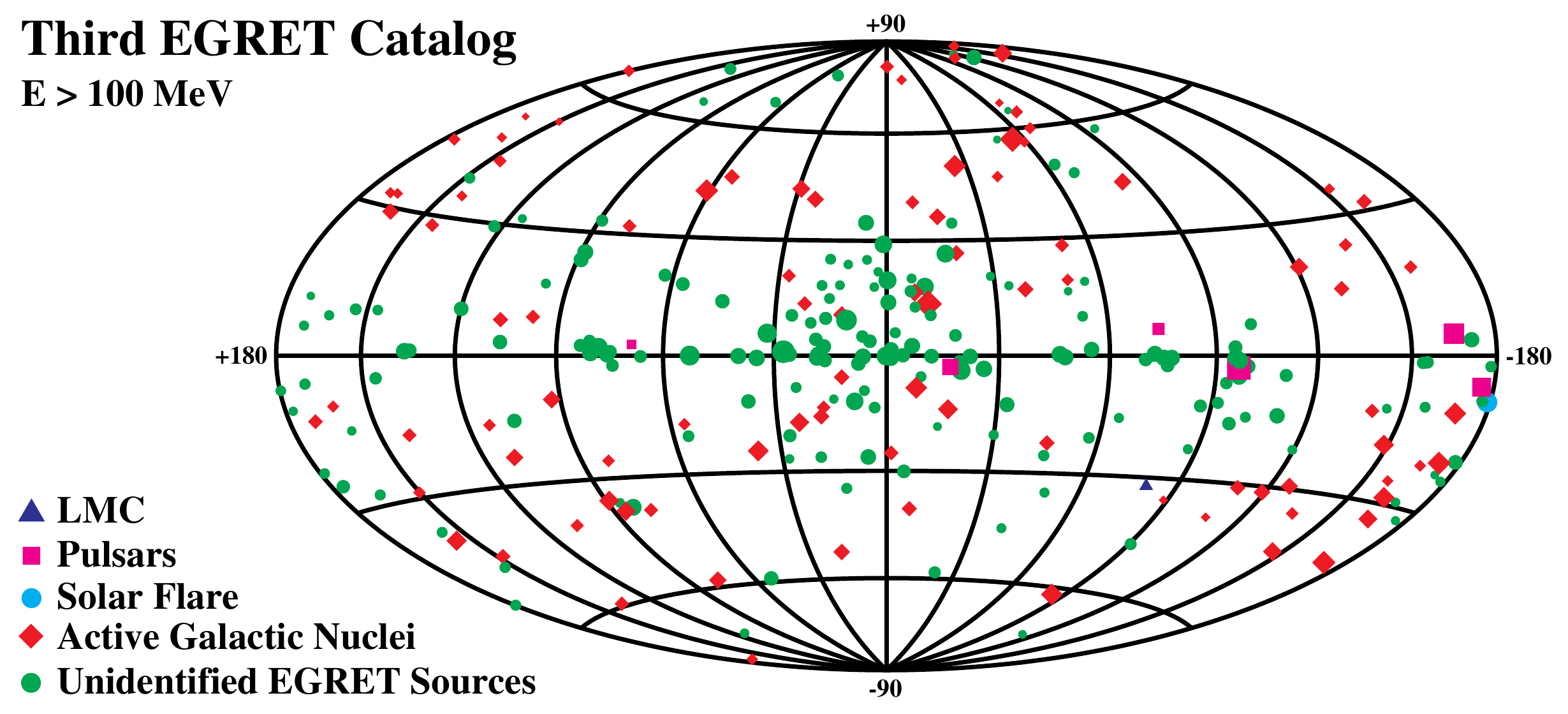}
  \caption{EGRET all-sky map of $\g$-ray sources \cite{har99}.}
  \label{fig:egret}
\end{figure}

The Third EGRET (3EG) catalog \cite{har99}, made from data accumulated
between 1991 April 22 and 1995 October 3,
 consists of 271 sources of $> 100$ MeV emission, including a Solar flare in 1991, 
the Large Magellanic Cloud, five pulsars, one probable radio galaxy, namely
Centaurus A, and 66 high-confidence detections of blazars.
The catalog also lists 27 lower confidence potential blazar detections, 
and contains 170 unidentified EGRET sources lacking associated sources. 
Five GRBs were detected with the spark chamber on 
EGRET. Figure 1 shows a skymap of the 3EG sources.

The EGRET Field of View (FoV), defined roughly by the solid angle within which the effective area
is greater than $1/2$ of the on-axis effective area, was $\approx 0.5$ sr, 
or $\approx 1/24^{th}$ of the full sky. EGRET operated in a pointing mode, and 
targeted one region of the 
sky for two weeks, representing $\sim 10^6$ s after time for Earth occultation and time spent
in the South Atlantic Anomaly are subtracted.  The Point Spread Function (PSF) 
at 100 MeV was $\approx 5.7^\circ$, 
with the PSF improving roughly as $\approx E^{-1/2}$ \cite{tho86,tho93}. 
The first 18 months of the mission 
were devoted to a full-sky survey.

Important results from analysis of the full 6 years of data (degradation of spark chamber gas
led to smaller apertures after 4 years into the mission) include the following:
\begin{enumerate}
\item Diffuse extragalactic background with intensity  $\approx 1.5$ keV/cm$^{2}$-s-sr,
corresponding to an energy density 
$u_{\gamma} \approx 10^{-17}$ erg/cm$^{3}$;
\item The galactic flux is as much as $50\times$ brighter than the extragalactic flux, and much softer, but 
not as soft as expected if the $\gamma$ rays are formed by secondary nuclear production by Galactic cosmic rays with the same spectrum as observed locally (the so-called ``EGRET excess");
\item Typical fluxes of EGRET sources are between $\sim (10^{-7}$ and $10^{-6}$) ph$(> 100$ MeV)/cm$^{2}$-s, with a typical 2-week
on-axis limiting flux at $\approx (15$ -- 25) $\times 10^{-8}$ ph$(> 100$ MeV)/cm$^{2}$-s;
\item Galactic sources, including young radio pulsars;
\item Sources with significant flux variability.
\end{enumerate}
Temporal variability is an essential characteristic of  GRBs and blazars, Solar flares, and now  V407 Cyg. 
The large FoVs of the LAT and Swift, and larger still with BATSE and GBM,
are crucial for study of $\gamma$-ray transients.

\subsubsection{Point source sensitivity of EGRET}

The best description of results leads to the question of units in GeV astronomy.  
Because medium-energy and high-energy $\g$-ray astronomy is
challenged by limited signal counts, an integral photon flux is the most
natural unit. For the EGRET experiment, units
of $10^{-6}$ ph$(> 100$ MeV)/cm$^{2}$-s are suitable, 
as this value roughly separates signal-dominated
 and noise-dominated detection in EGRET, as we now show.
Units $10^{-8}$ ph$(> 100$ MeV)/cm$^{2}$-s are more suitable 
for Fermi sources.

As measured with EGRET, the diffuse, or at least unresolved intensity 
of photons at 100 MeV, is measured to be $\epsilon I_\epsilon 
\approx 1.5$ keV/cm$^2$-s-sr at $E \equiv \e m_ec^2  \approx 100$ MeV.
Writing
\beq
\e I_\e = 1.5\;({\e\over\e_{100}})^{2-\Gamma_\gamma}\;\;{{\rm~keV}\over{\rm cm}^2\mbox{-}{\rm s}\mbox{-}{\rm sr}}\;,
\label{eIe}
\eeq
with photon number index $\Gamma_\gamma$ and $\epsilon_{100} = 100$ MeV/$m_ec^2$,
implies 
\beq
{dN(>\epsilon )
\over dAdt d\Omega}= \int_\e^\infty d\ep\;{\ep I_{\ep}\over m_ec^2\e^{\prime 2}} = {1.5\times 10^{-5}\over {\Gamma_\gamma-1}}\;({\e\over\e_{100}})^{1-\Gamma_\gamma}\;{{\rm ph(}>100{\rm ~MeV})\over{\rm cm}^2\mbox{-}{\rm s}\mbox{-}{\rm sr}}\;.
\eeq
EGRET measured a hard spectrum, with $\Gamma_\gamma \approx 2.1$ \cite{sre98}, whereas LAT measures a softer spectrum, with 
$\Gamma_\gamma \approx 2.4$ \cite{abd10id}, and with a 100 MeV intensity only 60\% as large as the EGRET intensity.\footnote{A number index $\Gamma_\gamma =2.41\pm 0.05$ 
 and intensity normalization 
$I(>100$ MeV) $ =(1.03\pm 0.17)\times 10^{-5}$/cm$^2$-s-sr 
for the intensity spectrum of the isotropic diffuse $\gamma$-ray background 
are measured 
from the first year Fermi data \cite{abd10id}. The lower diffuse extragalactic flux measured with Fermi compared 
to EGRET is 
partially but not entirely due to resolving more point sources out. } 
For the EGRET intensity, therefore, $dN(>\e_{100} )/dAdtd\Omega \approx 1.5\times 10^{-5}$ ph$(>100$ MeV)/cm$^2$-s-sr.

The EGRET PSF at 100 MeV, as previously noted, is $\approx 5.7^\circ$. In comparison, 
the LAT PSF (68\% containment radius) at 100 MeV is $\approx 3.5^\circ$, with the PSF dropping to $\approx 0.6^\circ$
for 1 GeV photons, and $\approx 0.15^\circ$ for 10 GeV photons (for conversion in the
thin layers) \cite{atw09}. The EGRET PSF
at 100 MeV represents about $\pi(5.7^\circ)^2\approx 100$ square degrees, or $\approx 1/400^{th}$ of the 
full sky. Thus the flux from each patch corresponding to the EGRET PSF 
is $\approx 5\times 10^{-7}$ ph ($>100$ MeV)/cm$^2$-s.
A $\gamma$-ray source is {\it signal dominated} for EGRET when its flux is $\gtrsim 10^{-6}$ ph ($>100$ MeV)/cm$^2$-s.
By contrast, it is {\it noise dominated} when its flux is $\ll 10^{-6}$ ph ($>100$ MeV)/cm$^2$-s.

The time $\Delta t$ needed to accumulate 100 photons with EGRET, consisting of 50 signal $S$ and 50 background 
$B$  photons, and to give a detection at the $\approx S/\sqrt{2B} \cong 5\sigma$ level, is found through an expression
for the integral photon flux $F(>\epsilon )$ of the source. For a source at the level of 
\beq
F_{-8}\equiv {F(>\epsilon_{100})\over 10^{-8}{\rm~ph}(>100{\rm~MeV})/{\rm cm}^2\mbox{-}{\rm s}}\;,
\eeq
the number of detected photons is
$
\approx 10^{-8} F_{-8} \times \Delta t\times 1000 {\rm~cm}^2 
$,
so that $\approx 50$ ph can be detected from a bright source at the level of $F_{-8}\sim 10^2$
during an EGRET observation period of $\Delta t\approx $ one day (during half this time, the Earth is occulted).
A nominal 2-week observation period with $\Delta t\approx 10^6$ s on source gives a 
5$\sigma$ limiting sensitivity for sources with  $F_{-8} \approx 2\times10^{-7}$ ph($>100$ MeV)/cm$^2$-s.

These sorts of arguments can be used to estimate the time needed to make a detection and 
resolve temporal variability with EGRET, LAT, and counter detectors with broad FoVs, including
neutrino telescopes. The Fermi LAT becomes noise dominated at much lower flux levels
than EGRET, with sources regularly detected at the $F_{-8}< 1$. 
Note that the background is smaller at higher energies as a result of the smaller 
PSF, but the smaller flux usually reduces significance for a given observation 
time except for hard-spectrum sources.
A better  approach for characterizing 
detection significance is likelihood analysis \cite{mat96} (described below), 
but the detection significance can 
be simply estimated as outlined above \cite{der07}.

The above estimates apply to high galactic latitude, $|b|>10^\circ$ sources where the $\gamma$-ray sky is dominated by extragalactic sources and
 unresolved isotropic $\gamma$-ray background. At lower galactic latitudes, the diffuse $\gamma$-ray emission from 
cosmic-ray interactions with gas and dust makes source detection more difficult. Subtraction of the diffuse emission and 
nonuniform and variable  background
requires  a Galactic model for cosmic-ray/gas interactions. The sensitivity of source detection to background model is seen in 
the alternate 3EGR analysis of the EGRET data by Casandjian \& Grenier \cite{cg08}. They do not confirm 107 3EG sources, most 
in the vicinity of the Gould Belt, and find 30 new sources in the full 9-year data set.

Long after EGRET's effective lifetime had expired (though BATSE, OSSE, and COMPTEL were still collecting valuable data),
{\it CGRO} was deorbited into the Pacific Ocean in 2000. The launch of {\it INTEGRAL} in October 2002, Swift 
on November 20, 2004, and {\it AGILE} on April 23, 2007, helped fill the $\gamma$-ray gap leading to {Fermi}.

\subsection{Fermi Gamma-ray Space Telescope}

The Fermi Gamma-ray Space Telescope consists of two major detector systems, namely
\begin{enumerate}
\item The Large Area Telescope (LAT), a pair-conversion telescope sensitive between $\approx 20$ MeV and $\gtrsim 300$ GeV, with the higher energy
limit  a result of the vanishing small detection rate given Fermi's $\sim 1$ m$^2$ aperture. 
The Fermi-LAT has opened the previously unexplored $\approx 10$ GeV -- 100 GeV window, as self-vetoing due to particle shower backsplash in the anti-coincidence detector may have 
reduced EGRET's effective area above $\approx 5$ GeV \cite{shk08}. 
Fermi nominally operates in a survey mode, and  given its large FoV, scans the entire sky every 3 hours.

\item The Gamma-ray Burst Monitor (GBM), sensitive in the 8 keV -- 40 MeV range, 
consisting of 12 NaI detectors sensitive between 8 keV and 1 MeV, and 2 BGO detectors sensitive in the 0.15 MeV -- 40 MeV range. The scintillator 
detectors surround the base of the LAT, and view the entire unocculted sky which, at the nominal 565 km altitude of Fermi, represents $\approx 2/3^{\rm rd}$ of the full sky. 
\end{enumerate}

Some of the science questions that Fermi was designed to answer are:
\begin{itemize}
\item How do supermassive black holes in Active Galactic Nuclei (AGN) create powerful jets of material moving at nearly light speed?  What are the jets made of? 
\item What are the mechanisms that produce GRB explosions?  What is the energy budget?    
\item What is the origin of the cosmic rays that pervade the Galaxy? 
\item How does the Sun generate high-energy $\gamma$ rays in flares?
\item How has the amount of starlight in the universe changed over cosmic time? 
\item What are the unidentified $\gamma$-ray sources found by EGRET? 
\end{itemize}

\subsubsection{GLAST becomes Fermi}

After a number of delays, a Delta II 7920-H(eavy) rocket carrying the Gamma-ray Large Area Space Telescope---GLAST---payload 
was launched from Cape Canaveral Air Station on 2008 June 11 at 12:05 pm EDT into a 565 km altitude circular orbit with $25.6^\circ$ inclination and
a 96 minute period. GLAST completed a 60 day checkout period by early August, and released its first-light image on 26 August 2008, when it was renamed 
the Fermi Gamma-ray Space Telescope, after the Italian-American physicist Enrico Fermi (1901 -- 1954).
 
The first-light image,  based on only 4 days of observation, already reveals dozens of sources to the eye, with the blazar 3C 454.3 almost as bright as the Vela pulsar, which is the brightest persistent $\gamma$-ray source in the sky  ($F_{-8} \cong 1060$). Limiting fluxes of EGRET sources are at the level of $F_{-8} \approx 20$ -- 30 for a two-week observation, with a corresponding all-sky flux limit of $F_{-8} \approx 15$ -- 30 for a year of observing (given the FoV of EGRET). For a source with a flat $\nu F_\nu$ SED, or a number index $\Gamma_\gamma=-2$, Fermi reaches $F_{-8}\sim 1$ in one year over the entire high Galactic latitude $|b| > 10^\circ$ sky. Due to its energy-dependent effective area and PSF, limiting fluxes are strongly dependent on both source and background spectrum. Hard-spectrum sources with number index $\approx -1.5$ are detected with comparable significance as soft spectrum sources, but at integral photon fluxes as low as $F_{-8}\approx 0.1$. 

The LAT operates in a nominal scanning mode whereby the spacecraft rocks about the zenith. The rocking angle was equal to 
39$^\circ$ in the first part of the {Fermi} mission, and then increased to $50^\circ$ after 3 September 2009. The larger rocking angle gives a more uniform exposure, 
but with the loss of data due the increased fluorescence $\g $-ray emission from cosmic-ray bombardment of the Earth's atmosphere.\footnote{\label{footnote:albedo}This is also called, colloquially and inaccurately, ``$\gamma$-ray albedo," or just ``albedo".}
The LAT observes the entire sky every two orbits, or $\approx 3$ hours,  by rocking 
north and south of the zenith on alternate orbits, with each point in the sky receiving  $\approx 30$ min exposure during this time. Onboard triggering has also been enabled for the Fermi spacecraft to autonomously slew.  As of October 2011, more than 30 autonomous repoint requests (ARRs) have taken place, resulting in 5-hr pointing mode observations in response to bright GRBs detected with the GBM  (the length of the ARR repoint was reduced  in the third year of the mission). Several dedicated Targets of Opportunity (ToO) pointings have been executed, including one to 3C 454.3, two to the flaring Crab, one to Cyg X-3, and two Solar pointings.\footnote{fermi.gsfc.nasa.gov/ssc/observations/timeline/posting/}  A regular schedule of nadir pointings to look for $\gamma$ rays  from terrestrial $\gamma$-ray flashes, or TGFs, is underway. 

The Fermi/GLAST LAT Collaboration is an international organization originally comprised of institutions in France, Italy, Japan, Sweden, and the United States. There are about 120 full Collaboration members, and a total of about 200 scientific members, including affiliated scientists, postdocs, and students. The principal investigator of the project is Professor Peter Michelson of Stanford, and the project scientist during the development, commissioning, and early science phase was Dr. Steven Ritz, now professor at UC Santa Cruz. Dr.\ Julie McEnery of NASA's Goddard Space Flight Center is now Fermi LAT project scientist.  A review of the first-year LAT results is given by Michelson, Atwood, and Ritz \cite{mar10}, and the detailed instrument description can be found in  the paper by Atwood et al.\ (2009) \cite{atw09}.

In recognition of the importance of the Fermi mission to high-energy astronomy, 
Peter Michelson, Bill Atwood, and the Fermi  Gamma Ray Space Telescope LAT team were awarded the 2011
Rossi Prize of the High Energy Astrophysics Division of the American Physical Society,
``for enabling, through the development of the Large Area Telescope, 
new insights into neutron stars, supernova remnants, cosmic rays, 
binary systems, active galactic nuclei, and gamma-ray bursts."
UCSC physicist Bill Atwood was also recently awarded the  2012 W.K.H.\ Panofsky Prize in Experimental Particle Physics by the American Physical Society.

\subsubsection{LAT instrument description}

The LAT is a pair-conversion telescope composed of 16 tracker-converter towers, a calorimeter, an anticoincidence shield, and an electronics system to filter signal from background. The conversion of $\gamma$-ray photons to electron-positron pairs takes place when a $\gamma$ ray produces a pair in one of the 16 high-Z tungsten planes in each tower. The photon direction is measured with a precision silicon-strip tracker with a 228 $\mu$ pitch (separation between strips), totaling $8.8\times 10^5$ channels distributed over 18 tracker planes in each of 16 tracker modules. The Si tracker follows the e$^+$-e$^-$ pairs to give direction, and the photon energy is measured in a by a cesium-iodide calorimeter in a hodoscopic assembly that images the shower, allowing for additional rejection of non-photon-like events. A segmented anticoincidence detector (ACD) rejects background of charged cosmic rays. Segmentation of the ACD  helps prevent self-vetoing at high energies.

\begin{figure}
\centering
\includegraphics[width=7cm,height=7cm]{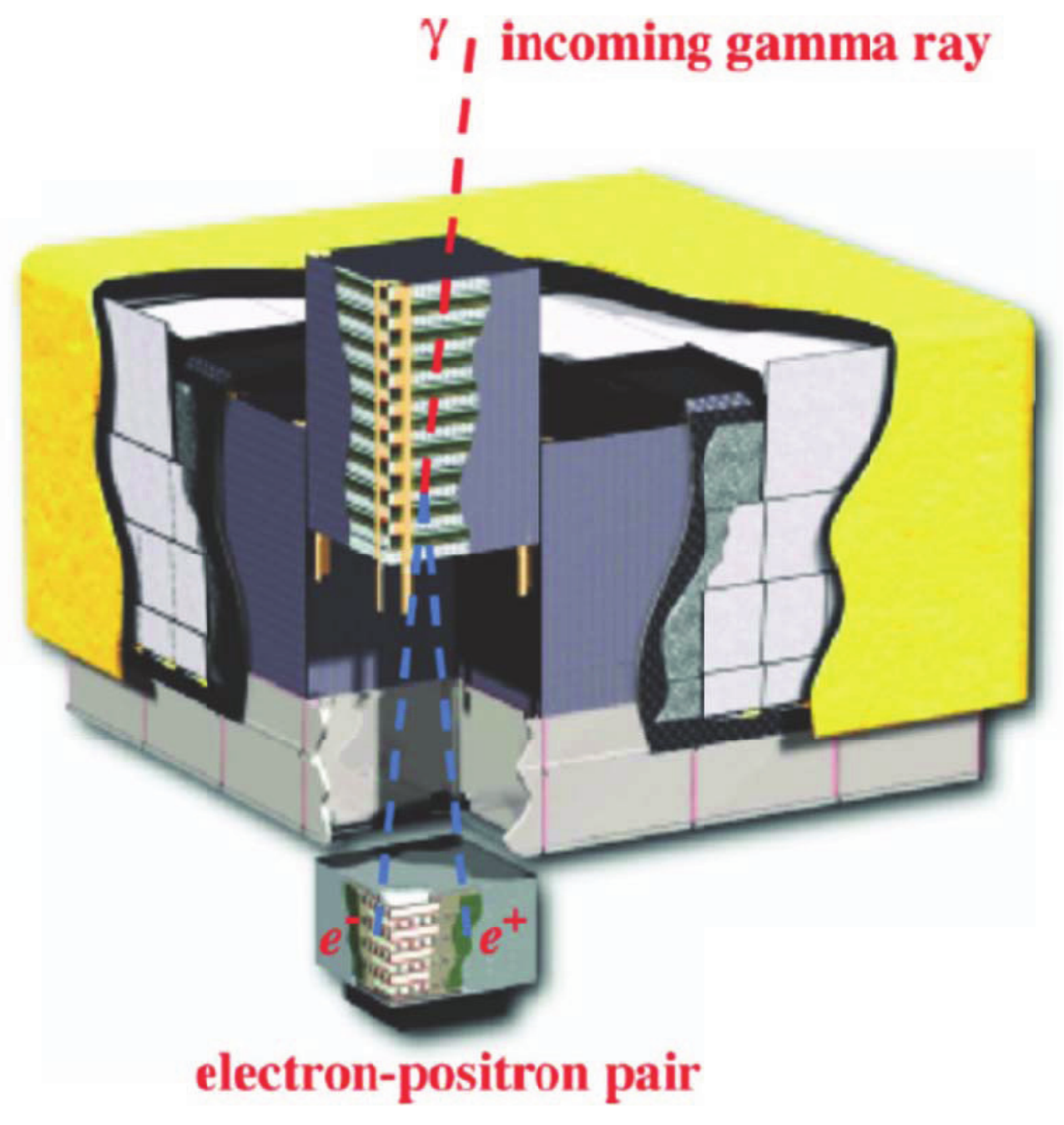}
\caption{Schematic of the Large Area Telescope on Fermi (from Atwood, et al., 2009 \cite{atw09}).}
\label{fig:3}       
\end{figure}

Table \ref{tab:1} gives the performance characteristics of the Fermi LAT. Depending on the source type, different classes of events are adopted to maximize sensitivity. Three hierarchical 
analysis classes based upon analysis cuts and 
event selections are generally used for Fermi analysis. For the most stringent {\it diffuse} event class used to identify $\gamma$ rays, the on-axis effective area increases from $\approx 1500$ cm$^2$ at 100 MeV to $\approx 8000$ cm$^2$ at 1 GeV, and is roughly constant at higher energies. The {\it transient} event class is used for the study of GRBs and brief transients where the amount of background can be greater due to the smaller time over which data is taken. The {\it source} class, intermediate to the other two, is where 
the residual instrumental plus environmental background is comparable to the extragalactic diffuse
emission measured with EGRET. 

\begin{table}
\centering
\caption{Performance Characteristics of the Fermi LAT \cite{atw09}}
\label{tab:1}       
%
%
\begin{tabular}{ll}
\hline\noalign{\smallskip}
Parameter & Value or Range  \\
\noalign{\smallskip}\hline\noalign{\smallskip}
Energy range & 20 MeV -- 300 GeV  \\
Effective area at normal incidence (60$^\circ$ off-axis)&  \\
~~~~100 MeV  &   $3700$ (700) cm$^2$ \\
~~~~1 GeV  &   $\leq 8300$ (2900) cm$^2$ \\
~~~~10 GeV  &   $\leq 8400$ (3100) cm$^2$ \\
Energy resolution ($1\sigma$, on-axis): &   \\
~~~~100 MeV -- 1 GeV &   15\% -- 9\% \\
~~~~1 GeV -- 10 GeV &   15\% -- 9\% \\
~~~~10 GeV -- 300 GeV &   8.5\% -- 18\% \\
Single photon angular resolution (68\% containment radius): &   \\
~~~~100 MeV &   $3.5^\circ$ \\
~~~~1 GeV  &   $0.6^\circ$ \\
~~~~$> 10$ GeV &   $<0.15^\circ$ \\
Field of View (FoV): & 2.4 sr  \\
Timing Accuracy & 300 ns  \\
Event readout time (deadtime) & 26.5 $\mu$s \\
\noalign{\smallskip}\hline
\end{tabular}
\end{table}

\subsubsection{LAT instrument response}

Even though two tracker towers and three 
calorimeter modules of the LAT were beam tested, the original science
tools provided for instrument response were the Pass 6\_v3 response functions based on 
extensive GEANT4 Monte Carlo simulations of the satellite
that followed secondary particles and photons through the digitization
and software filters to determine---assuming tracker azimuthal symmetry---effective area,
energy uncertainty, and PSF (or containment radius). The event reconstruction, filters,
and different event classes are described in \cite{atw09}.

The P6\_v11 corrections to the P6\_v3 response functions using a few months of on-orbit data 
\cite{2009APh....32..193A} showed that the earlier response functions gave smaller angles for 
68\% and 95\% containment radii above $\approx 5$ GeV than given by 
the on-orbit calibration. The PSF is best fit
with a King model profile of the form
\begin{equation}
f_{\rm King}(\theta,\sigma,\g) = (1-{1\over g}) (1+{\theta^2\over 2\sigma^2 g})^{-g}\;,
\label{fKing}
\end{equation}
where $\theta$ is the angle between the incident, ``true" photon direction and the reconstructed direction, 
and $g$ is a fitting parameter. 
The wings of the PSF follow a power-law rather than exponential behavior, and
are well fit with the sum of two King functions.

The first-year analyses and the 1FGL, and second-year analyses leading to the 2FGL, generally employ
the P6 and P7 instrument response functions, respectively.  
Updated instrument performance that improves on pre-flight Monte Carlo and muon 
calibrations by using inflight corrections to the instrument response functions are found at \footnote{www.slac.stanford.edu/exp/glast/groups/canda/lat\_Performance.htm}. 

Utilization of {Fermi}-LAT to energies as low as 30 MeV is possible with
the LAT Low-Energy (LLE) technique \cite{2010arXiv1002.2617P}. This approach was developed for transients, such as Solar flares and GRBs,
where a direction is known and a short time window can be defined. During this period, 
less discriminating event selections are used, the analysis includes all photons with reconstructed arrival
directions within $20^\circ$ of the target, and  the criteria for $> 100$ keV 
vetoing in the ACD is relaxed.

\subsection{Energy, flux, and luminosity}

Consistent with the notation of \cite{dm09}, 
we define the dimensionless photon energy, in units
of electron rest mass, as 
\beq
\e = {h\nu\over m_ec^2} = {E_\gamma\over m_ec^2}
\eeq
Flux density $F_\nu$ is usually reported in units of Jansky
(1 Jy = $10^{-23}$ erg/cm$^{2}$-s-Hz), so that the
quantity $\nu F_\nu$ is an energy flux $F$ (units of erg/cm$^{2}$-s, 
or Jy-Hz, noting that $10^{10}$ Jy-Hz $ =
10^{-13}$ erg/cm$^{2}$-s). The luminosity distance $d_L$ for
a steady, isotropically emitting source is defined so that the energy
flux $F$ is related to the source luminosity $L_*$ (erg/s) 
according to the Euclidean expression
\begin{equation}
F = {L_*\over 4\pi d_L^2 }\;.
\label{PhiE_intro}
\end{equation}

If $\phi(\e)$ is the measured spectral photon flux (units of photons
cm$^{-2}$ s$^{-1} \e^{-1}$), then $\nu F_\nu = m_ec^2 \e^2 \phi(\e )$.
Henceforth we use the notation
\begin{equation}
f_\e = \nu F_\nu 
\label{fe_define}
\end{equation}
for the $\nu F_\nu$ flux. From the definitions of $F$ and $f_\e$,
\begin{equation}
F = \int_0^\infty d\e\; {f_\e\over \e}\;.
\label{PhiE_fe}
\end{equation}
Considering eq.\ (\ref{PhiE_intro}),
the luminosity radiated by a source between measured photon energies
$\e_1$ and $\e_2$, or between source frame photon 
energies $\e_1(1+z)$ and $\e_2(1+z)$, is therefore given by 
\begin{equation}
L_*[\e_1(1+z),\e_2(1+z)] = 
4\pi d_L^2 m_ec^2\int _{\e_1}^{\e_2} d\e \;\e\;\phi(\e )
= 4\pi d_L^2 \int _{\ln \e_1}^{\ln \e_2} d(\ln \e )f_\e\;.
\label{L(e1,e2)}
\end{equation}
Eq.\ (\ref{L(e1,e2)}) shows that if the $\nu F_\nu$ spectrum is flat
with value $f_\e^0$, corresponding to a photon flux $\phi(\e) \propto \e^{-2}$, 
then the apparent power of the source over one decade
of energy is $\approx (\ln 10) L_0 \approx 2.30 L_0$, where $L_0 =
4\pi d_L^2 f_\e^0$. 

From these relations, one obtains the
mean apparent isotropic $\gamma$-ray energy release 
\begin{equation}
{\cal E}_*
= {4\pi d_L^2 F \langle \Delta t\rangle \over (1+z)}\;.
\label{totalenergyestar}
\end{equation} 
for a source at redshift $z$ that releases average energy flux $F$ during 
observing timescale $\Delta t$. The apparent versus the absolute energy
releases and luminosities depends on the jet structure and variability 
behavior. For a steady, two-sided top-hat jet with uniform emission 
within angle $\theta\leq \theta_j$ of the jet axis,
the absolute luminosity 
\beq
L_{abs} = f_b L_{iso}\;,
\label{Labsfb}
\eeq
where the beaming factor $f_b = 2\times 2\pi \int^1_{\mu_j}d\mu_j/4\pi = 1-\mu_j$. For $\theta_j = 0.1$ (5.7$^\circ$), 
$f_b \cong 1/200$, whereas if $\theta_j = 0.01$ (0.57$^\circ$), $f_b \cong 1/20,000 = 5\times 10^{-5}$.  What is reported is the apparent isotropic luminosity $L_{*,iso} = 4\pi d_L^2 F$, with absolute luminosity implied by arguments for the jet opening angle and beaming factor.

\subsubsection{Variability information}

The Schwarzschild radius of a black hole of mass $M$ is
\begin{equation}
R_{\rm S} = {2G M\over c^2} \cong 3.0\times 10^5\;\left({M\over
M_\odot}\right)\;{\rm cm}\;\cong 10^{-4}\;M_9\;{\rm pc},
\label{Rschwarz}
\end{equation}
defining $M_9 = M/(10^9\,M_\odot)$. 
 Variations in the
source flux by a large factor ($\gtrsim 2$) over a time scale $\Delta
t$ must, from causality arguments for a stationary source, 
originate from an emission region
 of size $R\lesssim c\Delta t/(1+z)$.  Incoherent superpositions of emission from larger size scales 
and from regions that are not in causal contact would usually (though not always) wash out
 large-amplitude fluctuations.  For high-quality data from bright
flares, large amplitude variations in source flux on timescale $\Delta
t$ would, from this argument, imply a black-hole mass
\begin{equation}
M_9\lesssim {(\Delta t/10^4 {\rm ~ s})\over 1+z }\;.
\label{mdeltat}
\end{equation}

This relation is even preserved in systems with relativistic outflows,
unless specific model-dependent provisions are made on the method of wind dissipation. 
Variability timescales far shorter than the light-crossing time 
across an $\approx 10^9 M_\odot$ black hole have been measured in the 
TeV BL Lac objects PKS 2155-304, Mrk 501, and Mrk 421, 
discussed in more detail below.
Another interesting fiducial is the ratio of the Eddington luminosity to the 
light-crossing time across a Schwarzschild black hole \cite{es74}, 
namely
\begin{equation}
{L_{\rm Edd}\over t_{\rm S}} = 4\pi\;{m_pc^4\over \sigma_{\rm T}}\cong 2.5\times 10^{43}\;{\rm erg~s}^{-2}\;.
\label{LEDDtS}
\end{equation}
We take this criterion as separating the  extreme universe 
from the moderate universe.
 During intense 3C 454.3 flaring activity  \cite{abd11}, 
the ratio of the apparent isotropic luminosity and the source variability timescale 
strongly violate this limit, making this an extreme event.

\begin{figure}[t]
\center
\includegraphics[scale=0.5]{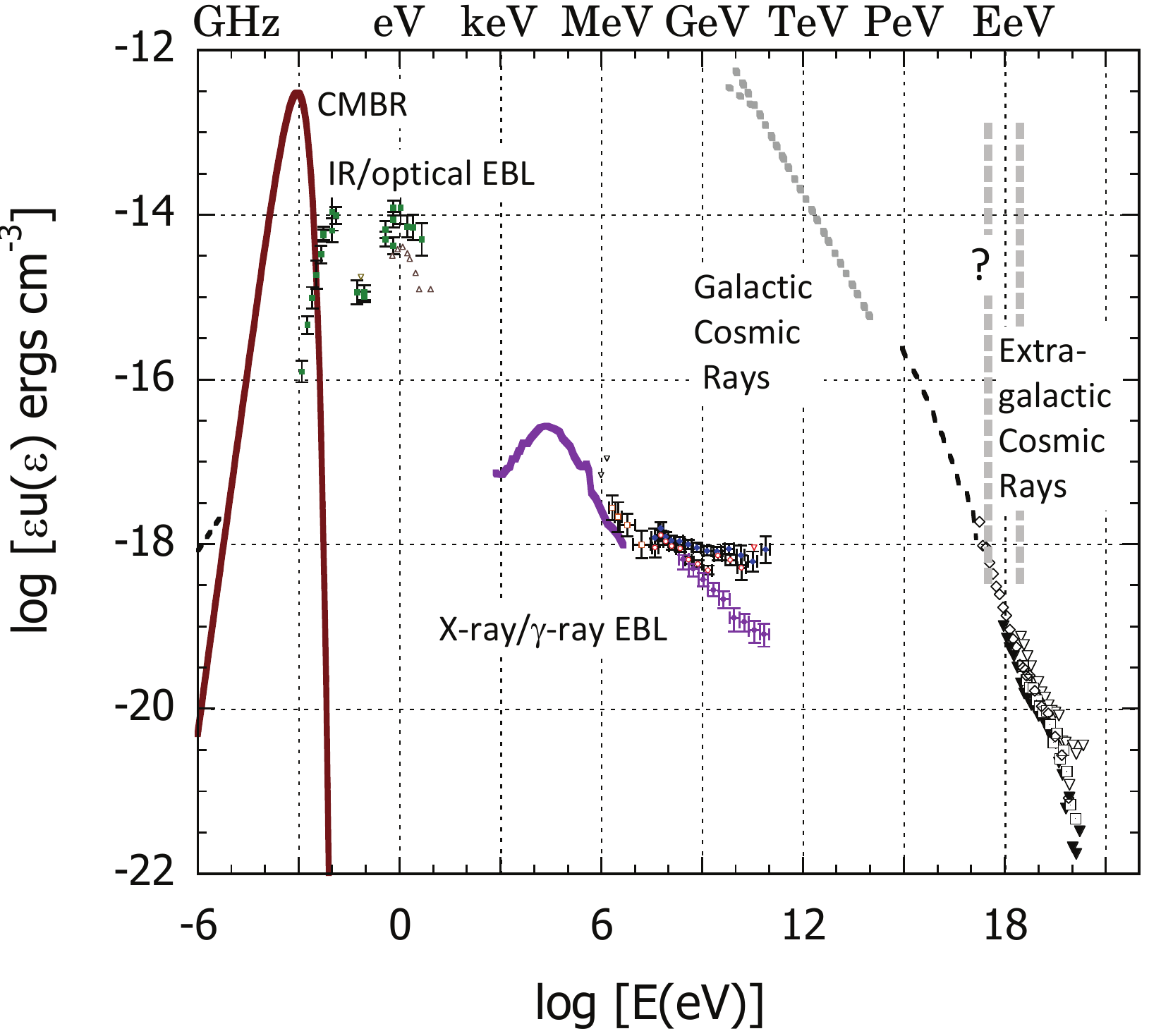}
\caption{Spectral energy densities in intergalactic space of various radiations,
including the CMB, the infrared (IR) and optical,  X-ray,
 $\gamma$ ray, and the extragalactic cosmic ray energy density. Also shown is the 
energy density of cosmic rays measured near Earth; the transition energy between the 
galactic and extragalactic component
remains uncertain.}
\label{f1-3}
\end{figure}

\subsubsection{Extragalactic background light (EBL)}

In intergalactic space, the energy density of the EBL
is dominated by that of
the cosmic microwave background radiation (CMBR),
with present temperature $T_{CMB} = 2.72$ K
and energy density $u_{CMBR} \cong 0.25$ eV/cm$^3 \cong 4\times 10^{-13}$
erg/cm$^3$.
The intensity of the EBL at infrared frequencies is difficult
to measure directly because of foreground radiations, such as 
Galactic electron synchrotron radiation and zodiacal light 
scattered by dust in our Solar system. Photons observed at 10 - 100 GeV
interact primarily with EBL photons at optical and UV energies.
The energy density of the dust and stellar components of the EBL
is $\approx 10$\% of the CMBR energy density at the present 
epoch. 

Figure \ref{f1-3} shows the energy density of photons in intergalactic space
and cosmic rays in outer space 
in the Solar cavity near Earth. 
{Fermi} measurements \cite{abd10id} give a lower intensity 
for the diffuse extragalactic 
$\gamma$-ray background (EGB) than EGRET data  \cite{sre98,smr04}, probably
due to EGRET miscalibration above $\approx 5$ GeV \cite{shk08}.\footnote{The spectral energy density of the isotropic background radiation field at 100 MeV as measured
with EGRET is
\begin{equation}
\epsilon u(\epsilon ) = {4\pi\over c}\;\epsilon I_\epsilon \cong \;{4\pi\over c}{{\rm~keV}\over {\rm cm}^2\mbox{-}{\rm s}\mbox{-}{\rm sr}} \cong 10^{-18}{\rm ~erg~cm}^{-3}.
\label{uepsilon}
\end{equation}
Because the $\epsilon I_\epsilon$ spectrum is essentially flat, the total energy density over the 100 MeV -- 100 GeV
energy range gives a factor $\ln 10^3 \cong 6.9$; thus the EGRET energy density of the EGB is nearly 5 orders of magnitude below the present CMBR energy density.}
The cosmic-ray particle 
spectrum is modulated at low, $\sim$GeV/nuc 
energies by changing Solar activity. Cosmic-ray 
source origin changes from a Galactic to 
extragalactic origin at an energy scale currently 
under debate. The case for an extragalactic origin of the UHECRs can be 
made by comparing Galactic size with the Larmor radius 
\beq
r_{\rm L} = {E\over QB} = ({A\over Z})\;{m_pc^2 \gamma\over eB}\;\cong\;
100\,{E/10^{20}{\rm~eV}\over Z B_{\mu{\rm G}}}\,{\rm kpc}
\label{rlarmor}
\eeq 
of an ion with energy $E$, Lorentz factor $\gamma$, 
atomic charge $Z$ and atomic mass $A$. The characteristic magnetic field 
$B=10^{-6}B_{\mu {\rm G}}$ G 
of the Milky Way is 
a few $\mu$G in the disk and probably much less in the Galactic halo.
  
\subsection{Limits to the extreme universe}

The largest luminosity is obtained if the entire rest mass energy of the 
object with mass $M$ is transformed into energy on a 
light-crossing time scale for the gravitational radius of 
the object, so that
\begin{equation}
L_{max} = { Mc^2\over GM/c^3} = {c^5\over G} = 3.6\times 10^{59} \;{\rm erg~s}^{-1}\; 
\label{Lmax}
\end{equation}
\cite{mtw73,tl11}. The most luminous blazar source yet detected with the Fermi LAT is
3C 454.3, with apparent isotropic $\gamma$ ray luminosity 
reaching $L_\gamma \approx 2\times 10^{50}$ erg/s. The most luminous GRBs detected
with the Fermi-LAT, namely GRB 080916C and GRB 090510A, reached $L_\gamma  \approx 10^{53}$ erg/s.

\section{Fermi Gamma-ray Source Catalogs and  Fermi Pulsars}

Now we take a global view, and look at the Fermi Large Area Telescope First Source
Catalog (1FGL) \cite{abd101FGL} as a {\it catalog}, or 
systematic collection of objects, before focusing on $\gamma$-ray emission from neutron stars
and the first Fermi pulsar catalog \cite{abd10PSR}.
The 1FGL catalog, taken with 11 months of data between 2008 August 4 and 2009 July 4, 
expands the 3 month bright source 
list (BSL, or 0FGL) \cite{abd09BSL}, just as the First LAT  AGN Catalog (1LAC) \cite{abd101LAC}, 
the subject of the next lecture, expands the LBAS (LAT Bright AGN Sample) \cite{abd09LBAS}. 
The 2FGL and 2LAC build on the 1FGL and 1LAC, respectively. The 1FGL and 2 FGL catalogs are 
described in this section.

Source detection significance for the 1FGL and 2FGL catalogs is based on the likelihood test statistic ($TS$). 
The likelihood ratio compares the likelihood of a null hypothesis
with the likelihood of the model. According to Wijk's theorem, $-2$ times the logarithm of this ratio
approaches $\chi^2$ for large $\chi^2$. The likelihood ${\cal L}$ is the probability of observing
$n_{ij}$ counts in pixel $i,j$ given the expected number of counts $\lambda_{ij}$ in pixel $i,j$.
Poisson statistics gives 
\beq
{\cal L} = \prod_{ij}p_{ij} = \prod_{ij}{\lambda_{ij}^{n_{ij}}\exp(-\lambda_{ij})\over n_{ij}!}\;.
\eeq
Dropping a term independent of model, the logarithm of the likelihood is
\beq
\log{\cal L} = \sum_{ij} n_{ij}\log (\lambda_{ij})  - \sum_{ij} \lambda_{ij} \;,
\label{logL}
\eeq
and $TS = -2 \log({\cal L}/{\cal L}_0)\rightarrow \chi^2$, where ${\cal L}_0$ is the likelihood of the null hypothesis
for the given data set. Eq.\ (\ref{logL}) gives a prescription 
for calculating $TS$ when analyzing $\gamma$-ray data.
Mattox and colleagues \cite{mat96} originally applied this method to EGRET data.

The Fermi catalogs are based on a limiting test statistic, 
which is $TS>25$ in the 1FGL, 2FGL, 1LAC and 2LAC, and $TS>100$ in the BSL and the LBAS. This introduces systematic effects
that should be taken into account when interpreting the 1FGL. For example, a cut on $TS$ produces 
integral flux thresholds strongly dependent on the source spectral hardness. 
 Because of the steeply falling Galactic background at high galactic latitudes, 
high-latitude hard spectrum
sources will have a large $TS$ for much smaller integral fluxes than a soft-spectrum source
detected at the same significance level.

\subsection{First Fermi catalog of gamma-ray sources: 1FGL}

The 1FGL comprises 1451 sources with $TS > 25$, corresponding to a significance
$\gtrsim 4\sigma$, and
 represents 21.22 Ms (or $\approx 73.3$\% livetime) of data. Most of the deadtime loss is due to 
passage through the SAA South Atlantic Anomaly ($\approx 13$\%) 
and to readout deadtime (9.2\%). The exposure is uniform to within a 
factor $\approx 1.25$ between north and south and the detections are based 
on integrated data rather than 
shorter bright periods or flares.
An improved background and calibration model is used in the 1FGL 
compared to the 0FGL.  

The 1FGL is not a flux-limited survey. Although the exposure is relatively uniform---an 
excellent feature of the GLAST design in nominal scanning mode---there are large differences in 
integral photon fluxes of sources depending on source location and the strong Galactic diffuse emission. 
Consequently, the 1LAC catalog is drawn from sources above $10^\circ$ latitude where the 
Galactic diffuse intensity is low.\footnote{The Galactic diffuse flux at $10^\circ \leq |b| \leq 20^\circ$, 
averaged over longitude, is still a factor of 2 -- 3 greater 
than the extragalactic diffuse at higher latitudes \cite{abd09gd}.} 
Any conclusions are weakened when source 
identifications are not complete, for example due to the limits of a counterpart catalog
used to make associations, or to uncertainty in distance or lack of redshift,
as this incompleteness can bias results.

The 1FGL catalog gives source location regions and association criteria 
defined in terms of elliptical
fits to the 95\% confidence regions,
 and power-law spectral fits as well as flux measurements in five energy
bands for each source. In addition, monthly light curves are provided. 
Firm identifications with sources found in other astronomical
catalogs are based on correlated variability, e.g., rotation or orbital period 
in the case of pulsars and X-ray binaries, or morphological similarities with counterparts at other wavelengths,
as in the case of SNRs.

For the catalogs and association criteria used, 630 of the sources in the 1FGL are
unassociated. Due to the sensitivity of the 
results to the model of interstellar diffuse
$\gamma$-ray emission used to model the bright foreground, 
161 sources at low Galactic
latitudes towards bright local interstellar clouds 
are flagged as having properties that are strongly dependent
on the Milky Way gas model.

\begin{table}
\centering
\caption{LAT 1FGL and 2FGL Source Classes \cite{abd101FGL,abd122FGL}}
\label{tab:2}       
%
%
\begin{tabular}{llll}
\hline\noalign{\smallskip}
Description & Designator & Assoc. (ID) & Assoc. (ID)\\
 &  & 1FGL & 2FGL \\
\noalign{\smallskip}\hline\noalign{\smallskip}
Pulsar, X-ray or radio, identified by pulsations & psr (PSR) & 7 (56)   &   0 (83)\\
Pulsar, radio quiet (LAT PSR, subset of above) & PSR & 24 --  &  25 --\\
Pulsar wind nebula & pwn (PWN)&  2 (3)   & 0 (3) \\
Supernova remnant&  snr$^a$ (SNR) & 41 (3)   & 62$^c$ (6) \\
Globular cluster & glc (GLC) & 8 (0)   &  11 (0) \\
High Mass X-ray binary & hxb (HXB)&  0 (2)  & 0 (4)\\
Micro-quasar object: X-ray binary (black hole & mqo (MQO) & 0 (1)   &    0 (1) \\or neutron star) with radio jet & & & \\
Nova	& nov(NOV)	&	& 0 (1) \\BL Lac type of blazar & bzb (BZB)&  295 (0)   & 428 (7)\\
FSRQ type of blazar&  bzq (BZQ)&  274 (4)   & 353 (17)\\
Non-blazar active galaxy & agn (AGN) & 28 (0) & 10 (1)\\
Radio galaxy	&	&	&	10 (2) \\
Active galaxy of uncertain type & agu (AGU)&  92 (0) & 257 (0) \\
Normal galaxy & gal (GAL) & 6 (0) & 4 (2)\\
Starburst galaxy & sbg (SBG)&  2 (0) & 4 (0)\\
Seyfert galaxy & sey (SEY) &   & 5 (1)\\
Unassociated & & 630 & 576 + 1$^d$ \\
Total  &  &   1478$^b$  & 1873$^{e}$  \\
\noalign{\smallskip}\hline
\end{tabular}
{\noindent $^a$Indicates a potential association with a SNR or PWN.}\\
{\noindent $^b$ $779 + 630+(69)=1409+(69) = 1478.$ Greater than 1451 because of multiple class assignments.}
{\noindent $^c$Some of the 62 sources may also be associated with PWNe.}
{\noindent $^d$576 unassociated plus one with uncertain class.}
{\noindent $^e$$1169 + 577+(127)=1746+(126) = 1873.$}
\end{table}

The principal source classes found in the 1FGL and listed in Table \ref{tab:2} are

\begin{itemize}
\item Sun, moon, Earth
\item 3 high mass X-ray Binaries
\item Rotation-powered and millisecond pulsars
\item Supernova remnants
\item Globular clusters
\item $>$ 600 blazars	
\item 28 Non-blazar AGNs and Radio Galaxies
\item Dozens of AGNs of uncertain type
\item 3 Starburst Galaxies
\end{itemize}
In addition, hundreds of GRBs and more than a dozen LAT GRBs had been 
detected by the time of publication of the 1FGL.

Association depends on the underlying catalog and the 
degree of confidence assigned in the probability of association. 
For blazars, four catalogs with substantial overlap were used, 
giving 689 1FGL  associations with sources found 
in at least one of these catalogs. 
A fuller discussion of the AGN 
catalog is deferred to the next lecture. Though no extended radio lobes of 
radio galaxies were reported in the 1FGL, the lobes of Centaurus A have 
since been imaged \cite{abd10f}.

Regarding other source classes in the 1FGL, 
41 Fermi sources are associated with 
SNRs or non-pulsed $\gamma$-ray emission, 
and three are sufficiently physically extended that
their morphology counts as an identification (W44  W51C, and
 IC 443; see Lecture 5). Fermi reported three high-mass
X-ray binary systems in the 1FGL, namely LS 5039, LSI+61$^\circ$303, and Cyg X-3 (Lecture 7), 
and a fourth was discovered later. 
In the 1FGL, it was unclear whether $\gamma$ rays were made by massive O stars,
but since then $\eta$ Carinae has been been established as a $\gamma$-ray source \cite{2010ApJ...723..649A}.
Ten BSL candidates out of 205 in the 0FGL, do not show up in the 1FGL, illustrating
the highly variable nature of some $\gamma$-ray sources (though note
that all these sources are found in the Galactic ridge, $|l| < 60^\circ$, 
where  background modeling is especially important). On top
of this, 630 sources were unassociated in the 1FGL. The unidentified 
Fermi sources are clustered towards the plane of the Galaxy where
diffuse background Galactic emission and higher source density makes 
associations more tentative, though there are still many high-latitude 
unidentified Fermi sources. 

\subsection{Second Fermi catalog of gamma-ray sources: 2FGL}

The 2FGL catalog \cite{abd122FGL} was released at the time of writing.
This source catalog was derived from data taken during the first
24 months of the science phase of the mission, which began on 2008 August 4.
Source detection is based on the average flux over the 24-month period. The
2FGL includes source location regions and fits to model spectral  forms. Also included
are flux measurements in 5 energy bands and light curves on monthly intervals
for each source. Twelve sources in the 2FGL are found to be spatially extended.

 The 2FGL  contains 1873
sources detected and characterized in the 100 MeV to 100 GeV range, of which
 127 are firmly identified and 1170 are reliably associated
with counterparts of known or likely 
$\gamma$-ray source classes. Although the diffuse Galactic and isotropic
models used in the 2FGL analysis are improved compared to the 1FGL catalog,
 caution flags for 162 sources  indicate possible confusion, given
the uncertainty in the underlying diffuse model. Table \ref{tab:2} lists the number of sources 
of various types in the 
2FGL. 

Some important improvements compared to the 1FGL catalog are:
\begin{enumerate}
\item The 2FGL catalog is based on data from 24 months of observations.
\item The data and Instrument Response Functions use Pass 7 event selections,
rather than the Pass 6 event selections used in the 1FGL.
\item The 2FGL employs a new, higher-resolution model of the diffuse Galactic and isotropic
emissions.
\item Spatially extended sources and sources with spectra other than power laws are incorporated
into the analysis.
\item The source association process has been refined and expanded.
\end{enumerate}

\subsection{Fermi pulsars}

In the 1FGL, 56 pulsars are identified by their $\gamma$-ray pulsations, 
and another 7 associations are based on expectations 
from pulsar catalogs, e.g., $\dot E/d^2$ ranking.
Six 1FGL sources are associated with pulsar wind nebulae (PWNe) 
that lack known pulsars, and 8 1FGL sources are associated with 
globular clusters illuminated, most likely, by the superposition
of millisecond pulsar (MSP) $\gamma$-ray emissions. 
The number of Geminga-like pulsars that
lack detectable  radio emission, has grown by a large factor.
There are 83 $\gamma$-ray pulsars, now all identified, in the 2FGL.

A catalog of 46 $\gamma$-ray pulsars
is presented in the First LAT Pulsar Catalog, Ref.\ \cite{abd10PSR}.
This catalog includes 16 ``blind-search" pulsars 
discovered by searching for pulsed $\gamma $-ray emission 
at the position of bright LAT sources. 
Pulsed $\gamma$-ray emission from 24 known pulsars
were discovered using ephemerides derived
from monitoring radio pulsars, of which 8
are MSPs. The remaining 6 $\gamma$-ray pulsars were known
 previously.

\subsubsection{EGRET pulsars}

The Crab  and Vela 
pulsars were known  prior to {\it CGRO}, and Geminga was known
as a bright $\gamma$-ray point source. The pulsar nature
of Geminga was only established by detection of X-ray pulsations
in ROSAT data early in the EGRET era \cite{hh92}.
The EGRET pulsars, with EGRET fluxes above 100 MeV
and 2FGL fluxes between 1 GeV and 100 GeV, ordered by brightness, 
are listed in Table \ref{tab:EGRETpulsars}.

\begin{table}
\centering
\caption{$\gamma$-ray Fluxes of EGRET Pulsars \cite{har99,abd10PSR,abd122FGL}}
\label{tab:EGRETpulsars}       
%
%
\begin{tabular}{lccccc}
\hline\noalign{\smallskip}
Pulsar & Period (ms) & Age (kyr)  & EGRET  &  Pulsar Catalog & 2FGL \\ 
& $P$ & $P/2\dot P$ & $F_{-8}$ & $F_{-8}$ & F(1 -- 100 GeV)$^a$ \\
\noalign{\smallskip}\hline\noalign{\smallskip}
0833$-$45, Vela & 89.3 & 11.3  &   $834.3\pm 11.2$ & $1061\pm 7.0$ & $135.8\pm 0.4$  \\
J0633$+$1746, Geminga & 237 & 340  &   $352.9\pm 5.7$ & $305.3\pm 3.5$ &  $72.9\pm 0.3$  \\
0531$+$21,$^b$ Crab & 33 & 1.25  &   $226.2\pm 11.2$ & $209\pm 4$ & $18.3\pm 0.15$  \\
1706$-$44 & 102 & 17.6  &   $111.2\pm 6.2$ & $149.8\pm 4.1$ & $19.1\pm 1.7$  \\
1055$-$52 & 197 & 540  &   $33.3\pm 3.82$ & $30.45\pm 1.7$ &  $5.0\pm 0.09$  \\
1951$+$32$^c$ & 39.5 & 110  &   $16\pm 2$ & $17.6\pm 1.9$ & $2.1\pm 0.07$  \\
1509$-$58,$^c$ Circinus & 88.9 & 150 &   -- &  $8.7\pm 1.4$ & $1.45\pm 0.08$  \\
\noalign{\smallskip}\hline
\end{tabular}
{\noindent $^a$Also in units of $10^{-8}$ ph cm$^{-2}$ s$^{-1}$}\\
{\noindent $^b$Associated with SN 1054}\\
{\noindent $^c$Pulsars not reported in the 3EG \cite{1995ApJ...447L.109R}}
\end{table}

PSR 1951+32, located in the confusing Cygnus arm region, 
is associated with the SNR CTB 80 discovered by Kulkarni et al. \cite{1988Natur.331...50K}.
It has a phase-averaged  $\nu F_\nu$ spectrum rising $\propto \epsilon^{0.2}$  
and peaking at $\approx 2$ GeV, with a slower than exponential decline at higher
energies  \cite{2010ApJ...720...26A}. 

The Circinus pulsar was detected with COMPTEL but not EGRET, and 
is unusual in having an inferred polar magnetic field 
$B_p = 15.77$ TG, compared to $\sim $TG ($10^{12}$ G) fields for the others.
Geminga is unusually close, at $d \cong 160$ pc \cite{car96}, whereas the others
are more likely to be at $\sim $kpc distances. 

\subsubsection{Elementary pulsar physics}

Over 1900 pulsars are 
known today, mostly through radio surveys.
Pulsars are rapidly rotating, highly magnetized 
neutrons star. The neutron stars themselves are formed by 
explosions made by the collapsing cores of massive stars
(core-collapse supernovae), or through accretion-induced
collapse of white dwarfs. Misalignment of the magnetic axis and rotation 
axis makes in the simplest configuration a rotating dipole-field geometry. Emission 
beamed along favorably defined field lines naturally makes a wide 
range of pulse structures for different obliqueness and inclination
and gap sizes.
The additional range of parameters associated with 
period and period derivative, along with poorly understand radiation 
physics, allows for purely empirical and kinematic pulse profile fitting.

 Neutron stars are  predicted to have masses
near $1.4 M_\odot$ and radii $\sim 15$ km. The two 
properties of the pulsar that can be very precisely measured are the 
period $P$ and period derivative $\dot P$. Besides these
observables, theory gives the mass of the neutron star, $M_{NS}$, 
and its radius $R_{NS}$ \cite{st83}. The uncertain equation of state of nuclear 
matter determines whether neutron stars with given masses can exist, 
whether collapse to a black hole occurs, or whether other degenerate
quark phases exist.

Electron degeneracy pressure in a white dwarf cannot
support the gravitational force of a degenerate core 
exceeding the Chandrasekhar-mass  $M_{\rm C} = 1.4 M_\odot$. Either through
nuclear burning or accretion, neutron stars can  
be formed. Neutron stars with 
masses $M>M_{\rm C}$ can test black-hole formation theory, but the uncertainty in orbital inclination usually makes
for large uncertainties in the neutron-star mass.  
By means of radio timing measurements of the Shapiro delay\footnote{The Shapiro 
delay is a regular general-relativistic change in the light travel 
time from the pulsar as the radio photon travels 
through the gravitational field of  the binary system; see \cite{wei72}.} 
of the binary millisecond pulsar J1614-2230, a $1.97 \pm 0.04 M_\odot$ pulsar mass was recently measured  \cite{dem10}, 
which rules out some exotic hyperon and boson condensate equations of state.

Noting that velocity ${\vec v} = {\vec \Omega}\times {\vec R}$, and the angular
speed $\Omega = 2\pi/P$, then the light-cylinder radius at which the speed $v =c$ is
\begin{equation}
R_{LC} = Pc/2\pi\;.
\label{RLC}
\end{equation}
Even the rotation of a simple misaligned dipole field leads to unusual
geometric features, for example, the footprint on the neutron star surface
of the field lines that open into the light cylinder  \cite{2004ApJ...606.1125D}.

An expression for the surface polar magnetic field can be obtained 
by equating the rotational spindown energy loss rate with the 
magnetic dipole radiation power. The former is
\begin{equation}
-{dE_{rot}\over dt} = {d\over dt}\left({1\over 2} I\Omega^2\right)= {4\pi^2 \over P^3}I\dot P\;,
\label{dErotdt}
\end{equation}
with the moment of inertia $I$ depending on the mass and equation of state of neutron star matter.
This implies a characteristic age
\begin{equation}
\tau = P/2\dot P\;,
\label{tau}
\end{equation}
for $P$ much longer than the initial spin period.
The magnetic dipole power can be estimated from the Poynting flux of a dipole radiation
at the light cylinder, namely,
\begin{equation}
-{dE_{md}\over dt} = {B^2(R_{LC})\over 8\pi}\left(4\pi R_{LC}^2 c \right)= 
{1 \over 2}B^2_{NS}\left({R_{NS}^6\over R_{LC}^4}\right) c \propto {B_{NS}^2\over P^4}\;.
\label{dEmddt}
\end{equation}
Thus
\begin{equation}
B_{NS} \propto \sqrt{P\dot P}\;.
\label{BNS}
\end{equation}
The magnetic field $B_{LC}$ at $R_{LC}$ is therefore
\begin{equation}
B_{LC} = \left({24\pi^4 I \dot P\over c^3 P^5}\right)^{1/2}\;.
\label{BLC}
\end{equation}

We can imagine a time-independent configuration for a spinning, magnetized neutron star if we 
wait long enough until all current flows have asymptotically relaxed to their steady-state values. This defines 
the condition of a force-free magnetosphere, where the vanishing of the Lorentz
force ${\vec F} = Q[({\vec v}/c)\times {\vec B} + {\vec E} )] = 0$ implies the existence 
of a magnetosphere filled with plasma with Goldreich-Julian \cite{gj69} density 
\begin{equation}
\rho_{\rm GJ} = -{{\vec\Omega}\cdot {\vec B} \over 2\pi }\;.
\label{rhoGJ}
\end{equation} 

Pulsar models build on these elementary concepts. For generic {\it polar cap} models,
charge depletion in the  polar field lines that open to the light cylinder, compared with the 
Goldreich-Julian density, generates strong electric fields
that induce vacuum breakdown and current flow. In {\it outer gap} models, strong fields are generated
at the gaps created by the surface defined by the condition  ${\vec \Omega} \cdot {\vec B} = 0$. The primary $\gamma$-ray production
and attenuation processes are curvature radiation, Compton radiation, and magnetic pair production. 
{\it Slot gap} models study a more sophisticated realization of the pair-starved region and how
a slot gap forms along field lines with different acceleration potentials. 
Besides discriminating between magnetospheric models, pulsar studies help disentangle the geometry of a pulsar.
For more on pulsar models, see \cite{2001astro.ph..6161B,2004ApJ...606.1125D,2010ApJ...714..810R}.

\subsubsection{Properties of Fermi pulsars}

Two different approaches are taken to discover pulsars, 
either by period folding with data from a pulsar 
established at other frequencies, or to perform
 a blind search for the pulsation. 
In the first method, the timing parameters of pulsars already known at 
radio or X-ray frequencies
are used to search for evidence of $\gamma$-ray pulsations in the $\gamma$-ray data.
For the blind-search technique, spectral power is calculated from 
time-tagged $\gamma$-ray photon event data by searching through likely values of
 $P$ and $\dot P$ values. 
This is extremely computationally intensity, and not feasible for MSPs found
in binary systems. More computationally efficient techniques
 compute the arrival time differences
of  $\gamma$ rays, which cluster at values corresponding
to multiples of the pulse period \cite{2006ApJ...652L..49A}. For $\gamma$-ray astronomy, 
with limited photon statistics, only the time-differencing technique for blind searches
is effective at finding pulsars.

The first blind-search detection with Fermi
was a pulsar with period $P=0.317$ s in the shell of the SNR CTA1, 
coincident with unidentified X-ray and 3EG J0010+7309 $\gamma$-ray sources
\cite{abd08CTA1}. The spin-down power of this pulsar is  $\approx 4.5\times
10^{35}$ erg s$^{-1}$, and the pulsar's inferred age
is 14,000 yr, which is consistent with the SNR age.
Of the  46 Fermi pulsars reported in the First Fermi Pulsar Catalog, 
16 of them were found using blind-search techniques in sources that
are positionally coincident with 
 unidentified EGRET sources 
and supernova remnants \cite{abd09psr}. These 16 pulsars are all
young and highly magnetized, with inferred magnetic fields between 
$\approx 1$ -- 10 TG.

\begin{figure}[t]
\begin{center}
 \includegraphics[width=4.4in]{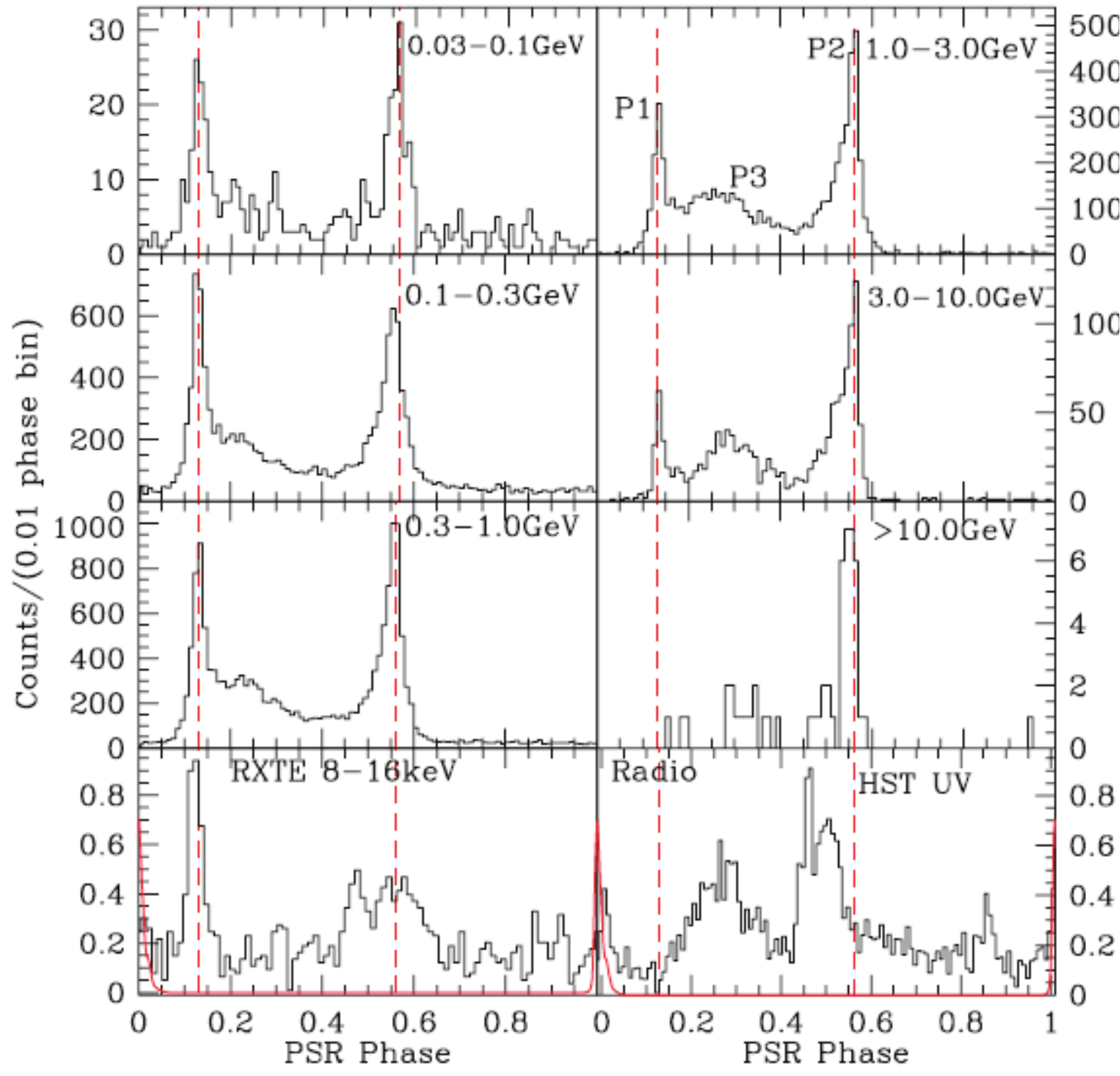} 
 \caption{Vela  light curves at optical, X-ray, and $\gamma$-ray energies \cite{2009ApJ...696.1084A},
binned to 0.01 of the pulsar phase. The main peaks P1, P2 and P3 are labeled in the top right panel. 
The bottom left panel shows the 8 -- 16 keV
{\it RXTE} light curve \cite{har02} along with the radio pulse profile (dashed lines). 
At lower right, the 4.1 -- 6.5 eV {\it HST}/STIS NUV light curve \cite{rom05} is shown.}
\label{vela_phase}
\end{center}
\end{figure}

The rotational energy-loss rates of
pulsars in the First Fermi Pulsar Catalog range from $\sim 3\times 10^{33}$ 
erg s$^{-1}$ to $5\times 10^{38}$ erg s$^{-1}$, with the young, highly magnetized pulsars
typically having rotational energy-loss rates exceeding 
$\gtrsim  10^{35}$ erg s$^{-1}$. 
Comparing with the phase-averaged apparent isotropic $\gamma$-ray luminosity  implies 
efficiencies for the conversion of spin-down power to $\gamma$-ray energy
in the range from $\sim 0.1$\% to $\approx 100$\%.
About 75\% of the $\gamma$-ray pulses have two peaks, but a third emission 
structure, P3, shows up in the Vela pulse profile and moves to later phases with increasing
photon energy \cite{2009ApJ...696.1084A}.
As Fig.\ \ref{vela_phase} shows, the main peak, P1, which is dominant from at least UV to GeV energies, 
becomes less intense compared to P2 at multi-GeV energies.

\begin{figure}[t]
\begin{center}
 \includegraphics[width=3.4in]{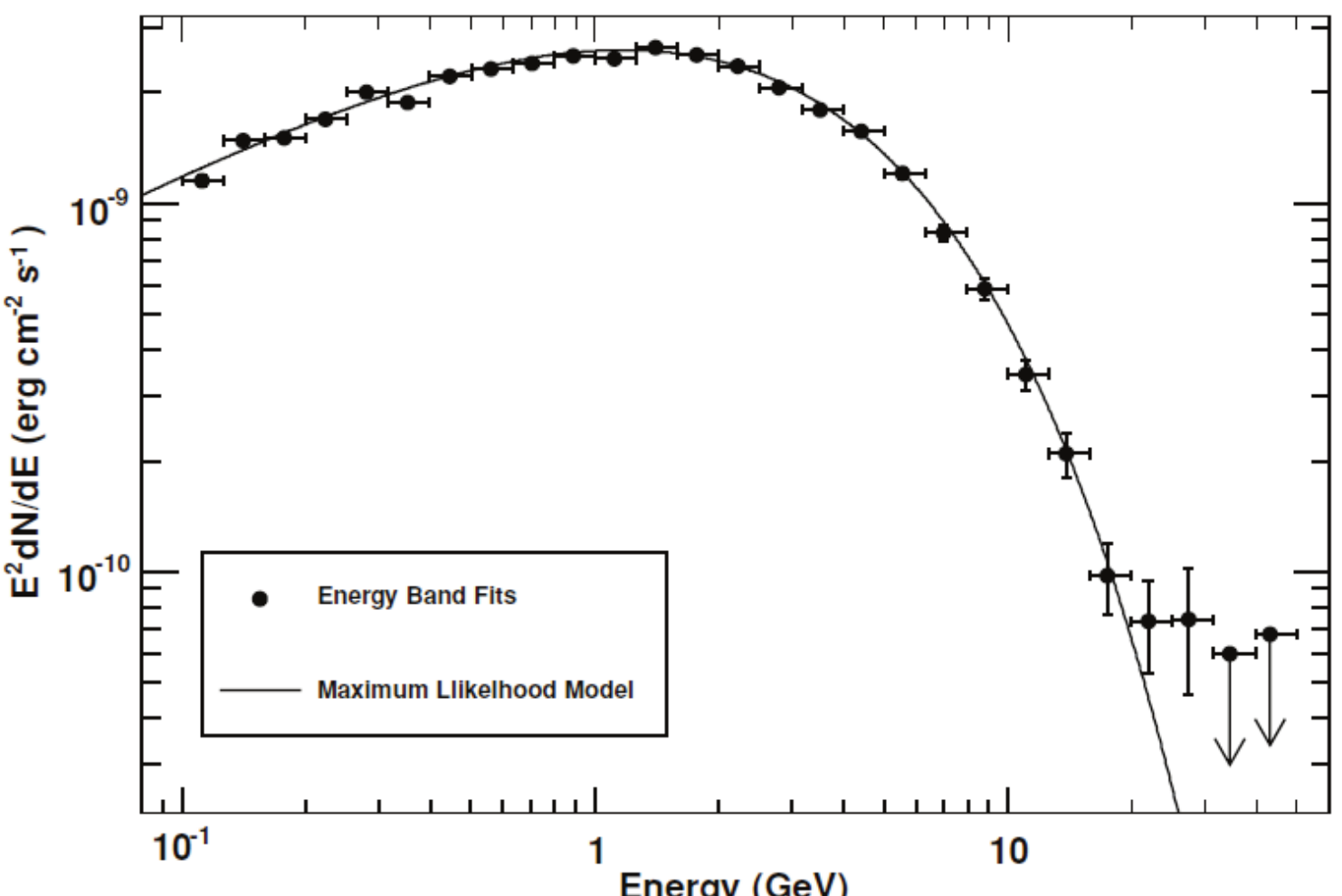} 
 \caption{Phase-averaged spectrum for $0.1 < E({\rm GeV}) < 60$ \cite{2010ApJ...713..154A}. }
\label{vela_spectrum}
\end{center}
\end{figure} 

The Fermi LAT phase-averaged spectrum of $\gamma$-ray pulsars can be well fit by a  generalized exponentially cutoff power law, given by 
\begin{equation}
N(E)\propto E^{-\Gamma_\gamma}\exp[-(E/E_c)^b]\;,
\label{N(E)}
\end{equation} 
with hard photon number indices $\Gamma_\gamma$ 
generally in the range  
$1 \lesssim \Gamma_\gamma \lesssim 2$, and cutoff energies $E_c$ 
between $\approx 1$ and $5$ GeV. This form has been found to apply to all types of pulsars, whether 
radio or $\gamma$-ray selected, or normal or millisecond pulsars.
For Vela itself, as seen in Fig.\ \ref{vela_spectrum}, $\Gamma_\gamma = 1.38\pm 0.03$, $E_c = 1.36\pm 0.15$ GeV,
and $b = 0.69\pm 0.02$ (quoting statistical errors only) \cite{2010ApJ...713..154A}. The sub-exponential ($b<1$) rather than super-exponential 
($b>1$) cutoff that would be produced if the $\gamma$ rays were made deep within the 
polar cap of the neutron star magnetosphere, and the detection of emission at tens of GeV, 
is evidence against a polar-cap model. A phase-averaged spectrum fitted with equation (\ref{N(E)}) is the composite phase-resolved spectra that
individually can be fit by simple exponential behaviours \cite{2010ApJ...713..154A,2010ApJ...720...26A}. Significant variations of cutoff energy with phase
are needed to reproduce the phase-averaged spectra.

\subsubsection{Millisecond pulsars and globular clusters}

Pulsed $\gamma$-ray emission
was detected with the Fermi LAT from J0030+0451, 
making it the first firm detection of a MSP in $\gamma$ rays 
\cite{abd0911FMSP}, although there was a marginal EGRET detection 
of PSR J0218+4232 \cite{kui00} that has been confirmed 
with the LAT \cite{abd0911FMSP}. Nine months into science operations, the 
Fermi-LAT Collaboration reported 8 $\gamma$-ray MSPs 
\cite{abd098FMSPs}, establishing a new population of $\gamma$-ray emitters.
 As noted above, it is not computationally feasible to 
perform blind searches for binary MSPs, which introduces too many possibilities into the 
timing solutions. This makes it harder to know what 
fraction of the unidentified sources are MSPs, or if there are any radio-quiet MSPs.

Globular clusters are quasi-spherical stellar systems consisting
of hundreds of thousands of 
old (ages of $\approx 10^{10}$ yr) metal-poor stars. The high-stellar density 
at the centers of these systems makes for unusual multi-body encounters.
According to the most common scenario,
binary neutron-star systems in low-mass X-ray binaries can be spin 
up by accretion over long time spans to periods of a few ms, provided that 
the polar surface magnetic field is $\lesssim 10^9$ G, which 
is $\sim 3$ orders of magnitude smaller than 
the typical TG fields of young $\gamma$-ray pulsars.
Even though they have very different magnetic field and rotation rates,
the $\gamma$-ray spectra of MSPs and young pulsars are similar, and well-represented
by a hard power-law with a modified exponential cutoff given by
equation (\ref{N(E)}).

The Fermi-LAT collaboration has reported the 
detection of the famous globular cluster 47 Tucanae at
a significance of $17\sigma$  \cite{abd0947Tuc}. The
$\gamma$-ray spectral shape 
is consistent with a large populations of MSPs.
 With a typical luminosity of $\lesssim 3\times 10^{33}$ erg s$^{-1}$ in $\g$-rays, 
 $\sim 20$ -- 40 MSPs are required to give the measured $\gamma$-ray luminosity.
Using 587 days of data, the number of Fermi sources associated with globular clusters has grown
to 8 \cite{abd10GC}, and now 11 in the 2FGL \cite{abd122FGL}.

Both $\gamma$-ray detected  ``garden-variety" pulsars and MSPs
 have the highest values of magnetic field at the light cylinder, $B_{LC}$,
suggesting that similar emission mechanisms operate.
The 
$\gamma$-ray luminosity grows with spin-down energy
 $L_\g \propto \dot E$ at $\dot E \lesssim 10^{35}$ erg s$^{-1}$ and 
$L_\g \propto \sqrt{\dot E}$ at $\dot E \gtrsim 10^{35}$ erg s$^{-1}$,
with large scatter in this relation due, at least, to distance uncertainties
(Fig.\ 6 in \cite{abd10PSR}).

\subsubsection{Pulsar wind nebulae}

Given the accuracy of timing parameters for a pulsar,  the rotational energy-loss rate is known to the uncertainty of the moment of inertia of the neutron 
star. The spin energy not coming out as $\gamma$ rays, which usually accounts for no more than $\sim 10$\% of the spin-down energy, must come out in a 
different form, e.g., as field energy in the form of a relativistic Poynting wind with such small baryon loading.
The wind Lorentz factors attains values of $\sim 10^8$. The interaction of the outflowing MHD wind with the ISM makes a termination shock where additional particle acceleration can take place.
The termination shock separates the cold upstream MHD wind with a turbulent flow downstream into the ISM. The boundary between the wind and ISM is highly structured because of various fluid instabilities.  Cold particles can be directly injected into the ISM with the wind Lorentz factor, or accelerated at the termination shock.

Identification of pulsar wind nebulae (PWNe) with the Fermi LAT depends on finding steady extended GeV radiation. Three PWNe were reported with Fermi LAT using 16 months of survey data, namely the Crab nebula, Vela-X, and the PWN inside MSH 15-52  \cite{ack11pwn}.  Searching in the off-pulse emission of the pulsar  yields a candidate PWN related to PSR J1023-5746 and coincident with HESS J1023-575. The sources with GeV-detected PWNe have the highest spin-down flux and tend to have the highest spin-down power, with $\dot E_{rot} \gtrsim 10^{37}$ erg s$^{-1}$.

\subsubsection{Crab nebula and flares}

\begin{figure}[t]
\begin{center}
 \includegraphics[width=3.4in]{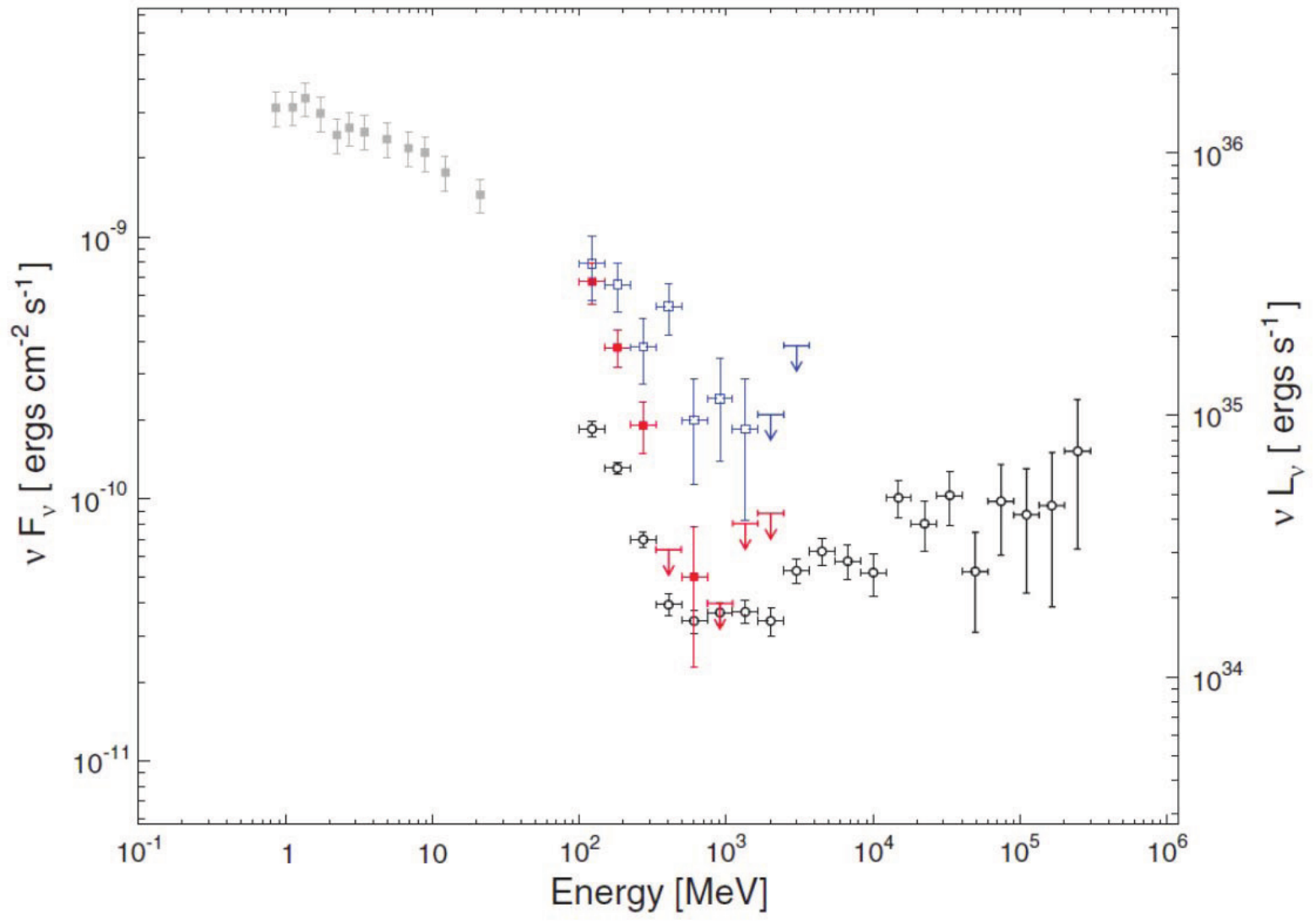} 
 \caption{COMPTEL 0.7 -- 30 MeV data (gray filled squares)  and long-term average
 {Fermi}-LAT $100$ MeV -- 300 GeV  $\gamma$-ray data (open circles), 
and February 2009 (open squares) and September 2010 (filled squares) flares of the Crab nebula \cite{abd10crabflares}.
}
\label{crab_flares}
\end{center}
\end{figure}

The Crab pulsar and nebula is associated with a supernova that took place in 1054 CE at a distance of 
$\approx 2$ kpc. It is the best example of a center-filled plerionic SNR, where the 
interior is illuminated by synchrotron emission powered by the wind of a young, 33 ms neutron star. 
The spin-down energy of the Crab pulsar is $4.6\times 10^{38}$ erg s$^{-1}$, of
which $\approx 30$\% is converted into the radiant power of the nebula.
The mean spectral index of the nebula in the LAT range is $\cong -4$ between $\approx 100$ MeV and 400 MeV, 
hardening to an index of $\cong -1.64$ at higher energies \cite{abd10crab} (see Fig.\ \ref{crab_flares}). The harder, high energy component is 
the synchrotron self-Compton component (predicted by Gould \cite{1965PhRvL..15..577G}), and well-explained as SSC if the mean magnetic field
is $\approx 200 \,\mu$G \cite{1996MNRAS.278..525A,abd10crab}.

Year-scale variations by a several percent are 
found in Crab nebular fluxes measured with RXTE, Swift BAT, and INTEGRAL.
The fluxes also declined during the first two years of the {Fermi} mission, 
consistent with measurements of an $\approx 7$\% decrease between
12 and 500 keV as measured with the GBM \cite{2011ApJ...727L..40W}.
In the period between 2008 August and 2009 April, no variations in the 100 MeV -- 30 GeV flux
were found with the LAT \cite{abd10crab}. 

Although some evidence for flux variations had been noted during the EGRET era \cite{1996ApJ...457..253D},
{Fermi}-LAT 
detected$\gtrsim 100$ MeV flaring episodes from the Crab  lasting for 16 d 
 in February 2009 and 4 d in September 2010  \cite{abd10crabflares}.
The September 2010 flare was first announced by the AGILE team \cite{2010ATel.2855....1T}, 
and they also reported an earlier October 2007 flare \cite{2011Sci...331..736T}. Typical $\gamma$-ray 
flare luminosities are $\approx 4\times 10^{36}$ erg/s, and reach an integral photon number  flux 
above 100 MeV that is  $\approx 10\times$ brighter 
than the mean flux.

The PSF of front events of the {Fermi}-LAT is $\approx 0.12^\circ$ -- 0.15$^\circ$ at
$E\gtrsim 10$ GeV, so imaging can be achieved to $\approx 0.12^\circ/8 \sim 1^\prime$
(at 2 kpc, $1^\prime$ and $1^{\prime\prime}$ correspond to $\approx 0.6$ pc and $\approx 3\times 10^{16}$ 
cm, respectively). The extent of the radio nebula is $\approx 10^\prime$, and the extent of the 
Chandra X-ray torus from which perpendicular jets emerge is $\approx 1.3^\prime$. A bright inner X-ray 
ring in the Chandra image is $\approx 0.5^\prime$ in extent. Zooming into the central region 
with both Chandra and HST\footnote{chandra.harvard.edu/photo/2002/0052/animations.html}
reveals a dynamic wind, wisps, and  $\sim 1^{\prime\prime}$ knots that brighten and dim. For comparison,
one light day corresponds to $\approx 3\times 10^{15}$ cm, far smaller than the smallest resolvable
knots, yet the $\gamma$-ray flares radiate $\approx 1$\% of the spin-down energy.

The crucial feature of the Crab  
spectrum may well be that the quiescent spectrum displays a strong softening at $\approx 20$ MeV. This 
value is a factor $\approx 10$ smaller than the 
maximum synchrotron frequency at $\approx 200$ MeV obtained by balancing the timescale for electron synchrotron losses 
with the Larmor timescale on which particles gain energy through Fermi particle acceleration mechanisms (see Section 5.5), {\it for a
nonrelativistic flow}. A mildly relativistic flow can enhance the emission by large factors
and could be compatible with changes in the flow profile at the termination shock when the ultra-relativistic wind  
slows to mildly relativistic speeds.  Considerable theoretical interest has focused on perturbations of 
the flow induced by statistical fluctuations \cite{2011ApJ...730L..15Y},
by the generation of large amplitude MHD waves \cite{2011arXiv1109.1204L}, and by outflowing knots formed 
at an oblique termination shock 
 \cite{2011MNRAS.414.2017K}.
No evidence for the flare energy generation is found in the pulsar spin-down behavior \cite{abd10crabflares}.

\subsubsection{Pulsar physics with {Fermi}}

Three years after launch, over 100 $\gamma$-ray pulsars are known, and $\gamma$-ray emission is 
associated with 8 globular clusters counting 47 Tuc \cite{abd10GC}. In NGC 6624, a single MSP dominates the 
$\gamma$-ray energy output \cite{newref}.  The $\gamma$-ray pulsars are
divided into 31 radio-selected MSPs, 37 blind search $\gamma$-ray pulsars, and 38 young 
radio-selected  $\gamma$-ray pulsars. Of the 37 blind search pulsars, extensive follow-up
observations reveal pulsed radio emission in only 3 cases. Radio followup of Fermi pulsars
is described, e.g., in \cite{ray2011}.
 
 Some of the interesting open questions relate to the field geometry and radiation mechanisms for 
 pulsar $\gamma$ rays. The significant fraction of Geminga-like blind-search pulsars is generally
interpreted in terms of a larger $\gamma$-ray than radio cone. The detection of pulsed emission
in the Crab to $\approx 125$ GeV \cite{2011Sci...334...69V} means that this emission has to be made high in the magnetosphere.
Emission to these energies requires extreme parameters for a curvature radiation origin, 
and may imply Compton-scattering in the Crab's magnetosphere. (The curvature radiation mechanism
remains a feasible mechanism for pulsars older than the Crab.)
The  $L_\gamma\propto \dot E^a$ behaviour is not well-explained from the lowest MSP powers to the highest
pulsar powers. Pulsar $\gamma$-ray emission can 
make a significant, tens of \%, fraction of the Galactic diffuse emission. This fraction could
be very different in early-type galaxies with only MSPs now active.  

For reviews of Fermi pulsars, see \cite{rsp10} and \cite{rom11}.

\section{ Fermi AGN Catalogs }

The discovery of the $\gamma$-ray blazar class is one of EGRET's lasting legacies. 
 Of the 66 high-confidence sources associated with AGNs in 
the 3EG catalog \cite{har99}, all except one---the nearby radio galaxy Centaurus A---were 
associated with blazars.  

Blazars are sources that exhibit  violent optical
variability (e.g., changing in flux by $\sim 50$\% in one day), high optical 
polarization (exceeding several per cent), and flat radio spectra
with radio spectral index $\alpha_r < 0.5$ at GHz frequencies ($F_\nu \propto 
\nu^{-\alpha_r}$). Superluminal motion at radio frequencies and 
highly luminous and variable $\gamma$-ray emissions are also typical blazar properties.
Blazars themselves are interpreted to be relativistic jet sources powered
by a supermassive black hole, like radio galaxies, though with the observer looking
nearly along the jet axis.
The variety of multiwavelength SEDs
displayed by blazars and their misaligned populations can often, though 
not exclusively, be attributed to orientation effects amplified by the 
Doppler boosting of the jetted radiation (Section 4).


\subsection{LAT Bright AGN Sample (LBAS) and First LAT AGN Catalog (1LAC)}

Three lists of AGNs detected with Fermi have now been published by the Fermi Collaboration. 
These are the LAT Bright AGN Sample (LBAS) \cite{abd09LBAS}, and  the First and Second LAT AGN Catalogs, 1LAC  \cite{abd101LAC} and 2LAC  \cite{2011arXiv1108.1420T}, respectively. 

The LBAS is based on 3 months of science observations taking place between 2008 August 4 and 2008 October 30, and consists of 106 high Galactic latitude ($|b|>10^\circ$) sources associated with AGNs. These sources have a test statistic $TS>100$, corresponding to $\gtrsim 10\sigma$ significance, and are a subset of the 205 sources listed in the BSL \cite{abd09BSL}.

By comparison, the 3EG \cite{har99} and  the EGR \cite{cg08}  list 31 sources with significance $>10\sigma$, of which 10 are at high latitude. Remarkably, 5 of the $>10\sigma$ EGRET sources are not found in the BSL. These are the flaring blazars NRAO 190, NRAO 530, 1611+343, 1406-076 and 1622-297, the latter of which is the most luminous blazar detected with EGRET \cite{mat97}. 

The 1LAC \cite{abd101LAC} is a subset of the 1451 sources in the 1FGL \cite{abd101FGL} derived from analysis of the first 11 months of LAT science data. There are 1043 1FGL sources at high latitudes, of which 671 are associated with 709 AGNs, with the larger number of AGNs than 1LAC sources due to multiple associations. Associations are made by comparing the localization contours with counterparts in various source catalogs, for example, the flat-spectrum 8.4 GHz CRATES (Combined Radio All-Sky Targeted Eight GHz Survey; \cite{hea07}) and the Roma BZCAT blazar catalog \cite{mas09}. The probability of association is calculated by comparing the likelihood of chance associations with catalog sources if randomly distributed. 

\begin{table}
  \begin{center}
  \caption{Classes of $\gamma$-ray emitting AGNs and galaxies in the 1LAC and 2LAC ``clean" samples}
  \label{tab3}
 {\scriptsize
  \begin{tabular}{|l|c|c|c|}\hline 
{\bf Class} & {\bf Number in } & {\bf Characteristics} & {\bf Prominent Members}  \\  
& {\bf 1LAC (2LAC)} &  &   \\  \hline
All & 599 (885) &  &   \\ \hline
BL Lac objects & 275 (395) & weak emission lines &  AO 0235+164  \\
 \dots LSP  & 64 (61) & $\nu^{\rm syn}_{pk}< 10^{14}$ Hz & BL Lacertae \\ 
 \dots ISP  & 44 (81) & $10^{14}$ Hz $<\nu^{\rm syn}_{pk}< 10^{15}$ Hz& 3C 66A, W Comae \\
 \dots HSP  & 114 (160) & $\nu^{\rm syn}_{pk}> 10^{15}$ Hz &  PKS 2155-304, Mrk 501 \\ \hline
FSRQs & 248 (310) & strong emission lines & 3C 279, 3C 354.3   \\
 \dots LSP  & 171 (221) & & PKS 1510-089 \\ 
 \dots ISP  & 1 (3) & & \\
 \dots HSP  & 1 (0)&  & \\ \hline
New Classes$^1$ & 26 (24) &  &   \\
 \dots Starburst  & 3 (2) & active star formation & M82, NGC 253 \\ 
 \dots MAGN  & 7 (8) & steep radio spectrum AGNs & M87, Cen A, NGC 6251  \\
 \dots RL-NLS1s  & 4 (4) & strong FeII, narrow permitted lines & PMN J0948+0022   \\
 \dots NLRGs  & 4 (--)$^3$& narrow line radio galaxy & 4C+15.05 \\ 
 \dots other sources$^2$  & 9 (11) & &  \\ \hline
Unknown & 50 (156)&  &     \\ \hline
  \end{tabular}
  }
 \end{center}
\vspace{1mm}
 \scriptsize{
$^1$Total adds to 27, because the RL-NLS1 source PMN J0948+0022 is also classified as FSRQ in the 1LAC\\
$^2$Includes PKS 0336-177, BZU J0645+6024, B3 0920+416, CRATES J1203+6031, CRATES J1640+1144, 
CGRaBS J1647+4950, B2 1722+40, 3C 407, and 4C +04.77 in 1LAC Clean Sample\\
$^3$Class designation deprecated in 2LAC}
\end{table}

Of the 671 associations, 663 are considered ``high-confidence" associations due to more secure positional coincidences. The ``clean" sample is a subset of the high-confidence associations consisting of 599 AGNs with no multiple associations or other analysis flags, for example, evidence for extended emission.  As listed in Table \ref{tab3}, these subdivide into 275 BL Lac objects, 248 flat spectrum radio quasars, 26 other AGNs, and 50 AGNs of unknown types. The ``New Classes"  category contains non-blazar AGNs, including starburst galaxies and various types of radio galaxies, e.g., narrow line and broad line. An AGN is classified as an ``unknown" type either because it lacks an optical spectrum, or the optical spectrum has insufficient statistics to determine if it is a BL Lac objects or a flat spectrum radio quasar (FSRQ). In comparison with the 671 AGNs in the 1LAC, EGRET found 66 high-confidence ($>5\sigma$) and another 27 lower-confidence detections with significance between $4\sigma$ and $5\sigma$, as noted earlier. Thus the 1LAC already represents an order-of-magnitude increase in the number of AGNs over EGRET. There are $\approx 300$ unassociated and therefore unidentified high-latitude Fermi sources in the 1LAC.

\subsection{Classification of radio-emitting AGNs and unification}

Different classes of extragalactic AGNs are defined according to observing frequency. We have already noted the association of Fermi sources with BL Lac objects and FSRQs, which are based on an optical classification.  The precise definition used by the Fermi team is that an AGN is a BL Lac object if the equivalent width of the strongest optical emission line is $<5$ \AA, and the optical spectrum shows a Ca II H/K break ratio  $< 0.4$ in order to ensure that the radiation is predominantly nonthermal (the Ca II break arises from old stars in elliptical galaxies).  The wavelength coverage of the spectrum must satisfy $(\lambda_{max} - \lambda_{min})/\lambda_{max} > 1.7$ in order that at least one strong emission line would have been detected if present. This helps guard against biasing the classification for AGNs at different redshifts where the emission lines could be redshifted out of the relevant wavelength range. For sources exhibiting BL Lac or FSRQ characteristics at different times, the criterion adopted is that if the optical spectrum conforms to BL Lac properties at any time, then it is classified as a BL object.  

The criterion for classification of radio galaxies according to their radio properties stems from the remarkable correlation between radio morphology and radio luminosity \cite{fr74}. The twin-jet morphology of radio galaxies is seen in low-power radio galaxies, whereas the lobe and edge-brightened morphology is found in high-power radio galaxies, with a dividing line at $\approx 2\times 10^{25}$ W/Hz-sr at 178 MHz, or at a  radio luminosity of $\approx 2\times 10^{41}$ erg s$^{-1}$ for the current cosmology. Besides a radio-morphology/radio-power classification, radio spectral hardness can also be used to characterize sources as flat-spectrum and steep-spectrum sources. Furthermore, radio galaxies can be subdivided according to the widths of the optical emission lines into broad- and narrow-line radio galaxies. Correlations between radio-core dominance and $\gamma$-ray luminosity supports a scenario where the jet $\gamma$-ray flux is greatest along the jet direction  \cite{lh05}.
 
Blazars and radio galaxies can also be classified according to their broadband SED when there is sufficient multiwavelength coverage to reconstruct a spectrum from the radio through the optical and X-ray bands and identify a $\nu F_\nu$ peak frequency $\nu^{\rm syn}_{pk}$ of the lower energy, nonthermal synchrotron component of the spectrum (see Section 4). When the peak frequency $\nu^{\rm syn}_{pk}$ of the synchrotron component of the spectrum is $< 10^{14}$ Hz, then a source is called low synchrotron-peaked (LSP), whereas if the SED has $\nu^{\rm syn}_{pk}> 10^{15}$ Hz, then it is referred to as high synchrotron-peaked (HSP). Intermediate synchrotron-peaked (ISP) objects have $10^{14}$ Hz $<\nu^{\rm syn}_{pk}< 10^{15}$ Hz. SEDs of the bright Fermi LBAS sources are constructed in \cite{abd10c}. Essentially all FSRQs are LSP blazars, whereas BL Lac objects have large numbers in all LSP, ISP, and HSP subclasses.

According to the standard unification scenario for radio-loud AGNs \cite{up95}, radio galaxies are misaligned blazars, and FR1 and FR2 radio galaxies are the parent populations of BL Lac objects and FSRQs, respectively. To establish this relationship requires a census of the various classes of sources that takes into account the different beaming properties for the Doppler-boosted radiation of blazars. Even if analysis of data of radio galaxies and blazars supports the unification hypothesis, this paradigm still does not explain the reasons for the differences between radio-quiet and radio-loud AGNs, or between BL Lac objects and FSRQs.

New classes of extragalactic Fermi sources found in the 1LAC include starburst galaxies (Section 6.1), narrow line radio galaxies (NLRGs), radio-loud narrow-line Seyfert 1s (RL-NLS1s), and radio-quiet AGNs.  Five NLRGs are reported in the 1LAC. These objects have narrow emission lines in their optical spectrum, suggesting that they are observed at large angles with respect to the jet direction, with the surrounding dust torus obscuring the broad line region (BLR). 

RL-NLS1s have also been recently established as a $\gamma$-ray source class \cite{abd09c}. These objects show narrow H$\beta$ lines with FWHM line widths $\lesssim 1500$ km s$^{-1}$, weak forbidden lines ($[OIII]/H\beta < 3$) 
and an Fe II bump \cite{pog00}. By comparison with the $\sim 10^9 M_\odot$ black holes in blazars, the host galaxies of RL-NLS1s are spirals that have nuclear black holes with relatively small ($\sim 10^6$ -- $10^8 M_\odot$) mass that accrete at a high Eddington ratio. The detection of these objects challenges scenarios (e.g., \cite{san88}) where radio-loud AGNs are hosted by elliptical galaxies that form as a consequence of galaxy mergers.

The 1LAC includes 10 associations with radio-quiet
AGNs. In 8 of these cases, at least one 
blazar, radio galaxy, or CRATES source is also found close
to the $\gamma$-ray source. In the remaining two cases, the
association probabilities are weak. Thus none appear in 
the 1LAC ``clean" sample. 
In some of these candidate radio-quiet $\gamma$-ray sources, such as the Sy 2 galaxies 
NGC 4945 or NGC 1068 \cite{len10}, which are also starburst galaxies, the $\gamma$ rays could be made by 
cosmic-ray processes rather than from a radio-quiet Sy nucleus.
More extensive searches for $\gamma$ rays from Swift-BAT AGNs
has not established, however, that radio-quiet AGNS are GeV $\gamma$-ray emitters \cite{2011arXiv1109.4678T}.

\subsection{Properties of Fermi AGNs}

\begin{figure}[t]
\begin{center}
 \includegraphics[width=3.8in]{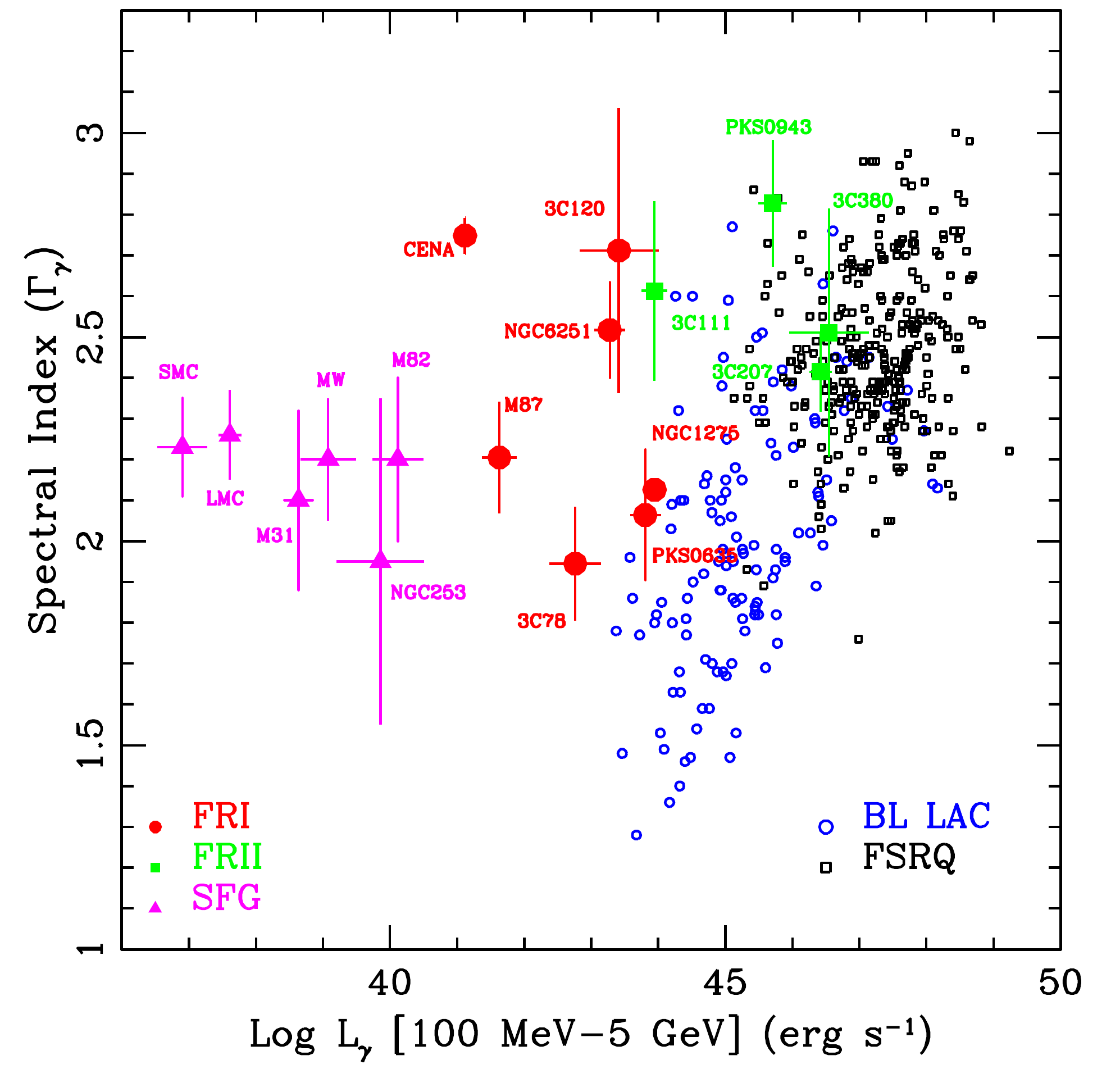} 
 \caption{Gamma-ray spectral slope $\Gamma_\gamma$ of BL Lac objects (open blue circles),  FSRQs (open black squares), FR1 radio galaxies (red circles), FR2 radio sources (green squares), and star-forming galaxies (magenta diamonds), are plotted as a function of their 100 MeV -– 5 GeV $\gamma$-ray luminosity $L_\gamma$.}
\label{GammavsL}
\end{center}
\end{figure}

Various correlations are found by comparing $\gamma$-ray properties of Fermi AGNs according to their radio, optical, or SED classification. Probably the most pronounced correlation is between the $> 100$ MeV $\gamma$-ray spectral index $ \Gamma_\gamma $  and optical AGN type. FSRQs have significantly softer spectra than BL Lac objects, with $\langle \Gamma_\gamma \rangle \cong 2.40\pm 0.17$ for FSRQs and $\langle \Gamma_\gamma \rangle \cong 1.99\pm 0.22$ for BL Lac objects in the LBAS \cite{abd09LBAS}. The SED classification shows that the mean $\gamma$-ray spectral index $\langle \Gamma_\gamma \rangle \cong 2.48, 2.28, 2.13,$ and 1.96 when the class varies from FSRQs to LSP-BL Lacs, ISP-BL Lacs, and HSP-BL Lacs, respectively. The progressive hardening from FSRQs to BL Lac objects can be seen in Figure \ref{GammavsL}, which also compares with values for radio galaxies and star-forming galaxies  \cite{abd10e}.

Fermi data reveal complex $\gamma$-ray blazar spectra. FSRQs and LSP-BL Lac objects, and most ISP blazars  with sufficiently good statistics, show breaks in the $\approx 1$ -- 10 GeV range \cite{abd10d}. This was already apparent from the first observations of the bright blazar 3C 454.3 \cite{abd09c}, to be discussed in more detail below. 
The HSP blazars, though, are generally well-described by a flat or rising $\nu F_\nu$ SED in the GeV range, with $\nu F_\nu$ peak frequencies between $\approx 100$  GeV -- TeV energies implied by VHE data. 

Only 121 out of 291 BL Lac objects had measured redshifts at the time of publication of the 1LAC.  For sources with measured redshift, BL Lac objects are mostly found at low redshift, $z\lesssim 0.4$, with only a few HSP BL Lac objects at higher redshifts. By contrast, the FSRQs span a wide range from $z\approx 0.2$ to the highest redshift 1LAC blazar with $z = 3.10$.

This significant redshift incompleteness hampers interpretation of AGN properties, in particular, Fig.\ \ref{GammavsL}, which can only display sources with known redshifts.  For these sources, the hard-spectrum BL Lac objects typically have much lower $L_\gamma$ than the FSRQs. This divide has been interpreted as a change in the accretion regime at $\approx 1$\% of the Eddington luminosity \cite{ghi09}. In addition, the nearby radio galaxies with $z\lesssim 0.1$ inhabit a separate portion of the $\Gamma_\gamma$ vs.\ $L_\gamma$ plane, and are characterized by lower $\gamma$-ray luminosities than their parent populations. The two more distant steep spectrum radio sources, 3C 207 ($z = 0.681$) and 3C 380 ($z = 0.692$), and the FR2 radio galaxy PKS 0943-76 ($z = 0.27$) fall, however, within the range of $\gamma$-ray luminosities measured from FSRQs. Indeed, steep spectrum radio sources are thought to be slightly misaligned FSRQs.

\subsubsection{3C 454.3 and FSRQs }

The FSRQ 3C 454.3, at $z = 0.859$, underwent giant flares and became the brightest $\gamma$-ray source in the sky for a week in 2009 December and 2010 April  \cite{ack10}, and again in 2010 November \cite{abd11}. The  outburst in 2010 April triggered a pointed-mode observation by Fermi. During the December outburst, its daily flux reached $F_{-8}= 2200 (\pm 100)$, corresponding to an apparent isotropic luminosity of $L_{iso}\approx 3\times 10^{49}$ erg s$^{-1}$, making it the most luminous blazar yet detected with Fermi. In its 2010 November outburst, it reached $F_{-8} \cong 6000$ and $L_{iso} \approx 10^{50}$ erg s$^{-1}$, becoming $\approx 5$ times brighter than the Vela pulsar.  

\begin{figure}[t]
\begin{center}
\includegraphics[width=4.5in]{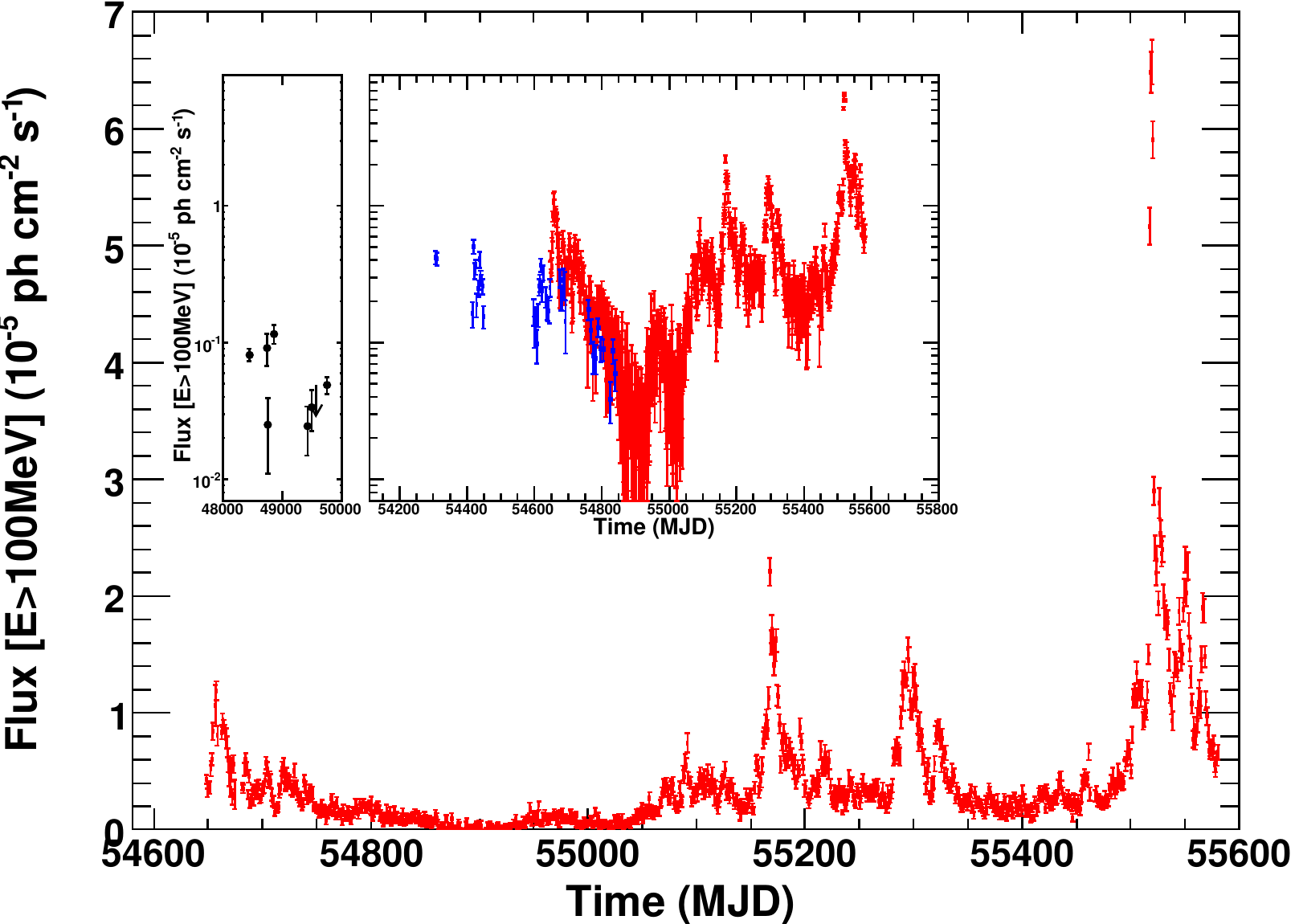} 
 \caption{Fermi LAT daily light curve of 3C 454.3  \cite{abd11}, showing the giant flares in December 2009, April 2010, 
and November 2010  (MJD 55200 corresponds to 4 January 2010). Inset shows the light curve on a logarithmic scale, with black points from EGRET \cite{har99}, and 
blue points from AGILE \cite{str10}.  }
\label{3c454ltcv}
\end{center}
\end{figure}

Figure \ref{3c454ltcv} shows the light curve of 3C 454.3 \cite{abd11}, which can also be found
at the public website of Fermi monitored sources.\footnote{fermi.gsfc.nasa.gov/ssc/data/access/lat/msl$\_$lc/} 
The fluxes are plotted in durations of one day over the course of the Fermi mission.  The flux rises to a plateau level preceding $\gamma$-ray outbursts, with the 2008 July outburst showing strong resemblance to those in 2009 August and December, and 2010 April and November. Intense flaring occurs during periods of enhanced activity, which is to say that the flares are not isolated events, but seem to occur during periods of enhanced accretion activity.

\begin{figure}[t]
\begin{center}
\includegraphics[height=3.0in]{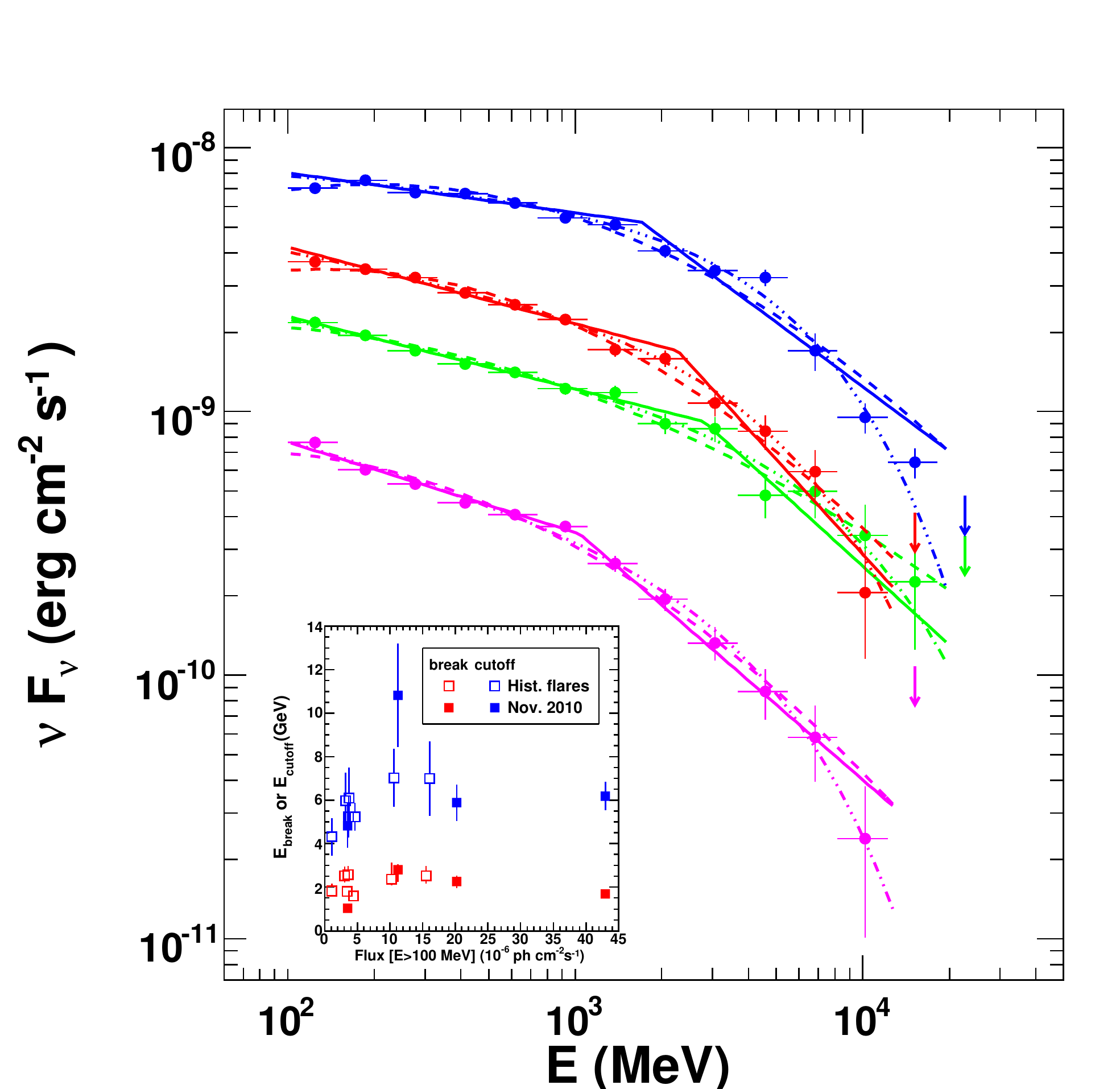} 
 \caption{Fermi LAT $\nu F_\nu$ SEDs for four different time periods
corresponding to preflare (magenta; lower spectrum), plateau (green; third from top), flare (blue; top spectrum), 
and post-flare (red; second from top) periods, along with the fitted functional forms, including
a broken power law (solid), log-parabola (dashed), and power-law with exponential
cutoff (dashed-triple-dotted) functions \cite{abd11}.
The inset displays $E_{\rm break}$ (red) and $E_{\rm cutoff}$ energies (blue) as a function of flux for the
different periods (filled symbols) and for historical flares (open
symbols).}
\label{3c454spct}
\end{center}
\end{figure}

As noted already in the initial Fermi report \cite{abd09c}, the spectrum of 3C 454.3 breaks strongly by $1.2(\pm 0.3)$ units at $E_{br}\approx 2$ GeV. Such a break is inconsistent with simple radiative cooling scenarios, which predict a break by 0.5 units. The more recent analyses of data from the major outbursts of 3C 454.3 \cite{ack10,abd11} confirm the strong spectral break and finds that  $E_{br}$ is very weakly dependent on the flux state, even when the flux changes by more than an order of magnitude (see Fig.\ \ref{3c454spct}). No consistent pattern expected in acceleration and cooling scenarios \cite{krm98} is found in the spectral index/flux plane. 

The origin of the spectral break in 3C 454.3 bears on several important issues in FSRQs: the location of the $\gamma$-ray emission site; the source of soft target photons in Compton-scattering models; and the relation of FSRQs and BL Lac objects in view of the disappearance of such breaks in ISP and HSP blazars. Such a break would be readily understood if the target field was sufficiently strong to attenuate the blazar radiation by $\gamma\gamma$ absorption processes, but the intense Ly $\alpha$ radiation field at 10.2 eV observed with GALEX  \cite{bon10}  would make a break at $E_{br}\gtrsim 30$ GeV \cite{rei07}. Models employing photon attenuation deep within the BLR by He II recombination and Ly $\alpha$ photons with $E> 54.4$ eV  \cite{ps10} have been proposed to explain this break, as have nonthermal electron scattering models with accretion-disk and BLR photon targets \cite{fd10}. 
 The spectral break could also be due to Klein-Nishina effects in scattering, 
as has been proposed to explain the SED of PKS 1510-089 \cite{abd10h}. The KN break due to upscattered Ly $\alpha$ radiation occurs at a few GeV, and the observed break energy is insensitive to the Doppler factor. This scattering problem is treated in Section 4. 

\subsubsection{PKS 2155-304, Mrk 501, and BL Lac objects }

PKS 2155-304, an X-ray selected BL Lac object at $z = 0.116$ and an EGRET source \cite{1995ApJ...454L..93V}, is one of the most 
prominent representatives of the HSP blazar population.  During a period of extraordinary flaring on 2006 July 28, PKS 2155-304 exhibited a succession of $\gamma$-ray flares varying on time scales as short as $\approx 5$ min with apparent luminosities $> 10^{46}$ erg/s---larger still when EBL absorption is included \cite{aha07}. 
This is an extreme, hypervariable source in the sense of equation (\ref{LEDDtS}).

An 11 day campaign between 2008 August 25 and 2008 September 6 was organized early in the Fermi mission  to measure the  SED at
optical (ATOM; Automatic Telescope for Optical Monitoring), X-ray (RXTE and Swift), and $\gamma$-ray (Fermi and HESS) frequencies. 
It turned out to be in a low state, well fit by a standard one-zone synchrotron/SSC model, 
with  Doppler factor $\delta_{\rm D} = 32$,  magnetic field $B^\prime = 0.018$ G, and comoving 
size $R^\prime = 1.5\times 10^{17}$ cm (corresponding to a variability time of 2 d; \cite{aha09}).\footnote{The reader is assumed to be familiar with the synchrotron/SSC model; see, e.g., \cite{1998ApJ...509..608T,1996A&AS..120C.503G,fin08,dm09}.}
When flaring, however, the one-zone synchrotron SSC model for PKS 2155-304 completely fails, or at least requires $\delta_{\rm D} \gtrsim 100$ \cite{fin08}.

\begin{figure}[t]
\begin{center}
 \includegraphics[width=3.4in]{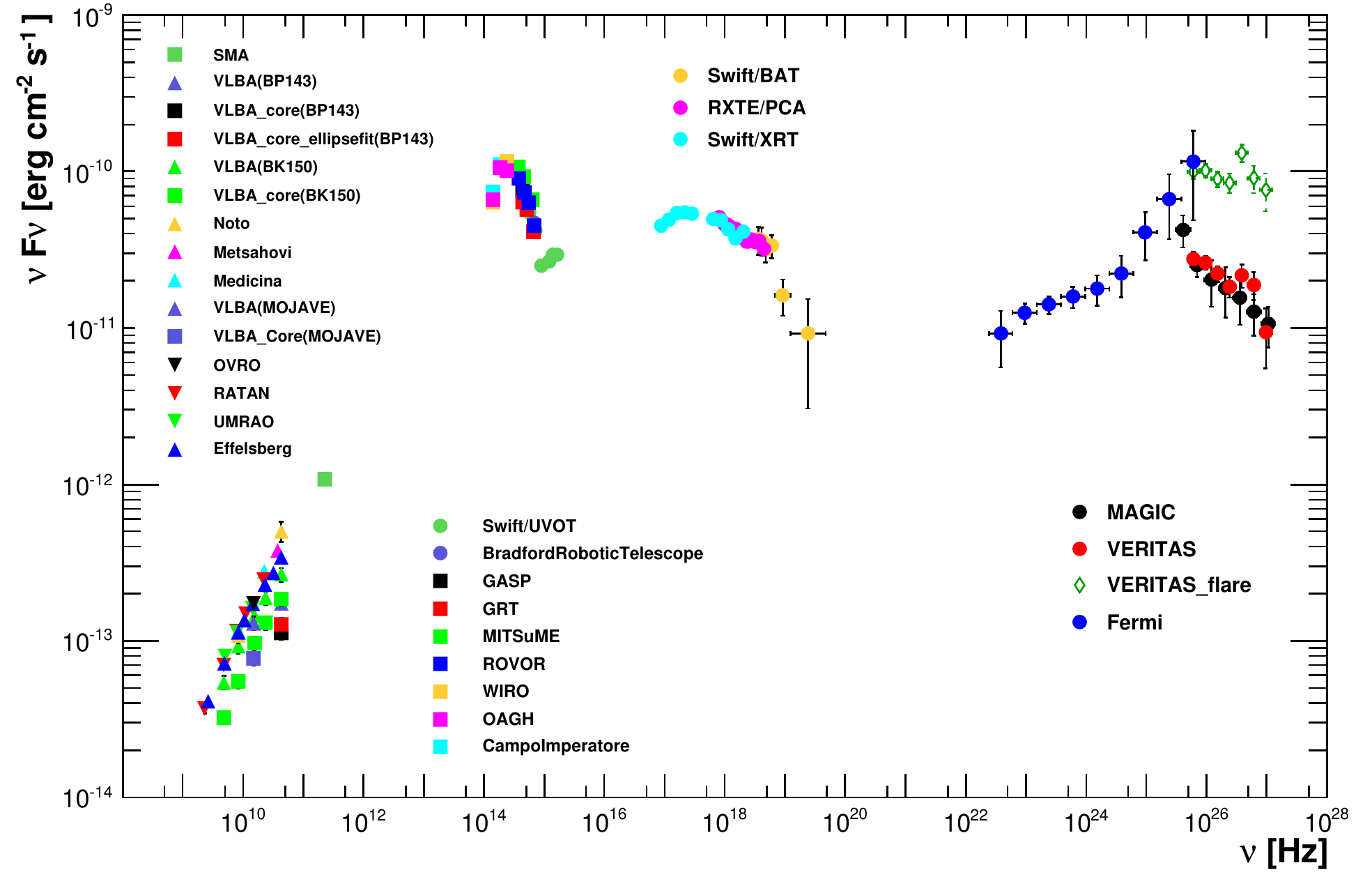} 
 \caption{SED for Mrk 501 averaged over all observations taken during the multifrequency campaign performed between 2009 March 15 (MJD 54905) and 2009
August 1 (MJD 55044) \cite{abd11mrk501}. The legend reports the correspondence between the instruments and the measured fluxes.  }
\label{mrk501fig}
\end{center}
\end{figure}

One-zone synchrotron/SSC models with $\delta_{\rm D} \gtrsim 10$ give good fits to the long-term average spectra of other HSP BL Lac objects such as Mrk 421 and Mrk 501. The spectacular multiwavelength SED of Mrk 421 shown in Figure \ref{mrk501fig} represents one of the most detailed multiwavelength blazar SEDs yet assembled  \cite{abd11mrk501}. 
The optical and X-ray data have been corrected for Galactic extinction, but the host galaxy, which is clearly visible at IR/optical frequencies, has not
been subtracted. The MAGIC and VERITAS data have been corrected for EBL absorption (Section 7.3). A single-zone synchrotron/SSC model with $12\lesssim \delta_{\rm D} \lesssim 22$, $B^\prime \approx$ 15 -- 30 mG, and $R^\prime \approx$ 8 -- 50 lt-day, gives a good fit to the nonflaring SED shown in Figure \ref{mrk501fig}. By comparison, radio galaxies are fit by synchrotron/SSC models with Doppler factors of order unity. We return to this point in Section 7.


\subsection{Second LAT AGN Catalog (2LAC)}

The 2LAC \cite{2011arXiv1108.1420T} is based on the first two years of scientific data taken with the Fermi Gamma ray Space Telescope. It contains 1016 $|b|>10^\circ$ sources that are associated at high confidence with AGNs (1319 of the 1873 1FGL sources are at $|b|>10^\circ$).  The 2LAC clean sample comprises 885 sources, consisting of 395 BL~Lac objects, 310 FSRQs, 156 blazars of unknown type, 8 misaligned AGNs, 4 RL-NLSy1 galaxies, 10 AGNs of other types, and 2 starburst galaxies (NGC 4945 has fallen out of the list); see Table \ref{tab3}.  The photon index distribution of the blazars of unknown type suggests that they comprise roughly equal numbers of BL Lacs and FSRQs. Of the 395 BL Lac objects, 220 ($\sim ˜55$\%) lack measured redshifts, and this fraction is roughly the same for LSP, ISP, and HSP BL Lac objects.

As shown in the 2LAC, threshold sensitivity in terms of photon flux is strongly dependent on source spectral index, whereas energy flux is not, and detection of FSRQs and BL Lac objects is complete to an energy flux of $\approx 5\times 10^{-12}$ erg/cm$^2$-s. Because more and deeper surveys have taken place in the northern hemisphere, the smaller fraction of southern-hemisphere BL Lac objects indicates that some 60 unassociated sources at negative declination are expected to be BL Lac objects. 

Besides validating the results of the 1LAC with additional data and improved analysis techniques, the 2LAC shows that the mean fractional variability of FSRQs is greater than that of BL Lac objects. 
The duty cycle, as defined by the fraction of time when the average flux exceeds 1.5 standard deviations above the mean flux, is $\approx 5$ -- 10\% for bright blazars. The number of BL Lacs and FSRQ have increased by 44\% and 35\% from the 1LAC to 2LAC clean samples, respectively, with the future increases in the number of FSRQs expected to be even more modest. This is due to cosmological effects, as reflected in the flattening in the $\log N$-$\log S$ distribution, and K-corrections that pushes the GeV peak of the $\nu F_\nu$ spectra to lower energies where the Fermi-LAT effective area rapidly declines and sensitivity to FSRQs gets worse.

Blazars make up $\gtrsim 97$\% of the extragalactic $\gamma$-ray sources detected with the Fermi-LAT. The number of misaligned sources has stalled, with the same number of radio galaxies---eleven---in the 2LAC as in the misaligned AGN paper. But these are not the same objects. The radio galaxies 3C 78, 3C 111, and 3C 120 are not found in the 2LAC, evidently because a variable jetted emission component that contributed to the $\gamma$-ray emission in the past has gone quiet.  Now found in the 2LAC are the FRI radio galaxies Fornax A and Cen B, and the head-tail radio galaxies IC 310 \cite{2010A&A...519L...6N}, the latter of which is also a MAGIC source \cite{2010ApJ...723L.207A}. 
The radio/$\gamma$-ray connection \cite{2011ApJ...741...30A} and GeV-TeV synergy \cite{2009ApJ...707.1310A} resulting from the {Fermi}-LAT for AGN studies, let alone Galactic sources, cannot be adequately reviewed here; see  \cite{2011arXiv1108.1420T} and \cite{abd122FGL}.

\section{ Relativistic Jet Physics }

Here we develop and apply relativistic jet radiation physics \cite{dm09} to some puzzles arising from the 
Fermi observations.

\subsection{GeV spectral break in LSP blazars}

As already described in Section 3.3, an important new result in blazar physics not anticipated 
before the launch of Fermi is the prevalence of a spectral 
softening in the $\gamma$-ray spectra of low synchrotron-peaked (LSP) blazars, including both FSRQs and LSP BL Lac objects.
The spectral softening occurs generally between 1 and 10 GeV
\cite{abd09LBAS,abd101LAC}. 
For 3C 454.3 \cite{abd09d,ack10}, the break is $\gtrsim 1$ unit, 
and the energy of the break
is rather insensitive to flux (Fig.\ \ref{3c454spct}). 
A softer spectral break, possibly consistent with radiative 
cooling, is found at
multi-GeV energies in the Fermi/MAGIC spectrum of  FSRQ 4C +21.35 \cite{2011ApJ...733...19T}.
If a GeV spectral softening and  significant GeV flux variability
accompanied by modest spectral variability
is typical of blazars, then some physical process should be 
able to explain this behaviour.
 
\subsubsection{$\gamma$ rays from external Compton scattering}

We treat the problem that blazar $\gamma$ radiation arises from the scattering of a 
quasi-isotropic target photon field \cite{1994ApJ...421..153S}
in the Klein-Nishina regime. Compton scattering takes place in the Thomson limit when
$4\gamma\epsilon^\prime\ll 1$ (eq.\ 6.123 in \cite{dm09}), where  $\gamma$ and $\epsilon^\prime$ are the electron Lorentz factor
and photon energy in the comoving frame. 

Consider an external isotropic, monochromatic photon field with photon energy $\epsilon_\star$ (in $m_ec^2$
units). The average energy in the comoving frame is $\langle \epsilon^\prime\rangle \cong  \Gamma \epsilon_\star$. 
The upscattered photon energy in the Thomson regime is
$\epsilon^\prime_C \cong (4/3)\gamma^2\epsilon^\prime$. Hence 
$\epsilon^\prime\epsilon^\prime_C= (4/3)\gamma^2\epsilon^{\prime 2} = (4\gamma\ep)^2/12$. Thus scattering is in the Thomson regime provided
$12 \epsilon^\prime\epsilon^\prime_C \lesssim1$, or
$\epsilon_\star\lesssim (\delta_{\rm D}/\Gamma)/[{12 \epsilon_C(1+z)]}$, implying 
$ E_C({\rm GeV})\lesssim 12/[{E_\star({\rm eV})]}$ for $\delta_{\rm D} \cong \Gamma$. 
Away from the endpoints, therefore, the scattered photon energy marking the 
transition to the KN 
regime  is very weakly dependent on Doppler
factor in highly beamed relativistic jets. 

The  spectrum of an isotropic
nonthermal power-law electron distribution, when transformed to the stationary frame,
 remains a power-law with the same index as in 
the comoving frame, but with a strong angular dependence. The endpoints 
of the distribution are the 
low-and high-energy electron Lorentz cutoffs, $\gamma_1^\prime$ and $\gamma^\prime_2$ respectively,
 boosted by $\delta_{\rm D}$ \cite{gkm01}.
Consequently, the shape of the scattered spectrum cannot depend on $\dD$ provided
that $\gamma_1 = \dD \gamma^\prime_1 \ll \gamma_{\rm KN}\cong (4\epsilon_*)^{-1}\ll\dD \gamma^\prime_2 = \gamma_2$.

If the break energy observed in 3C
454.3 at $\approx$ 2 GeV is due to the transition to scattering in
the KN regime, then the underlying target
photon energy $E_*\approx$ 6 eV. This is
close to the energy of the Ly $\alpha$ photon at 10.2 eV, or
H$\beta$ at 2.55 eV. If the origin of the spectral break is due
to scattering of nearly mono-energetic line radiation, then 
this would (1) place the scattering site within the BLR, and
(2) explain the near constancy of the spectral
break, independent of flux state.


We use the method of Georganopoulos et al.\ (2001) \cite{gkm01}, 
applied to isotropic radiation fields
in the stationary frame of the supermasssive black hole and BLR.
The differential Compton-scattered spectrum when isotropic, monoenergetic electrons
Compton upscatter  isotropic, monochromatic target photons, is
given for the full Compton cross section in the head-on approximation by
\begin{equation}
df_{\e}^{\rm C}\;=\; {3\over 4}{c\sT u_0\over d_L^2}
\left({\e_s\over \e_* }\right )^2\;{N_e(\g,\Omega_s )d\g\over \g^2} F_{\rm C}(q,\Gamma_e )
H\left[{\gamma\over 1+(1/\Gamma_e)}-\e_s \right]\;
\label{argequation16}
\end{equation}
(eq.\ 6.129; \cite{dm09}), with 
\begin{equation}
 N_e(\gamma, \Omega_s) = \dD^3\;
{N^\prime(\gp)\over 4\pi }\;,\;\gp = \g/\dD \;
\label{Nprime(gp)}
\end{equation}
(eq.\ 6.124),
\begin{equation}
\e_s = (1+z)\e \equiv \e_z\;\;,\;\;\Omega_s = \Omega_*\;,
\label{esezdef}
\end{equation}  
\begin{equation}
 F_{\rm C}(q,\Gamma_e)= 2q \ln q +(1+2q)(1-q) +{1\over 2}
{(\Gamma_e q)^2\over (1+\Gamma_e q)}(1-q)\;,
\label{FC}
\end{equation}
\begin{equation}
q \equiv {\e_s/\gamma \over \Gamma_e
(1-\e_s/\gamma )}\;, \; \Gamma_e \equiv 4\gamma\e_*\;,
\label{jesCq}
\end{equation}
and $q$ is restricted to the range $(4\gamma^2)^{-1} \leq q \leq 1$
\cite{jon68,bg70} (eqs.\ 6.74-6.76, 6.125--6.127; cite{dm09}). 
Restriction to 
the Thomson regime occurs when $\Gamma_e \ll 1$ or $4\g\e_* \ll 1$. 
The final term in eq.\ (\ref{FC}) dominates for
scattering in the KN regime.
(A simpler form for analytic work is the isotropic Thomson cross section,
\begin{equation}
F_{{\rm T},iso}(\hat \e) = {2\over 3}\left ( 1 - \hat \e\right )\;
\label{FTredux1}
\end{equation}
(eq.\ 6.71 \cite{dm09}), with $\hat \epsilon = \epsilon_s/4\gamma^2\epsilon_\star$ (eq. 6.70), 
which assumes isotropic scattering in the electron rest frame.)

\begin{figure}
\begin{center}
 \includegraphics[width=4.64in]{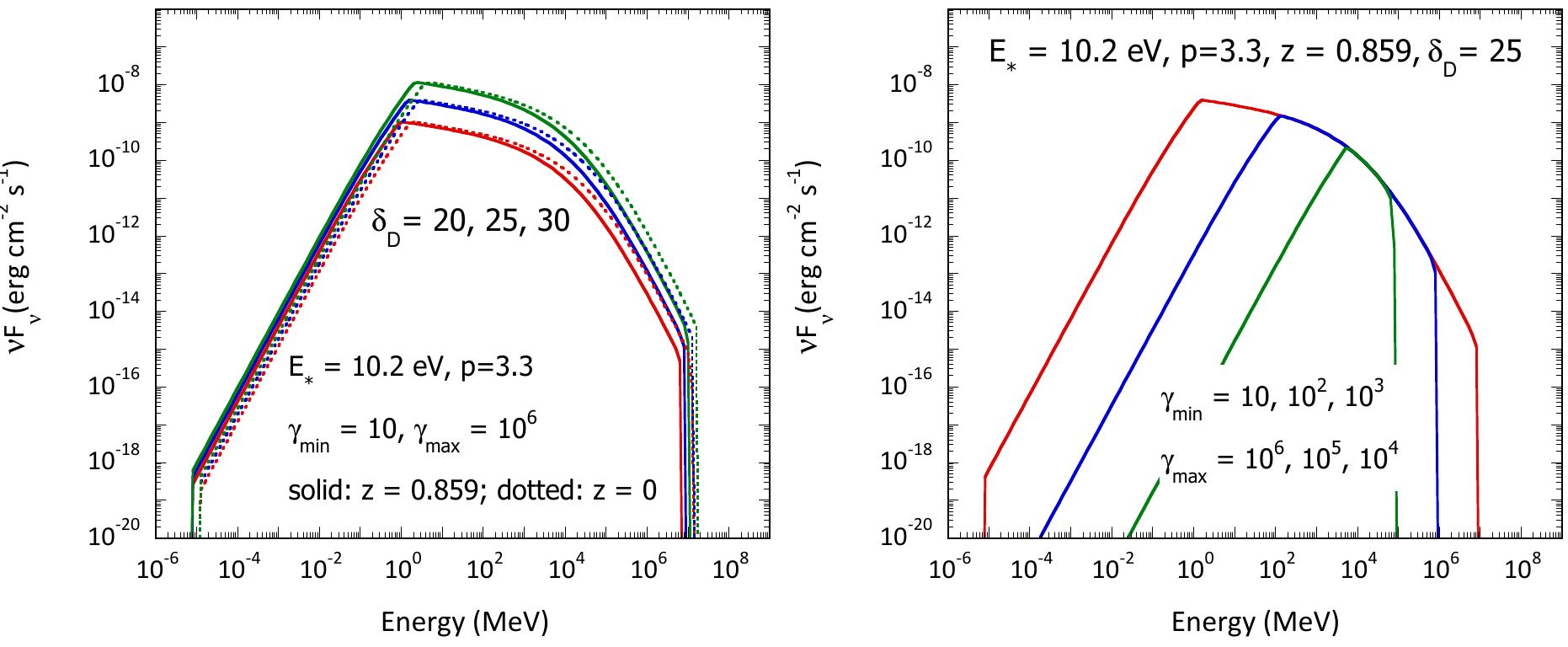} 
 \caption{Spectrum of Compton-scattered line radiation. {\it Left:} Dependence on Doppler factor $\delta_{\rm D}$ for 
sources at different redshifts. {\it Right:} Dependence on endpoints. Within the endpoints,
the spectral shape of the function is independent of $\delta_{\rm D}$.}
\label{compLya}
\end{center}
\end{figure}

Thus an isotropically distributed nonthermal electron distribution in the comoving frame
gives a Compton-scattered spectrum
\begin{equation}
f^{{\rm C},iso}_{\e} = {3\over 4}\;{c\sigma_{\rm T} \e_s^2\over  4\pi d_L^2}\;\delta_{\rm D}^3
\;
\int_0^\infty d\e_*\;{u_*(\e_* )\over\e_*^2}
\int_{\gamma_{min}}^{\infty} d\gamma\;
{N^\prime_e(\g /\delta_{\rm D} )\over \gamma^2 } \; F_{\rm C}(q,\Gamma_e)\;,
\label{jesCpowerlaw19}
\end{equation}
(eq.\ 6.123; considering only upscattering), with
\begin{equation}
\g_{min} = {\e_s\over 2}\left( 1 + \sqrt{1 + {1\over \e_s\e_*}}\right)\;
\label{gmin}
\end{equation}
(eq.\ 6.90). 
Substituting eq.\ (\ref{Nprime(gp)}) for an isotropic power-law electron spectrum in the 
comoving frame, so that $N^\prime(\gp) = N_{eo}\gamma^{\prime~-p}H(\g;\g_1,\g_2)$,\footnote{The Heaviside functions
$H(x-a)$ and $H(x;a,b)$ are defined such that $H(x-a) = 1$ if $x>a$ and $H(x-a) = 0$ otherwise, 
and $H(x;a,b) = 1$ if $a \leq x \leq b$ and $H(x;a,b) = 0$ otherwise.}
scattering
an external monochromatic line spectrum $u_*(\e_*) = u_*\delta(\e_*-\e_{*o})$
gives
\begin{equation}
f^{{\rm C},iso}_{\e} = {3\over 4}\;{c\sigma_{\rm T}u_* \over  4\pi d_L^2}\;\big( {\e_s
\over \e_* }\big)^2\;\dD^{3+p}\;N_{eo}
\;
\int_{\min[\gamma_{min},\g_1]}^{\g_2} d\gamma\;
\g^{-(p+2)} \; F_{\rm C}(q,\Gamma_e)\;.
\label{jesCpowerlaw20}
\end{equation}
This is numerically solved to give the results shown in Fig.\ \ref{compLya}, which 
were compared in \cite{ack10} with the rapidly falling spectrum of 3C 454.3.

\begin{figure}
\begin{center}
\includegraphics[width=4.0in]{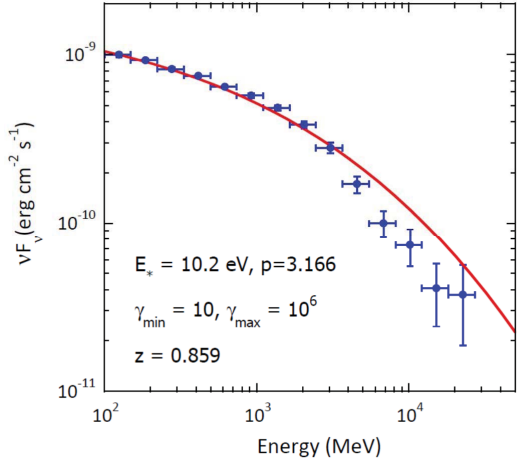} 
 \vspace*{-1.8 in}
 \caption{Model for the $\gamma$-ray spectrum of 3C 454.3 when 
isotropic power-law electrons with index $p = 3.17$ in 
a relativistic jet Compton scatter Ly $\alpha$ photons \cite{ack10}. }
\label{3c454lya}
\end{center}
\end{figure}

Fig.\ \ref{3c454lya} shows a model where
the $\gamma$-ray continuum observed from 3C 454.3 with 
the Fermi LAT \cite{ack10} are formed by power-law nonthermal jet electrons,
with number index $p = 3.17$, that Compton-scatter Ly $\alpha$ line photons.
The model is insensitive to values of 
lower and upper comoving electron Lorentz factors $\gamma_{min}$ and $\gamma_{max}$ 
provided $\gamma_{min} \lesssim 10^2$ and $\gamma_{max}\gtrsim 10^4$. The Klein-Nishina
softening from a power-law electron distribution gives a poor fit to the data,
but if one assumes that the underlying electron distribution is curved, 
as in the case of a log parabola function \cite{2008A&A...478..395M}, for example, it may be possible to obtain 
a good fit to the 3C 454.3 data for this model.

For integrations over blackbody spectra, 
substitute \begin{equation}
u_{*,bb}(\e_*) = {2m_ec^2\over \lambda_{\rm C}^3}{\e_*^3\over\exp(\e_*/\Theta) - 1}\;
\label{ustarestarredux}
\end{equation}
into eq.\ (\ref{jesCpowerlaw19}). The dimensionless temperature of the radiation 
field is $\Theta = k_{\rm B}T/m_ec^2$.
The CMBR can be approximated as an isotropic, monochromatic radiation field 
$u_*(\e_*) = u_0\delta(\e_*-\e_0 )$, with
$u_0 = 4\times 10^{-13}(1+z)^4$ ergs/cm$^{3}$ and
$\e_0 = 2.70\Theta_{CMBR} \cong 1.24\times 10^{-9}(1+z)$. The condition $q < 1$ implies
$\e_s < \g/(1+\Gamma_e^{-1})$.

\subsubsection{Compton emissivity and electron energy-loss rate}

In the calculation of time-dependent blazar spectra with evolving
electron distributions, or in pair cascade calculations, it is necessary
to consider the full Compton cross section in scattering and 
electron-energy loss calculations. When $\gamma\gg 1$,  
the head-on approximation, where the incident photon is 
assumed to be directed opposite (``head-on" to) the direction of the electron 
when transformed to the electron rest frame,  can be employed. 

For general treatments, the time-dependent emissivity 
$\e_s j_{\rm C}(\e_s,\Omega_s;\vec r,t)$ is calculated.
For external isotropic photon spectra, the angle-dependent emissivity when 
jetted 
nonthermal, isostropic electrons with electron spectrum $n_e(\gamma,\Omega_e)$ 
Compton-scatter photons of an isotropic monochromatic photon field is 
given, in the head-on approximation, by the expression
\begin{equation}
\e_s j_{\rm C}(\e_s,\Omega_s) = {3\over 4} c\sigma_{\rm T} u_0 \big({\e_s\over \e_\star}\big)^2
\int_1^\infty d\gamma\;{n_e(\g,\Omega_e)\over \gamma^2} F_{\rm C}(q,\Gamma_e)H(q;{1\over 4\gamma^2},1)\;
\label{esjCesOs}
\end{equation}
(eq.\ 6.74).
Here the electron spectrum is written in the stationary frame where the radiation field 
is isotropic. 
The $\nu L_\nu$ Compton luminosity for a single electron is given by
\begin{equation}
\e_s L_{\rm C}(\e_s) = {3\over 4} {c\sigma_{\rm T} u_0\over \g^2} \big({\e_s\over \e_\star}\big)^2
F_{\rm C}(q,\Gamma_e)\;H(\e_s;{\e_\star\over 1+\e_\star/\g},{\gamma\over 1+\Gamma_e^{-1}})\;.
\label{esLCesOs}
\end{equation}

The electron energy loss rate by Compton scattering is given by
$$-\dot\g_{\rm C} = \int_0^\infty d\e_s\;L_{\rm C}(\e_s) = {3\over 4} {c\sigma_{\rm T} u_0\over m_ec^2\gamma^2 \e_\star^2}\;
\int_{\e_\star}^{\gamma/(1+\Gamma_e^{-1})}d\e_s \;\e_s\; F_{\rm C}(q,\Gamma_e )$$
\begin{equation}
= {12c\sigma_{\rm T} u_0\gamma^2\over m_ec^2}\;
\int_{0}^1 dq \;{q\over (1+\Gamma_eq)^3} \; F_{\rm C}(q,\Gamma_e )\;.
\label{-dotgC}
\end{equation}
 Using the isotropic Thomson kernel, equation (\ref{FTredux1}) becomes in the Thomson regime, $\Gamma_e \ll 1$,
\begin{equation}
\e_s L_{\rm C}(\e_s) \cong  {c\sigma_{\rm T} u_0\over 2\g^2} \big({\e_s\over \e_\star}\big)^2
\big[ 1 + {3\over 4}{x^2\over (1-x)} \big]\; \big[ 1 - {x\over \Gamma_e(1-x)}\big]\;H(\e_s;\e_\star,{\gamma\over 1+\Gamma_e^{-1}})\;.
\label{esLCesOs1}
\end{equation}
Here  $x = \e_s/\gamma$ so that $q = x/[\Gamma_e(1-x)]$. 
To lowest order, 
\begin{equation}
-\dot\g_{\rm T} \cong  { c\sigma_{\rm T} u_0\over 2\e^2}\;
\int_{0}^{\Gamma_e}dx\;x\;\big( 1  - {x\over \Gamma_e}\big)\;= 
{ c\sigma_{\rm T} u_0\Gamma_e^2\over 12\e^2}= {4\over 3} c\sigma_{\rm T} u_0\g^2\;,\;
\label{-dotgT}
\end{equation}
recovering the familiar result in the limit $\g\gg 1$, $\e_\star/\gamma \ll 1$.
Expanding the energy-loss rate expression, again using the isotropic Thomson kernel, 
gives  the expansion
\begin{equation}
-{\dot\gamma}_{{\rm C},i = 1} \rightarrow {4\over 3}c\sigma_{\rm T} u_0 \gamma^2\;(1-{3\over 2}\Gamma_e+
{163\over 40}\Gamma_e^2+{\cal O}[\Gamma_e^3])\;
\label{dotg(i=1)}
\end{equation}
correct to ${\cal O}(\Gamma_e^2)$.


\subsection{Leptonic jet models}

A photon flux of $F_{-8}  = 2200$ from 
3C 454.3 at $z = 0.859$ implies an apparent $> 100$ MeV isotropic $\gamma$-ray luminosity of 
L$_{\gamma}= 3.3\times 10^{49}$ erg s$^{-1}$  \cite{ack10}. The flux reached a factor $\approx 6$ larger in the 
November 2010 outburst  \cite{abd11}, corresponding to an apparent luminosity of $\approx 2\times 10^{50}$ erg/s, 
which is the record-holder for all blazars, including PKS 1622-297 from the EGRET era \cite{mat97}. 

Using the measured flux and a one-day variability timescale at the time that the most energetic photon with $E\approx 20$ GeV was detected implies a minimum Doppler factor of $\delta_{\rm D,min}\approx 13$, as we show. Assuming that the outflow Lorentz factor $\Gamma \approx 20$, consistent with the inferred value of $\delta_{\rm D,min}$ and with radio observations at a different epoch \cite{jor05}, then simple arguments suggests a location $R \lesssim c \Gamma^2 t_{\rm var}/(1+z)\approx 0.2 (\Gamma/20)^2 (t_{\rm var}/{\rm day})$ pc, which is at the outer boundary of the BLR. Flux variability on time scales as short as 3 hr was measured at another bright flux state \cite{abd11}, which suggests that the $\gamma$-ray emission site would be even deeper in the BLR. This stands in contrast to inferences based on coherent optical polarization changes over timescales of weeks prior to a $\gamma$-ray flare in 3C 279  \cite{abd10g}, which places the emission site at much larger distances. The MAGIC detection of VHE emission from PKS 1222+21 (4C +21.35) \cite{2011ApJ...730L...8A} is even stronger evidence for $\gamma$-ray production at the pc scale or farther from the central engine.

\subsubsection{Jet Doppler factor}

Emission from bulk magnetized plasma in relativistic motion
provides the best explanation for the large apparent isotropic 
luminosities   $L_{iso}$, energy releases ${\cal E}_{iso}$,
and short variability times $t_{\rm var}$
from cosmological $\gamma$-ray sources. 
The comoving 
emission-region size scale $R^\prime$ is related to $t_{\rm var}$\
through the expression $R^\prime \lesssim c\Gamma t_{\rm var}/ (1+z)$.
Combined {Fermi} LAT and GBM observations give the best GRB data 
from which to determine the minimum bulk outflow Lorentz factor $\Gamma_{min}$
through $\gamma\gamma$ opacity arguments. For blazars, multiwavelength
X-ray measurements combined with {Fermi}-LAT data are important for
inferring the minimum Doppler factor.

It is simple to derive $\Gamma_{min}$ in a blast-wave formulation,
noting that the internal photon energy density 
\begin{equation}
u^\prime_\gamma \approx {4\pi d_L^2 \Phi\over \Gamma^2 4\pi R^2 c} \approx
{(1+z)^2d_L^2\Phi\over \Gamma^6 c^3 t_{\rm var}^2}\;,
\label{eq1}
\end{equation}
using $R\approx \Gamma^2 ct_{\rm var}/(1+z)$. The optical depth for 
$\gamma\gamma\rightarrow e^+e^-$ processes is $\tau_{\gamma\gamma}(\epsilon_1^\prime)
\cong R^\prime \sigma_{\rm T} (\epsilon_1^\prime/2)u^\prime_\gamma(2/\epsilon_1^\prime)/(m_ec^2)$, where 
$R^\prime = R/\Gamma$ and $\epsilon^\prime = 2/\epsilon_1^\prime$ from the threshold 
condition. The condition $\tau_{\gamma\gamma}(\epsilon_1^\prime)< 1$ 
with the relation $\Gamma\epsilon^\prime_1/(1+z) = \epsilon_1$ 
implies 
\begin{equation}
\Gamma \gtrsim\Gamma_{min} \cong \left[ {\sigma_{\rm T} d_L^2 (1+z)^2 
f_{\hat \e}\e_1\over 4t_{\rm var} m_ec^4}\right]^{1/6}\;,\;
\hat \e = {2\Gamma^2\over (1+z)^2\e_1}\;
\label{eq2}
\end{equation}
\cite{ack10_grb090510}.
Here $f_\e $ is the $\nu F_\nu$ flux at photon energy $m_ec^2
\epsilon$, which is evaluated at $\e = \hat \epsilon$ due to 
the peaking of the $\gamma\gamma$ cross section near
threshold. 

The minimum Doppler factor $\delta_{min}$ ($\delta_D$ = [$\Gamma(1-\beta \cos \theta)$]$^{-1}$) 
defined by the condition $\tau_{\gamma\gamma}(\epsilon_1)= 1$,
can be estimated to $\approx 10$\% accuracy compared to results of 
more detailed numerical calculations through the expression
\begin{equation}
\delta_{min}\cong \left[\frac{\sigma_T d_L^2(1 + z)^2 f_{\hat{\epsilon}} \epsilon_1}{4t_vm_ec^4}\right]^{1/6}\;,
\label{deltaD}
\end{equation}
where $f_\epsilon$ is the $\nu F_\nu$ spectrum measured at frequency $\nu = m_ec^2 \epsilon/h$, $t_v$ is the variability time, and $E_1= m_e c^2\epsilon_1$ is the maximum photon energy during the period in which $f_\epsilon$ is measured.  
The $\nu F_\nu$ flux $f_\epsilon$ in eq.\ (\ref{deltaD}) is evaluated at 
$\epsilon = \hat{\epsilon}={2\delta_{\rm D}^2}/{(1+z)^2\epsilon_1}$ from the pair-production threshold condition.

For 3C 454.3, writing $f_{\epsilon} = 10^{-10}f_{-10}$ erg cm$^{-2}$ s$^{-1}$ in eq.\ (\ref{deltaD})
gives 
$\delta_{min}\approx 16.5\left[{f_{-10} E_1(10{\rm~GeV})}/{t_v (6{\rm~hr})}\right]^{1/6}$
and $\hat{E}({\rm ~keV}) \cong 3.4 (\delta_{min}/15)^2/E_1$(10 GeV).
Contemporaneous Swift XRT observations give the $\nu F_\nu$ 
flux between 0.2 and 10 keV. From Swift public data \cite{bon10},
$\nu F_\nu(4 {\rm~keV}) \approx 6\times 10^{-11}$ erg/cm$^{2}$-s, so that $\delta_{min}\approx 15$. Because $\delta_{\rm D} >\delta_{min}$, we write $\delta =20\delta_{20}$
and $\delta_{20}\approx 1$. This value compares favorably with 
$\delta_{\rm D} =24.6\pm 4.5$, bulk Lorentz factor $\Gamma = 15.6\pm 2.2$,
and observing angle $\theta(^\circ) = 0.8\pm 0.2$ 
obtained from superluminal observations \cite{jor05}, derived from measurements made at a different time.

\subsubsection{Variability time scale}

In the case of 3C 454.3,   the emission region for one
of its major flares is constrained in a colliding shell geometry to be at distance
$R\lesssim 2c\Gamma^2t_v/(1 + z)\approx 0.1\Gamma^2_{20}t_v(6 {\rm~hr})$ pc, 
assuming $\Gamma_{20}  \equiv  \Gamma/20 \cong 1$. This is within
the BLR \cite{2010MNRAS.405L..94T}. Ways to avoid this conclusion are to assume
that $\Gamma\gg 20$, which would be associated with large jet powers, or for
only a small portion of the emission region to be active,
which would lower the jet's radiative efficiency given that only a 
small fraction of the emitting
surface is radiating. This might be implausible in view of
the already large apparent isotropic $\gamma$-ray luminosity
of 3C 454.3. Recollimation shocks (e.g., \cite{bl09}) at the pc scale might reduce the characteristic size of the
emission region, though the contrast ($c\delta t_v/[(1+z)R_{pc}({\rm pc})]\sim
2\times 10^{-3} t_v{\rm pc}/R_{pc}$) between the size of the emission region and location $R = R_{pc}$ pc 
seems unexpectedly small even for a narrow jet. Alternately, flaring episodes
with short variability times might take place within the BLR, whereas the
more slowly varying emissions could be radiated by jet plasma 
at larger distances, but then we would expect large spectral
variations due to the different target photon sources.

The uses of variability studies to infer properties of blazars, GRBs, or other sources
is an ongoing concern, because flow properties are inferred from the variability timescale.
The auto-correlation function can be used to infer a characteristic variability timescale, 
as can breaks in the power density spectrum of sources \cite{2010ApJ...722..520A}. Other techniques
for variability analysis of the extensive compilation of blazar light curves are currently in development.

\subsubsection{Equipartition field and jet power}

Here we perform an equipartition power analysis for
the giant outbursts from 3C 454.3 \cite{ack10,abd11}.
In the blob framework, where a
spherical emitting region with radius $R^\prime$ in the comoving
frame moves with Lorentz factor $\Gamma$ at an angle $\theta$ to the
line of sight, the absolute jet power is \cite{1993MNRAS.264..228C,2010MNRAS.402..497G} (Section 5.5, \cite{dm09}), 
including the photon power,
\begin{equation}
L_j = 2\pi r^{\prime 2}\beta c\frac{\Gamma^2}
{\delta_D^2}\frac{(\delta_D B_{eq})^2}{8\pi}(\chi^2+\frac{4}{3\chi^{3/2}})
+4\pi d^2_L \frac{\Gamma^2}{2\delta_D^4}\Phi\;.
\label{jetpower}
\end{equation}
Here $\Phi = 10^{-9}\Phi_{-9}$ erg cm$^{-2}$ s$^{-1}$ is the measured bolometric
photon energy flux ($4\pi d^2_L \Phi = 3.6 \times 10^{48}$ erg s$^{-1}$
for $\Phi_{-9}$ = 1; $\Phi_{-9} \cong$ 10 and 50 during the December 2009 and November 
2010 outbursts, respectively),
and the factor $\Gamma^2/2\delta_D^4$
in the last term of this expression
recovers the increased energy requirements due to debeaming
when $\theta \gg 1/\Gamma$¸ or $\delta_D \ll\Gamma$, in the limit
that the jet opening angle is $\ll 1$. The factor $\chi$ is the
deviation from equipartition, so that $\delta_D B=\chi (\delta_D B_{eq})$ and
\begin{eqnarray}
 \delta_D B_{eq}= B_{cr}\left[\frac{9\pi d_L^2 f_{\epsilon_{pk}} m_e c^2 (1+\zeta_{pe})\ln(\epsilon_2/\epsilon_1)}{4\sqrt{(1+z)\epsilon_{pk}} U^2_{cr} c \sigma_T V^\prime_b}\right]^{2/7}\;
\end{eqnarray}
(eq.\ 7.80, \cite{dm09}). Here  $f_{\epsilon_{pk}}$ is the $\nu F_\nu$ peak of the synchrotron spectrum,
$U_{cr} = B^2_{cr}/8\pi$, $B_{cr} = 4.414 \times 10^{13}$ G, and $\zeta_{pe}$ is the
relative energy in protons to nonthermal electrons which
emit synchrotron radiation with a spectrum $\alpha $ = 0.5 between
observed photon energies $\epsilon_1$ and $\epsilon_2$. 

Letting $(1 + \zeta_{pe}) \ln(\epsilon_2/\epsilon_1)$ = 10$^2 \Lambda_2$, the peak synchrotron
frequency $\nu_{pk}$ = $10^{13}\nu_{13}$ Hz, and  $f_{\epsilon_{pk}}=
10^{-10} f_{-10}$ erg cm$^{-2}$ s$^{-1}$, substitution of parameters for 3C 454.3 gives
\begin{equation}
B_{eq}({\rm G}) = 3.25 \frac{(f_{-10}\Lambda_2)^{2/7}}{\delta^{13/7}_{20} \nu^{1/7}_{13} t^{6/7}_v ({\rm d})}
\label{Beq}
\end{equation}
 i.e., an equipartition field of a few G. The corresponding
jet power is
\begin{eqnarray}
L_j ({\rm erg~s}^{-1}) \cong \frac{\Gamma^2}{\delta_D^2}[2.5\times 10^{46}\frac{(f_{-10}\Lambda_2)^{2/7}\delta^{2/7}_{20} t^{2/7}_v ({\rm d})}{\nu^{1/7}_{13}}(\chi^2+\frac{4}{3\chi^{3/2}})\end{eqnarray}
$$+\;5\times 10^{45}\frac{\Phi_{-9}}{\delta^2_{20}}]\;.
$$
The Eddington luminosity for a 10$^9 M_9$ Solar mass
black hole is 1.26 $\times 10^{47}$ erg s$^{-1}$. Estimates
for the black-hole mass in 3C 454.3 range from $M_9\approx 0.5$ \cite{bon10}
to $M_9\approx 4$ \cite{gu01}. Thus we see that the jet power from 3C 454.3
would be super-Eddington if $\chi$ departs from its equipartition
value by more than a factor $\approx$ 4. Moreover, the system
cannot be severely debeamed, though this is already
unlikely from other arguments, e.g., core dominance parameter
and superluminal motion observations \cite{jor05}.

By taking the ratio of the synchrotron and Compton $\nu F_\nu$ peaks, using the relations
$\epsilon_{pk,syn} \cong  (3/2)\gamma_{pk}^2 (B/B_{cr})\delta/(1 + z)$ and $\epsilon_{pk,C} \cong (4/3)\gamma_{pk}^2 \Gamma\epsilon_\star\delta/(1 + z)$, we have 
$
\epsilon_\star \cong ({\epsilon_{pk,C}}/{\Gamma\epsilon_{pk,syn})}({B}/{B_{cr})}$
or
\begin{equation}
E_\star({\rm eV}) \cong 4.6\;
{E_{pk,C}(100 {\rm~MeV})B({\rm 3\,G})\over \nu_{13}\Gamma_{20}}\;.
\end{equation}
This suggests that the soft photon energy scattered to make the GeV emission is a few eV, which would correspond to either atomic 
line radiation or scattered Shakura-Sunyaev accretion-disk radiation.

The energy density of the external radiation field can also be estimated
from the 3C 454.3 observations. 
Let $A_C$ represent the ratio of the energy fluxes in the
Compton and synchrotron components. If u$_{\star}$ denotes the
energy density of the target radiation field in the stationary
frame, and $u´_{B}$ denotes the comoving magnetic-field
energy density, then for $\Gamma\gg 1$, $4u_\star\Gamma^2/3 \approx A_CU^\prime_B$, or
$
u_{\star}({\rm erg~cm}^{-3}) \cong  0.007 {(A_C/10)B^2({\rm 3G})}/{\Gamma_{20}^2}
$
This can be compared to energy densities in the BLR using
relations between the BLR radius and line luminosity
\cite{bon10}. The implied energy densities
are an order of magnitude larger, suggesting that either $B >$ 3
G or $\Gamma< 20$.


\subsection{Hadronic jet models}

Measurements of air showers induced by cosmic rays impacting the atmosphere
give the best evidence for the existence of powerful accelerators of UHECRs with $E\gtrsim 10^{18}$ eV, 
with Larmor radii so large that they almost certainly originate from beyond the Galaxy.

\subsubsection{Adiabatic losses, and  photopair and photopion losses on the CMBR} 

The radial scale factor 
\begin{equation}
R = {R_0\over 1+z}\;
\label{R}
\end{equation}
at redshift $z$. In an adiabatic expansion process, $dQ = 0 
= dU +pdV$ and the energy content $U = uV$, where
$u$ is the energy density and V is the volume. 
Thus $udV + pdV = - Vdu$, or
$(u+p)dV = -V du$, 
so
\begin{equation}
-{dV\over V} = {dx\over x} + {dy\over y} + {dz\over z} = 3{dR\over R} = {du\over u+p}\;.
\label{dVV}
\end{equation}
For a relativistic gas with $p = u/3$,
\begin{equation}
-{3dR\over R} = {du\over ({4u\over 3})} =-{3d\ln R} = {3d\ln u\over 4}\;,
\label{3RR}
\end{equation}
which implies $u\propto R^{-4}$. Because the energy density $U = uV\propto R^{-1}$, 
$\gamma \propto R^{-1}$, so $-\dot \gamma \propto -(1/R^2) dR/dt_*$ (see Section 9.4 in \cite{dm09}).

With $R$ given by equation (\ref{R}), 
\begin{equation}
-\dot \gamma = (1+z)^{-1} |{dz\over dt_*}|\,\gamma= H_0\sqrt{\Omega_m(1+z)^3+\Omega_\Lambda}\;,
\label{dotgamma}
\end{equation}
for a flat cold dark matter cosmology with cosmological constant $\Lambda$. 
Thus 
\begin{equation}
-{d\ln \gamma\over dt} = {1\over (1+z)} |{dz\over dt_*}|= H_0\;,
\label{dlngammadt}
\end{equation}

The mean-free path for energy losses due to photopair production is given in Section 9.3.2  of \cite{dm09} 
by the expression 
$$r_{\phi e} (E_{20}) = {cE\over |dE/dt|} = {c\over |d\ln \gamma/dt|}\;\cong \;
$$
\begin{equation}
{1.0 [E_{18}(1+z)]^{1/2}\over (1+z)^3 {\cal F}_{\phi e}[E_{18}(1+z)]}{\rm Gpc}\equiv 
{\sqrt{y}\over (1+z)^3 {\cal F}_{\phi e}(y)}{\rm Gpc}\;,
\label{rphie}
\end{equation}
where
\begin{equation}
{\cal F}_{\phi e}(y)\;= 0.74 + 1.78\ln(y/2) + {0.69\ln(y/2)\over (y/2)^{3/2}}\;,
\label{calFy}
\end{equation}
taking the asymptote $y\gg 1$ or $E_{18} \gg (1+z)^{-1}$. If the ankle in the spectrum
is a consequence of photopair losses of ultra-high energy protons, as advocated by Berezinskii and 
colleagues \cite{2005PhLB..612..147B},
then the energy of the ankle would reflect the redshift epoch of greatest source production 
of the UHECRs, noting that $r_{\phi e}\propto (1+z)^{-3}$ at the  maximum
energy loss of protons due to photopair losses on the CMB.

Approximately, then
\begin{equation}
\big[ {d\ln E\over dt}\big]_{\phi e} = -{(1+z)^3 [0.74 + 1.78 \ln(y/2)]\over \sqrt{y} {\rm (Gpc/c)}}\;,
\label{dlnEdtphie}
\end{equation}
 when $y = E_{18}(1+z) \gg 1$. The conversion (Gpc/c) $= 1.03\times 10^{17}$s $\cong 3.2\times 10^9$ yr.

The mean-free-path for a proton to lose energy through photopion losses on the CMBR
is given by Ref.\ \cite{dm09}, Section 9.2.4, by
\begin{equation}
r_{\phi\pi}(E_{20})= {c\over |d\ln E/dt|_{\phi \pi}}\cong {13.7 \exp[4.0/E_{20}(1+z)]
\over (1+z)^3 [1+4.0/E_{20}(1+z)]}\;{\rm Mpc}\;,
\label{rphipi}
\end{equation}
which, though derived in the limit $E_{20}\ll 4/(1+z)$, gives a good 
approximation even at higher energies.

\subsubsection{Photopion production efficiency}

From the relations between time elements in the stationary frame (starred),
the comoving frame (primed), and the observer frame (unscripted), $R = \beta c t_* 
= \beta c \Gamma t^\prime$, so $t^\prime \cong R/\beta\Gamma c$ for relativistic flows, and 
$t \cong (1+z) R/\beta \Gamma^2 c$, since $t \cong (1+z)t^\prime/\Gamma$. The comoving dynamical time scale
is therefore
given by $t^\prime_{dyn} = R^\prime/c = R/\Gamma c$.

The energy flux
\begin{equation}
\Phi = {d{\cal E}\over dA dt} = {L_*\over 4\pi d_L^2}\;,\;L_* = {d{\cal E}_*\over dt_*}=
\Gamma^2 {d{\cal E}^\prime\over d\tp}\;,
\label{Phi}
\end{equation}
noting $\Gamma{\cal E}^\prime = {\cal E_*}$ and $dt_* = d\tp/\Gamma$. Therefore
$\e L(\e) = 4\pi d_L^2 f_\e = \Gamma^2 \ep L(\ep )$, and
\begin{equation}
n(\e ) = {\e L(\e)\over 4\pi R^2 c m_ec^2\e }\;,\;{\rm so}\; 
n^\prime(\ep ) = {\ep L^\prime(\ep)\over 4\pi \Gamma^2 R^2 m_ec^3\e }\;.
\label{ne}
\end{equation}
The causality constraint for rapidly variable emissions from  a relativistic
blast wave is 
\begin{equation}
\Delta r^\prime \cong {\Delta r\over \Gamma} \lesssim {\Delta \tp \over c} = {\Delta t_*\over \Gamma c} \cong {\Gamma \Delta t\over (1+z)c}\;,
\label{Deltarprime}
\end{equation}
and $t_{var} \cong (1+z) R/\Gamma^2 c$. 

We write the target $\nu F_\nu$ photon spectrum as
\begin{equation}
f_\e = f_{\e_{pk}}\big[\big({\e\over\e_{pk}}\big)^a H(\e_{pk}-\e)
+
\big({\e\over\e_{pk}}\big)^b H(\e - \e_{pk})\big]
\;
\label{fetarget}
\end{equation}
The threshold for photopion ($\phi\pi$) processes is $\gp_p\ep \gtrsim m_\pi/m_e \cong 400$,
using the approximation of \cite{ad03} that the product of the photopion inelasticity 
and cross section is $K_{p\gamma}\sigma_{\phi\pi} = 70\,\mu$b above
comoving photon energy $\ep \cong 400$.  Because $\gp_p = E_p/\Gamma m_pc^2$ and 
$\ep = (1+z)\e/\Gamma$, the threshold {\it escaping} proton energy is
\begin{equation}
E_p^{thr} \cong {400 m_pc^2\Gamma^2\over (1+z)\e}\;\;{\rm and}\; \;
E_p^{pk} \cong {400 m_pc^2\Gamma^2\over (1+z)\e_{pk}} \cong 
{3.7\times 10^{17}\Gamma_3^2\over (1+z)\e_{pk}}\;{\rm eV}\;
\label{Epthr}
\end{equation}
where 
$$\Gamma_3 \equiv {\Gamma\over 1000}\;.$$

The timescale for energy loss due to photohadronic processes is
\begin{equation}
t^{\prime-1}_{p\gamma}(E_p) \cong \;c(K_{p\gamma}\sigma_{p\gamma})
\;\int_{\ep_{thr}}^\infty d\ep\;n^\prime(\ep)\;.
\label{tprime-1}
\end{equation}
Thus
\begin{equation}
t^{\prime-1}_{p\gamma}(E_p) \cong \;{c(K_{p\gamma}\sigma_{p\gamma})d_L^2
\over 
R^2 m_ec^3 \Gamma^2}
\;\int_{\ep_{thr}}^\infty d\ep\;{f_\e\over \e^{\prime~2} }\;.
\label{tprime-2}
\end{equation}
Using eq.\ (\ref{fetarget}) gives
\begin{equation}
t^{\prime-1}_{p\gamma}(E_p^{pk}) \cong \;{(K_{p\gamma}\sigma_{p\gamma})d_L^2f_{\e_{pk}}
\over 
R^2 m_ec^2 \Gamma\e_{pk}(1-b)(1+z)}\;,
\label{tprime-3}
\end{equation}

The photopion production efficiency 
\begin{equation}
\eta_{p\gamma}(E_p) = {\tp_{dyn}\over \tp_{p\gamma}(E_p)}
\;.
\label{etapgamma}
\end{equation}
At $E_p = E_p^{pk}$, 
\begin{equation}
\eta_{p\gamma}^{pk} = \eta_{p\gamma}(E^{pk}_p) = {(K_{p\gamma}\sigma_{p\gamma})d_L^2f_{\e_{pk}}
\over 
\Gamma^4 m_ec^4 t_{var}\e_{pk}(1-b)}\;.
\label{etapgamma1}
\end{equation}
Thus
\begin{equation}
 \eta_{p\gamma}(E_p) \cong \eta_{p\gamma}^{pk} 
\cases{({E_p\over E^{pk}_p})^{1-b} \;, &  $E_p < E^{pk}_p$
 $~$ \cr\cr 
 \cases{({E_p\over E^{pk}_p})^{1-a} \;, &  $a<1$
 $~$ \cr\cr 
 1\;, & $ a>1$ \cr}, & $ E^{pk}_p < E_p$ \cr}\;.
\;,
\label{etapgamma2}
\end{equation}

To illustrate these results, we apply them to Fermi observations of 
GRBs. For GRB090510 at $z = 0.903$ and $d_L \cong 1.8\times 10^{28}$ cm, 
$E_p^{pk} \cong 2\times 10^{17}(\Gamma_3^2/\e_{pk})\;{\rm eV}$ and
$\eta_{p\gamma}^{pk} \cong 0.03 f_{-5}/\Gamma_3^4 t_{-2} \e_{pk}(1-b)$, and
is thus at the $\sim 1$ -- 10\% level. For GRB 080916C at $z = 4.35$ 
and $d_L \cong 1.25\times 10^{29}$ cm \cite{abd09science}, 
$E_p^{pk} \cong 7\times 10^{16}(\Gamma_3^2/\e_{pk})\;{\rm eV}$ and
$\eta_{p\gamma}^{pk} \cong 0.0015 f_{-6}/\Gamma_3^4 t_{var}({\rm s}) \e_{pk}(1-b)$.

\begin{table}
  \begin{center}
  \caption{Band function fits to GRB 090926A$^1$
}  \label{tab4}
 {\scriptsize
  \begin{tabular}{ l c c c c }\hline 
{\bf Time interval} & {\bf $f_{-6}$}$^2$ & {\bf a} & {\bf b} & {\bf $\e_{pk}$} \\  \hline
(a): 0.0 -- 3.3 s & 3.5 & $1.58\pm 0.03$ & $-0.64^{+0.07}_{-0.09}$ &  0.66 \\ 
(b): 3.3 -- 9.7 s & 6 & 1.45 & $-0.46$ &  0.56 \\ 
(c): 9.7 -- 10.5 s & 5 & 1.41 & $-1.69$  &  0.41 \\ 
(d): 10.5 -- 21.6 s & 1 & $0.30$ & $-0.80$ &  0.36 \\ 
\hline
  \end{tabular}
  }
 \end{center}
\vspace{1mm}
 \scriptsize{
$^1$Intervals (c) and (d) are best fit with additional cut-off power-law/power-law component\\
$^2$Units of $10^{-6}$ erg cm$^{-2}$ s$^{-1}$}
\end{table}

For GRB 090926A at $z = 2.1062$ and $d_L = 16.54$ Gpc,
the estimate above gives 
 $E_p^{pk} \cong  10^{17}(\Gamma_3^2/\e_{pk})\;{\rm eV}$ and
$\eta_{p\gamma}^{pk} \cong 0.01 (f_{-6}/6)/\Gamma_3^4 \hat t_{var} \e_{pk}(1-b)$.
The spectral parameters to derive the photopion efficiency are given
for the Band function in Table \ref{tab4} \cite{ack10c}.
Here the variability time scale $t_{var}$ is scaled to 
$0.15 \hat t_{var}$ s, which is the characteristic FWHM time of the 
well-resolved pulse. One percent efficiency can easily become 100\%
if $\Gamma_3\approx 0.3$, and more for higher energy protons, making escape
difficult. 
If long GRBs accelerate UHECRs, then the blast-wave Lorentz
factor must be close to that given in the simple $\gamma\gamma$ opacity limit,
otherwise the GRB would be highly radiative by hadronic processes. With the strong dependence on $\Gamma$, photohadronic interactions
become 100\% efficient when $\Gamma\approx 300$, so if GRBs accelerate UHECRs, an accompanying photohadronic 
$\gamma$-ray spectral component is predicted \cite{rmz04,mn06,der07a,drl07}.

\subsubsection{Proton models}

The comoving synchrotron cooling timescale of an
ion with atomic mass $A$ and  charge $Z$ is given by (eqs.\ 7.49 -- 7.50, \cite{dm09})
\begin{equation}
t^\prime_{syn} = {A^3\over Z^4} \left({m_p\over m_e}\right)^3 {6\pi m_ec\over \sigma_{\rm T} B^{\prime 2}\gp}\;.
\label{tprimesyn}
\end{equation}
Balancing the synchrotron energy-loss timescale with the shortest energy-gain timescale allowed in standard first- or second-order Fermi acceleration processes gives the relation
\begin{equation}
B^\prime \gamma^{\prime 2}= {A^2 \over Z^3} \left({m_p\over m_e}\right)^2 \;{6\pi m_e c\over \sigma_{\rm T} \phi }\;,
\label{BprimeFermi}
\end{equation}
where $\phi \gtrsim 1$ is the number of radians over which a particle gains $\sim e$ of its original energy. 
The observed peak synchrotron photon energy is 
$$\epsilon_{syn} \cong {\Gamma \epsilon^\prime_{syn}\over (1+z)} \cong {3\Gamma\over 2(1+z)}{Z\over A}\left({m_e\over m_p}\right) {B^\prime\over B_{cr}}\,\gamma^{\prime 2}$$
\begin{equation}
\cong{\Gamma\over (1+z)\phi} {27\over 8\alpha_f} {m_p\over m_e}  \approx {10^8\Gamma_3\over(1+z) (\phi/10)}{A\over Z^2} .
\label{epsilonprimesyn}
\end{equation}

Overlooking other limitations on particle acceleration, equation (\ref{epsilonprimesyn}) shows that proton synchrotron radiation from GRB jets and blazars can reach $\approx 50$ TeV  ($\Gamma\approx 10^3$) and $\approx 5$ TeV ($\Gamma\approx 100$), respectively. Proton synchrotron models have been developed in Refs.\ \cite{aha00,mp01} for $\gamma$-ray blazars, and in \cite{rdf10} for GRBs. We consider application of this type of model to these two source classes.

\vskip0.1in
{\bf GRBs}

Eliminating $\gp$ from these equations using the observer time $t_{syn} = (1+z)t^\prime_{syn}/\Gamma$
for the emission to be radiated at measured energy $m_ec^2\epsilon_{syn} = \Gamma m_ec^2 \epsilon^\prime_{syn}/
(1+z)= 100 E_{100}$ MeV implies a comoving magnetic field \cite{wan09,rdf10}
$$B^\prime({\rm G}) = {3\over 2^{5/3}}\left({A^{5/3}\over Z^{7/3}}\right)\;
\left({1+z\over \Gamma B_{cr}\e_{syn}}\right)^{1/3}\;\left({8\pi m_e c\over \sT t_{syn}}\right)^{2/3}\;$$
\begin{equation}
\cong {1.3\times 10^5\over t_{syn}^{2/3}({\rm s})}\;\left({A^{5/3}\over Z^{7/3}}\right)\;  \left({ 1+z\over \Gamma_3 E_{100}}\right)^{1/3} \;,
\label{BprimeG}
\end{equation}
and an isotropic jet power, dominated by magnetic-field energy,  of
$$L_* >   L_{*,B} \cong {1\over 2} R^2 c\Gamma^2 B^{\prime 2} \cong {1\over 2}{c^3\Gamma^6 t_{var}^2 B^{\prime 2}\over (1+z)^2} = {3^2\over 2^{13/3}}\;{ A^{10/3}\over Z^{14/3} }\;{c^3\over (1+z)^{4/3}}
\left({m_p\over m_e}\right)^{10/3}$$
\begin{equation}
\times {\Gamma^{16/3} t_{var}^2 \over B_{cr}^{2/3} \e_{syn}^{2/3}}\;\left({8\pi m_ec\over t_{syn}\sT }\right)^{4/3} \cong 2.5\times 10^{59}\;{ A^{10/3}\over Z^{14/3} }\;{\Gamma_3^{16/3}t_{var}^2({\rm s})
\over [(1+z)t_{syn}({\rm s})]^{4/3} E_{100}^{2/3}   } \;
{\rm erg/s},\label{LB}
\end{equation}
letting the blast-wave radius $R\cong \Gamma^2 c t_{var}/(1+z)$. For Fe ($A = 56, Z = 26$), the
power requirements are reduced by a factor $\approx 0.17$.

Eq. (\ref{LB}) shows that $L_B \propto \Gamma^{16/3}$. The absolute energy release is
\begin{equation}
{\cal E}_{*,abs} = {\Delta t\over 1+z}\;f_bL_{*} \gtrsim 
{2.5\times 10^{59}\over (1+z)^{7/3}}\;{ A^{10/3}\over Z^{14/3} }\;{\Gamma_3^{16/3}t_{var}^2({\rm s})
\over E_{100}^{2/3}  t^{4/3}_{syn}({\rm s}) }\,\Delta t({\rm s}) \,f_b\;{\rm erg},
\label{calE*abs}
\end{equation}
where $f_b$ is the beaming factor.
Using a 100 kG magnetic field as a fiducial,
so $B = 10^5 B_5$ G, then the absolute GRB energy release is
${\cal E}_{*,abs}\cong 1.3\times 10^{59} B_5^2 \Delta t({\rm s})t^2_{var}({\rm s})f_b \Gamma_3^6/(1+z)^2$ erg.
Such large energy requirements pose a problem for highly magnetized GRB models. 

In the model of Razzaque et al.\ (2010) \cite{rdf10},
taking a lower limit $\Gamma_{3} \cong 0.5$ implied by naive $\gamma\gamma$ 
opacity arguments,
gives the absolute energy requirements for GRB 080916C \cite{abd09science}
of
$${\cal E}_{*,abs}\cong 2\times 10^{52}E_{100}^{-2/3}(\Gamma_3/0.5)^{16/3}(\theta_j/1~{\rm deg})^2 t^{5/3}_{syn}({\rm s})\;{\rm erg},$$
after multiplying by a two-sided jet beaming factor 
$f_b \cong 1.5\times 10^{-4}(\theta_j/1~{\rm deg})^2$. 
The precise values of $t_{var}$, $t_{syn}$ and $\Delta t$ depend on model interpretation. 
For the external shock model \cite{dm99}
applied to GRB 080916C, the rapid variability may be made by irregularities in the 
complex surrounding medium, so $t_{var} \cong 0.1$ s, the proton synchrotron cooling timescale to make 
the delayed onset is $t_{syn} \cong4$ s, and the GRB duration $\Delta t \cong 10$ -- 50 s. 


From equation (\ref{calE*abs}), strong proton/ion synchrotron radiations can be
emitted in the Fermi range near 1 GeV or even 100 MeV due to strong cooling in 
an intense magnetic field, but the absolute energies are large without
a very small beaming factor. 
The jet break time  with apparent isotropic energy release $\approx 2\times 10^{57}$ 
ergs is $t_{br}\cong 0.3 (\theta_j/1~{\rm deg})^{16/3}n_0^{-1/3}$ d. For such a narrow jet, the jet break would
have taken place before Swift slewed to GRB 080916C at $\approx T_0 + 17.0$ hr \cite{raz10}. 

More complicated geometries might, however, relax the bulk Lorentz factor requirement further \cite{gra08}. 
If the inner engine makes the prompt MeV radiation 
and residual shell collisions at larger radii make LAT $\gamma$-ray photons, then $\Gamma$ could
be as low as $\sim 300$ \cite{li10}.  In this case, the absolute energy release from GRB 080916C 
could easily be $\lesssim 10^{52}$ erg.

For a proton 
synchrotron model of GRBs to be viable \cite{rdf10}, 
a narrow jet opening angle of order $1^\circ$ along with a value of 
$\Gamma \lesssim 0.5$ gives 
${\cal E}_{*,abs}\cong$ few $\times 10^{52}$ ergs, within ranges
implied by interpretation of radio and $\gamma$-ray observations and beaming \cite{cen10}. 
The detection of distinct
components in GRB spectra suggests that a cascading interpretation be more 
carefully considered \cite{asa09}.
The local rate density of Type Ib/c supernova progenitors of long GRBs like 
GRB 080916C or low-luminosity GRBs like GRB 980425 or GRB 060218 
cannot exceed $\approx  300$ Gpc$^{-3}$ yr$^{-1}$  \cite{sod06},
and the local 
beaming-corrected rate density of long duration GRBs is $\approx$ 10 -- 50 Gpc$^{-3}$ yr$^{-1}$ 
\cite{gpw05,ld07}. 
The local Type 1b/c rate is $9^{+3}_{-5}\times 10^3$ Gpc$^{-3}$ yr$^{-1}$ \cite{sod06}.
This agrees with 
the star-forming galaxy SN Ib/c rate of $\approx 0.28$ century$^{-1}$
per 300 Mpc$^{3}$ per $L_*$ galaxy $\approx 10^4$ Gpc$^{-3}$ yr$^{-1}$.
These imply a beaming rate of $\approx (10$ -- 50)/9,000 $\approx 
 10^{-3}$ -- 0.006. With the $\approx 1^\circ$ opening angle required to explain GRB 080916C,
then all Type 1b/c SNe would have to make GRBs, presenting another challenge for strong magnetic-field
jet models of GRBs.  

\vskip0.1in
{\bf Blazars} 

Proton synchrotron and photohadronic models for blazars have been developed, as noted,
in \cite{aha00,mp01}. The protons and ions
make a $\gamma$-ray component that originates from 
pion-induced cascades, and the synchrotron radiations of 
pions and muons make additional $\gamma$-ray emissions. Primary 
electron synchrotron radiation is still usually 
required to make the nonthermal radio/optical/X-ray synchrotron blazar emission. 

Anita Reimer gives  \cite{abd11Mrk421} a detailed fit to the famous HBL Mrk 421 at $z = 0.031$ at luminosity distance $d_L = 134$ Mpc. 
Mrk 421 is one of the first Whipple TeV blazars and the first one with sub-hour \cite{gai96}
variability detected. For the  composite SED of Mrk 421 averaged between
2009 January 9 (MJD 54850) to 2009 June 1, Reimer makes 
a proton fit with the parameters shown in Table \ref{tab:14}. 

\begin{table}
\centering
\caption{Proton blazar model for Mrk 421 \cite{abd11Mrk421}}
\label{tab:14}       
%
%
\begin{tabular}{ll}
\hline\noalign{\smallskip}
Parameter & Value   \\
\noalign{\smallskip}\hline\noalign{\smallskip}
$B^\prime$ & 50 G  \\
$\delta_{\rm D}\approx \Gamma$ & 12  \\
$r_b^\prime$ & $4\times 10^{14}$ cm  \\
$E_{max}$ & $7.2\times 10^{19}Z \Gamma_{12} B^\prime_{50}$ eV$^a$ \\
\noalign{\smallskip}\hline
\end{tabular}
{\noindent$^a$ $\Gamma_{12} = \Gamma/12$, $B^\prime_{50} = B^\prime/50$ G, 
$\delta_{12} = \dD/12$}
\end{table}

The Hillas criterion \cite{hil84} for a single-zone blob model
requires the comoving Larmor radius to be smaller than the 
comoving blob radius, that is, 
\begin{equation}
r_{\rm L}^\prime < r_b^\prime\;.
\label{hillas}
\end{equation}
With eq.\ (\ref{rlarmor}) giving the maximum escaping particle
energy $E_{max} \approx \Gamma E^\prime_{max}\approx ZeB^\prime r_b^\prime$, 
we see that this model blazar can accelerate super-GZK ($E\gtrsim 5\times 10^{19}$ eV)
 particles if charged ions, but this is only  possible for protons with 
slightly larger magnetic fields. These protons have maximum comoving 
Lorentz factors
$$\gamma^\prime_{p,max} \approx 6.4\times 10^9 B^\prime_{50}\;.$$ 
For particles well-trapped by the magnetic field of the plasma, 
$\gamma^\prime_{p,max} = 10^9 \gp_9$ with $\gp_9 \ll 1$.

For the 2-sided jet power as defined in eq.\ (\ref{jetpower}), taking $\delta_{\rm D} \approx \Gamma$ 
gives
\begin{equation}
L_j = 2\pi r_b^{\prime 2}\beta c\Gamma^2 \sum_i u_i^\prime\;,\; \;i = p, e, B\;,
\label{Lji}
\end{equation}
and the total Poynting power dominated by magnetic field energy is 
\begin{equation}
L_{j,B} = {1\over 4} r_b^{\prime 2}\beta c\Gamma^2 B^{\prime 2} \cong
4.3\times 10^{44} \Gamma_{12}^2 B_{50}^{\prime 2}\;{\rm erg/s}\;,
\label{LjB}
\end{equation}
which is not excessive, considering that the Eddington luminsoity 
is $L_{\rm Edd} \approx 10^{47}M_9$ erg/s, and $M_9 \approx 0.2$ -- 0.9 for
the supermassive black hole power Mrk 421. 

We still need to calculate the particle power and the 
efficiency for hadronic production of $\gamma$-ray emission.
In the case of  proton synchrotron radiation amounting
to total energy flux $\Phi$ (erg/cm$^2$-s), the 
particle power is 
\begin{equation}
L_{j,p} = 2\pi r_b^{\prime 2}\beta c\Gamma^2 \big( {N_{p0}m_pc^2 \gp_{p,max}\over V_b^\prime}\big)
=
{3\over 2}{m_pc^3 N_{p,0} \gp_{p,max}\Gamma^2\over r_b^\prime } \;,
\label{Ljp}
\end{equation}
for a spherical comoving emission region.

According to our standard approach \cite{dm09}, 
$$\nu F_\nu  = f_\epsilon \cong \delta_{\rm D}^4\;{L^\prime\over 4\pi d_L^2}\;\sim \Phi\;.$$
The proton-synchrotron energy loss rate is
\begin{equation}
-\left( {dE^\prime \over d\tp }\right)_{p,syn} = {4\over 3} \;\left({m_e\over m_p}\right)^2 c\sigma_{\rm T}
U^\prime_{B^\prime } \gamma^{\prime 2} \;,\;
\label{dEdtsyn}
\end{equation}
where $U^\prime_{B^\prime } = B^{\prime 2}/8\pi$. 

For a large range of proton spectra, the $\gamma$-ray power from photohadronic processes can be 
approximated by the number $N_{p,0}$ of protons with Lorentz factor $\gp\approx \gp_{p,max}$, 
from which follows that 
\begin{equation}
N_{p,0} = \left({m_p\over m_e}\right)^2 \;{3\pi d_L^2 \Phi\over \dD^4 c\sT U^\prime_{B^\prime } \g_{p,max}^{\prime 2}}
\label{Np0}
\end{equation}
and
\begin{equation}
L_{j,p} = \left({m_p\over m_e}\right)^2 \;{9\pi m_pc^3 d_L^2 \Gamma^2 \Phi\over 2r_b^\prime \dD^4 c\sT U^\prime_{B^\prime } \g_{p,max}^{\prime}}\;\approx \; 3.2\times 10^{44} \; {\Gamma_{12}^2 \Phi_{-10}\over 
\delta^4_{\rm D} B_{50}^{\prime 2} \gp_9}\;{\rm erg/s}\;,
\label{Ljp1}
\end{equation}
where the bolometric $\gamma$-ray energy flux supplied by the proton synchrotron process is 
$10^{-10}\Phi_{-10}$ erg/cm$^2$-s. The particle power is reasonable, even if $\gp_9\rightarrow 0.1$.
The proton synchrotron frequency, from eq.\ (\ref{epsilonprimesyn}),  
\begin{equation}
\e_{p,syn} \cong {3\dD\over 2(1+z)}\;{m_e\over m_p}\;{B^\prime \over B_{cr}}\g_{p,max}^{\prime 2}\;
\cong 1.1\times 10^4 \g_9^{\prime 2} B_{50}^\prime \delta_{12},
\label{epsyn}
\end{equation}
so
\begin{equation}
E_{p,syn} \cong 5.6 \g_9^{\prime 2} B_{50}^\prime \delta_{12}\;{\rm GeV}\;.
\label{Epsyn}
\end{equation}

The equipartition field is, from eq.\ (\ref{Beq}), or eq.\ (7.81) \cite{dm09}
\begin{equation}
B_{eq}({\rm G}) \cong 130 \;
{d_{28}^{4/7}(1+z)^{5/7}f_{-10}^{2/7} \Lambda^{2/7}\over t_{d}^{6/7} \dD^{13/7} \nu_{13}^{1/7}}
\;.
\label{Beq_2}
\end{equation}
In the case of Mrk 421, $\nu_{13} = 10^4 \nu_{17}$, $\Lambda = 10\Lambda_{10}$, 
$t_d = (1+z)r_b^\prime/c\dD = 1200/\delta_{12}$ s, implying 
that 
\begin{equation}
B_{eq}({\rm G}) \cong 4.5 \;
{f_{-10}^{2/7} \Lambda_{10}^{2/7}\over \delta_{12} \nu_{17}^{1/7}}
\;,
\label{Beq_421}
\end{equation}
so that this model is magnetically dominated with a magnetic field $\approx 10\times B_{eq}$.

The minimum bulk Lorentz factor from $\gamma\gamma$ constraints, using eq.\ (\ref{eq2}), is
$\Gamma_{min} = 20.3  f_{-10} [E_1/({10~{\rm TeV}})]^{1/6}=13.8  f_{-10} [E_1/({{\rm TeV}})]^{1/6}$, for a maximum 
measured photon
energy $E_1$ and using a variability timescale of$\approx 10^3$ s. Note that $f_{-10}$
represents the $\nu F_\nu$ flux for the target photons and, 
though both have the same units, is a differential quantity 
compared to the bolometric energy flux $\Phi$. The condition for $\gamma$-ray 
transparency is somewhat inconsistent with the detection of multi-TeV photons from Mrk 421.

The cooling regime for ultra-high energy protons through synchrotron cooling
is determined by the ratio $\rho \equiv \tp_{syn}/\tp_{dyn}$, with strong 
synchrotron cooling when $\rho \ll 1$ and weak cooling when $\rho \gg 1$.
From eq.\ (\ref{tprimesyn}) we have
\begin{equation}
\rho = {3(m_p/ m_e)^3\over 4\sigma_{\rm T}U^\prime_{B^\prime} \gp_p r_b^\prime}\rightarrow {140\over \gp_9}\;.
\label{rho}
\end{equation}
Thus this proton synchrotron model for Mrk 421 is in the weak cooling regime. Alternately,
we can use the expression \cite{rdf10} for the saturation Lorentz factor 
\begin{equation}
\gp_{sat} = \left({m_p\over  m_e}\right)\sqrt{B_{cr}\over B^\prime} \sqrt{9\over 4\alpha_f}\;= {3\times 10^{10}\over \sqrt{B_{50}^\prime}}\;,
\label{gpsat}
\end{equation}
for which particles with larger Lorentz factors are in the strong cooling regime (here $\alpha_f = e^2/\hbar c\approx 1/137$ is the fine structure constant).

We can use eq.\ (\ref{etapgamma2}) to determine the photopion production efficiency for this model.   
For $\e_{pk} = 10^{17}\nu_{17}/1.24\times 10^{20} = 8.1\times 10^{-4}\nu_{17}$, $a \cong 0$ and $b \cong -1$, we obtain
\begin{equation}
\eta_{p\gamma }^{pk} = {5.6\times 10^{-7}f_{-10}\over t_d \Gamma_{12}^4 \nu_{17}} \rightarrow 4.3\times 10^{-5} {f_{-10}\over \nu_{17}}\;,
\label{etapgammapblazar}
\end{equation}
taking $t_v = (1+z)r_b^\prime/c\dD = 0.013/\delta_{12}$ d. The comoving Lorentz factor of protons interacting with target photons with energy $\ep_{pk}$ is $\gp_{pk} \cong 6\times 10^6$. Protons with larger Lorentz factors interact primarily with the $a \cong 0$ portion of the $\nu F_\nu$ spectrum with frequencies less than $\nu_{pk}$. The efficiency of proton energy loss with $\gp\gg \gp_{pk}$ approaches 
\begin{equation}
\eta_{p\gamma}(E_{p}) \approx \eta_{p\gamma }^{pk} \;\left( {E_{p}\over E_{pk}}\right)
\cong 7.4\times 10^{-3} {f_{-10}\gp_9\over\nu_{17}}\;.
\label{etapgammaemax}
\end{equation}
For photopion losses (as for proton synchrotron losses), the energy loss is most efficient for the highest energy particles, reaching $\sim 1$\% for $\gp_9 \sim 1$. Somewhat larger efficiencies are allowed---provided these allowances are consistent with the variability data---if the protons are trapped on long times compared with the crossing time, and the emission region is slow to expand. 

In conclusion, hadronic models for blazars and GRBs face no insurmountable objections 
based on power or energetics.

\subsection{Cascade halos and the intergalactic magnetic field (IGMF)}

\label{halo}
The magnetic field $B_{\rm IGM}$ in the IGM
is bounded by  a primordial  field generated by quantum fluctuations in 
the early universe or decoupling transitions of the fundamental forces \cite{2009PhRvD..80b3010E}. 
Dynamo processes amplify the seed fields.
Gamma ray astronomy provides a method for measuring the intergalactic 
magnetic field (IGMF) by exploiting the effects of 
 the $\g\g\rightarrow$e$^+$ e$^-$ pair-production process
on $\gamma$ rays from  extragalactic TeV sources interacting with soft photons of the EBL.

When VHE 
$\gamma$ rays interact with ambient radiation fields and the EBL, 
then blazars, and therefore  radio galaxies (i.e., misaligned blazars) 
are surrounded by  anisotropic jets of relativistic leptons made
when  the $\gamma$ rays 
materialize into energetic e$^+$e$^-$ pairs. 
These leptons spiral 
in the ambient magnetic field to make extended synchrotron radiation halos \cite{acv94}, 
and  GeV $\gamma$ rays by Compton-scattering photons of the CMBR  \cite{2007A&A...469..857D}. 
Arrival time information in pair cascades generated
by impulsive sources of high-energy, multi-TeV $\gamma$ radiation can also 
be used  to infer the strength of the IGMF \cite{pla95,mur08}.

\begin{table}
\caption{Derived Limits on $B_{\rm IGM}$ for the source 1ES 0229+200 assuming persistent emission}
\begin{center}
\begin{tabular}{|l|l|l|}
\hline
1ES 0229+200 & $\theta_{\rm j}$ (rad)& $B_{\rm IGM}$(G) \\ 
\hline
Ref.\  \cite{nv10} & $\pi$ & $\gtrsim 3\times 10^{-16}$ 
  \\
Ref.\ \cite{tav10a} & 0.1 & $\gtrsim  5\times 10^{-15}$ \\
Ref.\ \cite{tav10b} & 0.03 & $\gtrsim  2\times 10^{-15}$ \\
Ref.\ \cite{dol10} & 0.1  & $\gtrsim  5\times 10^{-15}$  \\ 
\hline
\end{tabular}
\end{center}
\label{tableBIGMF}
\end{table}

Because the emission of 1 -- 10 TeV photons from a source at redshift $z\ll 1$ is attenuated
by, primarily, the IR EBL, then the cascade spectrum 
can be calculated for a given spectral model of the EBL and properties of 
of $B_{\rm IGM}$. For sufficiently weak magnetic fields, the pairs travel 
rectilinearly while Comptonizing CMB photons to $\g$ ray energies. 
The absence of this cascade signature in joint Fermi-HESS observations of candidate TeV blazars
implies a lower limit on $B_{\rm IGM}$   \cite{2007A&A...469..857D} under the assumption that 
the blazar persistently emits high-energy radiation for arbitrarily long time.
Neronov \& Vovk \cite{nv10} and Tavecchio et al.\ \cite{tav10a} 
 argued that nondetection of the TeV blazars 1ES 1101$-$232 ($z = 0.186 $), 1ES 0229$+$200 ($z = 0.14 $),  1ES 0347$-$121 ($z = 0.188$),
and H 2356$-$309 ($z = 0.165 $) by Fermi implies a lower limit $B_{\rm IGM}\gtrsim 3\times 10^{-16}$ G (Table \ref{tableBIGMF}).

The high-energy electrons and positrons also undergo deflections in the ambient magnetic 
field, so the emissions arriving latest generally come from leptons that
have been most severely deflected. This will cause steady sources to be surrounded by an extended angular halo 
formed by leptons deflected back into the line of sight. 
Ando \& Kusenko \cite{ak10} claim  that 
$\sim 30^\prime$ halos are found in stacked Fermi data for 170 AGNs that are 
bright at 10 -- 100 GeV. By fitting the angular distribution of the halo, they 
deduce that $B_{\rm IGM} \approx 10^{-15}(\lambda_{\rm coh}/{\rm kpc})^{-1/2}$, 
for magnetic-field correlation length $\lambda_{\rm coh} \sim 10$ -- 100 kpc.
This claim is disputed in \cite{ner10}, in part because of the use of P6\_v3 
response functions which are known to have inaccurate PSF from on-orbit
calibration data \cite{2009APh....32..193A}.

More realistic limits on $B_{IGMF}$ based on evidence that the blazar was operating 
only during the time frame over which it was observed, was proposed in \cite{der10}, and 
independently, in \cite{dol10}.

\subsubsection{Cascade radiation from EBL attenuation of TeV photons}

\begin{figure}[b]
\begin{center}
 \includegraphics[width=3.4in]{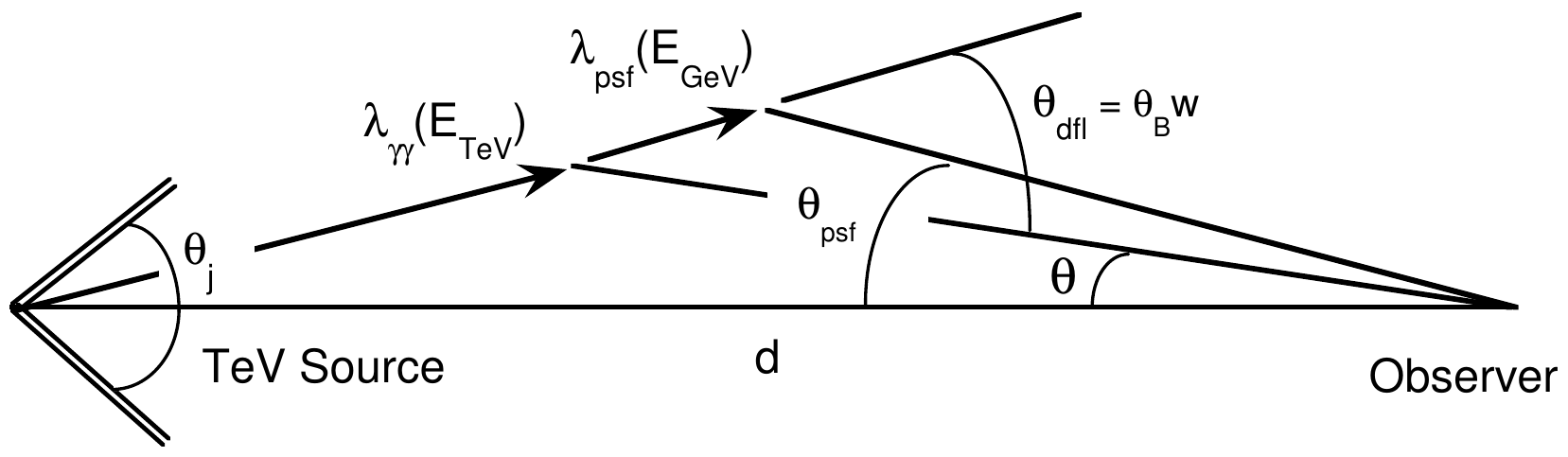}
 \caption{Sketch of the geometry of the event. A photon with energy $E_{\rm TeV}$ TeV emitted at angle $\theta_1\leq \theta_{\rm j}$ to the line of sight, 
where $\theta_{\rm j}$ is the jet half-opening angle,
interacts with an EBL photon to create an electron-positron pair. The electron is deflected by angle $\theta_{\rm dfl}$ and scatters a CMB photon to 
energy $E_{\rm GeV}$ GeV and angle $\theta$, which is detected as a source photon by the Fermi LAT when $\theta$ falls within the angular point spread function 
$\theta_{psf}({\rm GeV})$ from the source direction.}
\label{geometry}
\end{center}
\end{figure}

Consider a source and observer separated by distance $d$, as shown in Figure \ref{geometry}.
 Photons with dimensionless energy $\epsilon_1 = h\nu_1/m_ec^2$ emitted at angle $\theta_1$ to the line of sight between the source and observer, travel an average distance $\lambda_{\gamma\gamma}= \lambda_{\gamma\gamma}(\epsilon_1,z)$ before materializing into an electron-positron pair via $\gamma\gamma$ absorption with photons of the EBL. After production, the pairs cool by scattering CMB radiation, which  is detected at an angle $\theta$ to the line of sight to the source when the secondary electrons and positrons (hereafter referred to as electrons) are deflected by an angle $\theta_{\rm dfl}$. Thus $\theta_1 = \theta_{\rm dfl} - \theta$. For the purposes of the argument, we neglect redshift effects for the TeV blazar sources under consideration, which limits the treatment to sources at $z \lesssim 0.2$ (see \cite{ns09} for redshift corrections).

The time delay $\Delta t$ between the reception of photons directed towards the observer and those which undergo the process described above is given through the expression
$$c\Delta t = \lambda_{\gamma\gamma} + x - d = \lambda_{\gamma\gamma} + {d\sin(\theta_{\rm dfl}-\theta)\over \sin\theta_{\rm dfl}}
 - d = $$
\begin{equation}
\lambda_{\gamma\gamma}(1-\cos\theta_{\rm dfl}) - d(1-\cos\theta)\;,
\label{cDeltat}
\end{equation} 
noting that $x = d\sin\theta_1/\sin\theta_{\rm dfl}$ and $\lambda_{\gamma\gamma} = d\sin \theta/\sin \theta_{\rm dfl}$.
In the limit of small observing and deflection angles, equation (\ref{cDeltat}) implies
\begin{equation}
\Delta t \cong {\lambda_{\gamma\gamma}\over 2c}\;\theta_{\rm dfl}^2\,( 1 -  {\lambda_{\gamma\gamma}\over d})\;.
\label{Deltat}
\end{equation}
Moreover, this time delay is observed by photons received at 
\begin{equation}
\theta \cong \theta_{\rm dfl}\;{\lambda_{\gamma\gamma}\over d}\;
\label{theta}
\end{equation}
to the line of sight.

The deflection angle depends on the Lorentz factor $\gamma = 10^6\gamma_6$ of the produced electrons, and can be written in terms of the received photon energy $E =E_{\rm GeV}$ GeV. The average photon energy of the  CMB at low redshift is $\epsilon_{0}\approx 1.24\times 10^{-9}$ in $m_ec^2$ units, so that the mean Thomson-scattered photon energy is $\epsilon_{\rm T}\approx (4/3)\epsilon_{0}\gamma^2$. Thus, an electron with Lorentz factor $\gamma$ scatters CMB radiation to photon energy $E$ when $\gamma_6 \cong 1.1\sqrt{E_{\rm GeV}}$. The characteristic length scale for energy losses due to Thomson scattering is 
\begin{equation}
\lambda_{\rm T} = {3m_ec^2\over 4\sigma_{\rm T} u_{\rm CMB}\gamma} = 
({0.75\over\gamma_6}) {\rm ~Mpc}\;,
\label{lambdaT}
\end{equation}
where $u_{\rm CMB} \cong 4\times 10^{-13}$ erg cm$^{-3}$ is the CMB energy density at low redshifts. While losing energy, the electron is deflected by an angle $\theta_{\rm B}\cong \lambda_{\rm T}/r_{\rm L}$ in a uniform magnetic field of strength $B_{\rm IGM} = 10^{-15} B_{-15}$ G oriented perpendicular to the direction of motion of the electron, where the Larmor radius  $r_{\rm L} = m_ec^2\gamma/eB \cong 0.55 (\gamma_6/B_{-15})$ Mpc. Thus the deflection angle for an electron losing energy by scattering CMB photons to energy $E$ in a uniform field is $\theta_B = \lambda_{\rm T}/r_{\rm L}\cong 1.1B_{-15}/E_{\rm GeV}$. Introducing a coherence length $\lambda_{coh}$ that characterizes the typical distance over which the magnetic field direction changes by $\approx \pi/2$, then the deflection angle
\begin{equation}
\theta_{\rm dfl} \cong \theta_{\rm B}\cases{~~~1 & if $\lambda_{\rm T}< \lambda_{coh}$\cr
	\sqrt{\lambda_{coh}\over \lambda_{\rm T}},& if $\lambda_{\rm T}> \lambda_{coh}$.\cr} \;\equiv w\theta_{\rm B}\;,
\label{thetadlf}
\end{equation}
with $w \equiv H(\lambda_{coh}-\lambda_{\rm T}) + \sqrt{\lambda_{coh}/ \lambda_{\rm T}}H(\lambda_{\rm T}-\lambda_{coh})$.

The EBL model of \cite{frd10} for sources at $z = 0.14$ gives $\lambda_{\gamma\gamma}(E) \cong 200$ Mpc, 125 Mpc, and 70 Mpc at $E = 1, 3,$ and 10 TeV, respectively. A low EBL based on galaxy counts  \cite{kd10} gives values of 
 $\lambda_{\gamma\gamma}(E) \cong 280$ Mpc,
150 Mpc, and 85 Mpc at $E = 1, 3,$ and 10 TeV, respectively. Thus we write $\lambda_{\gamma\gamma} = 100\lambda_{100}$ Mpc, so that $\lambda_{100} \approx 1$ gives a minimum value of $\Delta t$ and  $\theta$ for the reprocessed TeV radiation.

For a source at distance $d= d_{\rm Gpc}$ Gpc, with $d_{\rm Gpc} \sim 1$ corresponding to $z \sim 0.2$,  the time delay for emission observed at angle 
\begin{equation}
\theta \cong 0.01\;{\lambda_{100}\over d_{\rm Gpc}}\;\big({B_{-15} w\over E/{\rm 10~GeV}}\big)\;
\label{theatnew}
\end{equation}
 from the line of sight is given from equation (\ref{Deltat}) by 
\begin{equation}
\Delta t ({\rm yr}) \cong 2\times 10^6\;\lambda_{100} \big({B_{-15} w\over E/{\rm 10~GeV}}\big)^2
(1- 0.1{\lambda_{100}\over d_{\rm Gpc}})\;
\label{Deltat1}
\end{equation}

Equation (\ref{Deltat1}) shows that small time delays are implied when  $\lambda_{\gamma\gamma}$ is small or $\lambda_{\gamma\gamma} \approx d$.
When $\lambda_{\gamma\gamma}\lesssim \lambda_{\rm T}$, an additional delay $\approx \lambda_{\rm T}\theta_{\rm dfl}^2/c$ arises  during the time that the electrons are losing energy and being deflected by the IGMF \cite{iit08,rmz04}. Such small values of $\lambda_{\gamma\gamma}\sim 1$ Mpc are only relevant at low redshifts to $\gtrsim 100$ TeV photons pair producing within $\approx 1$ Mpc of their source which, however, may be in the galaxy cluster environments where the magnetic field is not representative of the dominant volume of the voids. Thus we can dismiss such an origin of a short time delay without assuming special properties of the TeV sources. The case $\lambda_{\gamma\gamma}\sim d$ formally gives short time delays, but this corresponds to the case when the mean free path for $\gamma\gamma$ pair production is about equal to the source distance, which occurs for $\gamma$ rays with energies of several hundred GeV when $d \sim 1$ Gpc. In this case, the secondary electrons which take half the energy of the high-energy photon  scatter CMB photons to $\ll 100$ MeV.  Even so, the attenuation of the high-energy photons takes place over the entire distance $d$, so that the upscattered photons with short delay time comprise only a very small fraction of the incident flux attenuated close to the observer.  

The remaining alternative to avoid assuming that TeV blazars are steady on timescales of millions of years is to suppose that either $B_{\rm IGM} \ll 10^{-15}$ G or that $\lambda_{coh} \ll 1$ Mpc. If $\lambda_{coh} \sim 1$ Mpc, this contradicts the claim that the IGMF  has been measured to be $B_{\rm IGM}\gtrsim 10^{-15}$ G. If $\lambda_{coh} \ll 1$ Mpc, then the field must be even larger in order that the electrons are deflected away from the direction of the photon source. Thus TeV blazars must be assumed to be steady emitters on long timescales. Here we relax this assumption, and suppose most cautiously that the blazar has been operative over the last few years that the they have been observed. This reduces the implied lower limit by several orders of magnitude, as will be seen. Here we treat the blazar as a point source; see \cite{der10} when the {Fermi}-LAT PSF constraint is included (Fig.\ \ref{geometry}). 


\subsubsection{Derivation of the cascade spectrum}

Consider a source that emits TeV-scale photons from within a photon beam of half-angle 
$\theta_j$, and an apparent isotropic spectral luminosity $\e_* L_*(\e_*)$ within the
beam (``flat-topped" jet). If the TeV photons can escape from the nuclear environments
without $\gamma\gamma$ absorption, then they may still be attenuated by interactions
with EBL photons.  After $\g\g$ attenuation, e$^+$-e$^-$ pairs lose energy, primarily by
Compton scattering the CMBR. This upscattered radiation will be detected by an observer until
the cooling pairs are deflected out of the beam. 

We use the notation of Ref.\ \cite{dm09}. From eq.\ (\ref{fe_define}), denote the $\nu F_\nu$ spectrum by
$$f_\e = \nu F_\nu = {\e_*L_*(\e_*)\over 4\pi d_L^2}\; , \, \e = h\nu/m_ec^2\;,\;\e_* = (1+z)\e.$$
The correction due to EBL attenuation is 
\begin{equation}
f_\e = {\e_*L_*(\e_*)\over 4\pi d_L^2}\,\exp[-\tau_{\g\g}(\e;z)]\;=\;
 {m_ec^2 \e^2 \dot N(\e )\over 4\pi d_L^2 }\,\exp[-\tau_{\g\g}(\e,z)]\; ,
\label{fe13}
\end{equation}

 At low redshifts, $z\ll 1$, $\e_* \approx \e$, and $\e L(\e) = m_ec^2\e^2 \dot N(\e)$, 
where $\dot N(\e)$ is the photon injection function. The 
number of photons surviving to the observer is $\dot N(\e)\exp[-\tau_{\g\g}(\e;z)]$, 
so the number of absorbed photons is $\dot N_{abs}(\e )= \dot N(\e)[1-\exp[-\tau_{\g\g}(\e;z)]$,
and this also represents the electron injection function $\dot N_{inj}(\g_i)$, with normalization
\begin{equation}
\int_1^\infty d\g_i\;\dot N_{inj}(\g_i)=2\int_0^\infty d\e \;\dot N_{abs}(\e )\;.\;
\label{normNinj}
\end{equation}
because each photon makes two leptons with $\g_i \cong \e/2$. So 
$\dot N_{inj}(\g_i) = 4 \dot N_{abs}(\e )= 4 \dot N(\e )\{\exp[\tau_{\g\g}(\e,z)-1)]\}$.
Using equation (\ref{fe13}) gives the injection function
\begin{equation}
 \dot N_{inj}(\g_i) = {16\pi d_L^2 f_\e \over m_ec^2\e^2  }\,\{\exp[\tau_{\g\g}(\e,z)]-1\}\;,\;\e = 2\g_i\;.
\label{Ninjg1}
\end{equation}
This injection source of e$^+$-e$^-$ loses energy by Compton-scattering photons of 
the CMBR to GeV energies, and this cascade GeV component is considerably dimmed when the leptons 
are deflected out of the beam.

Photons with energy $\approx 1$ TeV make leptons with $\gamma\approx 10^6$,\footnote{At high energies, most of the energy is taken by one of the leptons. See, e.g., \cite{kal11}.} which scatter
the CMBR to $\gamma^2 \e_0\sim 10^3$, or to photon energies $\sim 500$ MeV. Scattering
is in the Thomson regime for $4\gamma\e_0\ll 1$ or $\gamma\ll 2\times 10^8$, that is, 
electrons with energies $\ll 100$ TeV. 

The Thomson energy-loss rate 
\begin{equation}
 -\dot \gamma_{\rm T} = - {d\gamma\over dt}|_{\rm T} = {4\over 3} c\sigma_{\rm T} {u_0\over m_ec^2}\gamma^2\equiv \nu_T\g^2\;,
\label{dotgammaT}
\end{equation}
with $u_0/m_ec^2 = 4.9\times 10^{-7}$ cm$^{-3}$. The solution to the steady-state electron continuity equation is 
\begin{equation}
N(\gamma )= {1\over \nu_{\rm T} \gamma^2}\;\int_{\gamma}^\infty d\gp\;\dot N(\gp )\;.
\label{Ngamma}
\end{equation}
Limiting $\gamma> \gamma_{dfl}$, where $\gamma_{dfl}$ is the deflection Lorentz factor where the 
lepton is deflected out of the jet opening angle by the IGMF.
Taking the luminosity spectrum from Compton scattering \cite{dm09},
eqs.\ (\ref{Ninjg1}) and (\ref{argequation16})  give 
$$\e_s L_{\rm C} (\e_s ) = 4\pi d_L^2 f_{\e_s} =$$
\begin{equation}
{12\pi d^2 c \sT u_0\over \nu_{\rm T}m_ec^2}\;\big({\e_s\over \e_0}\big)^2 
\int_{\g_{low}}^\infty d\g \;{F_{\rm C}(q,\Gamma_e)\over \gamma^4}\int_\gamma^\infty d\g_i\;
{f_\e\; \{\exp[\tau_{\g\g}(\e,z)] - 1\}\over \e^2}\;,
\label{esLCes}
\end{equation}
with $\e =2\g_i$. The interior integral is the injection function from high-energy $\gamma$ rays 
absorbed by photons of the EBL, and depends on a model of the optical-depth function
 $\tau_{\g\g}(\e,z)$ for 
a photon with detected energy $\e$ that was emitted by a source at $z$. 
In general, one uses the expression 
eq.\ (\ref{FC}) for the Compton kernel $F_{\rm C}(q,\Gamma_e)$, though
the isotropic Thomson kernel, eq.\ (\ref{FTredux1}), 
is sufficiently accurate our purposes here.

The lower limit $\gamma_{low}$ in eq.\ (\ref{esLCes}) is the maximum of various constraints 
given by kinematic factors, engine duration, and  $\gamma_{dfl}$.
The kinematic Lorentz factor $\gamma_{knm} $ given 
by eq.\ (\ref{gmin}). In the Thomson regime, $\gamma_{knm} \rightarrow (1/2)\sqrt{\e_s/\e_0}$. 

The deflection Lorentz factor is determined by the condition that
the energy-loss timescale is equal to the timescale for e$^+$-e$^-$ pairs to be deflected
out of the beam. The Thomson energy loss timescale $-\dot\gamma_{\rm T} = \nu_T\gamma^2$ implies
$t_{\rm T} = 1/\nu_{\rm T}\gamma$. The deflection timescale 
\begin{equation}
t_{dfl} = {\theta_j r_{\rm L}\over c}
= \theta_j({m_ec\gamma\over eB})\;,\;{\rm~when~}\lambda_{\rm T} < \lambda_{coh}\;,
\label{tdfl}
\end{equation}
 and $t_{dfl} 
=\theta_j(m_ec\gamma/eB)\sqrt{\lambda_{\rm T}/\lambda_{coh}}$
when $\lambda_{\rm T} > \lambda_{coh}$. Solving gives
\begin{equation}
\gamma_{dfl} =\cases{\sqrt{eB\over \theta_j m_ec\nu_{\rm T}}\;, &  $\gamma_{dfl} > {c\over \nu_{\rm T} \lambda_{coh}}$
 $~$ \cr\cr 
 \big({c\lambda_{coh}\over \nu_{\rm T}}\big)^{1/3}\;\ \big( {eB\over m_ec^2\theta_j} \big)^{2/3}\;, & $
 \gamma_{dfl} < {c\over \nu_{\rm T} \lambda_{coh}} \;$ \cr}\;.\;
\label{gammadfl}
\end{equation}
This constraint implies $\gamma_{dfl} = 3.7\times 10^6 \sqrt{B_{-15}/\theta_{-1}}$ for $\gamma > 7.5\times 10^5/\lambda_{coh}($Mpc), and $\gamma_{dfl} = 6.2\times 10^6 \lambda^{1/3}_{coh}({\rm Mpc})(B_{-15}/\theta_{-1})^{2/3}$ for $\gamma > 7.5\times 10^5/\lambda_{coh}($Mpc).

To avoid solving a time-dependent electron continuity equation, we introduce 
electron Lorentz factor limits $\gamma(\Delta t)$ to define the time the engine was operating.
Following eq.\ (\ref{jesCpowerlaw19}), noting that $f_{\e_s} = \e_sL(\e_s)/4\pi d^2$,
gives
$$
f_{\e_s}  = {3\over 2} \big({\e_s\over \e_0})^2    \int_{\max\big[\sqrt{\e_s\over 4\e_0} ,\g_{dfl},\gamma(\Delta t)\big]}^\infty
\;d\gamma\;\gamma^{-4} \big( 1-{\e_s\over 4\gamma^2 \e_0}\big)\;,
$$
\begin{equation}
 \times \int_\gamma^\infty
d\gamma_i\; {f_\e\; \{\exp[\tau_{\g\g}(\e,z)] - 1\}\over \e^2}\;,
\label{dgammai}
\end{equation}
with $\e = 2\gamma_i$.

The minimum Lorentz factor $\gamma$ related to period $\Delta t$ of activity of central engine is 
determined by equating the time delay with the extra pathlength followed by photons. Thus
\begin{equation}
\Delta  t \cong {\lambda_{\gamma\gamma}(\e_1)+ \lambda_{\rm T}(\gamma )\over 2c} \;\theta_{dfl}^2\;,
\label{Deltatlgg}
\end{equation}
and $\lambda_{tot} = \lambda_{\gamma\gamma}(\e_1)+ \lambda_{\rm T}(\gamma ) = 100\lambda_{100}$ Mpc, $\theta_{dfl} = w\theta_B$, $\theta_B = \lambda_{\rm T}/r_{\rm L}$. From this we derive
$$\gamma(\Delta t) =$$
\begin{equation}
\cases{ \sqrt{ eB\over m_ec\nu_{\rm T}  } \;\big({\lambda_{tot}\over 2c \Delta t }\big)^{1/4} \;\cong \;{9.9\times 10^9\lambda_{100}^{1/4} B_{-15}^{1/2}\over 
[\Delta t({\rm s})]^{1/4}}\;, &  $ {7.5\times 10^5\over \lambda_{coh}({\rm Mpc})}<\gamma$
 $~$ \cr\cr 
\left({ eB\over m_ec^2  }\right)^{2/3} \;\big({\lambda_{tot}\lambda_{coh}\over 2\nu_{\rm T}\Delta t }\big)^{1/3} \;\cong {2.3\times 10^{11}\lambda_{100}^{1/3} B_{-15}^{2/3}\lambda_{coh}^{1/3}({\rm Mpc})\over 
[\Delta t({\rm s})]^{1/3}}\;, & $
 \gamma <{7.5\times 10^5\over \lambda_{coh}({\rm Mpc})}$ \cr}\;.\;
\label{gammadfl2}
\end{equation}

\begin{figure}
\begin{center}
 \includegraphics[width=4.0in]{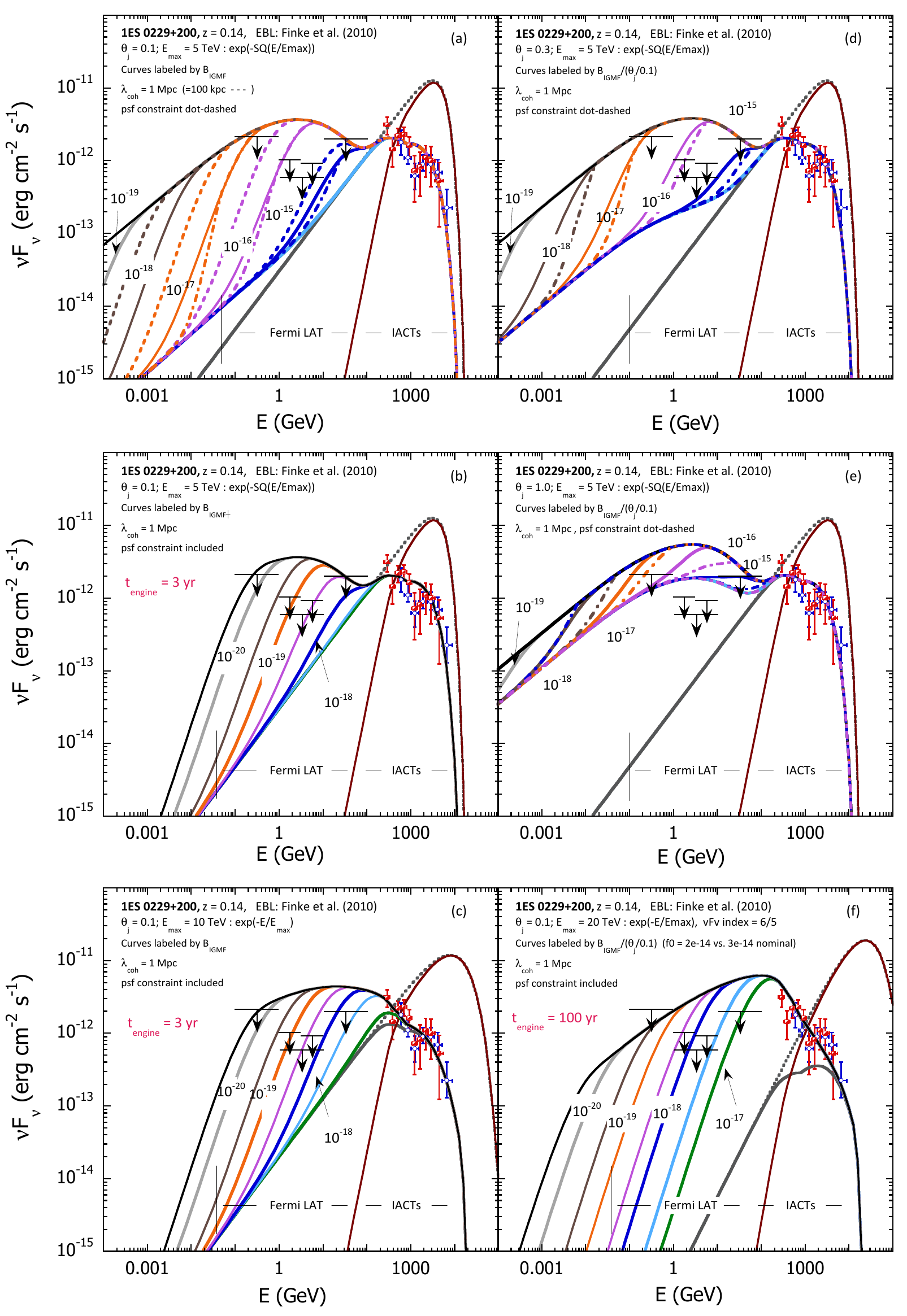} 
 \caption{
Model of cascade radiation spectrum, equation (\ref{fe13}), 
applied to HESS,
 VERITAS, 
and Fermi observations of 1ES 0229+200, using model spectra (solid curves) and
 EBL model of \cite{frd10} 
  to give attenuated source spectra.  (a) Cascade spectra for 
1ES 0229+200 assuming persistent TeV emission at the level observed with HESS and VERITAS, for
different values of $B_{\rm IGM}$ and $\lambda_{coh} = 1$ Mpc (solid) or $\lambda_{coh} = 100$ kpc (dot-dashed)
for a jet opening angle $\theta_j = 0.1$. The intrinsic TeV source spectrum is
given by power-law with $\nu F_\nu$ index $=4/5$ with super-exponential cutoff
$\propto \exp[-(E/5{\rm~TeV})^2]$.
The PSF constraint for the  $\lambda_{coh} = 1$ Mpc case is shown by the dot-dashed curves.
(d) Same as for (a), except that $\theta_j = 0.3$.  (e) Same as (a), except that $\theta_j = 1.0$. 
(b), (c), (d): Here the TeV engine operates for 3 yr, 3 yr, and 100 yr, respectively, but
the intrinsic TeV source spectrum differs. In (b), it is the same as (a). In (c),  it has an exponential cutoff
$\propto \exp(-(E/10 {\rm~ TeV})$.
In (f), the intrinsic source spectrum use parameters of Dolag et al. (2011) \cite{dol10}. 
In (b), (c), and (d), $\lambda_{coh} = 1$ Mpc and cascade spectra are calculated 
for  different values of $B_{\rm IGM}$, as labeled.}
\label{row4}
\end{center}
\end{figure}

Note that there is no separate constraint that the electrons have cooled long enough to scatter significant emission in the given band. In the stationary 
frame, one can define the cooling electron Lorentz factor 
$\gamma_{cool}$ giving the characteristic Lorentz factor of electrons that
have cooled in time $\Delta t$, expressed in terms of the GeV photon energy $E_{GeV}$ to which CMBR photons are Compton-scattered CMB photons.
At low redshifts,  
\begin{equation}
 t_{\rm T}= \Delta t = {3\over4c\sT u_0\gamma_{cool}(\Delta t)} = {\sqrt{3}\over 2c\sT u_0} {\sqrt{{\e_0}\over \e}}\;\cong {7.7\times 10^{19}{\rm s}\over \gamma }
\;\cong {2.2{\rm~Myr}\over \sqrt{E_{GeV}}},
\label{tT}
\end{equation}
we see that it takes nearly a million years for the generated electrons and positrons to cool and begin to 
make their strongest emissions in the GeV -- 10 GeV band of Fermi. But this applies for pair halo emissions made at large angles to the jet axis. The cooling timescale compared to the engine starting time, as measured by an observer within the jetted emission cone, is instead given by equation (\ref{Deltatlgg}), 
which we write as
\begin{equation}
\Delta t({\rm s})\;\cong\; 60\; \lambda_{100}\,{B_{-22}^2
\over E_{\rm GeV}^2}w^2\;.
\label{Deltat62}
\end{equation}
The lack of distinct GeV echoes from impulsive or flaring 
high-energy sources
can limit parts of the $B_{\rm IGM}$-$\lambda_{coh}$ parameter space. Better yet,
 the discovery of such echoes in GRB light curves, as originally proposed by 
Plaga (1995) \cite{pla95}, would finally reveal the primordial magnetic field. 
From equation (\ref{tT}), the Thomson cooling length of relativistic electrons scattering the CMBR is, at redshift $z$, given by 
\begin{equation}
 \lambda_{\rm T}= {2.3\times 10^{30}{\rm cm}\over \gamma (1+z)^4}
\simeq {0.7{\rm ~Mpc}\over \sqrt{E_{\rm GeV}}}\;,
\label{lambdaT}
\end{equation}
where the last expression applies at low redshifts.

Fig.\ \ref{row4} shows calculations with this semi-analytic model \cite{der10}. Here we use a photon-energy dependent expression for $\lambda_{\g\g}$ and perform an integration over the CMBR spectrum. If the jet is persistent on long time scales, then 
the jet opening angle is limited to be $\lesssim 0.4$, as can be seen in the middle panels of this figure. 
Restricting  TeV activity of 1ES 0229+200 to $\approx 3$ -- 4 years, during which the source has been observed, leads to a more robust lower limit of $B_{\rm IGM} \gtrsim 10^{-18}$ G, which can be larger by an order of magnitude if the intrinsic source flux above $\approx 5$ -- 10 TeV  from 1ES 0229+200 is strong.

If there were no intergalactic magnetic field at all, then the pairs made from a source at distance $d$, minus the distance $\lambda_{\g\g}$ over which they are made,
will have cooled to electron Lorentz factors $\gamma$ given by  $d - \lambda_{\g\g} = \lambda_{\rm T} \cong 750{\rm~Mpc}/(\g/10^3)$. It is interesting to think that the pair injection process will have seeded the voids of intergalactic space with an ultra-relativistic nonthermal electron-positron 
component that has already cooled to low energies. Given how active TeV sources are in our present declining phase of the universe (in terms of star formation activity), TeV sources must have been vigorous operating in the early universe.


Without making severe assumptions about the $\gamma$-ray duty cycle and radiative behavior of blazars like
1ES 1101-232 and 1ES 0229+200 on long timescales, the best limit to the strength of the IGMF is $\gtrsim 10^{-18}$ G for  $\lambda_{coh}\gtrsim 1$ Mpc \cite{der10,dol10}. 
Constraints on the value of $B_{\rm IGM}\lesssim 3\times 10^{-19}$ G can be 
obtained from a search for pair echos in the analysis of GRB data \cite{tak08,mur09}, 
so if the larger field is correct, then no GeV echo radiation is predicted from GRBs.

\section{ $\gamma$ Rays from Cosmic Rays in the Galaxy }

Cosmic rays are the most energetic
particles in the universe, and sources of
\begin{itemize} 
\item the light elements Li, Be, B;
\item the Galactic radio emission;
\item the Galactic $\gamma$-ray emission;
\item Galactic pressure;
\item collisional excitation of atoms and molecules;  
\item terrestrial $^{14}$C and $^{10}$Be, with half lifes of $\approx 5700$ yr and $\approx 1.5$ Myr, respectively; and
\item astrobiological effects.
\end{itemize}
Cosmic rays are composed mainly of protons and ions, but also include energetic 
electrons, positrons, and antiprotons.\footnote{High-energy neutrinos could also, depending on definition, be included.} They make up an important particle
background in the space radiation environment and contribute to the space weather. Cosmic-ray electrons emit 
radio and X-ray synchrotron radiation, X-ray and $\g$-ray bremsstrahlung radiation, and X-rays and $\g$ rays from (``inverse") Compton scattering \cite{2011ApJ...739...29B}. It is believed with good reason that Galactic GeV-PeV cosmic rays are accelerated 
by supernova remnants \cite{gs64,hay69}. The origin of the UHECRs, and its relation to the origin of the cosmic rays, is an open question; the hypothesis that their origin involves rotating black holes is developed in  \cite{dm09}.

The difficulty to solve the problem of cosmic-ray origin for the $\sim$GeV cosmic-ray protons and ions that carry the bulk of the 
cosmic-ray energy content is that, being charged particles,
they do not point 
back to their sources as a consequence of intervening magnetic fields that deflect them in transit. Sites of high-energy particle interactions are
identified by $\g$ rays, but ascertaining whether the emission is made by hadronic cosmic rays is 
complicated by 
the possible leptonic origin of most of the $\g$ rays. GeV-PeV neutrinos, by comparison with charged cosmic rays, 
unambiguously point to the sources of the cosmic-ray hadronic interactions, but are faint and difficult to detect.  

Cosmic-ray models use   strong nuclear interaction cross sections  and acceleration, loss and transport 
physics to  derive the cosmic ray intensity and spectrum, 
whether in supernova remnants, diffuse and extended clouds of gas, galaxy clusters, 
the Galactic Center region, or wherever there is a significant column of gas with an illuminating cosmic-ray 
flux. 
The Sun and Solar flares are especially instructive
for study of particle acceleration, transport, and radiation physics from $\gamma$-ray observations. 

\subsection{ $\gamma$ rays from Solar system objects}

Besides direct observations, the effects of cosmic rays
in the Solar system
are traced  via the cosmic-ray induced $\g$-ray flux of
the Sun, moon, and Earth. 
In the Solar cavity, the cosmic-ray intensity is 
modulated by the outflowing Solar 
wind plasma,  making an anti-correlated decrease and lag
in the $\lesssim 10$ GeV/nucleon cosmic-ray flux
reaching in the Solar cavity with a period 
of the 11-year sunspot cycle (one-half the 22 year Solar cycle).
Interaction of these cosmic rays with Solar system objects 
make GeV $\gamma$-ray flux that varies on  these timescales. With 3 years of data taken with {Fermi}, 
this only amounts to $\lesssim 30\%$ of the sunspot cycle, 
the first 2.5 yrs of which were taken  with the Sun in a deep
and extended Solar minimum.

Due to its proximity,  the Earth is the strongest $\gamma$-ray source for the {Fermi}-LAT, which is
why zenith-angle cuts are made on source spectral reconstruction to eliminate the 
interfering effects of the cosmic-ray induced $\gamma$-ray ``albedo" emission (recall footnote \ref{footnote:albedo}). The  
$\gamma$-ray spectrum of the Earth albedo depends on angle to the nadir. At {Fermi}-LAT's
$h = 565$ km orbit, it views the horizon at $\theta_{ndr}\approx  \arcsin(1+h/r_{\rm E})^{-1}\cong 66.5^\circ$; 
the Earth's radius is $r_{\rm E} = 6378.1$ km.
{Fermi} is therefore exposed to $(1/2)(1-\cos (\pi -\theta_{ndr}))\approx 70$\% of the full sky,
with the Earth occulting $\approx 30$\% of the sky. 

Observations of Earth albedo flux reveals a number of interesting things. 
One is the exposure bias toward the North Ecliptic Pole due to favored rocking to the North. 
Recall that Fermi rocks about zenith, early in the mission by $39^\circ$, which was increased to $50^\circ$ later.
Because of the intensity of the Earth albedo, a standard analysis cut is to accept photons only within $105^\circ$
of zenith. For studies of the albedo, then, an acceptance of $\theta_{ndr}\lesssim 80^\circ$ gathers mostly 
cosmic-ray induced terrestrial $\gamma$-ray emission, i.e., albedo. Knowing the albedo spectrum gives, empirically,
a better characterization of the $\gamma$-ray background, and is useful for {Fermi}-LAT calibration.
Deconvolving the $\gamma$-ray spectrum with an Earth atmosphere model and knowledge of the 
interaction cross sections would give the 
primary cosmic-ray spectrum and information about the deflection of primary
cosmic rays by the geomagnetic field.

In \cite{abdprd09}, two data sets are gathered: 
one during the first 90 days of the {Fermi}
mission, and a second for a two-orbit pointing at the Earth's
limb. The analysis covers a total  of $6.4\times 10^6$ events
giving the albedo spectrum in the 
$100$ MeV  -- TeV range, with  218 ph$(>$ 100 GeV), and 16 ph$(>$ 500 GeV).

\begin{figure}[tbp] 
  \centering
  \includegraphics[width=4in]{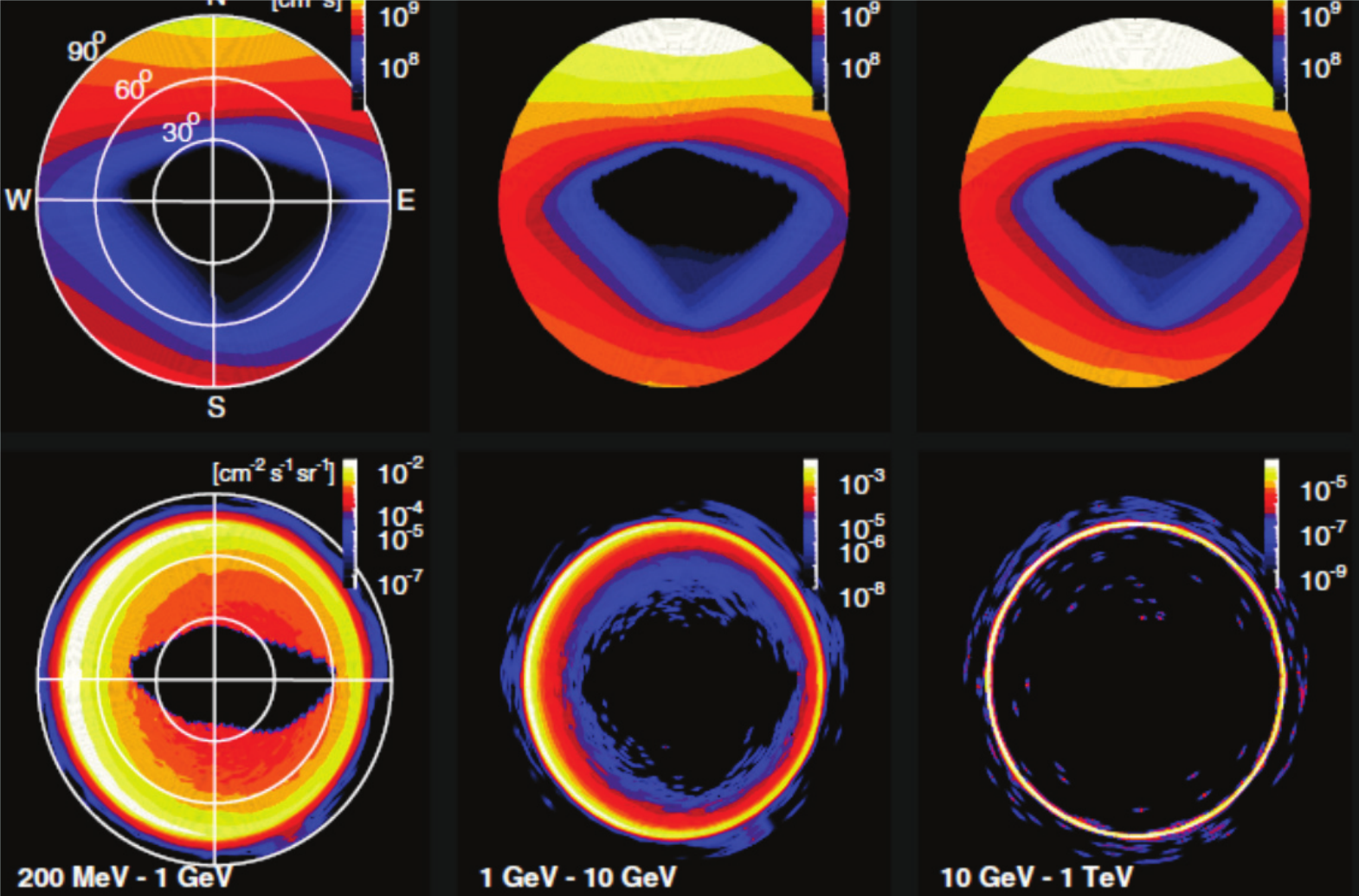}
  \caption{Exposure maps (top) and intensity maps (bottom) of terrestrial $\gamma$-ray albedo emission \cite{abdprd09}.}
  \label{albedo}
\end{figure}

From two-dimensional intensity maps 
with increasing energy range, a beautiful 
high-energy ring forms above 3 -- 10 GeV  due not to Fermi's
energy-dependent PSF but to beaming 
of the emission during formation (see Fig.\ \ref{albedo}).
The bright limb at the Earth's horizon
are $\gamma$-ray light showers made by grazing incidence
cosmic rays  coming directly towards
the LAT. 

The next interesting feature of albedo emission
is the angle-dependent spectrum. At the nadir,
the spectrum is soft, and generated by 
$\gamma$-rays that are backscattered by large angles,
which is considerably less likely due to threshold
and beaming effects at $\gtrsim 1$ -- 10 GeV/nuc, where 
the particle Lorentz factor $\gamma \gtrsim $ a few.
The deflection of cosmic rays by the geomagnetic
field makes an east-west effect that fades out
at high energies. The spectral intensity rises and hardens
until the Earth's limb is reached.  See 
\cite{abdprd09} for details.
 
For the moon, without an atmosphere, a steep secondary nuclear production spectrum with flux 
$F_{-8} = 110 \pm 20$ \cite{gig09} is made as cosmic-ray GeV protons and ions 
impact the surface of the moon, confirming the EGRET detection \cite{1997JGR...10214735T}.
A search for $\gamma$-ray emission from asteroid populations
and other Solar system rocks and dust \cite{2008ApJ...681.1708M,2009ApJ...692L..54M} is currently in progress.

 The $\gamma$-ray emission from the Sun consists
of two components: cosmic rays impacting the surface of the Sun to make an 
albedo-type emission, and cosmic-ray electrons Compton-scattering solar photons to $\g$-ray energies \cite{2006ApJ...652L..65M,os08}. 
Both components of emission are sensitive to the phase of the Solar cycle, and both components have been detected. In analysis of 18 months of data \cite{2011ApJ...734..116A}, 
the solar disk emission is found at the level of 
$F_{-8} \approx 46$. The measured integral flux of the extended non-disk emission from a region of 
$20^\circ$ radius centered on the Sun is $F_{-8}\approx 70$. 
So the Sun and moon are really bright $\gamma$-ray sources, and one has to be alert to data contamination
and spurious variations when the Sun or moon drift past.

The {Fermi}-LAT measurement of the Solar emission confirms Seckel's 
model \cite{1991ApJ...382..652S}, but at a much higher (by a factor $\approx 7$)
flux. The disk emission is practically flat in a $\nu F_\nu$ spectrum
 up to $\approx 8$ GeV, where it begins to fall off. 
The observed spectrum and angular profile is in
good agreement with theoretical predictions for the 
quiet Sun emission \cite{2006ApJ...652L..65M}.

The first {Fermi}-LAT as well as nuclear $\gamma$-ray line flare in Solar Cycle 24 is
the 2010 June 12 M3 flare \cite{ack11flare}. The LAT emission lasted for 
only $\approx 50$ s, compared to past long-duration Solar $\gamma$-ray flares. 
Combined {Fermi} GBM and LAT analysis reveals a rich $\gamma$-ray line 
spectrum superimposed on a continuum 
with a hard-energy tail that is consistent with either a separate 
nonthermal bremsstrahlung component, or pion production in energetic 
nuclear events. The hard bremsstrahlung X-rays and the $\approx 300$ MeV
$\gamma$ rays flare up within 3 s of each other, placing 
strong requirements on acceleration theory and 
target properties. As we enter the active
portion of the Solar cycle, a wealth of new {Fermi} data on particle 
acceleration physics of large Solar flares is anticipated.

Also interesting and timely is the use of {Fermi} for the study of 
terrestrial $\gamma$-ray flashes \cite{1994Sci...264.1313F} that accompany thunderstorms found
mainly in the sub-tropics and tropics. Experiments in nadir-pointing modes
to increase sensitivity to TGFs are currently underway. Interesting analysis 
effects having to do with the shortness of the pulse duration and the 
strong soft X-ray emission affecting the ACD, which is also important for 
Solar flares, reveals {\it Fermi's} capability for this science.

\subsection{GeV photons from cosmic rays}

\begin{figure}[tbp] 
  \centering
  \includegraphics[width=4.6in]{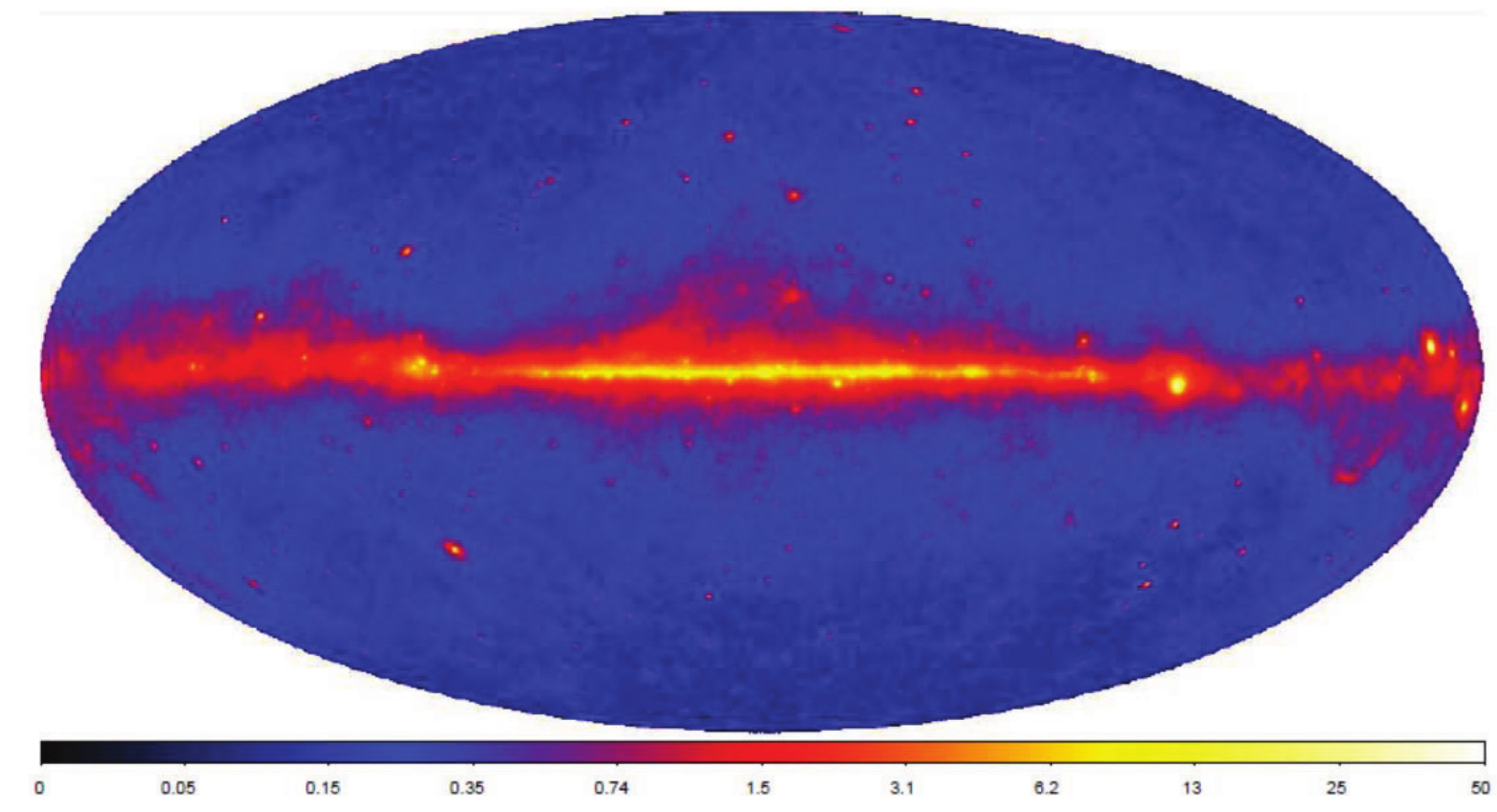}
  \caption{{Fermi}-LAT 2FGL 100 MeV -- 10 GeV all-sky map, using 2 years of sky survey data (2008 Aug 4 -- 2010 Aug 1) \cite{abd122FGL}.  Scale in units of $10^{-7}$ erg/cm$^2$-s-sr. The 2FGL consists of 1873 sources, with 1170 associations and 127 identifications, of which 1319 are at $|b|>10^\circ$.  }
  \label{fig:LATallskymap}
\end{figure}

The cosmic-ray induced $\gamma$-ray glow of the Milky Way is the most pronounced and distinctive feature of the Fermi sky (Fig.\ \ref{fig:LATallskymap}).
A dominant fraction of the Galactic $\gamma$-ray emission is believed to be truly diffuse, and made by cosmic-ray 
 bombardment of gas and dust in the interstellar medium
(ISM). The most important hadronic process for $\gamma$-ray 
production is secondary nuclear production from the collisions of cosmic-ray protons and ions with ISM particles. 
The most abundant secondaries are pions (others are kaons and heavier baryons and baryonic resonances).
The pions decay according to the scheme
\begin{equation}
\pi^0\rightarrow 2\gamma\;,\;\pi^+\rightarrow \mu^++\nu_\mu\;\;,\;\;
\pi^-\rightarrow \mu^-+\bar\nu_\mu\;\;,\;\;
\label{pidecayscheme}
\end{equation}
\begin{equation}
\mu^+\rightarrow e^++\bar\nu_\mu+\nu_e\;\;,\;\;
\mu^-\rightarrow e^-+\nu_\mu+\bar\nu_e\;.
\label{mudecayscheme}
\end{equation}
For the purposes of Galactic cosmic-ray physics, the decays are essentially 
instantaneous, and result in $\g$-ray, electron, and neutrino injection
emissivities (as defined below, eq.\ \ref{npHpi0}) proportional to the cosmic-ray intensity and gas density at
that location. The secondary electrons and positrons, as well as those 
accelerated at cosmic-ray sources,  lose energy during propagation mainly by synchrotron and Compton losses at high energies,
bremsstrahlung losses at intermediate energies, and 
 ionization and Coulomb losses at low energies, 
and can also in principle 
be reaccelerated. 

The diffuse Galactic $\gamma$-ray glow is the superposition of all the radiations made by $\pi^o$-decay $\gamma$ rays, $\g$ rays from cosmic-ray electrons that Compton-scattered the available radiation fields, and electron bremsstrahlung $\g$ rays. Electron synchrotron radiation would only contribute to the Galactic $\g$-ray emission from localized sources, such as pulsar wind nebula.
 
A simple expression for the demodulated cosmic-ray proton intensity in the local interstellar space inferred from measurements of
the  near-Earth
cosmic-ray intensity is 
\begin{equation}
J_p(E_p,\Omega_p) = 2.2E_p^{-2.75} \;{\rm CR~p/cm}^{2}{\rm \mbox{-} s \mbox{-}GeV \mbox{-}sr}\;
\label{JpEpOmegap}
\end{equation}
\cite{der86}, so that the GeV cosmic-ray flux  represents tens of cosmic-ray protons per cm$^{2}$ per s. The cosmic-ray kinetic-energy density is dominated by the kinetic energy $T_p (= E_p - m_pc^2$) of $\sim$GeV protons, and is given by 
\begin{equation}
u_{CRKE} = {4\pi\over c}\;\int_0^\infty dT_p \;T_p\; J_p(E_p,\Omega_p) \cong
0.7 \;{\rm eV/cm}^{3}\;,
\label{ucrke}
\end{equation}
to which ions contribute another factor $\sim$30 -- 50\%.  The dominant elementary hadronic process is $p+p\rightarrow \pi^{\pm,0}$, so astrophysical studies of secondary nuclear production focus on this process. For more detail on radiative processes and cosmic-ray propagation, see \cite{gai90,ber90,lon94,2007ARNPS..57..285S}.
 
\subsubsection{Background modeling}

Reliability of source detection is improved if the intensity of the diffuse background radiation is known, 
because a clumpy gaseous structure illuminated by cosmic-ray induced $\gamma$-ray fluorescence could be mistaken for a point source. 
 Moreover, knowledge of the background 
is required to recognize dim sources, because the background dominates for all but the brightest sources. 
The background model for {Fermi} analysis
takes into account diffuse Galactic $\gamma$ rays  from interactions of cosmic rays with material found in the various phases of the ISM, including
the neutral hydrogen, HI, molecular hydrogen, H$_2$, ionized hydrogen, HII or H$^+$, and the dark gas phase \cite{gct05}. 
The distribution of neutral atomic hydrogen (HI) 
is traced by 21 cm line surveys. The molecular hydrogen (H$_2$)
distribution is derived indirectly, most commonly by using 2.6 mm line observations of carbon monoxide (CO). (The ratio of H$_2$ to CO---the so-called `$X$'-factor, $X = N(H_2)/W_{\rm CO}$\footnote{$N(H_2)$ is the column density of molecular hydrogen, and
$W_{\rm CO}$ is the brightness temperature of CO integrated over velocity \cite{1983ApJ...274..231L}.
} is derived from $\gamma$-ray observations.)
The total atomic and molecular gas column density can also be traced indirectly from extinction and reddening by dust, 
depending on the relative fraction of dust and gas. 
Cosmic rays also interact with ionized hydrogen. The low-density ionized gas can be inferred from dispersion measures of pulsar signals in the radio band.

Cosmic rays move in large-scale galactic magnetic field and diffuse by scattering off magnetic turbulence. In the thick disk of the Galaxy, where the bulk of the diffuse $\gamma$-ray emission is made  at 
 $|b|\lesssim 5^\circ$ -- 10$^\circ$, the cosmic-ray intensity changes with galactocentric distance since the source distribution is peaked at star-forming arms at 4 -- 6 kpc. The cosmic-ray intensity also changes with the distance from the Galactic plane because of escape, and because the diffusion coefficient is energy dependent, with high-energy particles diffusing faster through the Galaxy. In addition to the CMBR, the different distributions 
of background optical and  IR fields,  and location-dependent Galactic magnetic fields, means that cosmic-ray electrons suffer position-dependent energy losses.

 For the electronic component, gradients in the particle distribution can be severe and contribute a Galactic background contribution to the $\gamma$-ray emission.
The  GALPROP (GALactic cosmic ray PROPagation) model \cite{sm98,por08,2004ApJ...613..962S,2000ApJ...537..763S}, started in  1996 and developed independently of Fermi,
determines cosmic-ray diffusion coefficients from fits to cosmic-ray data. The spatial and momentum diffusion equations for cosmic-ray transport are solved, taking into account source injection, energy and fragmentation losses, and energy changes for cosmic-ray protons, ions, and electrons. The stellar optical field, assorted IR and PAH lines in the 10$\mu$ valley, and a FIR dust peak at $\sim 100\mu$ \cite{por08} provide target photons to be Compton scattered by relativistic electrons. The GALPROP model is constrained by the energy dependence of the 
B/C and $^9$Be/$^{10}$Be ratio, from which predictions for the  e$^+$, e$^-$, $\bar p$, and  $\g$-ray spectra and intensity can be made.

Template modeling of $\gamma$-ray emission from Gould belt clouds in Cassiopeia and Cepheus \cite{2010ApJ...710..133A} shows a weak  Galactocentric gradient from the Gould belt to the Perseus arm, and an increase in the $X$-factor from $\cong 0.87\times 10^{20}$ cm$^{-2}$(K km s$^{-1})^{-1}$ in the Gould belt clouds to $\cong 1.9\times 10^{20}$ cm$^{-2}$(K km s$^{-1})^{-1}$ in the Perseus arm. The dark gas represents $\approx 50$\% of the  mass traced by CO.

\subsubsection{Diffuse Galactic $\g$ rays from cosmic rays}

There can be little doubt that cosmic-ray interactions make a large fraction of the $\g$ rays observed with Fermi. This is established most clearly by  {Fermi}-LAT  observations \cite{abd09diffuse} towards  a region in the third quadrant between 
Galactic longitude 200$^\circ$ -- 260$^\circ$ and latitude 22$^\circ$ -- 60$^\circ$ that contains no known molecular clouds. After subtracting point sources and Compton emission, 
the residual 100 MeV -- 10 GeV $\g$-ray intensity exhibits a very strong linear correlation with atomic gas column density.

According to the model of Cordes \& Lazio (2002) \cite{cl02},  the  N(HII) column
density is only $(1$ -– $2)\times 10^{20}$ cm$^{-2}$ and fairly smooth
in the third quadrant regions analyzed in Ref.\ \cite{abd09diffuse}. The contribution from ionized gas has a small effect on the measured emissivity, as  the N(HI) column density ranges from $(1$ -- $13)\times 10^{20}$ cm$^{-2}$.
The measured integral $\g$-ray emissivity measured with the Fermi LAT is $1.63\pm0.05\times 10^{-26}$ photons/s-sr-H-atom and $0.66\pm 0.02\times 10^{-26}$ photons/s-sr-H-atom above 100 MeV and above 300 MeV, respectively, with an additional systematic error of $\sim 10$\%. 

These numbers are explained in the first approximation if cosmic rays pervade the gaseous disk of the Milky Way with the same intensity, eq.\ (\ref{JpEpOmegap}), as observed locally. The pion production rate per unit volume, or differential photon emissivity, is 
\begin{equation}
\dot n_{pH\rightarrow \pi^0}(T_\pi ) = 4\pi n_H \int_0^\infty dT_p \; J_p(T_p,\Omega_p ) \; {d\sigma_{pH\rightarrow \pi^0}(T_p)\over dT_\pi }\;,
\label{npHpi0}
\end{equation}
and the bolometric $\pi^0$ $\g$-ray production rate per H atom is, therefore
\begin{equation}
q_\gamma\equiv{{dN_\gamma\over dtdVd\Omega}/ n_H }\cong 
2\zeta \int_{T_{p,thr}}^\infty dT_p \;J_p(T_p,\Omega_p)\sigma_{pp\rightarrow \pi^0}(T_p)\;,
\label{dndgdtdv}
\end{equation}
including a factor for two photons per $\pi^0$ and a metallicity correction 
$\zeta\approx 1.5$. Between $E_p\approx 1.3$ GeV, just above threshold, and 
$E_p \approx 10$ GeV, the inclusive $\pi^0$ production cross section is approximately linear, and can be written as 
$$\sigma_{pp\rightarrow \pi^0} ({\rm mb}) \cong 4{\rm~mb}(E_p/1.3{\rm~GeV})\;.$$
from which, with eq.\ (\ref{JpEpOmegap}), we find that 
\begin{equation}
q_\gamma \cong 2.2\times 10^{-26}\left({\zeta\over 1.5}\right)
\;[{\rm s}\mbox{-}{\rm sr}\mbox{-}{\rm H}\mbox{-}{\rm atom}]^{-1}
\;.
\label{qgamma}
\end{equation}
 roughly agreeing with the Fermi-LAT measurements  \cite{abd09diffuse} quoted above.

\begin{figure}
\begin{center}
 \includegraphics[width=2.2in]{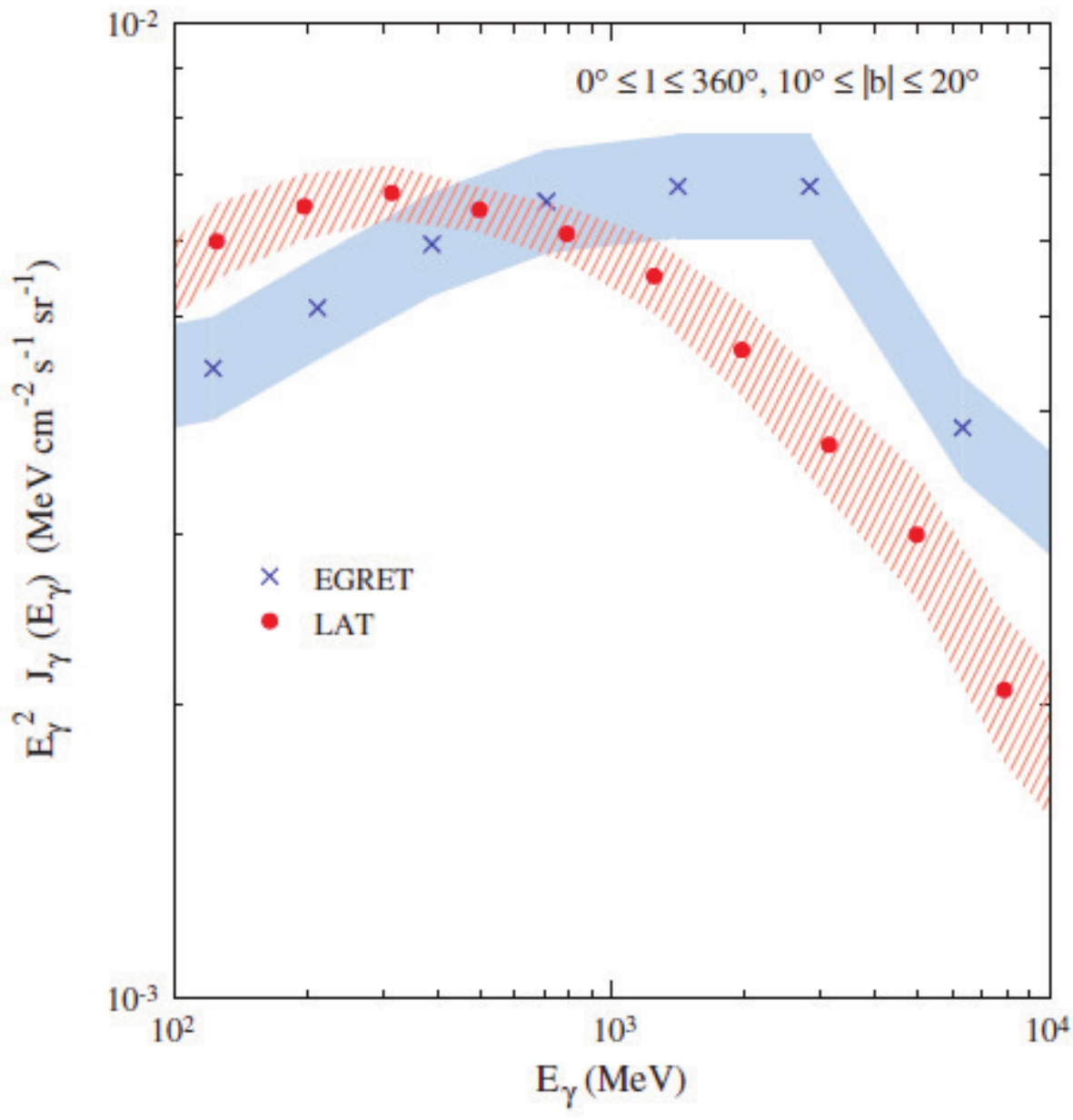}
 \includegraphics[width=2.2in]{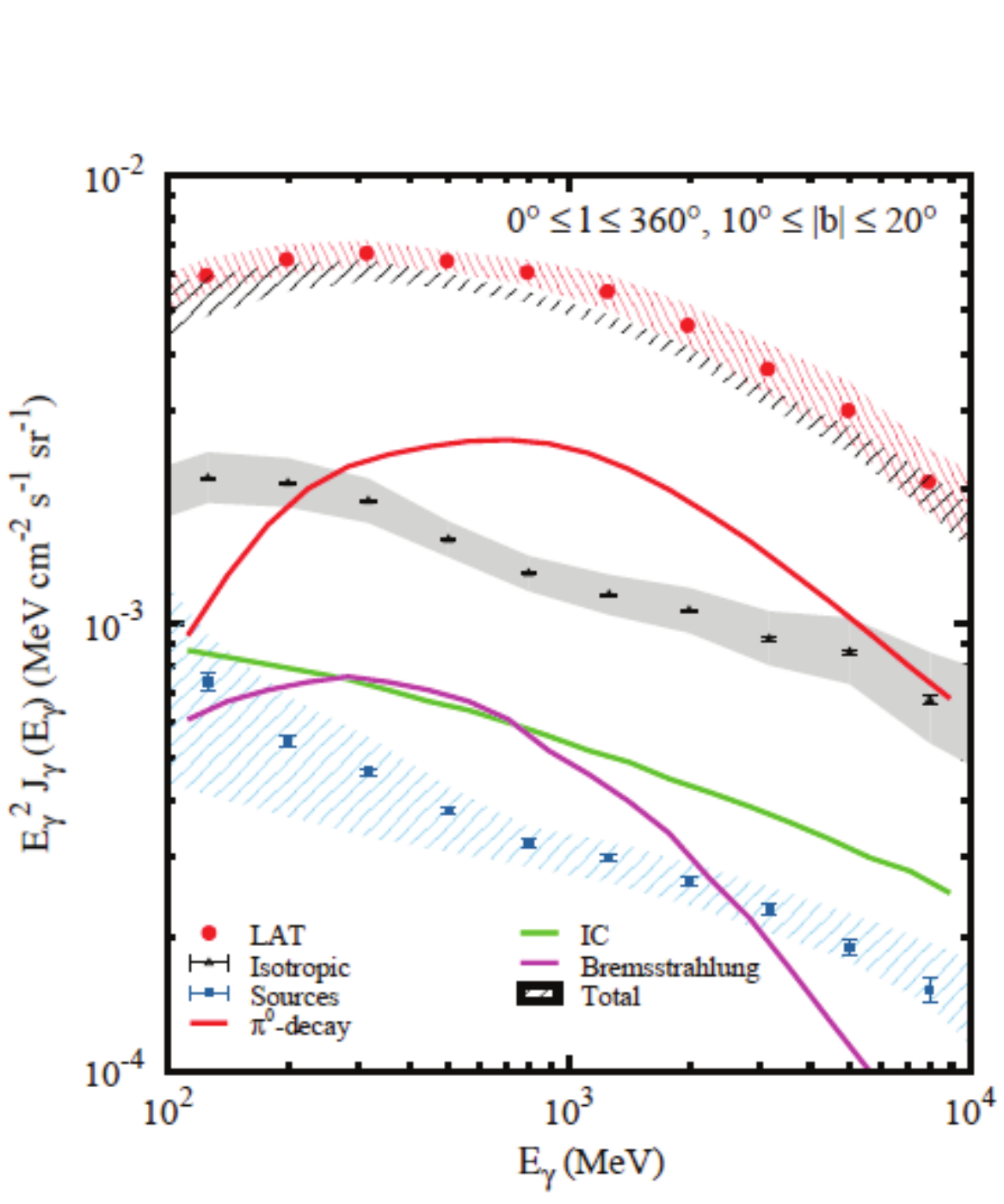}
 \caption{Galactic diffuse emission intensity as measured in the latitude range
$10^\circ \leq |b|\leq 20^\circ$ \cite{abd09gd}. ({\it left}) LAT Galactic diffuse $\gamma$-ray intensity data are given by red dots and 
red cross-hatched error regions; EGRET data are given by the blue crosses and 
blue shaded regions.  ({\it right}) LAT data compared to a model consisting of $\pi^0$ decay
$\gamma$ rays (red); bremsstrahlung from primary and secondary electrons and positrons (magenta);
Compton-scattered soft photons to $\gamma$-ray energies by cosmic-ray electrons (green); 
unidentified background consisting of point sources, isotropic diffuse Galactic, and the isotropic 
extragalactic $\gamma$-ray background. }
\label{diffgal}
\end{center}
\end{figure}

The intermediate latitude, $10^\circ \leq |b|\leq 20^\circ$, Galactic diffuse $\gamma$-ray emission intensity spectra  as 
measured with the {Fermi}-LAT and EGRET are shown in the 
left panel of Fig.\ \ref{diffgal} \cite{abd09gd}. The prominence of the $\pi^0$ decay feature is apparent, as also are the large systematic differences at both low, $\lesssim 300$ MeV,  and high, $\gtrsim 300$ MeV energies. 
In the right panel of Fig.\ \ref{diffgal}, the 
 LAT spectrum shown in the left panel
is compared with spectra from an {\it a priori} 
model (based only on local cosmic-ray data) 
for the diffuse Galactic $\gamma$-ray emission, updated from GALPROP.
 As the Fermi LAT accumulates data, and analysis becomes even more accurate below $100$ MeV for the diffuse class, we can foresee using Fermi data to determine the best interstellar cosmic-ray proton and He ion spectra rather than the other way around. 

\subsubsection{Other Fermi LAT cosmic ray results}

{\it CR e spectrum:} The Fermi LAT can be used as a cosmic-ray electron detector by 
using its anticoincidence dome to identify incoming charged particles, and distinguishing
between leptons and ions from the tracks in the tracker and the calorimeter. The Fermi telescope has an acceptance of $>2$ m$^2$-sr for combined cosmic ray e$^+$ and e$^-$ (CR e), since it cannot distinguish charge.

The extraordinary statistics of the Fermi enabled a precise measurement of the  
CR e spectrum, showing a featureless spectrum consistent with a power law of number index  $\cong -3.04$ between $\approx 25$ and 900 GeV \cite{abd09CRe}. The CR e spectrum is harder than a GALPROP model prediction where the diffusion coefficients tuned to the low-energy electron spectra are extrapolated to higher
energies. The measurement disagrees with the ATIC report \cite{cha08}. The PAMELA measurement of CR e$^+$ \cite{adr09} and the cosmic-ray electron spectrum inferred at TeV energies with HESS data \cite{aha08}, in addition to the Fermi measurement of the cosmic-ray electron spectrum, place constraints on cosmic-ray electron models that require either (or a combination of)
modifications to the propagation characteristics; local sources, most likely pulsars \cite{yks09}; or contributions from dark matter (see lectures by Prof.\ Bergstr\"om).  

 The  2010 Fermi analysis gives the CR e spectrum   from $\approx 7$ GeV to 1 TeV, with a slight hardening above 100 GeV and a softening above 500 GeV \cite{ack10CRe}.

{\it EGRET excess:} This term 
refers to EGRET measurements of the diffuse Galactic $\gamma$-ray intensity that were found, irrespective of direction,
to be in excess of that predicted using the local demodulated cosmic ray spectrum and measured target gas mass  \cite{hun97}. 
Possible explanations included an
\begin{enumerate}
\item unusual location and local cosmic-ray spectrum measured here at Earth;
\item nuclear physics wrong;
\item addition of $\gamma$ ray signal from annihilating dark matter; or
\item EGRET miscalibration.
\end{enumerate}
With the launch of the Fermi Gamma ray Space Telescope,  measurements \cite{abd09gd} of the diffuse Galactic emission 
have been found to favor the latter hypothesis, namely, that EGRET was poorly calibrated above $\approx 5$ GeV.  
The differences in the EGRET and Fermi LAT diffuse Galactic $\gamma$-ray emission are shown in the right panel of Fig.\ \ref{diffgal}.


The GALPROP model decomposes the Galactic plane emission into pionic, electron bremsstrahlung and Compton fluxes, point sources, and an isotropic diffuse background. 
The $\pi^0\rightarrow 2\gamma$ signature is clearly seen in the spectrum of the diffuse Galactic $\gamma$ radiation \cite{abd09gd}. Discrepancies between EGRET and Fermi spectra still remain at $\lesssim 100$ MeV, where systematics effects become severe for both the EGRET spark chamber and the LAT tracker; see Fig. \ref{diffgal}.
The latest analysis of diffuse Galactic $\gamma$ radiation in the Third Galactic Quadrant finds weak evidence at best for a Galactocentric gradient in the cosmic-ray intensity \cite{ack11CR}.

Fermi has also measured the positron flux and fraction using the Earth as a magnetic field
\cite{2011arXiv1109.0521T}.


\subsection{Fermi bubbles}

It is not possible to cover, or even mention, all  {Fermi}-LAT diffuse studies in this short review, 
e.g., the Galactic center region,  the Cygnus Loop \cite{2011ApJ...741...44K}, and $\gamma$-ray emission from massive star clusters and OB associations
like the cosmic-ray filled cocoon region in Cygnus \cite{abd11cyg}, but
we would be remiss not to mention the Fermi bubbles that D.\ Finkbeiner and colleagues 
 find \cite{2010ApJ...717..825D,2010ApJ...724.1044S}. In their analysis, backgrounds determined by template fitting are subtracted from the $\gamma$-ray data,
leaving (apparently) well-defined structures---the Fermi bubbles---at
$\approx 1$ -- 10 GeV, and which probably extend to much higher energies. They are symmetrically arranged
north and south of the Galactic Center with a width of $\approx 40^\circ$, and extend some $\approx 50^\circ$ in height.
They are correlated with the WMAP haze,  which is excess 20 -- 40 GHz emission 
found after correcting for the dipole anisotropy and subtracting out emission traced
by H$\alpha$ and soft synchrotron radiation extrapolated from the 408 MHz Haslam 
survey \cite{2004ApJ...614..186F}. The 1 -- 100 GeV luminosity of both bubbles, if emanating 
from the Galactic Center region, is $\approx 4\times 10^{37}$ erg/s. With a minimum lifetime 
of $\approx 10$kpc/c, its minimum energy content is $\gtrsim 4\times 10^{49}$ erg. The bubble 
luminosity represents $\sim 5$\% of the 100 MeV -- 100 GeV luminosity of the Galaxy \cite{2010ApJ...722L..58S}.

The haze itself, if not an artifact of 
template fitting or mis-extrapolation of the radio emission \cite{2010JCAP...10..019M},  could be formed by dust, spinning 
dust, or dark matter, but a synchrotron origin seems favored, 
especially given the Fermi bubbles. Indeed, a leptonic model for the bolometric SED of the bubbles,
with synchrotron radiation for the WMAP haze and Compton-scattered CMB $\gamma$ rays for 
the {Fermi}-LAT emission, can  be arranged by tailoring the electron distribution \cite{2010ApJ...724.1044S}. A detailed model including inverse Compton scattering off the CMB,
FIR, and optical/UV radiation fields fits the average LAT $\gamma$-ray spectrum \cite{2011PhRvL.107i1101M}, which is practically flat between 1 and 100 GeV.
There is furthermore no energetics problem in a leptonic model, but it requires reacceleration
and a low-energy cutoff in the electron spectrum. The sharp edges reported for the Fermi bubbles,
to be confirmed in ongoing {Fermi}-LAT analyses,  stand in contrast to the 
the WMAP haze. A specific hadronic model  \cite{2011PhRvL.106j1102C} suffers from  weak radiative efficiency in 
the dilute wind from the Galactic center, and lateral diffusion would smear the edges of the bubbles. 

More fundamental, perhaps, is what the existence of the Fermi bubble says 
about our Galaxy. Was it a giant Galactic explosion, the residuum of a
more dynamic period in the history of our Galaxy, a superwind from a past 
starburst episode, or something else?

\subsection{$\gamma$-ray supernova remnants}

 The cosmic-ray power $P_{CR}$ can be estimated as 
$$P_{CR} \sim  \left({1{\rm~eV/cm}^3\over t_{esc}}\right)V_{MW}
\sim {1.6\times 10^{-12}{\rm~erg}/{\rm cm}^3 \times 4\times 10^{66}{\rm cm}^3\over 20{\rm~Myr}}
$$
\begin{equation}
\sim 10^{40}{\rm~erg/cm}^3\;,
\label{lcr}
\end{equation}
using a residence time of $\approx 20$ Myr from analysis of $^{10}$Be abundance in cosmic rays  \cite{gai90}, and the Milky Way volume
$V_{MW}\sim \pi (200$ pc)(15 kpc)$^2\sim 4\times 10^{66}$ cm$^3$. Thus the sources of the Galactic cosmic rays need to supply 
$\gtrsim 10^{40}$ erg/s in the form of nonthermal 
particle power throughout the Galaxy.
One Galactic supernova every thirty years supplying $\approx 10^{51}$ erg in outflowing kinetic energy will inject with 10\% efficiency $\approx 10^{41}$ erg/s of cosmic-ray power, so significant inefficiency can be accommodated if SNRs accelerate the cosmic rays. 
For the cosmic-ray power implied by {Fermi}-LAT $\gamma$-ray observations of the Milky Way for leptonic and hadronic cosmic-ray models, see \cite{2010ApJ...722L..58S}.

The substantial time-averaged kinetic powers of SNRs make them the favored candidate source population for the  hadronic cosmic rays in our Galaxy. 
This power is greater than the  time-averaged power available from, for example, pulsars, galactic X-ray binaries, stellar winds, or novae. In the electromagnetic window, the confirming signature of enhanced cosmic-ray activity is the $\pi^o$ $\gamma$-ray bump \cite{gs64,hay69} peaking at $m_{\pi^0}/2 \approx 70$ MeV in a number spectrum, and at several hundred MeV (depending on the spectrum) in a $\nu F_\nu$ representation. The confirming signature of $\pi^0$ production at $\lesssim 200$ MeV competes with systematics and inaccuracies in the background model which rise due to the large LAT PSF in the lowest decade of its energy range (Table \ref{tab:1}). 

Nearly 300 SNRs are known, mostly through radio detections. The Chandra X-ray catalog
contains nearly 100 SNRs,\footnote{hea-www.harvard.edu/ChandraSNR/} and ROSAT Galactic SNRs must number several 
dozen. The poor EGRET PSF made it impossible to identify specific SNRs with sites of enhanced $>100$ MeV emission, though several associations could be made, including IC 443, W28, and W44 \cite{sd95,esp96,tor03}.    

In the 1FGL, 41 SNRs are associated with LAT $\gamma$-ray sources, and 3 are, through morphological features, 
identified with SNRs (Table \ref{tab:2}), namely W51C, W44, and IC 443.
By the time of 2FGL, 62 associations and 9 (morphological) identifications of
SNRs and PWN with Fermi LAT and other wavelengths have been made (Table \ref{tab:2}). 
The identifications represent 
6 SNRs, adding also W28, W30, and the Cygnus Loop \cite{2011ApJ...741...44K},
and three PWNe, namely J0835.3-4510 (Vela X), J1509.6-5850 (MSH 15-52; possibly the remnant of a supernova
in 185 CE), and J1826.1-1256 (HESS J1825-137). A fourth PWN, MSH 15-52, was found 
in a search of off-pulse emission of PSR B1509-58 \cite{2010ApJ...714..927A}.

  Some important questions to be answered are
\begin{itemize}
  \item  Are cosmic-ray protons and electrons accelerated at SNR shocks, and with what relative efficiencies?
\item  With what efficiency is shock kinetic energy converted to cosmic-ray energy? 
\item What is the spectrum of accelerated particles escaping from the shock acceleration site?
\item What is the maximum particle energy, and do SNRs accelerate CRs up to or beyond the knee of the cosmic-ray spectrum?
\item Does magnetic field amplification take place in the vicinity of SNRs?
\end{itemize}

\begin{table}
\centering
\caption{Historical Supernovae \cite{2003LNP...598....7G}}
\label{tab:histSNR}       
%
%
\begin{tabular}{lccccc}
\hline\noalign{\smallskip}
SN  & Type$^a$  & $~~~$Distance$~~~$  &  Size$^b$  & $~$$\gamma$-ray detection$^c$ & $~$ Pulsar$^d$   \\ 
&  & (kpc) &  & GeV/VHE &  \\
\noalign{\smallskip}\hline\noalign{\smallskip}
1987A  & II  &  55  & $0.8^{\prime\prime}$ &    &  \\
1680; Cas A$^e$  & II  & 3.4  &    $5.9^\prime \times 5.5^\prime$    & Y/y &  \\
1604; Kepler&    Ia  &  5  &      $4.5^\prime \times 3.7^\prime$    & & \\
1572; Tycho&    Ia  &   2.4   &  $8.7^\prime \times 8.6^\prime$  & Y/y & \\
1181; 3C58 (?) &  II  & 2.6  &  $6^\prime \times 3^\prime$   &  & J0205+6449  \\
1054; Crab &   II  &    2    & $2.3^\prime \times 2.1^\prime$ & y/Y  & Crab\\
1006 &  Ia  &   1.6  & $32^\prime \times 26^\prime$  &   ?/Y &  \\
393; RX J1713.7-3946$~~$  &  II (?) & 1  & $\approx 40^\prime\times 40^\prime$   & y/Y &  \\
386? &    &   &    &  &  \\
369? &    &   &    &  &  \\
185; MSH 15-52 (?) &  II  &   &    & y/Y &  1509-58  \\
\noalign{\smallskip}\hline
\end{tabular}\\

{\noindent $^a$\cite{2009ApJ...706L.106L,2011ApJ...732..114L} }\\
{\noindent $^b$Chandra X-ray sizes$^{16}$, except for SN 1987A (HST),
 RX J1713.7-3946 \cite{2006A&A...449..223A}, and 3C 58 (radio; \cite{1988srim.conf..331R}) }\\
{\noindent $^c$Flux measured at GeV or VHE energies; Y: larger $\nu F_\nu$ flux, y: smaller $\nu F_\nu$ flux}\\
{\noindent $^d$See Table \ref{tab:EGRETpulsars}}\\
{\noindent $^e$Detection by Flamsteed disfavored \cite{2003LNP...598....7G}}
\\
\end{table}

Fermi has now detected  historical/young  ($\lesssim 3000$ yr; see Table \ref{tab:histSNR}) SNRs, namely Cas A (1680 CE), Tycho (1572 CE), and RX J1713.7-3946 (393 CE),
 intermediate age ($\approx 10^4$ yr) SNRs, e.g.,  IC 443 and W51C, and 
middle-aged ($\gtrsim 2\times 10^4$ yr) SNRs, including W44, W28, the Cygnus Loop, and G349.7+0.2.  
All SNRs detected with Fermi, other than Cas A, show evidence
for molecular cloud interactions.

Detecting the $\pi^o$ decay feature in {Fermi}-LAT SNR 
spectra depends on good background modeling to expose the low-energy ($\lesssim 200$ MeV)
flux. Shock compression of pre-existing cosmic rays could be argued to account for apparent 
cosmic-ray production at SNR shocks, but this seems less plausible now that we have 
measured GeV $\gamma$-ray luminosities ranging from $\approx 10^{33}$ erg/s for the Cygnus loop, seen as 
an $\approx 1^\circ$ radial loop and the largest resolved $\gamma$-ray SNR, to  $\approx 10^{34}$ -- $10^{35}$ erg/s for young
SNRs, and $\approx 10^{35}$ -- few $\times 10^{36}$ erg/s for SNRs showing strong molecular cloud interactions \cite{2011arXiv1104.1197U}. 
This implies cosmic-ray kinetic energy on the order of ${\cal E}_{CR}({\rm erg}) \sim 10^{36} L_{36}t_{pp} \cong
10^{51} L_{36}/n({\rm cm}^{-3})$,  in accord with the SNR hypothesis for cosmic-ray origin.
This estimate is not, however, so strong, given that distance uncertainties
can easily lead to factor of 4 or more uncertainty in $L$. 


\begin{figure}
\center
\includegraphics[scale=0.6]{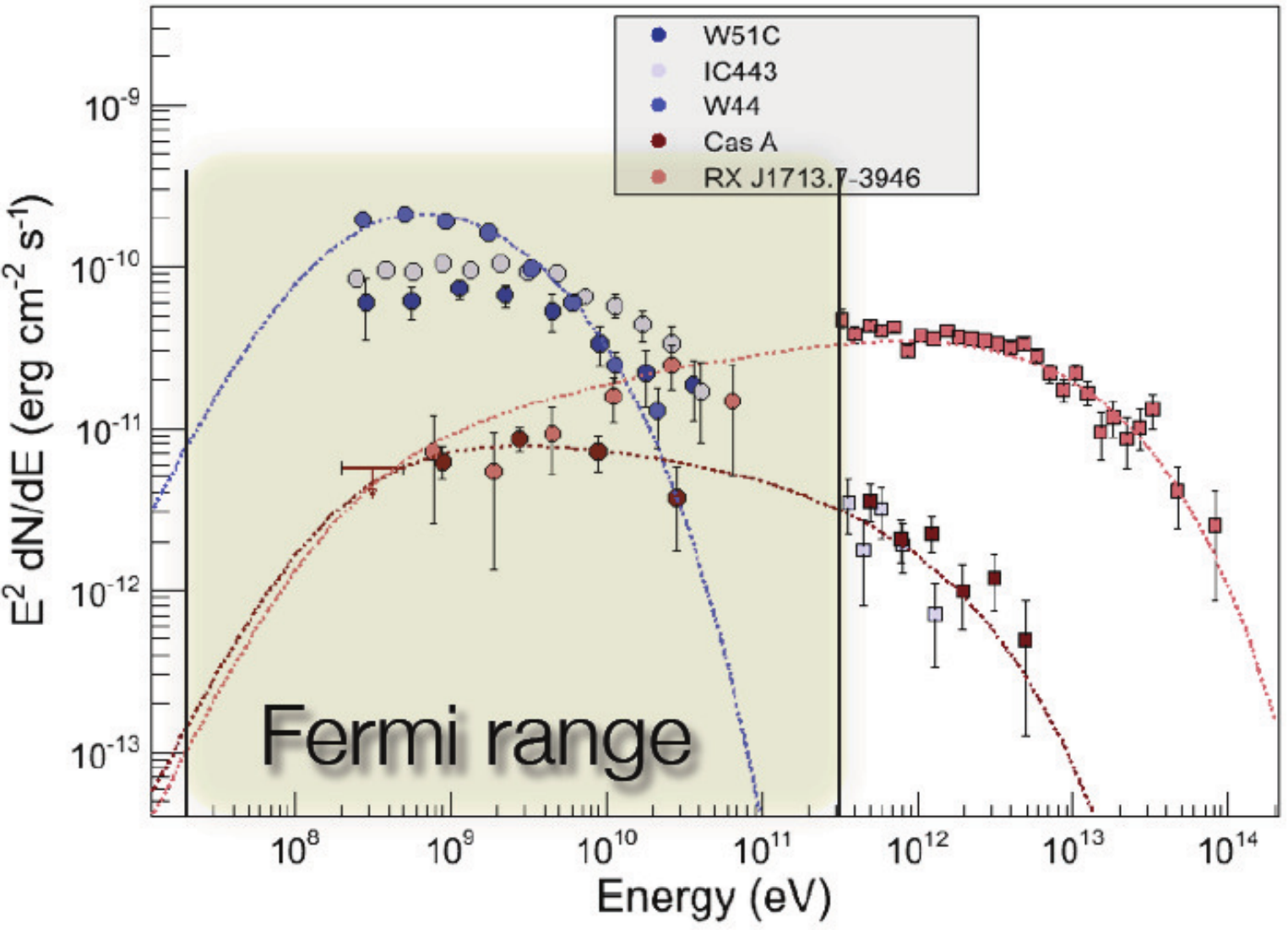}
\caption{Combined Fermi-LAT and TeV spectra of selected SNRs \cite{2010APS..APR.S3002F}.}
\label{Fermi_LAT_SNRs}
\end{figure}
 
For Cas A, one of the youngest SNRs detected with {Fermi} 
\cite{2010ApJ...710L..92A}, combined VHE (MAGIC, VERITAS, and HEGRA) data
show a $\nu F_\nu$ peak at a few GeV, with ambiguous evidence for 
a low-energy cutoff, but clear evidence for a TeV steepening (see Fig.\ \ref{Fermi_LAT_SNRs}). Cas A, 
with an extent of $\approx 6^\prime$, appears almost as a point source 
for the LAT.
The somewhat older SNR RX J1713.7-3946
is famous for being the first SNR for which a VHE (HESS) map was made \cite{2004Natur.432...75A}.
The TeV emission is extremely 
well-correlated with X-ray maps, e.g., HESS and Suzaku \cite{2008ApJ...685..988T}.
The {Fermi}-LAT map  \cite{2011ApJ...734...28A} displays an asymmetry towards the north, in the direction 
of a molecular cloud where the X-ray and TeV emission is enhanced.
The GeV spectrum itself is remarkably hard, with spectral index $= 1.5\pm 0.1$. 
The $\nu F_\nu$ spectrum of this remnant peaks between $\approx 200$ GeV and 1 or 2 TeV, 
with a rapid cutoff at higher energies.  Nevertheless, the spectrum of 
RX J1713.7-3946 extends to $\approx 100$ TeV, making it the source of the most energetic
$\gamma$ rays yet detected. Its hard spectrum cannot be made by conventional hadronic models.
One-zone leptonic models can fit the X-ray/$\gamma$-ray spectra of RX J1713.7-3946, as can two-zone 
models.

The prototypical Type Ia Tycho SNR has also been recently reported as a GeV source
with a GeV-- TeV $\gamma$-ray spectrum implying, if hadronic, a proton spectrum with a $-2.3$ number index \cite{2011arXiv1108.0265G}. Claims have been made that the combined LAT/VERITAS spectrum requires hadrons \cite{2011arXiv1105.6342M}, but a 
two-zone leptonic model with both bremsstrahlung and Compton processes
can fit the spectrum as well \cite{atoyan2011}. Because of the weakness of the GeV/TeV $\gamma$-ray flux, 
the $\nu F_\nu$ peak of Tycho can be constrained to be  $\lesssim $ few GeV, and the spectrum is not 
well-resolved in the crucial energy range $\lesssim 300$ MeV where the $\pi^0$ bump might be evident. 
For this purpose, a better SNR might be IC 443 \cite{2010ApJ...712..459A}, a 3 -- 4 kyr old core collapse SN (like
Cas A and RX J1713) surrounded by clouds of gas. With an 0.75$^\circ$ radio radius, it is resolvable with Fermi. 
The SED of IC 443 is a flat $\nu F_\nu$ spectrum between $\approx 300$ MeV and 5 GeV with a peak at 1 or 2 GeV.
The spectrum extends to VERITAS/MAGIC VHE energies like a power-law with $\nu F_\nu$ slope equal to $\approx -0.6$. 

The middle-aged shell SNR W44 \cite{2010Sci...327.1103A} has an age of $\approx 20$ kyr. 
The $\gamma$-ray shape closely following the 4.5 $\mu$ Spitzer IR image 
which traces shocked $H_2$. Its spectrum peaks at 1 GeV or less, and it falls off sharply at higher energies,
so is not detected with the VHE telescopes. Its physical extent precludes the $\gamma$-ray emission from 
originating from a pulsar, though a pulsar, B1853+01 with age $\approx 20$ kyr, is found in this SNR and could be associated with W44.
If so, W44 is probably the result of a core-collapse SNR, and therefore Type II.

As a final example of {Fermi}-LAT observations of SNRs, consider
W51C, the first resolved GeV SNR at GeV energies. It is an $\approx 10$ kyr old remnant with extended GeV emission compatible
with the location and shape of the ROSAT emission  \cite{2009ApJ...706L...1A}.
The spectrum of W51C is similar to that of W44, peaking below $\approx 1$ or 2 GeV, and 
falling off steeply at higher energies.  HESS weakly detects emission
at about 1 TeV from W51C, (whereas W44 is not detected at VHE energies), 
implying a  $\nu F_\nu$ spectral index $\approx -0.5$ between 10 GeV and 1 TeV.

In terms of cosmic-ray origin theory, a few remarks can be made. First, almost all {Fermi}-LAT
SNRs, other than Cas A, show evidence for interactions with molecular clouds, in particular, those
with OH(1720 MHz) maser emission from the OH hydroxyl molecule tracing dense, shocked gas \cite{2009ApJ...706L.270H,2011arXiv1108.4137F}. Out of 24
known maser SNRs reported in \cite{2009ApJ...706L.270H}, 10 have GeV and/or TeV associations, 
and 6 have both. Detection of illuminated molecular cloud complexes in front
of the SNR shock could reveal the existence of runaway cosmic rays that  
more likely would be protons than electrons.

Second, nonthermal cosmic-ray protons 
lose energy on the secondary nuclear production timescale $t_{pp} = (n\sigma_{pp} c)^{-1} \approx 35$ Myr/n(cm$^{-3}$), 
whereas nonthermal electrons lose energy through bremsstrahlung on the free-free timescale
$t_{ff} = (n\sigma_{ff} c)^{-1} \approx 6\times 35$ Myr/$n$(cm$^{-3}$), where $\sigma_{ff} \approx 
\alpha_f \sigma_{\rm T} \cong \sigma_{pp}/6$. In either case, the process is radiatively inefficient
unless the target density $n\gg 1$ cm$^{-3}$, but the nonthermal electron bremsstrahlung is less efficient
by no more than an order of magnitude than the nuclear production cross section. The 
nonthermal bremsstrahlung or  secondary nuclear
production model  face 
energetics problems if they  requires $ \gtrsim 10^{50}$ erg in cosmic rays.

Third, ionization/Coulomb losses can harden a nonthermal electron spectrum, making $\gamma$ rays from a leptonic 
bremsstrahlung model   masquerade as 
a $\pi^0$ feature. A full spectral model for multiwavelength production from SNRs requires, most generally,
particle acceleration at the forward and reverse shocks, zones of different magnetic field strength 
at which particle acceleration can occur, leptonic bremsstrahlung and Compton scattering, and secondary 
nuclear production. 

Perhaps there is a trend in the SEDs of SNRs that can reveal the likely 
$\gamma$-ray production mechanism (Fig.\ \ref{Fermi_LAT_SNRs}). For leptonic models, both  (i) nonthermal bremsstrahlung
 Compton scattering, are potentially feasible. For hadronic models, only (iii)
secondary nuclear/particle process is effective in the SNR environment.
Confrontation of these three nonthermal processes with  multiwavelength SNR data 
imply power and spectral and morphological constraints that can in principle
identify the dominant radiation process, though in practice, this has not been so simple.
But establishing a trend from 
hard to soft $\gamma$-ray spectra with age, even scaled to the Sedov age, is oversimplified,
as is evident from the rather soft spectrum of Tycho. 

For useful studies, a plot of the VHE (100 GeV -- 10 TeV) flux divided by the 
GeV (300 MeV -- 30 GeV) flux vs.\ SNR age (to avoid uncertainties in distance measurements) 
for Type Ia and Type II SNe might 
provide some insight on the evolution of particle acceleration with remnant age. 
At present, it still seems premature to claim that the problem of the origin of the Galactic 
cosmic rays is solved.

\subsection{Nonrelativistic shock acceleration of electrons}

First-order Fermi acceleration is highly developed as 
a mechanism to accelerate cosmic rays \cite{dru83,be87}. 
Here we illustrate this process by calculating the nonthermal spectrum of test particles accelerated
at a shock-discontinuity in density and velocity. 
We use a continuity equation approach,\footnote{The treatment of \cite{1979A&A....76..276A} assumes energy-independent escape timescale representing some second-order Fermi acceleration scenarios.} which complements the approach to 
shock acceleration using the convection-diffusion equation and probability 
arguments \cite{dm09}. 

\begin{figure}
\center
\includegraphics[scale=0.5]{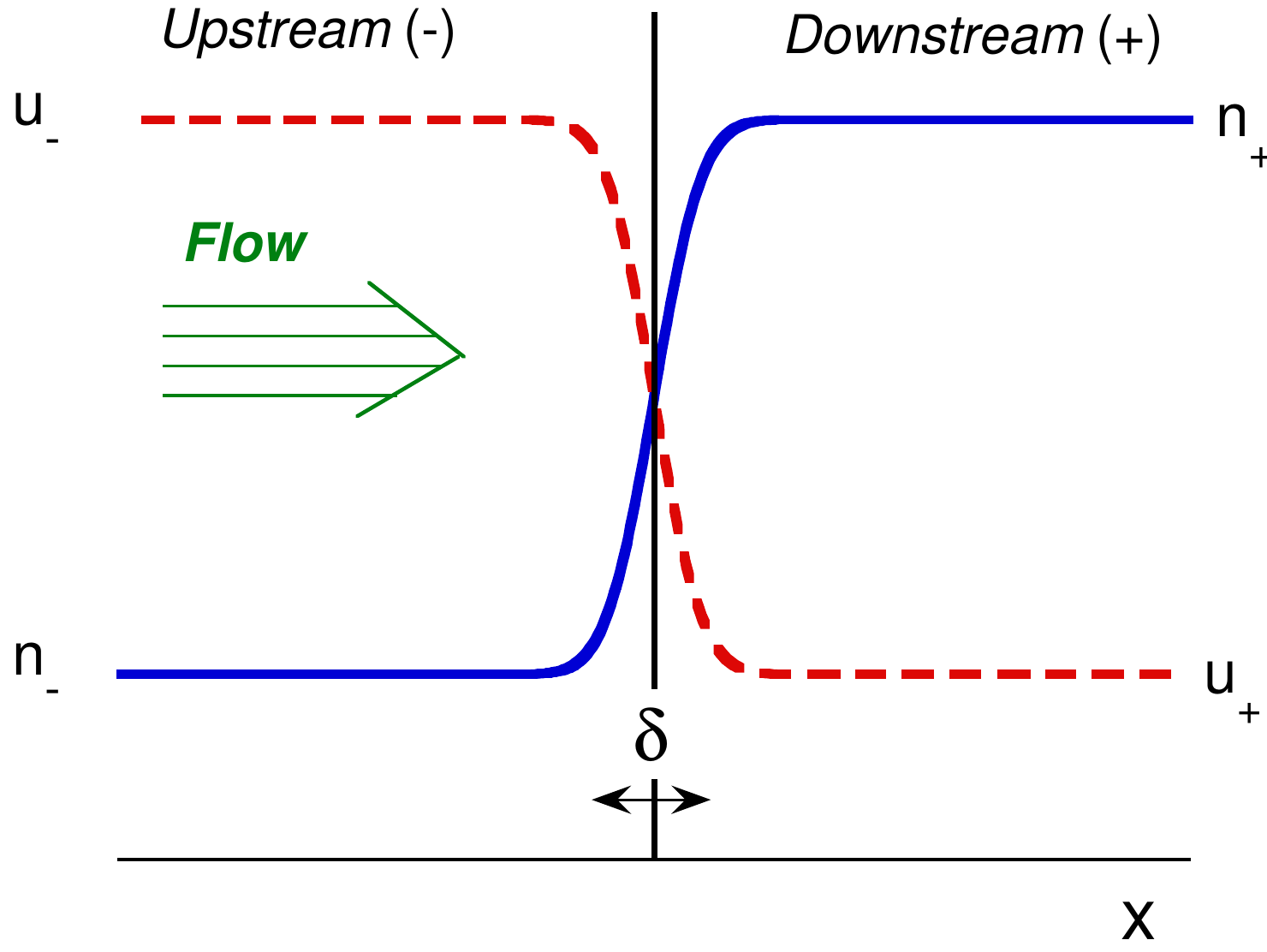}
\caption{Shock geometry in rest frame of 
shock. In this figure, $\delta$ is the shock width. }
\label{shock_structure}
\end{figure}

The geometry we consider is particle acceleration at 
a discontinuity in velocity and density, as illustrated in Fig. \ref{shock_structure}. The upstream
unshocked material has speed $u$ in a frame where the supersonic flow intercepts
gas at rest. For a strong nonrelativistic shock with compression ratio $\chi = 4$, 
 the upstream ($-$) flow approaches
with speed $u_- = \beta_-c = 4\beta c/3$ and the downstream ($+$) flow
recedes with speed $u_+ = \beta_+c = \beta c/3$ in the comoving primed frame
stationary with respect to the shock. Consequently  $u =
u_--u_+ = \beta c$.  Here we treat
the acceleration of relativistic nonthermal
particles with Lorentz factor $\gamma \gg 1$ and speed $\beta_{par}c \approx c$.

In the general case, the compression ratio
\begin{equation}
\chi = {u_1\over u_2} = {u_-\over u_+}\;.
\label{chidefinition}
\end{equation}
For a lossless adiabatic shock wave, 
\begin{equation}
\chi = {\h +1 \over \h - 1 + 2/{\cal M}_1^2} \;\mq {4 \over 1 + 3/{\cal M}_1^{2}} 
\label{chisolution}
\end{equation}
(eq.\ (13.8), \cite{dm09}), where $\hat \gamma$ is the adiabatic index, and  
$\h = 5/3$ for a nonrelativistic monatomic gas. 
When ${\cal M}_1 \gg 1$ and $\h = 5/3$, $\chi \rightarrow 4$.
It follows from eq.\ (\ref{chidefinition}) and the condition
$u = u_- - u+$ that 
\begin{equation}
u_- = {\chi u\over \chi-1 }\;,\;\; u_+ =  {u\over \chi -1}\;, 
\label{newu}
\end{equation}
so that in the limit ${\cal M} \gg 1$, $u_- \rightarrow 4u/3$ and $u_+ \rightarrow u/3$ in the frame 
comoving with the shock.


The change of energy of a relativistic particle with dimensionless momentum $p =  \beta \gamma =\sqrt{\gamma^{2}-1}\rightarrow 1, {\rm~when~}\g\gg 1,$
in a complete cycle of Fermi acceleration is, 
from eq.\ (12.10) \cite{dm09}, given by 
\begin{equation}
\big({\Delta \g\over \g}\big)_{\rm F1} \cong {4\over 3}\beta\;,
\label{deltapoverp}
\end{equation}
where $u = \beta c = u_--u_+$. For second-order Fermi acceleration, 
\begin{equation}
\big({\Delta \g\over \g}\big)_{\rm F2} \cong {4\over 3}\beta^2\;,
\label{deltapoverpF2}
\end{equation}
and both expresssions receive an added boost $\propto \Gamma^2$, at least in the 
first cycle of acceleration, for  shock acceleration by a relativistic
flow with speed $\sqrt{1-1/\Gamma^2} c$.

The rate of energy gain, or acceleration rate, in nonrelativistic first-order 
Fermi acceleration is given
for particles \begin{equation}
\dot \g_{\rm FI} \cong {\Delta \g\over t_{cyc}}\;,
\label{dotgFI}
\end{equation}
where the cycle time $t_{cyc}$ is given by 
$$t_{cyc} = {4\over v}\;
({\kappa_-\over u_-}+{\kappa_+\over u_+})\; = 
{4\over v u_- } (\kappa_-+\chi \kappa_+)
$$
\begin{equation}
={mc^2\over QB_-}\,{4p\over 3u}\,{\chi-1\over \chi}\left(\eta_-+\chi\eta_+{B_-\over B_+}\right)
\cong \;{mc^2\over QB_-}\,{4\gamma\over 3u}\,{\chi-1\over \chi}\propto \gamma\;.
\label{tcyc}
\end{equation}
\cite{lc83a}, 
where $v \cong c$ is the particle speed,
$\chi = u_-/u_+ = \rho_+/\rho_-$ is the compression ratio, and
$\rho = m n$ is the mass density. The spatial diffusion coefficient
$$\kappa_\pm \;=\; {1\over 3} \lambda_\pm v \;=\; {1\over 3}\eta_\pm r_{{\rm L}\pm} v \;=\; 
{1\over 3}\eta_\pm
r_{{\rm L}\pm}^opv\;,$$ 
here writing the diffusion coefficients in
terms of the parameters $\eta_\pm$ that give the particle
mean-free-paths scaled to the values in the local magnetic field. In the Bohm diffusion approximation,
the diffusion mean-free path is set equal to the Larmor radius, 
so $\eta_\pm = 1$.
Thus the Bohm approximation is 
$$\kappa_{{\rm B}\pm} = {1\over 3}r_{{\rm L}\pm} v\;,$$
where the Larmor radius
\begin{equation}
r_{{\rm L}\pm} = r_{{\rm L}\pm}^o p = {mc^2\over QB_\pm}\;p\;.
\label{rL}
\end{equation}
The acceleration rate of nonthermal relativistic particles by 
first-order shock acceleration can therefore be expressed as 
\begin{equation}
\dot \g_{acc} \;= \;{4\beta\over 3t_{cyc}}\gamma= {4\over 3}\,{u\over c t_{cyc}}\gamma\;.
\label{dotgaccF1}
\end{equation}
The acceleration rate is independent of $p$ or $\gamma$ because $t_{cyc} \propto \gamma$
from eq.\ (\ref{tcyc}). 

In nonrelativistic shock acceleration, the escape probability per cycle is
\begin{equation}
{\cal P}_{esc} \;=\;4\beta_+
\;=\; 4{u_+\over c} \;=\;{4u\over c(\chi - 1)}
\;.
\label{Pret}
\end{equation}
\cite{bel78,bel78a}.
The escape time is therefore
\begin{equation}
t_{esc} = {t_{cyc}\over {\cal P}_{esc}} = {t_{cyc}\over 4\beta_+}\;.
\label{tesc}
\end{equation}

The steady-state particle continuity equation with gains, losses, and escape takes the form (eq.\ (C10), \cite{dm09})
\begin{equation}
{\partial \over \partial \gamma} [{\dot \g}\;n(\gamma )] 
+
{n(\gamma)\over t_{esc}(\gamma)}
= \dot n(\gamma)\;.
\label{dgdtnnot}
\end{equation}
Eq.\ (\ref{dgdtnnot}) has solution, for 
an energy-gain process $\dot\g>0 $, given by (eq.\ (C11), \cite{dm09})
\begin{equation}
n(\gamma ) = \dot\g ^{-1} \int_1^\gamma d\gp \;\dot n(\gp )
\exp\left[-\int^\gamma_{\gamma^\prime} 
{d\gamma^{\prime\prime}\over t_{esc}(\gamma^{\prime\prime}) 
\dot\gamma(\gamma^{\prime\prime})}\right]\;.
\label{ngamma2}
\end{equation}

Eqs.\ (\ref{dotgaccF1}) and (\ref{tesc}) imply
\begin{equation}
t_{esc}(\g )\dot \g_{acc}(\g ) = {\beta \over 3\beta_+}\gamma = {\chi - 1\over 3}\gamma\;,
\label{escgdot}
\end{equation} 
so
\begin{equation}
n(\gamma ) = {3\dot n_0\over 4\beta\gamma}\,t_{cyc} \; 
\left({\g\over \g_0}\right)^ {-3/(\chi - 1)}\;.
\label{ngamma1}
\end{equation} 
noting that this represents the differential number density spectrum of particle accelerated
at the shock discontinuity. 

Taking a $\delta$-function source injection  $\dot n(\g) = \dot n_0 \delta (\gp - \gamma_0)$, 
the injection spectrum downstream where the particles are no longer subject
to acceleration at the shock is 
\begin{equation}
\dot n_{ds}(\gamma ) = {n(\g )\over t_{esc}} = {4\beta_+\over t_{cyc}}\,n(\g )
= {3\dot n_0\over (\chi - 1)\gamma_0} \; 
\left({\g\over \g_0}\right)^ {-A_{tp}}\;.
\label{ndsgamma}
\end{equation} 
where
\begin{equation}
A_{tp}\;=\; {u_- + 2 u_+\over u_- - u_+} = {2+\chi\over \chi -1}\;
\label{Achidef}
\end{equation}
is the test-particle number index well known in studies of nonrelativistic shock acceleration \cite{be87}.

Now consider shock acceleration with the addition of Thomson/synchrotron 
radiative losses, so
\begin{equation}
\dot \gamma = \dot \g_{acc} + \dot \g_{rad} = {4\over 3}\,{\beta\over t_{cyc}}\g -\nu \g^2\;.
\label{dotgammaaccrad}
\end{equation}
Following the approach above gives
\begin{equation}
n(\gamma ) = {\dot n_0\over \left({4\over 3}{\beta\gamma\over t_{cyc}}-\nu\g^2\right)} \; 
\exp\left[ -{3\over \chi -1}\int_{\gamma_0}^\g {dx\over x(1-kx^2)}\right]\;,
\label{ngamma4}
\end{equation} 
where 
\begin{equation}
k = {3\nu T\over 4\beta}\;,\;{\rm and}\;\;T = {m_ec^2\over QB_-}{4\over 3u} \left( 
{\chi - 1 \over \chi }\right)\;(\eta_- + \chi \eta_+{B_-\over B_+} )
\label{kK}
\end{equation} Therefore
\begin{equation}
\dot n_{ds}(\gamma ) = {n(\g )\over t_{esc}} 
= {3\dot n_0\over (\chi - 1)(1-k\g^2)^{3/2}} \; 
\left({\g\over \sqrt{1-k\g^2}}\right)^ {-A_{tp}}\;.
\label{ndsgamma1}
\end{equation} 
This is a pileup spectrum for $\chi > 5/2$, that is, when $A_{tp}$ is harder than $3$.
The maximum electron energy is given by $1 - k\g^2 = 0$, or 
\begin{equation}
\g_{max} = {1\over \sqrt{k}} = \sqrt{4\beta \over 3\nu T}\;.
\label{gammamax}
\end{equation}
Even though the spectrum piles up, the number of particles and the energy of 
the accelerated particles is convergent. 

The synchrotron emission made by this electron distribution has a maximum value 
at dimensionless photon energy 
\begin{equation}
\e_{syn,max} = {3\over 2} {B\over B_{cr}}\g^2 = {9\pi e\over B_{cr} \sT}
\left({\chi \over \chi -1 }\right) \,\beta^2 \;
{\left({B_-B_+/ B^2}\right)\over \left( 1+{u_{ph}\over u_{B}}\right)   ( \eta_- + \chi \eta_+{B_-\over B_+}   )}\;\label{esynmax}
\end{equation}
(cf.\ \cite{aa99}). Thus
\begin{equation}
\e_{syn,max} = {27\over 8\alpha_f}\left({\chi \over \chi -1 }\right) \,\beta^2 \;
{{B_-/ B_+}\over \left( 1+{u_{ph}\over u_{B}}\right)   ( \eta_- + \chi \eta_+{B_-\over B_+}   )}
\;.
\label{esynmax1}
\end{equation}
The apparent divergence due to the term $\chi - 1$ in the denominator of eq.\ (\ref{esynmax1}) can be seen 
not to arise, considering that $\chi u^2/(\chi - 1)= u_- u$. 
The leading term, $\e_{syn,max} \cong {27/8\alpha_f}\cong 462$, or $E_{syn,max}\cong 236$ MeV, represents a bound for nonrelativistic 
shock acceleration, remarkably close in value to the cutoff energy of the Crab pulsar wind nebula \cite{abd10crab}.
This indicates that the pulsar wind nebula of the Crab is formed by a wind termination shock moving out at
mildly relativistic velocities. Small changes in the bulk speed from a knot in the Crab pulsar wind, 
as imaged with Chandra, could produce the Fermi and AGILE flares from the Crab \cite{abd10crabflares}. 
Better imaging at $\approx 100$ MeV would localize the emission source, but is hardly possible with Fermi. 

Addition of a diffusion term will produce smoothed, realistic pile-up electron injection distributions
formed in first-order shock acceleration.  The addition of a diffusion term is under study 
in work with P.\ Becker. Note that these pileup functions differ from the Schlickeiser pile-ups \cite{sch84,sch85} where
escape is always independent of particle energy. At relativistic energies and with relativistic flows, 
the maximum dimensionless synchrotron energy is $\e_{syn,max} \cong (27/8\alpha_f)\Gamma/(1+z)$, which
is relevant when interpreting the maximum photon energies in GRBs as a consequence of synchrotron emission formed
by particle acceleration at an external shock.

\section{ $\gamma$ Rays from Star-Forming Galaxies and  Clusters of Galaxies, and the Diffuse Extragalactic $\gamma$-Ray Background }

\subsection{$\gamma$ rays from star-forming galaxies }

Galaxies with ongoing star formation, most notably the Milky Way in which we live, are illuminated at $\gamma$-ray eneregies by secondary  
products of  cosmic-ray interactions with gas and dust.  The characteristic $\gtrsim 100$ MeV $\gamma$-ray luminosities of normal star-forming galaxies are $\sim 10^{38}$ -- $10^{40}$ erg s$^{-1}$, 
some factors of $\sim 10^3$ -- $10^{10}\times$ smaller than those of active galaxies. Yet the star-forming galaxies vastly outnumber the AGNs. For example, the space density of a typical L$_*$ spiral galaxy like the Milky Way is $\approx 3\times 10^6$ -- $10^7$ Gpc$^{-3}$, by comparison with the space density of FR II radio galaxies, which is $\approx 2000$ Gpc$^{-3}$ \cite{gbw10}. (The volume of 1 Gpc$^3$ extends to about $z = 0.15$ from the present epoch.)

Besides the Milky Way, the Large Magellanic Cloud, detected earlier with EGRET with flux $F_{-8} \cong 19$, is now measured with Fermi LAT at the level of $F_{-8} \cong 26\pm 2$ \cite{abd10lmc}. At a distance of $\approx 50$ kpc, the LMC has about 10\% of the mass and a supernova rate $\approx $20\% of the Milky Way. The LAT resolves the LMC, and finds that 30 Doradus, its major star-forming region, is a bright source of $\gamma$ rays that does not consist of significant point source contributions. The $\g$-ray spectrum is consistent with an origin in cosmic-ray production.  Its $\gamma$-ray emission correlates well with massive star forming regions and an ionized H$^+$ template, but more poorly with neutral or molecular gas distribution.  The $\g$-ray emission morphology is surprisingly compact, and indicates that cosmic rays are accelerated in star-forming regions that are not very diffusive, thus accounting for the bright compact emission centered around 30 Doradus. 

Our other notable dwarf companion galaxy, the Small Magellanic Cloud, is also for the first time detected in $\gamma$ rays. The LAT measures a flux of $F_{-8} = 3.7\pm 0.7$ \cite{abd10smc}  from an extended, $\sim 3^\circ$ region. Unlike the LMC, the $\g$-ray emission from the SMC is  not clearly correlated with the distribution of 
massive stars or supernova remnants, though the emission may trace supergiant shells.

\begin{table}
  \begin{center}
  \caption{Properties of Star-Forming Galaxies$^a$
}  \label{tabstar}
 {\scriptsize
  \begin{tabular}{ l c c c c c c  c}\hline 
{\bf Galaxy} & { $d$} & { R$_{\rm SN}$} & { M$_{\rm Gas}$} &  $F_{-8}$$^b$ & $4\pi d^2 F_\gamma$ & L$_\gamma^c$ & Index \\  
& (kpc) & (century$^{-1}$) & $(10^9 M_\odot$) & & ($10^{41}$ ph/s)& ($10^{39}$ erg/s) &  \\
\hline
MW & -- & $2.0\pm 1.0$ & $~6.5\pm 2.0~$ &  -- & $~11.8\pm 3.4^d~$ &$1.2\pm 0.5$ & $2.2 \pm 0.15$ \\
LMC & $52\pm 2~$ & $0.5\pm 0.2$ & $~0.67\pm 0.08~$ &  $26.3\pm 2.0$ & $~0.78\pm 0.08~$ &$0.041\pm 0.007$ & $2.26 \pm 0.11$ \\
SMC & $61\pm 3~$ & $\approx 0.12^e~$ & $~\approx 0.45~$ &  $3.7\pm 0.7$ & $~0.16\pm 0.04~$ &$0.008\pm 0.003$ & $2.23 \pm 0.12$\\  
M31 & $780\pm 30$ & $1.1\pm 0.2$ & $7.7\pm 2.3$ & $0.9\pm 0.2$ & $6.6\pm 1.4$ & $0.43\pm 0.09$  &  $2.1 \pm 0.22$\\ 
M82 & $3600\pm  300~$ & $20\pm 10$ & $~2.5\pm 0.7~$ &  $1.6\pm 0.5$ & $~250\pm 90~$ &$13\pm 5$  &  $2.2 \pm 0.2$ \\ 
N253 & $3900\pm 400~$ & $20\pm 10$ & $~2.5\pm 0.6~$ &  $0.6\pm 0.4$ & $~110\pm 70~$ &$7.2\pm 4.7$    & $1.95\pm0.4$ \\ 
\hline
  \end{tabular}
  }
 \end{center}
\vspace{1mm}
 \scriptsize{
$^a$Refs.\ \cite{abd10starburst,abd10smc,2010ApJ...722L..58S}\\
$^b$$F_\gamma = 10^{-8}F_{-8}$ ph $(>100$ MeV)/cm$^{2}$-s\\
$^c$Fluxes and luminosities in $100$ MeV -- 5 GeV range\\
$^d$Value is strongly dependent on assumed size of Galactic halo used to model MW;  \cite{2010ApJ...722L..58S} find
$0.6 \lesssim F_{-8} \lesssim 1.0$\\
$^e$Ref.\ \cite{tls94}
}
\end{table}

Though not unexpected \cite{pf01}, the discovery of starburst galaxies at GeV \cite{abd10starburst} and TeV energies is important to test predictions of cosmic-ray origin. The two nearest starburst galaxies, M82 and NGC 253, each at a distance of $\approx 4$ Mpc, were detected at a level of $F_{-8} \cong 1.6$ and $F_{-8} \cong 0.6$, respectively \cite{abd10starburst}, with the LAT, soon after being reported as VERITAS \cite{acc09m82} and HESS \cite{ace09ngc253} sources, respectively. Properties of the brightest star-forming galaxies at $\gamma$-ray energies are shown in Table \ref{tabstar}. Although correlations with mass, SN rate, and their product can be extracted, the situation, as exhibited by the distribution of $\gamma$-ray emission in the LMC, is  far more complex. As can be seen from Fig.\ \ref{GammavsL}, starburst galaxies define a separate track in the spectral index vs. luminosity plane.

Now that the LAT sensitivity for star-forming galaxies has reached beyond our neighboring galaxies, we can expect the number to grow. The Andromeda galaxy, M31, 
has been detected at a level of  $F_{-8} \cong 0.9$ \cite{abd10m31}. Other starburst and infrared luminous galaxy detections from 
sources like Arp 220 \cite{tor04} or Mrk 273 are keenly anticipated as the LAT photon statistics accumulate.

\subsection{$\gamma$ rays from clusters of galaxies}

Clusters and superclusters of galaxies are the largest manifestations of 
the ongoing structure formation. Regions 
collapse by gravitation, driven primarily by dark matter fluctuations.
The magnitude of the density fluctuation determines the formation time, 
insofar as larger structures form by accreting smaller clumps
in a hierarchical merging scenario. The result, as seen in 
numerical simulation codes, is lumpy, continuous accretion forming 
filaments and webs of enhanced density and 
magnetic field. 

Clusters of galaxies are the most energetic 
events in the universe since the big bang,
releasing as much as $G M_1 M_2/ R \sim 10^{63}$ -- $10^{64}$ erg, here taking 
$M_1\sim 10^{15} M_\odot\sim 10 M_2$ and $R \sim 1$ Mpc. Yet they are hardly the most luminous, 
since the energy is released over a Hubble timescale $H_0^{-1} \sim 4.3\times 10^{17}$ s $
\sim 13.6$ Gyr, implying a power of $\sim 2\times 10^{45}$ -- $2\times 10^{46}$ erg/s. 
This energy, carried primarily in the gravitational potential of the dark matter, 
 goes into heating, turbulence, and gas motions. In the events of cluster mergers, 
turbulence is generated in wakes, and shocks are formed in the collision. Nonthermal 
particles are accelerated by shocks and turbulence.

Evidence for nonthermal particle acceleration in merging clusters of galaxies
is already known from the existence of radio halos and relics. 
The relativistic electrons and positrons making this radio emission are either accelerated directly or produced as secondary pairs by cosmic-ray protons and ions colliding with particles in the intracluster medium.  The unpolarized central {\it radio halos}, $\sim$ Mpc size, with a morphology similar to the X-ray bremsstrahlung, reflect particle acceleration at the merger shocks between two merging clusters, intermittent AGN activity, and dynamical friction of galaxies in the hot, $\sim$ keV, thermal plasma. By contrast, {\it radio relics}, also about $\sim 1$ Mpc in size, which lie on the cluster outskirts, display elongated morphologies, and are up to $\sim 50$\% polarized, may reflect a different origin from the radio halos. Most likely, they represent ongoing accretion of pristine gas from the big bang.   

The halo radio emission of, e.g., the Coma cluster of galaxies, is very soft above $\approx 1$ GHz, and can be modeled in a merging cluster framework if the system is observed soon after the cluater and subcluster have merged \cite{2003ApJ...594..709B}.  By scaling nonthermal cosmic-ray proton and ion energy to the electron energy required to make the observed radiation, predictions for the $\gamma$-ray emission can be made. The identification of a nonthermal hard X-ray feature as Compton-scattered CMBR would mean a larger nonthermal electron total energy, giving more optimistic predictions for $\gamma$-ray emission from clusters of galaxies. In any case, clusters of galaxies act as storage volumes for the cosmic-ray protons \cite{1997ApJ...487..529B}, which have radiative lifetimes exceeding the Hubble time. Because of the long, $\sim 10$  Gyr, crossing time of electrons, versus a radiative lifetime of $\sim 0.1$ Myr, in-situ acceleration of pre-existing relativistic electrons by turbulence \cite{bbcg04,bb05} is now favored to explain radio halos and relics. 

The Fermi LAT has many high-priority cluster candidates, including not only Coma and Abell 2256, but Hydra, Centaurus, and Fornax. The Perseus cluster  is dominated by the AGN 3C 84 in the central elliptical galaxy NGC 1275, and is detected with the Fermi LAT \cite{abd09e}. But this $\gamma$-ray emission is from a variable blazar core (though there may be steady   emission at the level of $F_{-8} \lesssim  4$). Radiation mechanisms involving the energetic leptons together with the decay of neutral pions produced by hadronic
interactions have the potential to produce abundant GeV photons. 
Using data from 2008 August to 2010 February, 
upper limits of 
33 galaxy clusters, selected according to their proximity,
mass, X-ray flux, temperature, and non-thermal activity 
were reported by the Fermi collaboration \cite{ack10_clusters}. 
The flux upper limits, in the 0.2 –- 100 GeV range, are typically at the level of $F_{-8}\cong 0.1$ -- 5. These results limit the 
 relativistic-hadron-to-thermal energy density ratio 
to be $\lesssim 5$\% -– 10\% in several clusters \cite{2009JCAP...08..002K,2011MNRAS.410..127B}.

As disappointing as it has been that $\gamma$ rays have not yet been detected from galaxy clusters, we are still learning about the injection conditions in the low Mach number merger shocks, the high Mach number accretion shocks, and the injection conditions of turbulence models. A purely secondary production model where all the radio emission results from secondary electrons formed in secondary nuclear production is likely ruled out \cite{ack10_clusters,bru11}, and dark matter annihilation cross sections are further constrained.

\subsection{Extragalactic $\gamma$-ray background and populations}

\begin{figure}[t]
\center
\includegraphics[scale=0.5]{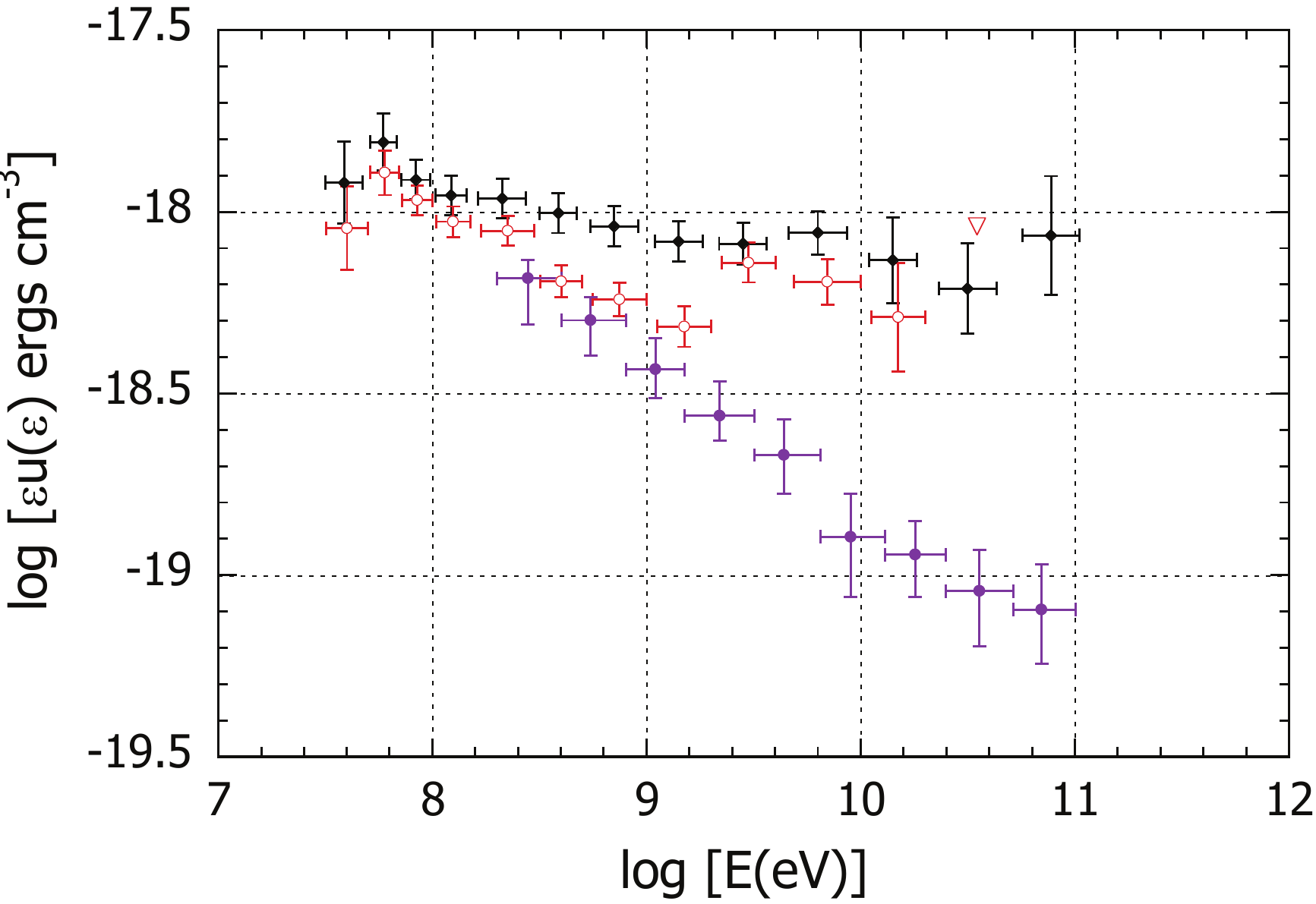}
\caption{Fermi measurement of the diffuse extragalactic gamma-ray background (EGB) \cite{abd10id} (purple filled circles). Also plotted is the EGRET result \cite{sre98} (filled black circles) and background intensity from GALPROP analysis of EGRET data \cite{smr04} (open red circles).}
\label{Fermiisodiff}
\end{figure}

Fig.\ \ref{Fermiisodiff} shows the spectrum of the diffuse extragalactic $\gamma$-ray background (EGB) obtained
in an analysis of thefirst 10 months of {Fermi}-LAT science data 
in the range 200 MeV -- 100 GeV based on the GALPROP model for the Galactic emission \cite{abd10id}, alongside the EGRET EGB \cite{sre98} and the EGRET EGB based on a  GALPROP analysis of
the Galactic diffuse EGRET emission  \cite{smr04}.
 The Fermi-LAT intensity extrapolated to 100 MeV
based on the power-law fit $I(> 100{\rm ~MeV}) = (1.03\pm 0.17)\times
10^{-5}$/cm$^2$-s-sr is significantly lower than
that obtained from EGRET data, namely $I_{\rm EGRET}
(> 100{\rm ~MeV}) = (1.45\pm 0.05)\times
10^{-5}$/cm$^2$-s-sr  \cite{sre98}. The GALPROP analysis of the EGRET data  \cite{smr04} agrees, however, with the {Fermi} flux extrapolated to 100 MeV.
Furthermore, the {Fermi}-LAT 
spectrum is compatible with a featureless power law
with index $-2.41\pm 0.05$  \cite{abd10id}. This is significantly softer
than the EGRET spectrum, with index $-2.13\pm 0.03$  \cite{sre98}. To check that the different spectra are not due
to the instrumental point-source sensitivities,
a threshold flux $F_{-8}^{thr} = 10$, comparable to the
average EGRET sensitivity, and an isotropic 
$\gamma$-ray intensity like that quoted above is measured.
{\it  Therefore, the
discrepancy cannot be attributed to a lower threshold for
resolving point sources.} Most likely the relatively hard
spectral slope for the EGRET EGB was due to underestimation of the 
EGRET effective area
above $\approx $ few GeV \cite{shk08}, or due to lack of a model for Compton $\gamma$ rays, which can make a significant contribution to the EGB at
high latitudes.

The origin of the EGB 
is an interesting open question. 
 Because of the 
preponderance of blazars in the Third EGRET catalog \cite{har99}, a large fraction of this 
emission was expected to be contributed by blazars, though some pre-{Fermi} studies \cite{der07,2009ApJ...702..523I} found  
that blazars made only a small fraction of this emission.
The detection of $\sim$1000 high Galactic latitude
sources \cite{abd122FGL}, most of which are blazars \cite{2011arXiv1108.1420T}, has allowed a precise characterization of the contribution of 
blazars to the EGB. 
Such studies have shown 
 that unresolved  blazars contribute 
$\lesssim$30\,\% of the EGB emission \cite{2010ApJ...720..435A,aje11}.

If not blazars, then other source classes must be invoked to account for 
the EGB. Possibilities include emission from star-forming 
galaxies, misaligned blazars (i.e., radio galaxies),
$\gamma$ rays made by particles accelerated by structure-formation
 shocks in clusters of galaxies,
\cite{loeb00, miniati00,gabici03},
and dark matter; see references in \cite{2010ApJ...720..435A}.
Star-forming galaxies such as our own Milky Way are known to be $\gamma$-ray 
emitters due to the interaction of cosmic rays with interstellar gas and
radiation fields \cite{abd10starburst}. The superposition of the numerous but individually $\gamma$-ray weak
star-forming galaxies can contribute a greater fraction of the EGB than the rare, individually 
bright $\gamma$-ray blazars  \cite{2010ApJ...722L.199F,2011ApJ...736...40S}.  {Fermi} detection 
 of several star-forming galaxies, including  NGC 253, M82, and the LMC  \cite{abd10starburst},
supports this possibility, though the exact percentage remains highly uncertain. 

\begin{figure*}[h]
\begin{center}
\begin{tabular}{cl}
 \includegraphics[scale=0.27]{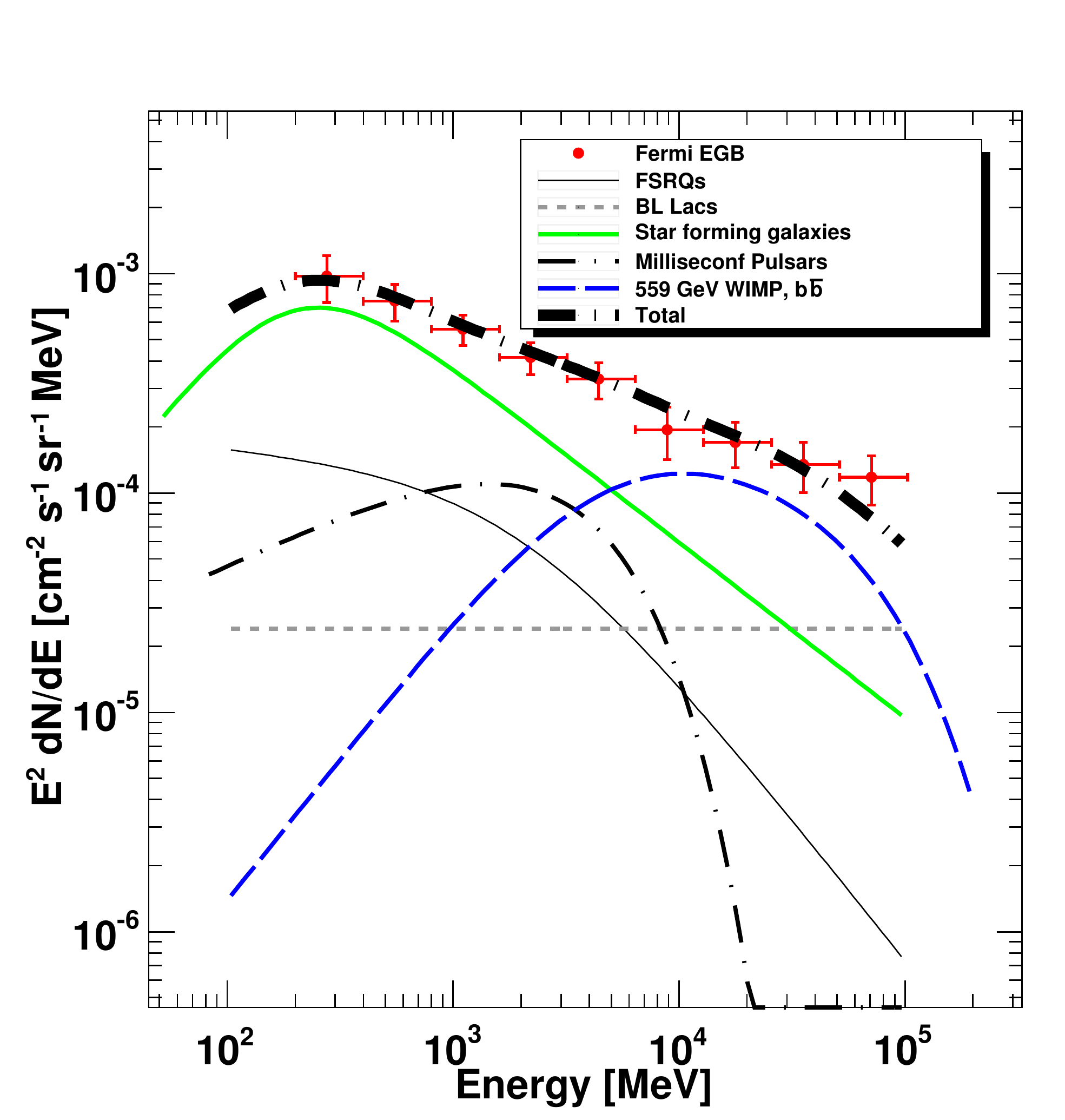} &
\includegraphics[scale=0.35]{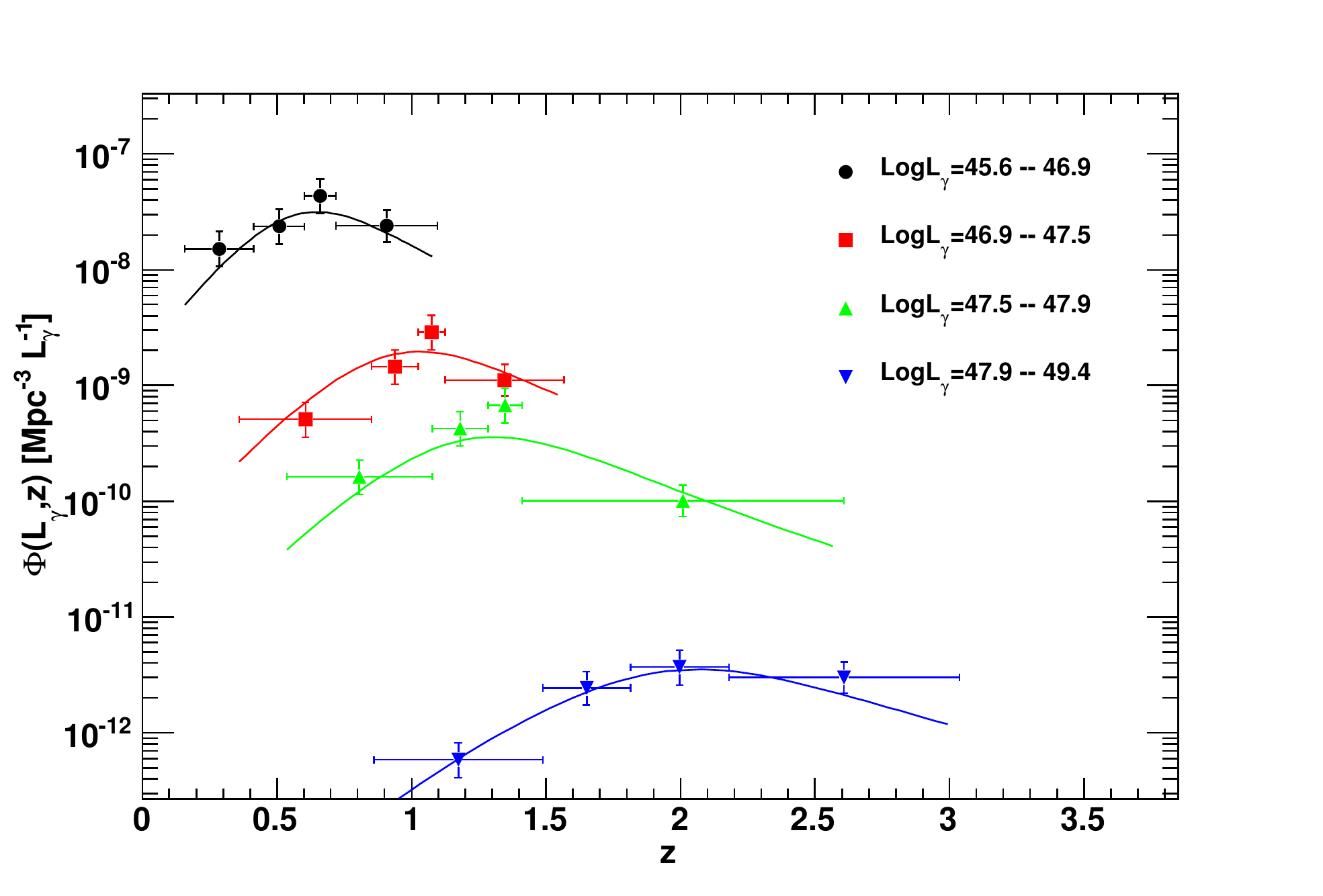}
\end{tabular}
\end{center}
\vspace{-0.5cm}
\caption{\footnotesize
{{\it Left panel:} 
{Fermi}-LAT measurement of the extragalactic diffuse background spectrum
(data points; see \cite{abd10id}), and model contributions from different source classes.
The curves represent the integrated contributions of:
1) FSRQs (thin solid line),
2) BL Lac objects (thin short dashed line),
3) star-forming galaxies,
4) MSPs (dot-long dashed),
5) WIMP $b\bar{b}$ annihilation (long dashed) and
6) sum of all the previous (hatched). 
{\it Right panel}: Luminosity function of $\gamma$-ray selected FSRQs \cite{aje11}. 
}}
\label{fig:edb}
\end{figure*}

Included in the EGB are pulsar contributions, including millisecond pulsars.
Because millisecond pulsars (MSPs) are also found in early-type galaxies, 
they might contribute a significant fraction of the EGB.
A way to determine their integrated diffuse emission would be
 through the integration
of their  flux (or luminosity) distribution in $\gamma$-rays. 
The typical spectrum of MSPs is hard (e.g. photon index
of $\sim$1.5) and shows an exponential cut-off around a few GeV 
\cite{2010ApJ...713..154A}.
Thus the diffuse emission arising from MSPs should show a bump around
a few GeV, similar to a  feature  found in
the Galactic diffuse emission that has been
ascribed to annihilating dark matter \cite{2009arXiv0910.2998G}. 
The left-hand panel of Fig.\ \ref{fig:edb} illustrates a possible decomposition of the EGB into different source classes, 
including a contribution from WIMP dark-matter annihilation.

A preliminary  $\gamma$-ray luminosity 
function for FSRQs detected with the {Fermi}-LAT 
is shown in the right-hand panel of Fig.~\ref{fig:edb} \cite{aje11}.
The luminosity function shows the 
change in the space density of FSRQs as a function of redshift and 
for different luminosity classes. What is apparent from
Fig.~\ref{fig:edb} is that the space density of powerful FSRQs 
peaked at redshifts $\approx 2$,
and then declined to the space density  now observed.
The peak of maximum growth occurs at different epochs 
for different luminosity classes, with the more luminous sources
reaching their maximum space density 
earlier in the history of the universe, while the bulk of the 
population (the lower luminosity blazars) are more abundant at
present times. This cosmic downsizing behavior cannot persist to very large ($z\gg 1$)
redshifts and early times, as may be reflected by the sudden turnoff of high-$z$ $\gamma$-ray blazar.
The largest redshift $\gamma$-ray blazar  in both
the 1LAC and 2LAC is at $z = 3.1$, despite many candidate higher-redshift blazars.

\section{ Microquasars, Radio Galaxies, and the EBL } 

\subsection{$\gamma$-ray binaries}

The literature on the subject of Galactic binaries is vast; 
a good X-ray review remains Ref. \cite{lph95}. 
Here we highlight only a few of those
aspects that make for an appreciation of GeV -- TeV
$\gamma$-ray binaries in the Galaxy,
of which 6 are well-established, namely
\begin{enumerate}
\item LS 5039 ($P =$ 3.9 d);
\item LSI +61$^\circ$ 303($P = 26$ d);
\item PSR B1259-63 ($P = 3.5$ yr); 
\item Cygnus X-3 ($P = 4.8$ hr);
\item 1FGL J1018.6-5856 ($P = 16.6$ d) \cite{cor11}; and
\item Cygnus X-1 ($P = 5.6$ d).
\end{enumerate}
These sources are exclusively identified with high-mass X-ray binary (HMXB) 
systems and are identified, not merely associated, with the 
named system by period folding or time differencing with the 
orbital period. With its long orbital period, this is harder for PSR B1259-53, 
but $\gamma$-ray flaring near periastron strengthens the identification. PSR B1259-63 is an example
of a Be X-ray binary system, where the compact object, a young pulsar with a 
spin period of 48 s,  is in a highly eccentric
orbit around a luminous B star that displays emission lines in its optical spectrum
(thus the designation ``Be").  Flares take place near periastron
at all frequencies, when the pulsar passes through the equatorial 
stellar wind \cite{joh96}, and now also at GeV energies \cite{abd11B1259}. 

High mass and low mass refer to the companion star, not the compact object.
HMXBs are  X-ray binaries composed
of a compact object---white dwarf, neutron star, or black hole---and 
a young massive O or B main-sequence or Wolf-Rayet star. 
The low-mass X-ray binaries (LMXBs) typically have
 older, less massive companions $\lesssim 2 M_\odot$. 
If the companion compact object is a pulsar, it can vary in age from 
young and energetic to old and slowed down (and/or up). 
The fuel illuminating galactic X-ray binaries is accretion onto 
the compact object, 
which is transferred, primarily, through stellar wind in HMXBs
 and Roche-lobe overflow in LMXBs. X-ray variability is modulated 
by orbital effects, the most important being the change 
in the binary orbital separation distance. In only the case of PSR B1259-63 are X-ray 
pulsations at the pulsar period seen, which
are also found for intermediate mass X-ray binaries like Her X-1
and other HMXBs at X-ray energies. 

Microquasars are X-ray binaries with radio jets. HMXBs number 
around 7, including SS 433 and V4641 Sgr to the list 
above. There are 9 LMXB microquasars in the Paredes list from 
2005 \cite{par05}. Some of the most famous
are GX 339-4, Circinus X-1 and Scorpius X-1.  

Like the radio structures seen in extragalactic AGNs (e.g., 3C 84), 
jetted structures have been found in unusual X-ray binary sources, 
e.g., the Hertz-Grindlay source 1E 1740.7-2942 lying $\sim 0.7^\circ$ 
away from the Galactic center \cite{mir92}.  The analogy with 
extragalactic radio sources was strengthened when this source \cite{mr94} and other 
LMXB sources such as GRS 1915+107 \cite{mr99} were shown to exhibit superluminal effects.
No LMXBs are yet detected at GeV or TeV energies.

Prior to Fermi, three Galactic binaries were detected at TeV energies, namely LSI + 61$^\circ$ 303 with MAGIC and VERITAS, and LS 5039 and PSR B1259-63 with HESS \cite{aha05B1259}. An isolated flare from Cyg X-1 was observed with MAGIC.  No solid EGRET identification with a Galactic binary source had been made, though both LS 5039 \cite{par00} 
and LSI + 61$^\circ$ 303, even in the COS-B days \cite{her77}, were suspected to be $\gamma$-ray sources.  

A central debate regarding $\gamma$-ray binaries is whether they are scaled-down 
quasars, as suggested by the microquasar label, or colliding stellar wind/pulsar wind systems. Leptonic microquasar models \cite{br06,db06,or07}
invoke electrons or hadrons in jets accelerated up to TeV energies with mildly relativistic outflows, accounting for the extended radio emission. Confirming evidence would be VHE emission from X-ray binaries with definite black holes, e.g., Cyg X-1, V 4641, or GRS 1915+105.  

The other class of pulsar/star model invokes the rotational energy of the neutron star 
and the interactions of the pulsar and stellar winds. \cite{mt81,dub06}.
This model surely applies to PSR B1259-63. Insofar as 
LS 5039 and LSI +61$^\circ$ 303 have compact objects with $M < 4 M_\odot$,
the question of whether the compact object is a black hole or neutron star
is unresolved. The confirming evidence for a neutron star is, of course, detection of pulsations.  

When considering the different sources, it is useful to have a theoretical 
picture in mind. Most developed and perhaps most likely is a leptonic model, 
where electrons, which are accelerated in the extreme environment near the 
compact object, Compton scatter the photons of the hot star. 
Those $\gamma$ rays are in turn attenuated by the stellar radiation. 
To first order, a source at superior conjunction should be GeV luminous
(more head-on collisions for Compton scattering) and less TeV luminous
(due to $\gamma\gamma$ attenuation). It should anti-correlate with the 
TeV emission, which has least $\gamma\gamma$ opacity
and therefore should be brightest at inferior conjunction.

\subsubsection{LS 5039}
Though sitting on the edge of the Galactic plane, LS 5039 is well-detected
and identified by phase-folding \cite{abd09LS5039}. Its {Fermi}-LAT light curve is 
remarkably anti-correlated with the HESS TeV emission. At superior 
conjunction, when the TeV flux  is in a low power-law state, the Fermi GeV spectrum
is in a high state. At inferior conjunction, the roles are reversed. 
Only quasi-simultaneity is possible, because Fermi will necessarily integrate over
longer times to achieve the same sensitivity as a TeV telescope. The joint
spectra are also remarkable, with the GeV high state fit as an exponentially cut-off power
law, and taking place at the same phase of the low-state power-law HESS spectrum. In reverse,
the low-state GeV power law takes place when the TeV flux is elevated.

\subsubsection{LSI +61$^\circ$ 303}
The Fermi LAT light curve of this source is a rather sharp-peaked
sinusoid, with
peaks of emission
around
periastron \cite{abd09LSI}.
The GeV radiation is modulated with 
a maximum close to periastron, and minimum close to apastron.
TeV emission peaks at different orbital phase, revealing
a flux anti-correlation with the MAGIC TeV emission that
might be expected in a leptonic model. 
The phase-averaged Fermi spectrum is well described by an exponentially
cutoff power-law with index $\approx 2.2$ and cutoff energy
$\approx 6.3$ GeV. This spectrum is not found to vary strongly with
phase. The non-simultaneous MAGIC and VERITAS data corresponding to different
phases look, if they correspond to the time of Fermi observations, 
 like a distinct spectral component, with the hard component extending
to TeV energies  with a spectra index $\approx 1.9$ overtaking the exponentially
decaying GeV spectrum. 

\subsubsection{Cygnus X-3}

Cyg X-3 is a HMXB microquasar with an orbital period of 4.8 hours---short for an HMXB---at  a distance of $\approx 7$ kpc from the Earth. 
It is found in a complicated region, the Cygnus arm, but its period 
readily identifies it in $\gamma$ rays. Its companion is the Wolf-Rayet star V1521 Cyg.
It exhibits bright radio flares and goes through active states. It is
not known whether it contains a black hole or neutron star binary.
Four episodes of flares were observed with AGILE \cite{tav09} a 
few days before radio/X-ray flares. The Fermi LAT light curve exhibits periods 
of scattered activity, and periods of quiescence that appear to be anti-correlated with 
the X-ray emission \cite{Abd09CygX3}. The radio and $\gamma$-ray activity are also correlated. 

\subsubsection{1FGL J1018.6-5856}

Period-folding analysis has resulted in the recent discovery of 
a new high-mass X-ray/$\gamma$-ray binary, 1FGL J1018.6-5856 \cite{cor11,corbet2}. 
The 100 MeV -- 200 GeV {Fermi}-LAT data is modulated with the $16.58\pm 0.04$ d 
orbital period.  
The Swift XRT finds an X-ray source coincident with the position of the Fermi source.
The digitized sky survey (DSS2) image of this region shows a star with spectrum very similar 
to that of the gamma-ray binary LS 5039, and the Australia Telescope Compact Array
detects a radio source at 9 GHz. Chandra and XMM-Newton observations \cite{2011ATel.3228....1P} find a 
coincident X-ray source from which the   
column density and reddening can be determined. With 2MASS optical observations, 
this implies a star of type of O6V at a distance of 6 -- 12 kpc, large by comparison
with the distances of LS 5039 and LSI.

\subsubsection{PSR B1259-63}

PSR B1259-63 may help us understand these systems best by 
virtue of the compact object being unambiguously a neutron star. 
The pulsar has a 47.76 ms rotation period and a massive, $\approx 10 M_\odot$ Be star companion, SS 2883.
The stars orbit with a 3.5 yr orbital period and eccentricity = 0.87,  guaranteeing 
episodes of strong interactions near periastron passage. 
Because of its inclination and the geometry of the equatorial stellar wind,
the pre- and post-periastron 
passages are not symmetrical. EGRET on CGRO failed to detect it during 7 weeks of observation
around the 1994 periastron passage \cite{tav96}, but flaring  at radio, 
optical, X-ray, and HESS TeV \cite{aha05B1259} has been monitored in previous 
passages.

\begin{figure}
\center
\includegraphics[scale=0.4]{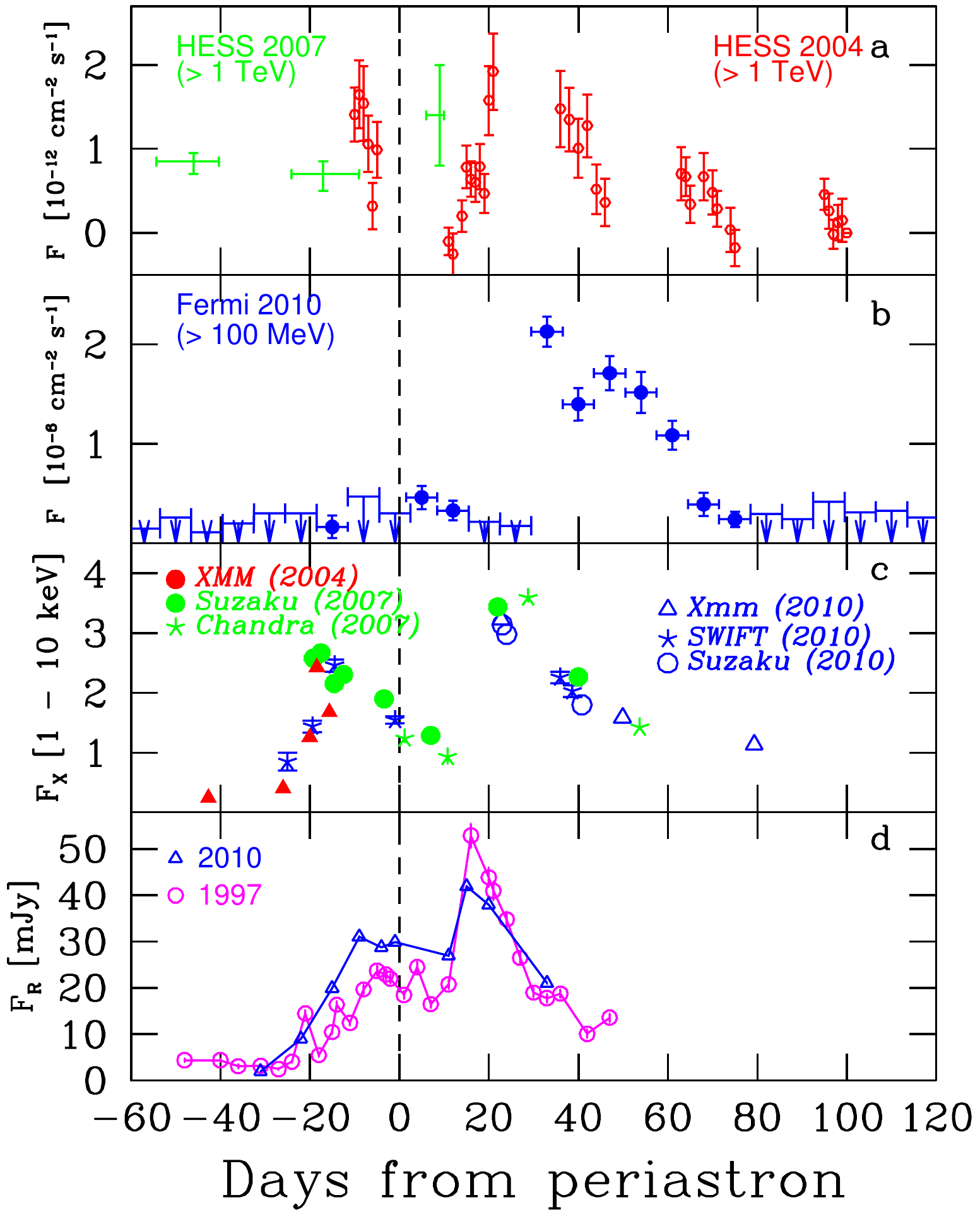}\includegraphics[scale=0.3]{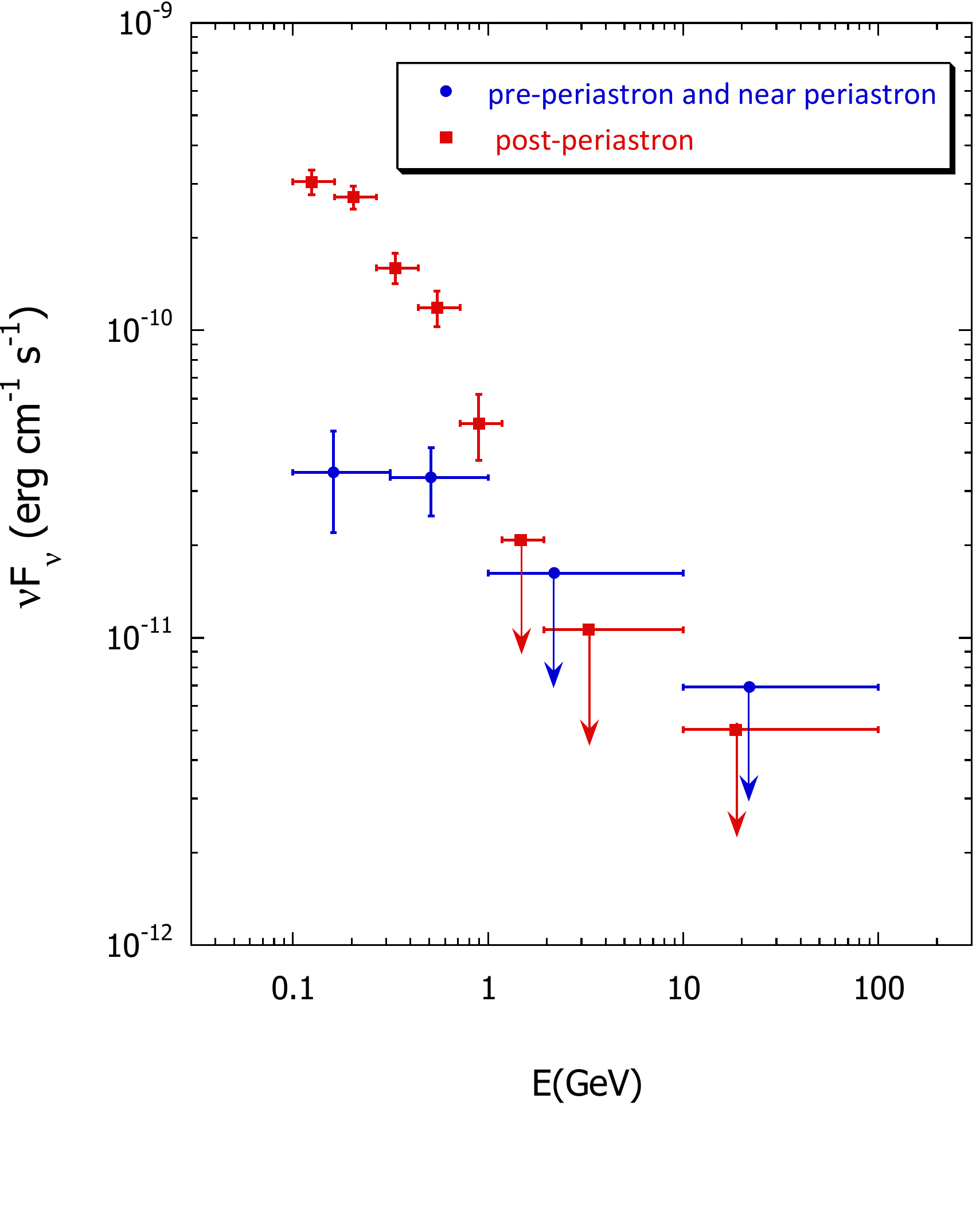}
\caption{{\it Left:} Radio, X-ray,  GeV, and  TeV light curves of PSR B1259-63 
at different epochs around the 2011 perastron passage on 2010 December 15 (MJD 55545), as labeled \cite{abd11B1259}.  
The 120 days shown represents $\approx 13$\% of the orbital period. 
{\it Right:} Fermi spectra of the system in the pre- and near-periastron period (blue  
circles) and in the post-periastron flare (red squares). }
\label{B1259_lc}
\end{figure}

So it was with great anticipation that observers at all wavelengths waited for 
the 2010 December passage.
PSR B1259-63 did not fail to perform: the light curves measured
by an array of instruments at different wavelengths and epochs are 
shown in Fig.\ \ref{B1259_lc}, and 
reveal the complexity of the encounter \cite{abd11B1259}.
It was  only weakly
 detected with the Fermi LAT on the 
approach to periastron and up to 20 d after (MJD 55517-55563), during which 
the pre- and near-periastron spectrum in Fig. \ref{B1259_lc} was measured. The system flared at GeV energies 
 starting at $\approx 30$ d post-periastron, with the Fermi LAT measuring 
during period MJD 55575 - 55624 shown as the post-periastron spectrum in Fig. \ref{B1259_lc}. The reason that the GeV flare seems to lag behind
the peak of the radio and X-ray emissions at $\approx 20$ d post-periastron is not understood. 

Though the time of 2010 periastron  passage was not ideal for TeV observations due to moonlight conditions, 
additional post-periastron TeV data for the most recent passage will help illuminate this system.  
We can reliably predict interesting events in the future.

\subsection{Misaligned blazars and radio galaxies}

The misaligned AGN (MAGN) population detected with Fermi using the first
15 months of science survey data consists of 11 sources (Fig.\ \ref{GammavsL};  \cite{abd10e}), including 7 FRI and 2 FRII radio galaxies, which are thought to be the parent population of BL Lac objects and FSRQs, respectively, in addition to 2 steep spectrum radio quasars that are believed to be slightly misaligned FSRQs. The MAGNs are associated with objects in the Third Cambridge (Revised) and Molonglo radio catalogs. The MAGN sources in the 3CRR catalog have large core dominance parameters compared to the general 3CRR source population, implying that the beamed component makes an appreciable contribution to the $\gamma$-ray flux. This is furthermore supported by the fact that three of these sources---3C 78, 3C 111, and 3C 120---do not appear in 2LAC, evidently due to variability. 

Why the ratio of measured $\gamma$-ray luminosities of FRI galaxies and BL Lac objects span a much larger range than the comparable ratio for FRII radio galaxies and FSRQs is an open question, if in fact not due to statistics. The simplest possibility is that there are a lot more nearby FRIs, and limitations on Fermi sensitivity will therefore favor detection of these nearby sources. At $z\ll 1$, the luminosity distance $d_L\approx 4200z(1+z)$ Mpc. As noted earlier, the {Fermi}-LAT reaches a limiting energy-flux sensitivity of $\approx 5\times 10^{-12}$ erg cm$^{-2}$ s$^{-1}$ for two years of observations that is, unlike integral photon flux, only weakly dependent on source spectral index \cite{2011arXiv1108.1420T}.  Fermi can thus only detect sources with luminosity $L_\gamma \gtrsim 10^{46}z^2(1+2z)$ erg s$^{-1}$. 

The number of nearby FRI and FRII galaxies depends on the space density of these objects which, in the study by Gendre, Best, and Wall \cite{gbw10}, is inferred from 1.4 GHz NVSS-FIRST radio observations (recall that the Fanaroff-Riley dichotomy is based on 178 MHz luminosities $P_{178~{\rm MHz}}$). From their Figure 12,  FRI and FRII radio galaxies have a local ($z<0.3$) space density of $\approx 35,000$ Gpc$^{-3}$ and $\approx 2,000$ Gpc$^{-3}$, respectively, at $P_{1.4~{\rm GHz}} > 10^{22}$ W Hz$^{-1}$ sr$^{-1}$ or $L_r \gtrsim 10^{40}$ erg/s, implying that the local space density of FRIs exceeds that of FRIIs by $\approx 20$.

The volume of the universe within $z<0.15$ ($d_L < 700$ Mpc) is $\approx $1 Gpc$^{-3}$, 
so there are thousands of FRIs and FRIIs within $z \cong 0.1$, yet other than 3C 111, all LAT-detected radio galaxies within this volume are FRI sources (see Figure \ref{figlumden};  \cite{2010ApJ...724.1366D}). Sensitivity limitations and the abundance of nearby FRIs could explain the MAGN population statistics, but low apparent-luminosity off-axis FRIIs, which are far more radio luminous than BL Lac objects and FRI galaxies, are lacking.  With only a few thousand randomly aligned sources within $z = 0.1$, a narrower $\gamma$-ray beaming cone in FSRQs, with a more rapid fall-off in off-axis flux, makes detection of these nearby sources far less likely than a broader $\gamma$-ray emission cone in BL Lac objects, as expected for the different beaming factors from SSC emission and external Compton processes \cite{1995ApJ...446L..63D,2001ApJ...561..111G}.  This could also reflect differences in jet structure between FSRQs and BL Lac objects \cite{2000A&A...358..104C,2005A&A...432..401G}, or extended jet or lobe emission in FRIs \cite{2008ApJ...679L...9B} that is missing in FRII galaxies.

Both core and lobes can significantly contribute to the measured $\gamma$-ray luminosities. In the case of Centaurus A, the values of $L_\gamma$ of the core and lobes are comparable, with the lobe emission primarily attributed to Compton-scattered CMBR
\cite{abd10f}. The significant or dominant lobe component means that the core luminosity of misaligned AGNs can be less than the measured $L_\gamma$ unless the lobe emission is resolved. 

One-zone synchrotron/SSC models with $\delta_{\rm D} \gtrsim 10$ give good fits to the long-term average spectra of other HSP BL Lac objects such as Mrk 421 and Mrk 501. 
The SEDs of radio galaxies, in contrast, are fit with much lower Doppler factors. The SED of the core of Cen A, for instance, can be fit with $\delta_{\rm D} \approx 1$ and bulk Lorentz factors $\Gamma \approx $ few \cite{abd10i}. Likewise, the SEDs of the FR1 radio galaxies M87 and NGC 1275 \cite{abd09e} and M87 \cite{abd09f} are well fit with $\delta_{\rm D}\approx 2$ and $\Gamma \sim 4$.   
 
The much larger values of $\Gamma$ for BL Lac objects than for their putative parent population, the misaligned FR1 radio galaxies, is contrary to simple unification expectations. Moreover, even though the $\gamma$-ray luminosities from FR1 radio galaxies are much smaller than that of BL Lac objects (Fig.\ \ref{GammavsL}), they are still larger than expected by debeaming the radiation of BL Lac objects with $\Gamma \gtrsim 20$.  Additional soft target photons that can be Compton scattered to high energies result in a reduction of the value of $\delta_{\rm D}$ compared to those implied by the one-zone synchrotron/SSC model. These target photons can  be produced in a structured jet, as in the spine and sheath model \cite{chi00}. Another soft photon source arises if blazar flows decelerate from the inner jet to the pc scale \cite{gk03}, seemingly in accord with the mildly relativistic flows at the sub-pc scale found in radio observations of Mrk 421 and Mrk 501.

\subsection{The EBL}

In the most general sense, the EBL refers to the cosmic microwave background radiation (CMBR),
the radiations from all the past stars and black holes,
 the glow from annihilating and decaying dark matter, and any residual emissions
from exotic particle decays. The EBL at infrared, optical and UV frequencies
 originates predominantly from stellar emissions, either directly or after
being absorbed and reradiated by dust. 
The present EBL is a consequence of the star formation history of the universe, 
light absorption and re-emission by dust, and different types of
dust extinction  in various classes of galaxies throughout cosmic time.
The IR and optical/UV EBL attenuates TeV and 10 -- 100 GeV $\gamma$-rays, respectively.
Knowledge of the absorption due to the EBL is therefore necessary to infer 
the intrinsic, escaping spectra of extragalactic $\gamma$-ray sources.

\begin{figure}
\center
\includegraphics[scale=0.5]{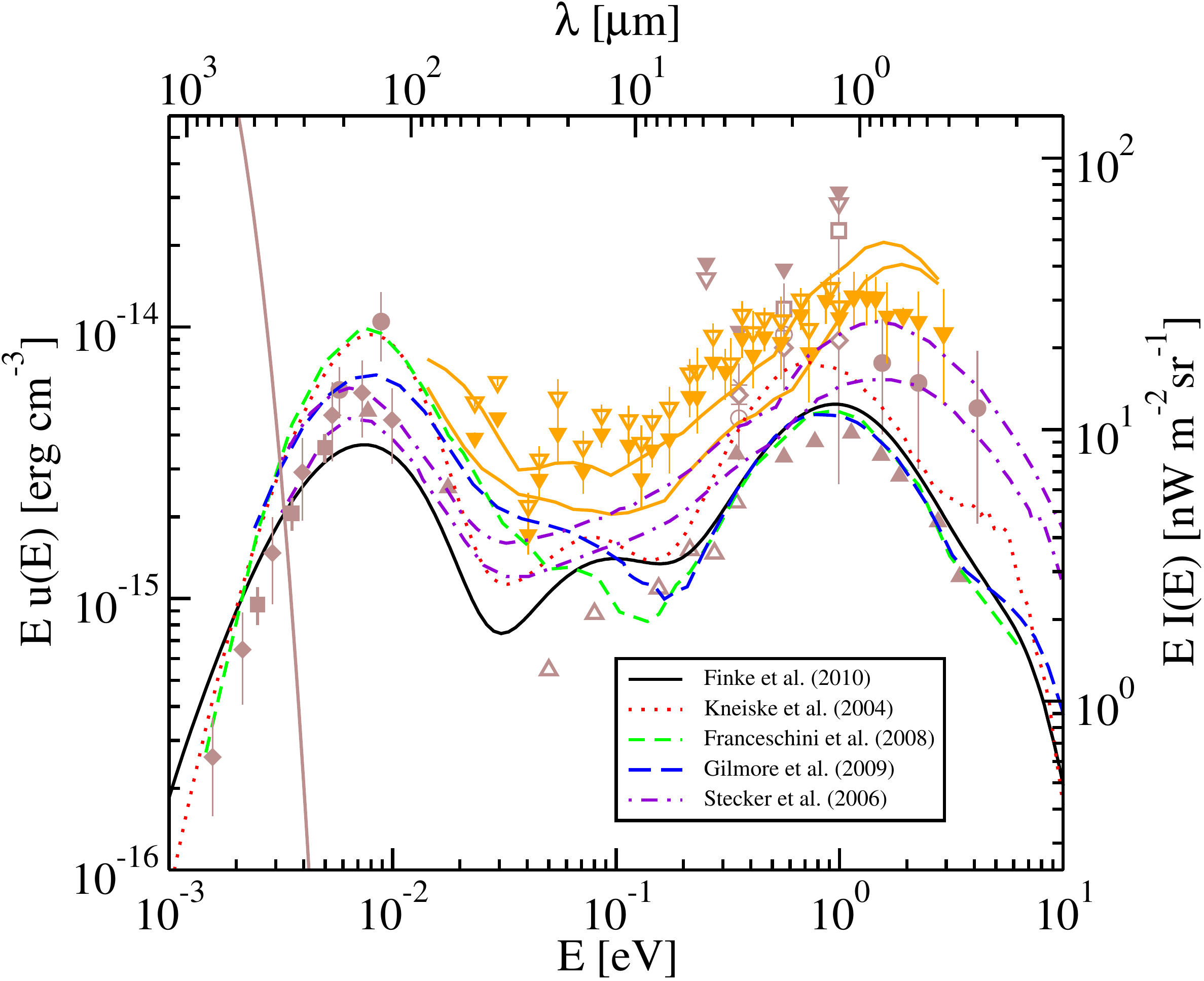}\\
\vskip 0.1in \includegraphics[scale=0.74]{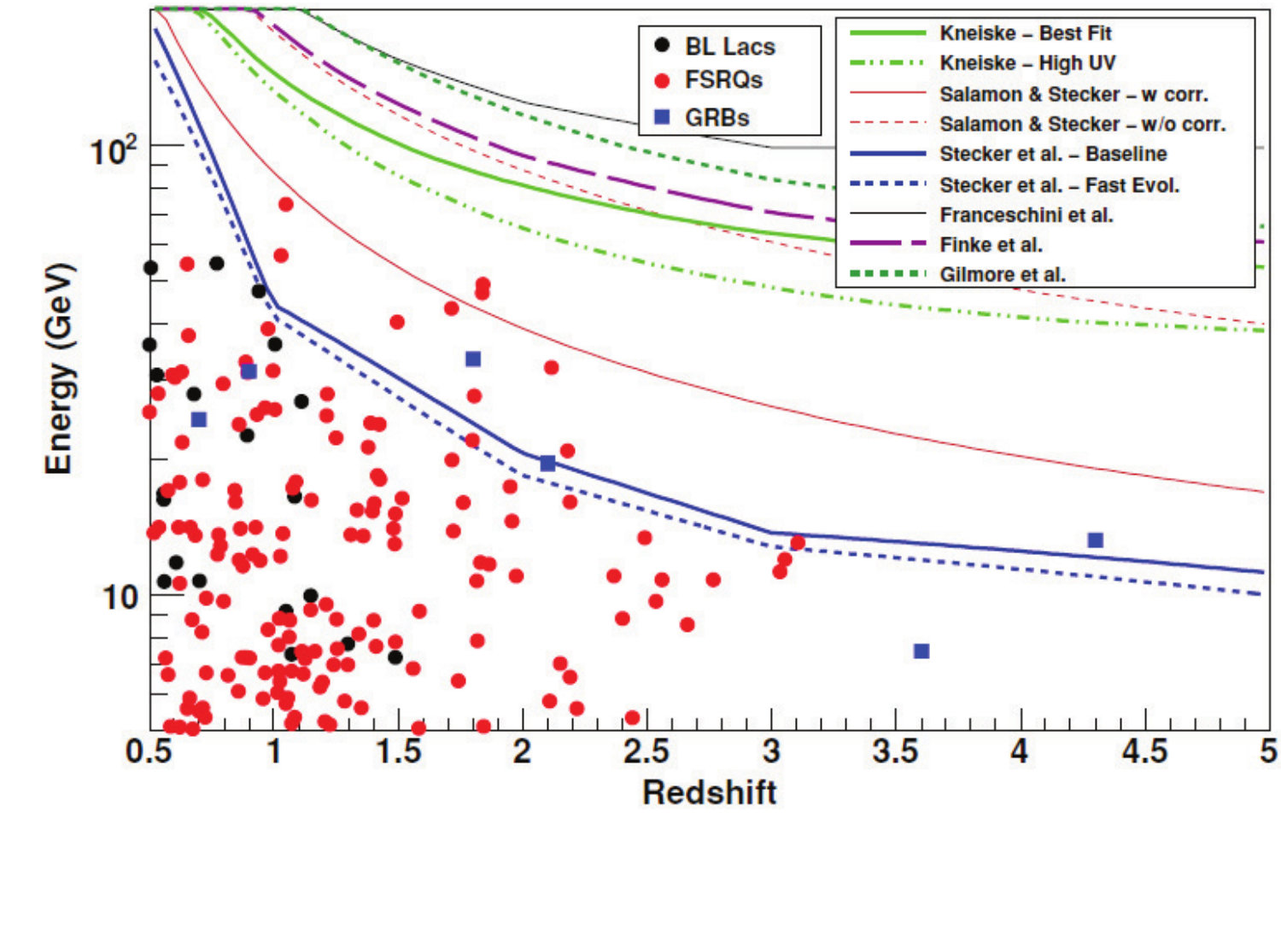}
\vskip-0.2in
\caption{{\it Upper:} EBL observations, models, and constraints \cite{frd10}. See text for references and details. 
{\it Lower:} Highest energy photons observed from blazars and GRBs with the {Fermi}-LAT as 
a function of the source redshift. Curves showing the $\tau_{\gamma\gamma} = 3$ optical depth 
for EBL attenuation are shown for different models as labeled \cite{2010ApJ...723.1082A}. }
\label{EBL+figs}
\end{figure}

Directly measuring the EBL at IR and optical wavelengths is difficult because of the 
interfering foreground zodiacal light and Galactic synchrotron radiation. 
Many attempts have been taken to predict the EBL intensity.
Empirical methods \cite{1998ApJ...493..547S,2006ApJ...648..774S,2008A&A...487..837F} 
 sum optical/IR emissions from galaxy-counts using, e.g., 
luminosity-dependent galaxy SEDs, and extrapolating to high redshift
where data is lacking. Semi-analytic models of the EBL
\cite{2005AIPC..745...23P,2009MNRAS.399.1694G,2011MNRAS.410.2556D}
model galaxy formation following mergers of dark matter halos, with
effects of supernova feedback, dust attenuation, and metal production 
 considered. Other models \cite{2002A&A...386....1K,2004A&A...413..807K,frd10} are based on 
integrating stellar light with dust absorption, employing arguments for 
the star formation rate and the initial mass function of the stars. 
A lower bound to the EBL intensity can be determined by galaxy counts at 
different frequencies and redshifts. The upper panel in Fig.\ \ref{EBL+figs}
shows a comparison of observations with models, and inferred constraints. 
 Curves are model C from \cite{frd10} (solid black curve); best fit model from \cite{2004A&A...413..807K} (red dotted curve); \cite{2008A&A...487..837F} (dashed green curve); the fiducial model of \cite{2009MNRAS.399.1694G} (dashed  blue curve); and the fast evolution and baseline models from \cite{2006ApJ...648..774S} (upper and lower dot-dashed violet curves, respectively).  Orange curves and inverted triangles are upper limits derived from blazar observations by \cite{2007A&A...471..439M} and \cite{2009ApJ...698.1761F}, respectively.  

Gamma-ray astronomy offers a special 
technique to probe the EBL at IR through UV frequencies, because 
photons of the EBL attenuate $\gamma$ rays via pair production through the
$\gamma\gamma\rightarrow {\rm e}^+  {\rm e}^-$ process.
The threshold condition implies that 400 GeV photons can pair produce on
soft EBL photons with 1 $\mu$ wavelength, and 4 TeV photons with 
$10\mu$ EBL photons, etc., so thatV 
$$\lambda (\mu )\Leftrightarrow {E_1\over 0.4{\rm~TeV}}\;.$$ 

Analysis of the effects of this process have already ruled out
high-intensity models of the EBL at IR/optical frequencies \cite{2010ApJ...723.1082A}. 
Ruling out other EBL models 
will require longer integration times to improve implementation of tests like the 
following:
\begin{enumerate}
  \item {\it Flux ratio method.} By assuming that intrinsic blazar spectra are independent of redshift,
then the ratio of high to low energy fluxes will decrease with redshift due to enhanced EBL attenuation \cite{2004ApJ...608..686C}. It is necessary to use separate blazar populations in this test since {Fermi} has established significant spectral differences between FSRQs and BL Lacs, but this test is not yet sensitive enough to identify any trend of spectral softenings due to EBL absorption. 
  \item {\it Deabsorbed $\gamma$-ray spectrum.} By extrapolating the {Fermi}-LAT spectrum of an individual source to high energies, and assuming that the spectrum does not harden with energy, then deabsorbing the spectrum limits the range of EBL models that are compatible with this underlying assumption \cite{2010ApJ...714L.157G}. Upper limits are also placed 
on the allowed EBL intensity by assuming that the intrinsic source spectrum is limited
in hardness by some theoretically determined value \cite{2005ApJ...628..617S,2007A&A...471..439M}, which depends on details of the particle acceleration mechanism
\cite{2007ApJ...667L..29S}. Specific approaches determine the likelihood of different EBL models to be compatible with the measured $\gamma$-ray spectrum by assuming an intrinsic (usually power-law) source spectrum. 
  \item {\it Distribution of highest energy photons.} The probability of detecting high-energy photons depends on the $\gamma\gamma$ optical depth associated with a given EBL model. 
The detection of such photons, though rare in number, can place severe demands on  the apparent power budget of a source. If the absolute power or energy  if separately determined, e.g., by using optical breaks to infer
the jet opening angle in GRBs, then the required energy release may be extreme and call the EBL model into question (see the lower panel in Fig.\ \ref{EBL+figs}).
\item {\it Detection of lobes of radio galaxies.}
Photons of the EBL are Compton scattered to GeV energies  by relativistic electrons in the radio lobes of nearby extended radio galaxies. Thus 
measurements of the GeV lobe flux can be used to determine the level of the EBL  \cite{2008ApJ...686L...5G}. The first radio galaxy
with radio lobes resolved by {Fermi} is Cen A 
\cite{abd10f}. 
Due to the low significance of the {Fermi} detection and the lack of data at energies $\sim$2\,GeV, the method cannot yet be applied. Longer exposures should reduce the error bars below 2\,GeV and detect photons in the critical regime above $\sim$2\,GeV where the emission from the up-scattered EBL is expected to dominate. Besides Cen A,  Cen B and Fornax A---two other radio galaxies with lobes---have recently been detected \cite{2011arXiv1108.1420T}. These objects, as well as NGC 6251 \cite{2011A&A...533A..72M}, are potential targets for resolving extended lobe emission to measure or constrain the EBL.
\end{enumerate}

Implicit in the above methods is the assumption that the  $\gamma$ rays are made at the source. This need not be the case if blazars or GRBs are sources of UHECRs.
UHECR protons that are able to escape from the blazar and structured regions without deflection can deposit energy in transit through the IGM via photopair production.
The cascade radiation can produce persistent TeV emission that would confuse the interpretation of the attenuation due to the EBL. This is less likely for UHECRs accelerated by GRBs, where deflection of the UHECRs would suppress the emission generated by the UHECRs. This is currently an active area of research; see, e.g., 
\cite{2010APh....33...81E,2010PhRvL.104n1102E,2011ApJ...731...51E,2011A&A...527A..54K,2011arXiv1107.5576M,2011arXiv1110.0853R}.

\section{Fermi Observations of Gamma Ray Bursts}

The effective lifetime for GRB studies 
using EGRET's spark chamber on the {\it Compton Gamma-ray Observatory}
ended $\approx 4.5$ yrs into 
mission, after 1996  \cite{kur97}.\footnote{Very bright GRBs like  
GRB 990123 \cite{bri99} could still be detected far off the  
COMPTEL and OSSE axes while making a signal in the Energetic Gamma Ray
Experiment Telescope's Total Absorption Shower Counter.
EGRET TASC and BATSE data  were used to made the discovery of the additional hard component in 
GRB 941017  \cite{gon03}.}
The depletion of spark-chamber gas was mitigated through 
the introduction of a narrow-field
mode suitable for pointed observations. This made 
the chance of catching a GRB, 
proportional to EGRET's FoV, too improbable
without rapid automated slewing, which was not possible for CGRO. Consequently
EGRET only detected a total of five spark-chamber GRBs, all 
early in the mission  \cite{din95}. These are GRB 910503, GRB 910601, 
the superbowl GRB 930131, the famous long-lived GRB 940217 \cite{hur94}, and GRB 940301. 
In the wide-field mode, EGRET was sensitive to 
$\approx 1/25^{th}$ of the full sky, 
which is $\approx 1/5^{th}$ as large as the FoV of {Fermi} \cite{atw09}.

\subsection{ Fermi LAT observations of GRBs }

Since {Fermi}  operations began, 
13 GRBs were reported as significantly detected in the LAT by the 
{Fermi} Collaboration
from early August 2008 through calendar year 2009.\footnote{See fermi.gsfc.nasa.gov/ssc/observations/types/grbs } The year 2010 
saw a dearth of bright LAT GRBs, but the rate has picked up in 2011 with 
the detection of a few remarkable events like GRB 110721A
and GRB 110731A, the former
of which displays classic ``fast-rise, exponential decay"-type GBM light curves with prompt LAT emission 
\cite{2011GCN..12187...1T,2011GCN..12188...1V},
and the latter of which is the first long-duration GRB
jointly detected with {Fermi}-LAT and Swift from the prompt phase 
into the afterglow \cite{2011GCN..12215...1O,2011GCN..12218...1B}. 
In the meantime, the development of the LLE technique (\cite{2010arXiv1002.2617P}; Section 1.2) has found LAT emission from 
several GRBs in the 30 MeV -- 100 MeV range. Given the addition of these GRBs, the rate of GRB
detection with the LAT is about 1 per month, and the number of {Fermi}-LAT GRBs has reached nearly 30.  Because analysis is 
in progress on 2011 GRBs, and there were so few GRBs in 2010, here we review GRB observations only through 2009.

\begin{figure}
\center
\includegraphics[scale=0.5]{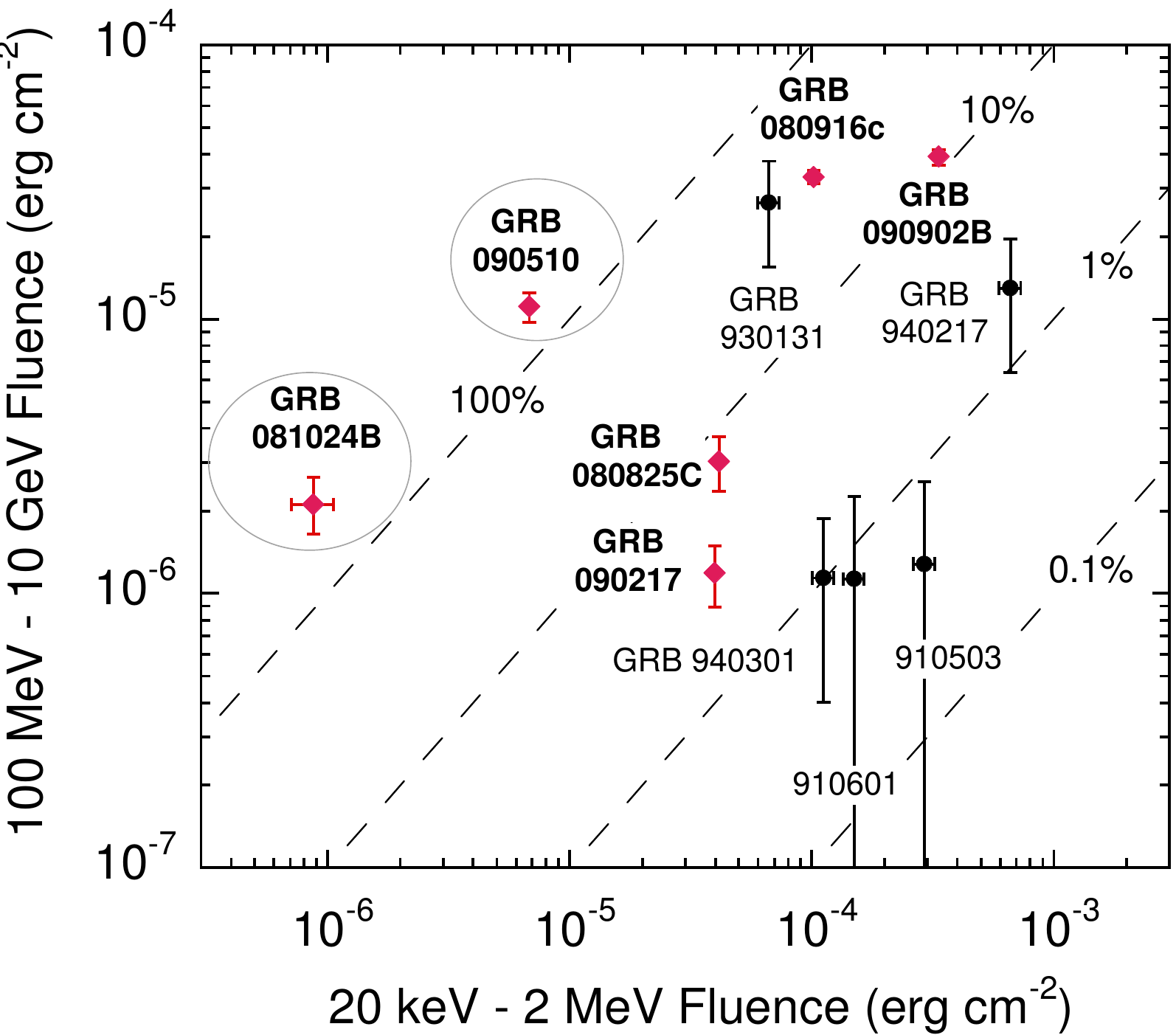}
\caption{Fluence-fluence diagram showing 6 {Fermi} \cite{abd10_grb081024b} GRBs (red data points) 
and 5 EGRET spark-chamber \cite{led09} GRBs (black data points). The EGRET fluence is measured from 100 MeV to 5 GeV, whereas the {Fermi} LAT fluence
is measured from 100 MeV to 10 GeV. Short hard GRBs are circled. }
\label{fluence}
\end{figure}

All LAT GRBs during this period are also 
GBM GRBs and comprise the brighter GBM GRBs, as already expected
from a comparison between EGRET and BATSE GRBs in terms of fluence \cite{led09}.
The 13 {Fermi} LAT GRBs observed through 2009 include 11 long GRBs
and 2 short bursts, namely
GRB 081024B and GRB 090510 ($z = 0.903$). 
The most studied---because they are brightest---GRBs are GRB 090902B ($z = 1.822$) \cite{abd09_grb090902b}, 
which provides the 
first strong evidence for a hard spectral component in long GRBs;
GRB 080916C ($z = 4.35$) 
\cite{abd09science}, the first bright long GRB; and GRB 090926A ($z = 2.106$), a burst with a narrow spike
from the lowest to highest energies in an SED that requires both 
a Band function and a hard power-law component to fit. 
GRB 090926A also reveals an extraordinary spectral softening at 
$\gtrsim 1$ GeV in its time-integrated spectrum when the hard LAT spectral 
component is bright. These long GRBs had 
GBM fluences ${\cal F}$  in the  20 -- 2000 keV range 
$\gtrsim  10^{-4}$ erg cm$^{-2}$ 
(Fig.\ \ref{fluence}). 
The bright, short GRB 090510, with ${\cal F} \cong 10^{-5}$ erg cm$^{-2}$, 
also shows (like GRB 090902B and GRB 090926A)
 a distinct high-energy power-law spectral component in addition to a Band component  
\cite{ack10_grb090510}. 
Its short duration, large distance, and the detection of a 
31 GeV photon permit strict tests on quantum gravity theories that predict a 
dependence of the speed of light in vacuo that is linear with energy
 \cite{abd09_grb090510}.

Besides these notable GRBs are the less well-known and also less fluent long duration
GRBs 090323 ($z = 3.57$), 090626 and 090328 ($z = 0.736$), 
the widely off-LAT-axis GRB 081215A, the first LAT GRB 080825C \cite{abd09_grb080825c}, the 
unusual GRB 090217 \cite{ack10_grb090217a} showing none of the typical properties of LAT GRBs,
and the unremarkable LAT GRBs 091003A and GRB 091031.  
The weakest fluence GRB of the sample is the first short GRB detected at LAT energies, GRB 081024B \cite{abd10_grb081024b},
with ${\cal F}\approx 4\times 10^{-7}$ erg cm$^{-2}$. The weakness of this 
GRB could be related to the high $E_{pk}\approx 2$ -- 3 MeV of its Band-function component, 
but the time-averaged $E_{pk}\cong 4$ MeV for GRB 090510 between 0.5 and 1  s after trigger is
even higher \cite{ack10_grb090510}.

For those GBM GRBs occurring within the LAT FoV,
detection of GRBs with the LAT is almost guaranteed when ${\cal F}\gtrsim 10^{-4}$ erg cm$^{-2}$. 
The detection rate slips to less than 50\% when ${\cal F}\approx 3\times 10^{-5}$ erg cm$^{-2}$, 
and becomes highly improbable for ${\cal F}\lesssim 10^{-5}$ erg cm$^{-2}$.
This behavior undoubtedly reflects a distribution in the ratios of $\gtrsim 100$ MeV LAT to 
GBM  energy fluence \cite{gpw10}.

In the first 16 months of {Fermi} science operations, $\lesssim 1$ GRB per month was detected with the {Fermi} LAT,
 or $\approx 9$ GRB/year, with LAT detecting short GRBs 
at $\approx 10$ -- 20\% of the rate of long GRBs. GRBs are detected with the GBM at a rate of 
250 GRB/yr, or $\approx 500$ GRB/yr (full-sky). 
When corrected for FoV, EGRET detected $\approx 25$ GRB/year (full sky), while the {Fermi} LAT detects 
$\approx 50$ GRB/yr (full sky). Given the much larger effective area 
of {Fermi} than EGRET, by a factor $\approx 6 ~[\approx (8000$ -- $9000 {\rm ~cm}^{2})/(1200$ -- $1500 $ cm$^{2}$)], 
this small rate increase is something of a surprise, compounded by the ongoing sparse period of
{Fermi} LAT detections of GRBs in 2010. Part of this difference
is the stronger detection criteria of {Fermi} LAT than EGRET. 
But an improvement in flux sensitivity by a factor $\approx 6$, with an accompanying rate increase by 
only a factor $\approx 2$ -- 3 suggests that LAT GRBs are being 
sampled in a portion of their $\log N-\log {\cal F}$ distibution that is flattened by 
cosmological effects. This is consistent with the known redshifts of LAT GRBs, which 
range from $\approx 0.7$ to $z = 4.35$, with a very rough average redshift of $\langle z \rangle=2$ for long
GRBs and $\langle z \rangle \approx 1$ for short GRBs (based only on GRB 090510). If typical,
both classes of GeV-emitting GRBs 
would be subject to strong cosmological effects on the fluence and flux distributions. 

\subsubsection{Fluence-fluence diagram}

Figure \ref{fluence} shows the fluence-fluence diagram for the 5 EGRET spark-chamber \cite{led09} 
and for 6 {Fermi} \cite{abd10_grb081024b} GRBs. Most GBM GRBs
have ${\cal F} \lesssim 10^{-5}$ erg cm$^{-2}$, 
and are only rarely detected with the LAT. 
Because of the small number of LAT GRBs, it is not yet 
clear whether there is a 
systematic difference between fluence ratios of EGRET and 
{Fermi} LAT-detected GRBs. The weakest {Fermi} LAT
GRBs in terms of GBM fluence are both short duration GRBs. 
This could indicate a preference for short GRBs to have a larger ratio 
of LAT to GBM fluences than long GRBs, depending on 
possible triggering biases, e.g., increased LAT 
background for long GRBs. 

The apparent isotropic energies ${\cal E}_{iso}$ of GBM and LAT emission for LAT GRBs
with known redshifts are in several cases $\gtrsim 10^{54}$
erg. For GRB 080916C, ${\cal E}_{iso} \approx 10^{55}$ erg. The LAT GRBs tend to 
have the largest energies of all measured GRBs, and as a result are good for radio studies
\cite{cen10}.

\subsubsection{Fermi  LAT GRB phenomenology}

Some distinct behaviors have been identified in {Fermi} GRBs, namely:
 
\begin{itemize}

\item Extended (long-lived) LAT (100 MeV -- GeV) emission compared to the GBM 
(20 keV -- 2 MeV) emission, known already from EGRET observations, especially GRB 940217 \cite{hur94}. 

\item Delayed onset of the LAT emission compared to the GBM emission, in both long
and short classes of GRBs.

\item Power-law temporal decay profiles of the LAT extended emission, decaying typically $\propto t^{-1.5}$ \cite{ghi10}. 

\item Appearance of separate power-law spectral components with photon number index harder than $-2$.

\item Delayed onset of the lowest energy GBM emission at $\approx 10$ keV, seen for example 
in GRB 090902B and GRB 090926A.

\item Quasi-thermal Band function components with steep Band $\beta$ 
found, e.g., in GRB 090902B at $E\gtrsim 1$ MeV \cite{abd09_grb090902b}.  

\end{itemize}

The onsets of the $> 100$ MeV emission appear to be delayed by $\sim 0.1 t_{90}$
compared to the 100 keV -- MeV emission (with $t_{90}$ measured, e.g., in the 50 -- 300 keV GBM/BATSE range). 
This is 
one of the key and unanticipated results on GRBs from {Fermi},
and it appears to operate equally for both the long- and  
short-duration LAT GRBs. There have as yet been no LAT detections of members of 
the low-luminosity/sub-energetic class of GRBs that includes GRB 980425
and GRB 030329, nor have any X-ray flashes or X-ray luminous GRBs been detected
with the LAT. Because GBM's primary triggering modes are similar to BATSE, 
high $E_{pk}$, relatively low-$z$ GRBs  are more likely to 
be detected compared to {\it Swift}. 

The luminosity function can be constructed from observations of GRBs. 
This function gives information about the rate density, the opening angle, and the total nonthermal
energy radiated by GRBs. A description of the luminosity function for GRBs, and how the 
luminosity density can be evaluated for different GRB models is now given.

\subsection{GRB luminosity function}

The luminosity function $Y(L)$ describes the distribution of apparent isotropic luminosities
of members of a common source class. For GRBs, one can construct 
the luminosity function for, at least, the long duration
(classical) GRBs, the short hard GRBs, and the X-ray flashes and sub-energetic 
(low-luminosity) GRBs. 
As defined by \cite{lia07}, the luminosity function
\begin{equation}
Y_{\rm LZVD}(L) = L\,{dN\over dL}\;= {dN\over d\ln L}\;
=Y_0\big[\big({L\over L_b}\big)^\alpha + \big({L\over L_b}\big)^\beta\big]^{-1}H(L;L_l,L_u)\;,
\label{PhiLZVD}
\end{equation}
with normalization
$$Y_0^{-1} = L_b\int_{L_l/L_b}^{L_u/L_b} dx(x^\alpha + x^\beta)^{-1}
$$

As defined by \cite{gpw05}, the luminosity function
\begin{equation}
Y_{\rm GPW}(L) 
=c_0\cases{({L\over L_*})^\alpha \;, &  $L_*/\Delta_1 < L < L_*$
 $~$ \cr\cr 
 ({L\over L_*})^\beta\;, & $L_* < L < \Delta_2 L_*$ \cr}\;\;
,
\label{PhiGPW}
\end{equation}
with normalization
$$c_0^{-1} = \alpha^{-1}(1-\Delta_1^{-\alpha}) + \beta^{-1}(\Delta_2^\beta - 1)\;.$$
Generally, the luminosity function depends on $z$, so that 
$Y(L) = Y(L;z)$.

\subsubsection{Luminosity density from the luminosity function}

The luminosity density $\ell = d{\cal E}/dVdt$ at redshift $z$, referred to comoving
volume, is given by 
\begin{equation}
{\ell}(z) = \dot n(z) \langle \Delta t \rangle
\int_0^\infty d(\ln L) \,L\,Y(L;z)\;,
\label{dndtdzdl}
\end{equation} 
where $\dot n(z)$ is the comoving rate density and the integration
over $dL$ or $d\ln L$ depends on the definition of the luminosity function $Y(L)$. 
The local luminosity density, which depends on the low-redshift luminosity function 
and the local rate density, is  
\begin{equation}
{\ell}_0 = \dot n_0 \langle \Delta t \rangle
\int_0^\infty d(\ln L) \;L\;Y(L)\;.
\label{ell0}
\end{equation} 
The relation of mean duration $\langle \Delta t\rangle$ to luminosity is normalized
so that the fluence ${\cal F} = L\langle \Delta t\rangle$. 

For the model of Ref.\ \cite{gpw05}, $Y(L)$ is dimensionless (per unit $\ln L$),
and the expression for the local luminosity density
is analytic, given by
\begin{equation}
{\ell}_{\rm GPW} = \dot n_0 c_0 \langle \Delta t \rangle
 L_*[\big(1-\Delta_1^{-(1+\alpha)}\big)/(1+\alpha)
+ (\Delta_2^{1+\beta} - 1)/(1+\beta)]\;,
\label{dndtdzdlGPW}
\end{equation} 
with $c_0^{-1} = \alpha^{-1}(1-\Delta_1^{-\alpha}) + \beta^{-1}(1-\Delta_1^{-\beta})$.

For the model of Ref.\ \cite{lia07}, $Y(L)$ has dimensions of $L^{-1}$, 
and the local luminosity density
requires a simple numerical integration of the expression
\begin{equation}
{\ell}_{\rm LZVD} = {\dot n_0 Y_0 \langle \Delta t \rangle
\over L_*}
\;\int_0^\infty dL\,L\,\big[\big({L\over L_*}\big)^\alpha + \big({L\over L_*}\big)^\beta\big]^{-1} \;,
\label{dndtdzdl1LZVD}
\end{equation} 
with 
$$
Y^{-1}_0\;=\; \left[ L_B \int_{L_{low}/L_b}^{L_u/L_b} dx\;(x^{\alpha_1} +x^{\alpha_2})^{-1}\right]^{-1}
$$

{\it Elementary Cosmology:} Following, e.g., \cite{lia07}, we write the 
differential rate of GRBs with redshift $z$ between
$z$ and $z+dz$ and luminosity $L$ in the range $L$ to $L+dL$ by
\begin{equation}
{dN\over dt dz dL} = {\dot n(z)\over 1+z}\;{dV(z)\over dz} Y(L)\;,
\label{dndtdzdl2}
\end{equation} 
The comoving volume element
\begin{equation}
{dV(z)\over dz} \;=\;{c\over H_0}\;{4\pi d_L^2\over (1+z)^2 \sqrt{\Omega_m(1+z)^3 + \Omega_\Lambda}}\;.
\label{dVdz}
\end{equation}
To derive this, note that $dV_* = cdt_*dA_*$, so $dV_*/dz = 
c|dt_*/dz| dA_*$. With $dA_* = (R_*\chi)^2 d\Omega_* =
d_L^2(z)d\Omega/(1+z)^4$ 
(Ref.\ \cite{dm09}, eq.\ (4.43); see below). 
The cosmology of the universe we inhabit is 
a flat $\Lambda CDM$ universe, that is, a dark-matter 
dominated flat universe with nonzero cosmological constant,
well-described by  
\begin{equation}
{dt_*\over dz} = {-1\over H_0(1+z)\sqrt{\Omega_m(1+z)^3 +\Omega_\Lambda}}\;
\label{dt*dz}
\end{equation}
This defines, noting that $V = (1+z)^3 V_*$, 
equation (\ref{dVdz}). Our standard values are $H_0 = 72$ km s$^{-1}$ Mpc$^{-1}$,
$\Omega_m = 0.73 = 1-\Omega_\Lambda$.

The proper distance, though not directly measurable, is
the distance between two objects that would be measured at the same
time $t$. 
The proper distance at the present epoch is just
the comoving coordinate, so
\begin{equation}
d_{prop} = \chi = ct = c\int_0^{\chi/c} dt =
 c\int_0^z d\zp\;|{dt_*\over d\zp}|(1+\zp )\;.
\label{dprop}
\end{equation}
The energy flux for a source isotropically radiating luminosity $L =
d{\cal E}/dt$ at proper distance $d_{prop}$ is related to the energy
flux from a source with isotropic luminosity $L_* = d{\cal E}_*/dt_*$
at luminosity distance $d_L$ through the relation 
\begin{equation}
\Phi = {d{\cal
E}\over dA dt} = {d{\cal E}/dt\over 4\pi d^2_{prop}} = {d{\cal
E}_*/dt_*\over 4\pi d^2_{L}}\;.
\label{dlum0}
\end{equation}
The fluence ${\cal F} = \int_{t_1}^{t_2} dt\; \Phi(t)$ measured over
some time interval $t_1$ -- $t_2$ is therefore related to the apparent isotropic
energy release 
through the expression
\begin{equation}
{\cal E}_* \;=\; {4\pi d_L^2 {\cal F}\over 1+z}\;.
\label{calE*fluence}
\end{equation}
For an expanding universe, $d{\cal
E}_* =
\e_* dN = \e(1+z) dN = (1+z) d{\cal E}$, and $dt_* = dt/(1+z)$. 
From the definition of $d_L$, the energy flux
\begin{equation}
\Phi = {d{\cal E}\over dA dt} = {L_*\over 4\pi d_L^2} = 
{dA\over 4\pi d_L^2} \;
|{d{\cal E}\over d{\cal}E_*}|\;|{dt_*\over dt}|{d{\cal E}\over dA dt}
= {(1+z)^2 dA\over 4\pi d_L^2}\;\Phi\; .
\label{phiE}
\end{equation} 
Therefore $dA = R^2\chi^2d\Omega = 4\pi d_L^2/(1+z)^2$.
At emission time $t_*$, $dA_* = R_*^2 \chi^2 d\Omega_*$, 
and $dA = R^2 \chi^2 d\Omega$, so that 
$
{dA_*/ dA} =  1/(1+z)\;.$
This implies from the definition of the luminosity distance
and the angular diameter $R_*\chi$, 
with $d{\cal E}/dt = (1+z)^{-2}(d{\cal E}_*/dt_*)$,
and letting 
$d\Omega = d\Omega_*$, 
$$
d_L = d_L(z) = (1+z) d_{prop}  =  ({R\over R_*})^2(R_*\chi)= (1+z)^2 d_{A} = 
$$
\begin{equation}
=\;{c\over H_0}
\;(1+z)\int_0^z dz^\prime\;{1\over \sqrt{\Omega_m (1+\zpr )^3 +
\Omega_\Lambda}}\rightarrow {cz\over H_0}[1+(1-{3\Omega_m\over 4})z + {\cal O}(z^2)\dots ]\label{dlum}
\end{equation}
$\cong 4170 z(1+0.8z)$ Mpc for a flat universe. 

Compare this derivation from Ref.\ \cite{dm09}.
The directional event rate, or event rate per sr, is
\begin{equation}
{d\dot N\over d\Omega} \;=\;{1\over 4\pi}\int
dV_*\; \dot n_*(t_*)\;|{dt_*\over dt}|\;=\;c\int_0^\infty dz 
\;|{dt_*\over dz}| {(R_*\chi)^2 \dot n_*(z)\over (1+z)}, 
\label{ndotOmega}
\end{equation}
where the burst emissivity $\dot n_*(z)$ gives the 
rate density of events at redshift $z$.
Volume density in comoving and proper 
coordinates is related by the expression 
$\dot n_*(\e_*;z) = \dot n_{co}(\e_*;z)(1+z)^3
$. Assuming separability of source emission properties
and the rate density, then $\dot n_{co,i}(z)
= \dot n_{0i} \Sigma_i(z)$, where $\Sigma_i(z)$ is the structure
formation history (SFH) of sources of type $i$, defined so
that $\Sigma_i(z\ll 1) = 1$, and $\dot n_{0i}$ is the local ($z\ll 1$)
rate density of bursting sources of type $i$.
Thus
\begin{equation}
{d\dot N\over d\Omega} \;=\;{c\over H_0}\int_0^\infty dz 
\;{d_L^2(z) \dot n_{i}\Sigma_i(z)\over (1+z)^3\sqrt{\Omega_m (1+z )^3 +
\Omega_\Lambda}}, 
\label{ndotOmega1}
\end{equation}
This expression can be easily generalized to accommodate spectral behavior.

{\it Star formation rate functions:} 
The redshift-dependent rate density by Porciani \& Madau (2001; SFR2) 
that has roughly constant star formation at $z>2$ \cite{pm01} is described by
\begin{equation}
\dot n_{\rm PM}(z) = 23 \dot n_0\;{\exp(3.4z)\over \exp(3.4z) +22}\;.
\label{dotrhoz}
\end{equation}
The analytic function
\begin{equation}
\Sigma_{_{\rm SFR}}(z) = \frac{1+a_1}{(1+z)^{-a_2} +
a_1(1+z)^{a_3}} \ , \label{ldeq14}
\end{equation}
\cite{wda04}, with $a_1 = 0.005$, $a_2 = 3.3$, and
$a_3 = 3.0$ for the lower star formation rate (LSFR)
and $a_1 = 0.0001$, $a_2 = 4.0$, and
$a_3 = 3.0$ for the upper star formation rate (USFR)
describes extreme ranges of optical/UV measurements without and with
dust extinction corrections, respectively.  

\begin{figure}[t]
\center
\includegraphics[scale=0.5]{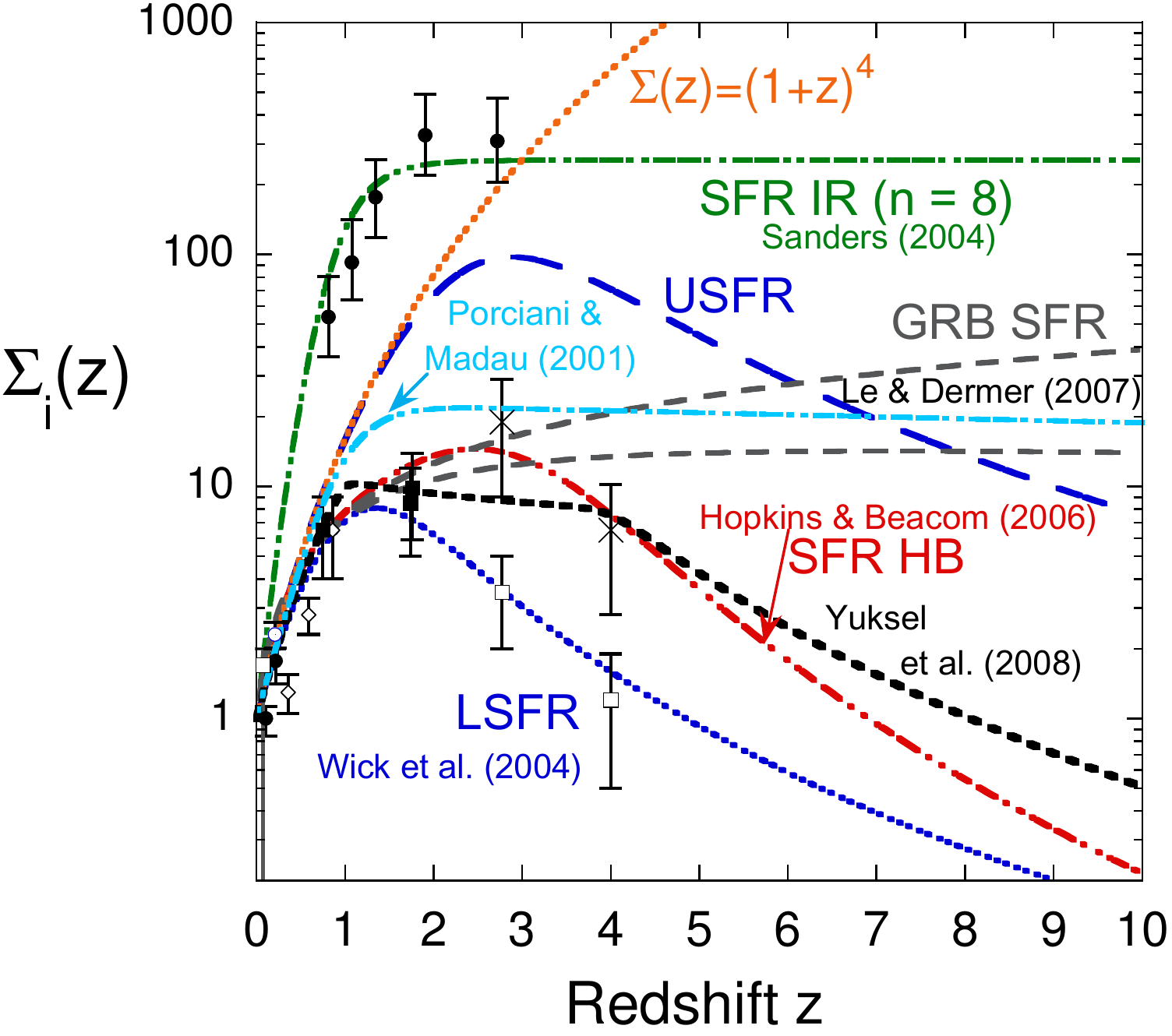}
\caption{Star formation rate factors \cite{wda04,hb06,ld07}.}
\label{fsfr}
\end{figure}

The SFR history (SFR3) of Hopkins and Beacom (2006) \cite{hb06}, which is intermediate
between the LSFR and USFR, is given by their analytic fitting profile
\begin{equation}
\Sigma_{_{\rm HB}}(z) = \frac{1 + (a_2 z/a_1)}{1+ (z/a_3)^{a_4}}
\ , \label{ldeq15}
\end{equation}
where $a_1 = 0.015$, $a_2 = 0.10$, $a_3 = 3.4$, and $a_4 = 5.5$
are best fit parameters (Figure~\ref{fsfr}).
An update of the SFH fit of \cite{hb06}
is contained in Ref.\ \cite{yuk08} in the form of 
a continuous broken power law,
\begin{equation}
\dot \rho_*(z) = \dot \rho_0\;\left[(1+z)^{a\eta}
+ \left({1+z\over B}\right)^{b\eta} 
+\left({1+z\over C}\right)^{c\eta} 
\right]^{1/\eta}
 \;, 
\label{dotnz}
\end{equation}
where $a = 3.4 (b = -0.3, c = -3.5)$ is the logarithmic slope
of the first (middle, last) piece breaking at $z_1 = 1$ and $z_2 = 4$, and the normalization is 
$\dot \rho_0 = 0.02 M_\odot$ Mpc$^{-3}$ yr$^{-1}$. Using 
$\eta = -10$ smooths the transitions,
while $\eta \rightarrow \infty$  recovers the kinkiness of the original form.
Here  
$B =  (1+z_1)^{1-a/b}\cong 5000$, and $C = (1+z_1)^{(b-a)/c}(1+z_2)^{1-b/c} \cong 9$.

Fig.\ \ref{fsfr} shows star formation rate factors used in different models for GRBs. For 
data, see \cite{san04,wda04}. An enhanced rate of GRBs compared to the global star formation rate
is required to fit Swift statistical data\ \cite{ld07}, as confirmed in other studies
\cite{yuk08,wp10}.

\subsubsection{Luminosity function in Le \& Dermer  model}

We approximate the temporally-evolving SED of, for example, a GRB or blazar
by the expression
\begin{equation}
\nu F_\nu \equiv \fluxeps(t) \cong f_{\epsilon_{pk}} S(x)
H(\mu; \mu_{j},1) \ H(t;0, \Delta t) \ , \\
\label{ldeq1}
\end{equation}
and
\begin{equation}
S(x) = x^a H(1-x) + x^bH(x-1)\;,\;x \equiv {\e_*\over \e_{*,pk}} ={\e\over \e_{pk}} = {\ep\over \ep_{pk}}\;. \\
\label{S(x)}
\end{equation}

The bolometric fluence of the model GRB for observers with $\theta
\leq \theta_j$ is 
\begin{equation}
{\cal F} = \int^\infty_{-\infty} dt \; \int^\infty_0 d\e\; \frac{\fluxeps(t)
}{\epsilon}  = \lambda_b\; f_{\epsilon_{pk}}\;\Delta t  \;,
\label{eq2F}
\end{equation}
where $\lambda_b$ is a bolometric correction to the peak measured $\nu
F_\nu$ flux. If the SED is described by eq.\ (\ref{S(x)}), then
$\lambda_b = (a^{-1} - b^{-1})$, and is independent of $\e_{pk}$. The
beaming-corrected $\gamma$-ray energy release $\Estarg$ for a
two-sided jet is 
\begin{equation}
\Estarg = 4 \pi d^2_L (1-\mu_j) \; \frac{{\cal F}}{1+z}  \ , \\
\label{eq3}
\end{equation}
where the luminosity distance is given in equation (\ref{dlum}).
Substituting eq.~(\ref{eq2F}) for ${\cal F}$
into eq.~(\ref{eq3}) gives the peak flux
\begin{equation}
f_{\epsilon_{pk}} = \frac{\Estarg}{4 \pi d^2_L(z) (1 - \mu_j)
\Delta t_* \ \lambda_b} \ . \\
\label{ldeq5}
\end{equation}
Finally, by substituting eq.\ (\ref{ldeq5}) into eq.\ (\ref{ldeq1}),
the energy flux becomes
\begin{eqnarray}
\fluxeps(t)  \; = \; f_{\e_{pk}}\;H(\mu; \mu_j,1) \, H(t;0,
\Delta t) S(x). \label{ldeq6}
\end{eqnarray}

The observed directional event rate for
bursting sources with $\nu F_\nu$
spectral flux greater than $\fluxthres$ at observed photon energy $\bar\e$
is given by
$$\frac{d\dot{N}(> \fluxthres)}{d\Omega} =  \frac{c}{H_0}
\int^\infty_{\fluxthres} df^{'}_{\epsilon} \int^\infty_0 d\Estarg
\int^\infty_0 d\epsilon_{pk*} \int^\infty_0 d(\Delta
t_*)\int^{\mu_{\rm jmax}}_{\mu_{\rm jmin}} d\mu_j \int^1_{\mu_j}
d\mu \nonumber \\ $$
\begin{equation}
\times  \int^\infty_0 dz\;\frac{d^2_L(z)
\dot{n}_{co}(z)Y(\Estarg,\mu_j,\epsilon_{pk*},\Delta t_*)
}{(1+z)^3 \ \sqrt{\Omega_m (1+z)^3 + \Omega_\Lambda}} \; \delta
\left(f^{'}_{\epsilon} - f_{\e_{pk}}S(x) \right) . 
\label{eq9}
\end{equation}
For discrete values of $\Estarg$, $\e_{pk*}$, and
$\Delta t_*$, the property distribution function%
\begin{equation}
Y(\Estarg,\mu_j,\epsilon_{pk*},\Delta t_*) = g(\mu_j) \
\delta(\Estarg- \Estargo) \ \delta(\epsilon_{pk*} -
\epsilon_{pk{*0}}) \ \delta(\Delta t_* - \Delta t_{*0}) \ ,
\label{eq10}
\end{equation}
where $g(\mu_j)$ is the jet opening angle distribution, and
$\fluxthres$ is the instrument's detector sensitivity. 
The detector threshold for \Swift~and
pre-\Swift~GRB detectors is taken to be
 $\sim 10^{-8}$ and $\sim 10^{-7} \ \rm ergs
\ cm^{-2} \ s^{-1}$, respectively.

In the model of Le \& Dermer (2007) \cite{ld07},
 the form for the jet opening angle $g(\mu_j)$ is chosen to be%
\begin{equation}
g(\mu_j) = g_0 \ (1-\mu_j)^s \ H(\mu_j;\mu_{\rm jmin},\mu_{\rm
jmax}) \;,
\label{ldeq12}
\end{equation}
where $s$ is the jet opening angle power-law index; for a
two-sided jet, $\mu_{\rm jmin} \geq 0$. Normalized to unity,
equation (\ref{ldeq12}) gives
\begin{equation}
g_0 = \frac{1+s}{(1-\mu_{\rm jmin})^{1+s} - (1-\mu_{\rm
jmax})^{1+s}} \;.
\label{ldeq13}
\end{equation}
Integrating over $\mu$ in eq.~(\ref{eq9}) gives the factor $( 1-
\mu_j)$ describing the rate reduction due to the finite jet
opening angle. Hence, eq.~(\ref{eq9}) becomes
$$\frac{d\dot{N}(>\fluxthres)}{d\Omega}  =  \frac{c}{H_0}
\int^\infty_{\fluxthres} df^{'}_{\epsilon} \int^{\mu_{\rm
jmax}}_{\mu_{\rm jmin}}d\mu_j \; g(\mu_j)(1-\mu_j) \;
$$
\begin{eqnarray}
\times \int^\infty_0 dz\; \frac{d^2_L(z) \ \dot{n}_{co}(z) \ \delta
\left(f^{'}_{\epsilon} -f_{\e_{pk}}S(x)\right)}{(1+z)^3 \ \sqrt{\Omega_m
(1+z)^3 + \Omega_\Lambda}} \;, \label{eq11}
\end{eqnarray}
where $f_{\e_{pk}}$ is given by eq.~(\ref{ldeq5}). The redshift, size, and opening angle distributions can then 
be calculated for comparison with data.

Fitting to redshift and opening-angle distributions of pre-Swift 
and Swift GRBs using this model
gives a good fit, assuming a flat $\nu F_\nu$ spectrum with 
bolometric factor $\lambda_b = 5$, for the following parameters:
\begin{equation}
\Delta t_* = 10{\rm s}, {\cal E}_{*\gamma}= 4\times 10^{50} {\rm~erg~s}^{-1},
s = -1.3,\mu_{min} = 0.765,\mu_{max} = 0.99875
\;,
\label{parameters}
\end{equation}
i.e., $\theta_{min} = 0.05$ (2.9$^\circ )$, $\theta_{max}
= 0.7$ (40$^\circ$). The ``{\it true}" event rate
$\dot n_{GRB} = (7.5$ -- $9.6)$ Gpc$^{-3}$ yr$^{-1}$.
This can be compared with beaming corrected rates of $ \cong 75\times 0.5$ 
Gpc$^{-3}$ yr$^{-1}\sim 40$ Gpc$^{-3}$ yr$^{-1}$
\cite{gpw05}.
The local emissivity in all forms of 
electromagnetic radiation from long GRBs is automatically implied
in this model, namely
\begin{equation}
\ell_{LD}= \dot n_{GRB}{\cal E}_{*\gamma} \;\approx (3 - 4) \times 10^{43}{\rm~erg~s}^{-1} {\rm~Mpc}^{-1}
\label{ellLD}
\end{equation}

{\it Luminosity function:}
We derive the luminosity function for the Le \& Dermer (2007) model, which 
can be checked in the limit $z \ll 1$ to agree with equation (\ref{ellLD}).
The isotropic luminosity $L_{*,iso} = 4\pi d_L^2 \Phi = 4\pi d_L^2 \int_0^\infty f_\e/\e$.
From equations (\ref{ldeq1}) and (\ref{ldeq5}),
$$f_{\e_{pk}}= L_{*,iso}(t) S(x)/\lambda_b\;,$$ where
$$L_{*,iso}(t) = L_{*,iso} \;H(\mu;\mu_j,1)H(t;0,\Delta t)\; = {{\cal E}_{*\gamma}\over \Delta t_* (1-\mu_j)}\;H(\mu;\mu_j,1)H(t;0,\Delta t)\;$$
is the apparent isotropic luminosity that would be measured within the beaming cone 
during time $t$ from 0 to $\Delta t$. Thus the relation between the apparent luminosity
$L = L_{*,iso}$ and $l_* = {\cal E}_{*\gamma}/\Delta t_* = 4\times 10^{40}$ erg s$^{-1}$ is, obviously, $L = l_*/(1-\mu_j)$ 
for a two-sided jet ($0 < \mu_j \leq 1$). Thus $dL/d\mu_j = L/(1-\mu_j)$, and 
\begin{equation}
{dN\over dV dt dL} \,dL \;=\; {dN\over dV dt d\mu_j} \,d\mu_j\,(1-\mu_j )\;,
\label{dNdVdtdL}
\end{equation}
noting the extra factor $(1-\mu_j)$ which accounts for the smaller opening angle of
the more apparently luminous jets.

From the formulation of the model,
\begin{equation}
{dN\over dV dt d\mu_j } = \dot n_i g(\mu_j ) = \dot n_i g_0(1-\mu_j)^sH(\mu_j;\mu_j^{min},\mu_j^{max})\;.
\label{dNdVdtdmuj}
\end{equation}
Therefore 
$${dN\over dV dt dL } = |{d\mu_j\over dL}|\; {dN\over dV dt d\mu_j }(1-\mu_j) = 
\dot n_i g_0{(1-\mu_j)^{s+2}\over L}\; H(\mu_j;\mu_j^{min},\mu_j^{max})\;$$
\begin{equation}
=\; {g_0\dot n_i\over l_*}\;\left({l_*\over L}\right)^{s+3} H\left(L;{l_*\over 1-\mu_j^{min}},{1\over 1-\mu_j^{max}}\right).
\label{dNdVdtdL1}
\end{equation}
The luminosity 
function in this model is 
\begin{equation}
\Phi_{\rm LD}(L) = {dN\over dL }=\; {g_0\over l_*}\;\left({l_*\over L}\right)^{s+3} H\left(L;1.7\times 10^{51}{\rm~erg~s}^{-1},3.2\times10^{53}{\rm~erg~s}^{-1}\right).
\label{dNdVdtdL2}
\end{equation}
Therefore $L^2 \Phi(L) \propto L^{-1-s}\propto L^{0.3}$.
From equation (\ref{ldeq13}), $g_0 = 0.051$. 

Integrating $\Phi(L)$ over $L$ gives the emissivity according to equation (\ref{ell0}).
We find that integration of equation (\ref{dNdVdtdL2}) recovers the luminosity density 
given by equation (\ref{ellLD}). The luminosity function in the form $\dot n_i 
L^2 \Phi_{\rm LD}(L)$ is plotted in Figure \ref{LGRBell}. The model is overconstrained, 
which problem can easily be relaxed by changing the functional dependence of the jet opening angle.

\subsubsection{Local luminosity density of GRBs}

The ``$\nu F_\nu$" intensity $\e I_\e$(erg/cm$^{2}$-s-sr) $=m_ec^2 \e^2 \phi(\e)/4\pi$
for unbeamed sources, where $\phi(\e)$ is the differential photon flux, is given by
\begin{equation}
\e I_\e = {m_ec^3\over 4\pi}\int_0^\infty dz\; |{dt_*\over dz}|
\;{\e_*^2 \dot n_{co}(\e_*;z)
\over 1+z}\;.
\label{eIe1}
\end{equation}
(eq.\ (4.57) in Ref.\ \cite{dm09}).
This equation also applies to beamed sources provided that the apparent comoving 
event rate density $\dot n_{co}(z) = \int_0^\infty d\e_* \dot n_{co}(\e_*;z)$ is
increased by the inverse of the beaming factor, $f_b$, to get the true source rate density
(see eq.\ ({\ref{Labsfb})). 

The photon luminosity density at redshift $z$ is given by
\begin{equation}
\ell (z) = m_ec^2 \int_0^\infty d\e_* \;\e_* \dot n_{co}(\e_*;z)
\label{callz}
\eeq
and the local luminosity density $\ell_0 \equiv \ell(z\ll 1)$.
Assuming that the only property of long-duration GRBs that is 
redshift dependent is the rate density, then $\ell(z) = \ell_0 \Sigma(z)$.\
The integrated intensity ${\cal I} = \int_0^\infty d\e I_\e$ is 
related to the mean photon flux $\Phi = d{\cal E}/dAdt$ 
through the relation ${\cal I}=\Phi/4\pi$, so
\beq 
\ell_0 = k\;{H_0 \Phi\over c}\;,
\label{ell}
\eeq
where
\beq
k^{-1} \equiv \int_0^\infty dz\; { \Sigma(z)\over (1+z)^2\sqrt{\Omega_m(1+z)^3 +\Omega_\Lambda}}\;.
\label{constant}
\eeq
Table \ref{tab:3} gives values for $k$ 
obtained by numerically integrating eq.\ (\ref{constant}), using 
the SFR factors shown in Fig. \ref{fsfr} with the standard 
$\Lambda CDM$ cosmology with Hubble constant $H_0 = 72$ km/s-Mpc, $\Omega_m = 0.27$ and 
$\Omega_\Lambda = 0.73$. The integration was truncated at a redshift $z_{max} = 10$.

\begin{table}
\centering
\caption{Constant $k$, from eq.\ (\ref{constant}), for different Star Formation Rate Factors}
\label{tab:3}       
\begin{tabular}{lll}
\hline\noalign{\smallskip}
SFR & k  \\
\noalign{\smallskip}\hline\noalign{\smallskip}
Constant comoving & 1.85  \\
LSFR \cite{wda04}  & 0.52\\
USFR \cite{wda04} & 0.13\\
HB  \cite{hb06} &  0.37\\
LD \cite{ld07} & 0.33 - 0.37 \\
\noalign{\smallskip}\hline
\end{tabular}
\end{table}

The mean flux $\Phi$ over the full sky 
can be estimated from the total fluence per year from 
GRBs, given the number of GRBs per year. For 1293 GRBs in 4B catalog
the total fluence (20-300 keV) is $63.80\times 10^{-4}$ erg/cm$^2$
and the total fluence $>20$ keV is $153.63 \times 10^{-4}$ erg/cm$^2$
(M. Gonzalez, private communication, 2003).
For the 4th BATSE Catalog, considering 666 Burst/year full sky with 1293 bursts 
implies 1293/(666 GRB/yr)= 1.94 yr = 61225300 s, so that
the mean 20 -- 300 keV flux 
$1.04\times 10^{-10}$ erg/cm$^2$-s or $3.3\times 10^{-3}$ erg/cm$^2$-yr.
Likewise, for the total $>20$ keV fluence was, we obtain
$2.54\times 10^{-10}$ erg/cm$^2$-s or $7.9\times 10^{-3}$ erg/cm$^2$-yr
Band (2002) \cite{ban02} obtains 550 GRBs/yr for BATSE exposure, which brings
the flux down to $6.5\times 10^{-3}$ erg/cm$^2$-yr. Subtracting 10 -- 20\% 
for short GRBs gives  (5 -- 6)$\times 10^{-3}$ erg/cm$^2$-yr.

From eq.\ (\ref{ell}), writing $\Phi_{-2} = \Phi/(10^{-2}{\rm ~erg/cm}^{2}\mbox{-}{\rm yr})$  we find
$\ell_0 = 2.3\times 10^{43} \; k \Phi_{-2}\;{\rm ~erg/Mpc}^{3}\mbox{-}{\rm yr}$.
With $\Phi_{-2}\approx 0.5$ and $k \cong 0.35$, then
\beq
\ell_0 = 0.5\times 10^{43} \; ({k\over 0.35}) ({\Phi_{-2}\over 0.6}){\rm ~erg/Mpc}^{3}\mbox{-}{\rm yr}\;,
\eeq
in agreement with the results of Eichler et al.\ (2010) \cite{egp10}.

This value is at least a factor of 5 smaller than values obtained through 
statistical treatments of GRB data  \cite{ld07,gpw05,lia07}, as can be seen from Table \ref{tab:4}.
Here we give the local luminosity densities for long GRB models derived either numerically or analytically
from the luminosity functions derived in these models.

\begin{table}
\centering
\caption{Local luminosity density of long GRBs for different models}
\label{tab:4}       
\begin{tabular}{cccc}
\hline\noalign{\smallskip}
Model & $\ell_0$  & Photon energy range \\
	& ($10^{44}$   erg/Mpc$^3$-yr) &			\\	
\noalign{\smallskip}\hline\noalign{\smallskip}
Schmidt (2001) \cite{sch01} & 0.03 & 50-300 keV\\
Guetta et al.\ (2005) \cite{gpw05}  & 0.05 -- 0.08 & 50-300 keV\\
Le \& Dermer (2007) \cite{ld07} & 0.30 -- 0.38  & all photon energies \\
Liang et al.\ (2007) \cite{lia07} & 2.0 & 1 keV - 10 MeV\\
Wanderman \& Piran (2010)  \cite{wp10} &  1.3$^a$ & 1 keV -- 10 MeV\\
\noalign{\smallskip}\hline
\end{tabular}
{~$^a$Normalizing to the Swift durations brings this value down by a factor of $\approx 2$.}
\end{table}

\begin{figure}[t]
\center
\includegraphics[width = 8.0cm,height=6.0cm]{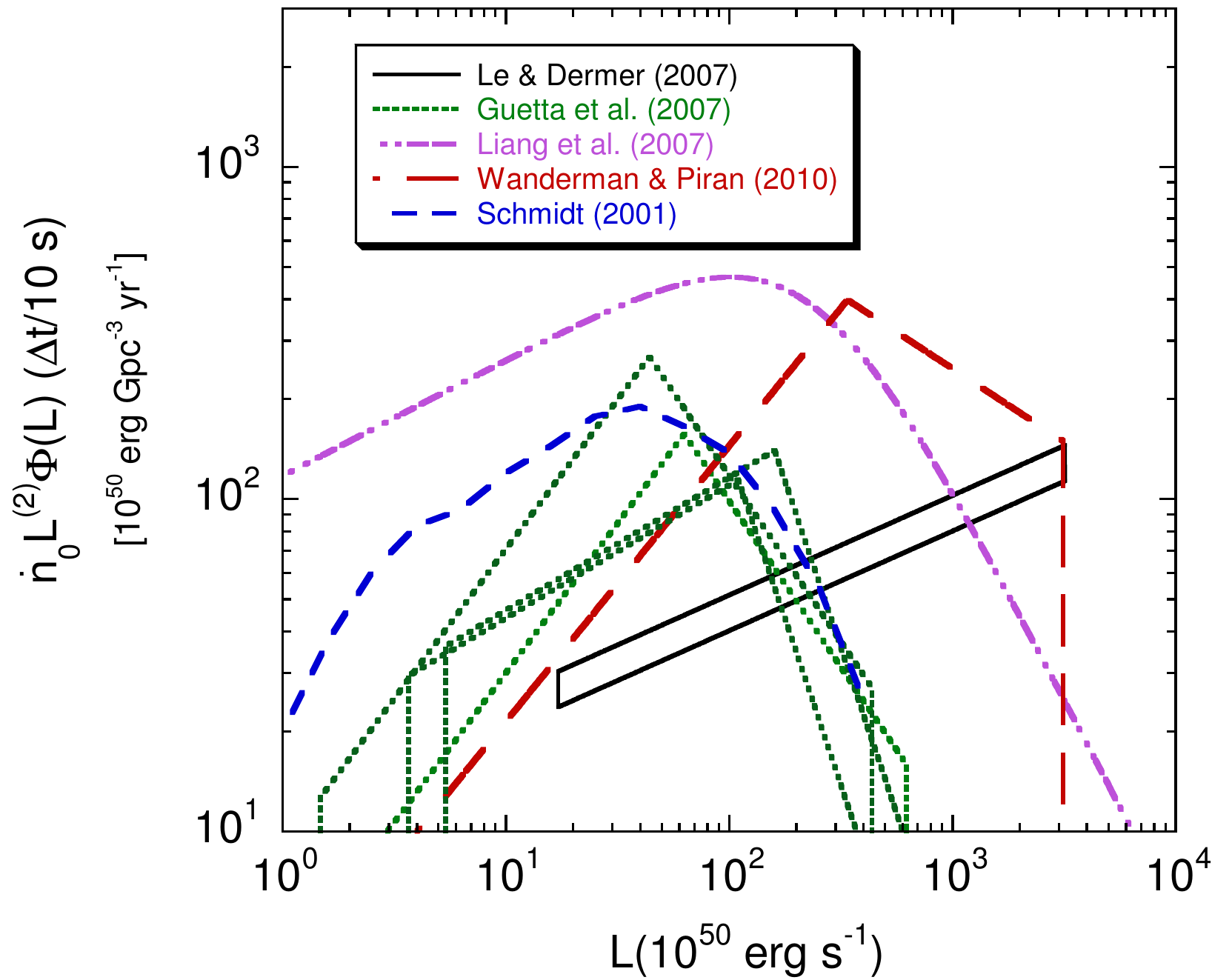}
\caption{Long duration GRB luminosity function, plotted in the form of a differential
luminosity density $\ell = \dot n_0 L\, Y(L) \Delta t$, for various models. }
\label{LGRBell}
\end{figure}

The reason for the discrepancy between the local emissivity of long duration GRBs derived
from BATSE data and inferred from Swift data is suggested by Fig.\ \ref{LGRBell}. 
This figure  compares different 
models for the local GRB luminosity functions, plotted 
in the form of a differential emissivity function 
$\dot n_0 \Delta t L\, Y(L)$, where 
$\dot n_0$ is the
local rate density, and the characteristic duration of 
a GRB in the explosion frame is $\Delta t = 10$ s.
Because the treatments of Schmidt (2001) \cite{sch01} and Guetta, Piran, \& Waxman
(2005) \cite{gpw05} are based on BATSE observations in the 50 -- 300 keV range, a
bolometric correction factor $= 6.3$ is applied to normalize them to the 1 keV -- 10 MeV range
(D.\ Wanderman, private communication, 2010). In this case, the implied luminosity density 
from the treatments of the luminosity function agree within factors of $\approx 3$.

This figure show that the integrated luminosity density depends sensitively 
on the maximum luminosity taken in the construction of the luminosity function.
When redshift information is available, very luminous GRBs are found with 
apparent luminosities reaching $10^{53}$ erg/s. To fit these luminous 
GRBs, the luminosity function must extend to large luminosities, and
these very rare, 
very luminous GRBs make a significant contribution to the 
integrated luminosity density. For such luminosity functions, the large fluence
GRBs make up a large, even dominant part of the total GRB fluence, but such 
high fluence events are so rare that the chances to now of such a GRB with 
large, $\gg 10^{-2}$ erg/cm$^2$, fluence having been detected
is low. 

\begin{figure}[t]
\center
\includegraphics[width = 12.0cm,height=5.0cm]{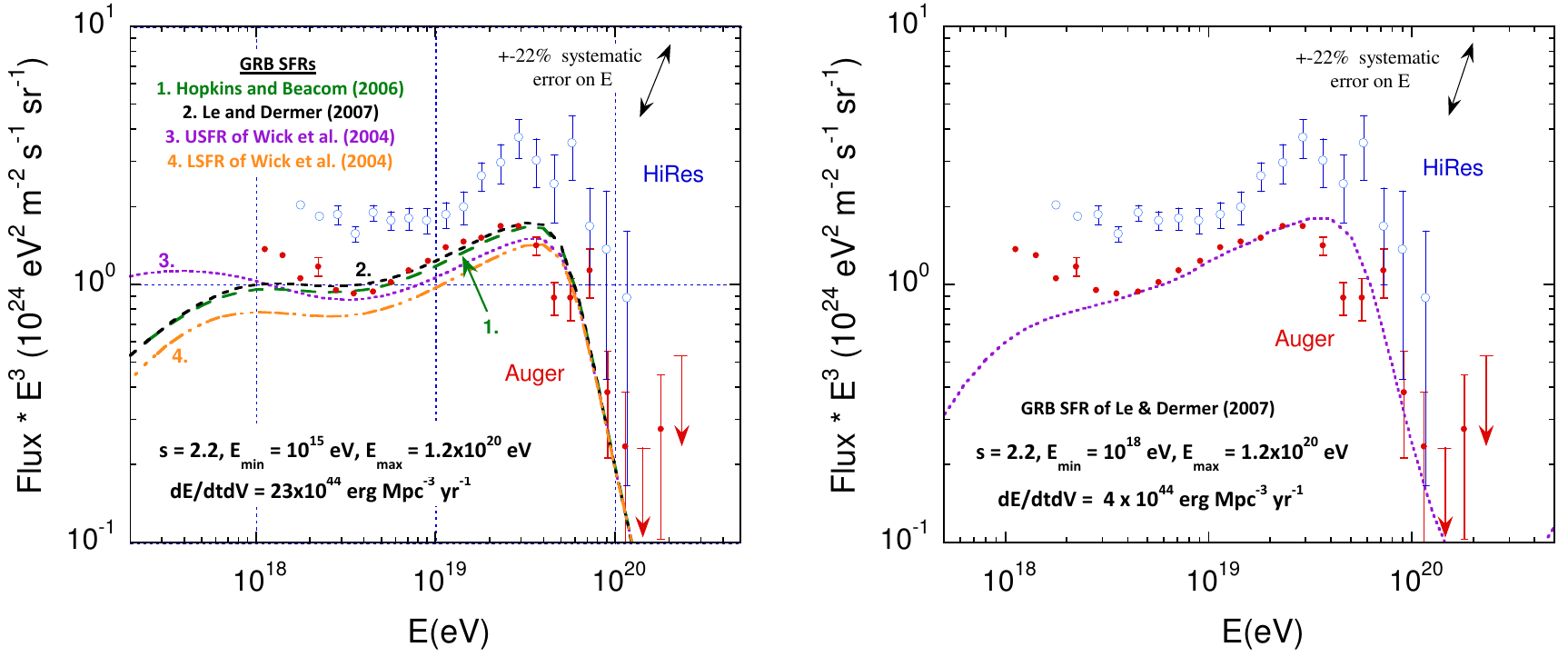}
\caption{Local luminosity density requirements on different models for UHECR production from long-duration GRBs
with rate density following various star-formation-rate functions shown in Fig.\ \ref{fsfr}. }
\label{uhecrvssfr}
\end{figure}

The local electromagnetic emissivity of long GRBs, given in Table \ref{tab:4},  is therefore found to be $\lesssim 10^{44}$ 
erg/yr-Mpc$^3$ \cite{2011NatCo...2E.175C}. 
This can be compared with the minimum local emissivity of $\ell_{UHECR} \approx 4 \times 10^{44}$ erg/s-Mpc required to 
power UHECRs, as seen in Fig.\ \ref{uhecrvssfr}, and discussed in more detail in the next section. 
Related estimates can be made for other source classes  \cite{lia07,2009ApJ...690L..14M}, including the short hard GRB class, the X-ray flashes, the sub-energetic GRBs, and engine-driven supernovae with relativistic outflows that lack GRB-type emissions. These classes, though they have much larger local rate densities than long-duration GRBs, are not much preferred over long-duration GRBs on the basis of their local luminosity density. The so-called low-luminosity GRBs, for example, have a large luminosity density in ejecta kinetic energy, $\gtrsim 10^{46}$ erg Mpc$^{-3}$ yr$^{-1}$ \cite{mur06,2008ApJ...677..432W}, but smaller or comparable nonthermal $\gamma$-ray luminosity densities than long-duration GRBs.

\subsection{Closure relations}

The relativistic blast wave theory for the afterglow has become the 
industry standard model. Familiarity with the derivation of the closure
relations is required for basic knowledge of GRB physics. 
The reader may see \footnote{heseweb.nrl.navy.mil/gamma/$\sim$dermer/notes/index.htm}
for the derivation, \cite{dm09} for review, and \cite{1998ApJ...497L..17S} for the
essential treatment.

Interpreting the delayed onset of the {Fermi}-LAT radiation 
as due external shock emission, one approach is to suppose that the blast wave decelerates
adiabatically in a uniform surrounding medium \cite{kb09}, with closure relation
$\nu F_\nu \propto t^{(2-3p)/4} \nu^{(2-p)/2}$, where $p$ is the electron injection 
index. A value of $p\approx 2.5$ gives a reasonable fit to the data.
Another regime to consider is a radiative GRB blast wave \cite{ghi10}, where the 
comparable closure relation is  $\nu F_\nu \propto t^{(2-6p)/7} \nu^{(2-p)/2}$, with
$p\approx 2$ giving a plausible fit to the data. The adiabatic model 
requires unusually low densities and magnetic fields for GRB 080916C, and the 
radiative model supposes pair loading can help achieve strong cooling. 

Alternate leptonic models for {Fermi} LAT GRBs 
include photospheric models with the 
photospheric emission passing through shocked
plasma in the colliding shells or external shocks
of the GRB outflow \cite{2011MNRAS.415.1663T}. A joint {Fermi}-{\it Swift} paper 
examines leptonic afterglow models for GRB 090510 \cite{pas09}.

\section{Fermi Acceleration, Ultra-High Energy Cosmic Rays, and Black Holes }

Fermi acceleration is intrinsically coupled to the presence of a magnetic field
that governs particle motions. The energy is gained stochastically in second-order Fermi scenarios, 
for example,  via resonant
pitch-angle scattering,  or the particle energy 
is gained systematically by those few particles that 
diffuse back and forth across a shock front multiple times before convecting downstream
(Section 5.5). In either case, the energy gain rate is related to the Larmor rate $v/r_{\rm L}$. 
The magnetic field and the size scale set a fundamental limit on the maximum particle 
energy through the Hillas criterion, eq.\ (\ref{hillas}).

\subsection{Maximum particle and synchrotron photon energy}

In Fermi acceleration scenarios, the maximum particle energy is obtained 
by balancing acceration rate with the energy-loss rate. 
For relativistic electrons, the maximum comoving electron Lorentz factor 
is $\gp_{max} = \sqrt{6\pi e/\phi\sT}B^\prime$. The maximum
synchrotron photon energy for a relativistic jet source 
at redshift $z$ is, from eq.\ (\ref{esynmax1}),
\begin{equation}
\e_{syn,max} \cong {27\over 8\alpha_f}{\Gamma\over \phi(1+z)} \;, 
\label{esyngamma}
\end{equation}
or 
\begin{equation}
E_{syn,max} \cong 240\;{(\Gamma/1000)\over \phi(1+z)} \;{\rm GeV}\;. 
\label{esyngamma1}
\end{equation}
The delayed appearances of the highest energy photons from Fermi LAT GRBs, which typically happens at times
after the GBM $t_{90}$,  calls into question a synchrotron origin 
for these photons, unless $\Gamma$ remains large at the end of the prompt
phase of GRBs.

The maximum energy for protons and ions is obtained from the Hillas criterion, 
written as $r^\prime_{\rm L} = E^\prime/\beta QB^\prime < R^\prime$. The maximum 
escaping particle energy $E_{max} = \Gamma E^\prime_{max} < \Gamma E^\prime
= Z\beta eB^\prime R$, noting that $R^\prime = R/\Gamma$ from length contraction 
of the stationary frame size scale as measured in the comoving frame. Relating 
the magnetic field energy density by a factor $\epsilon_B  <1$ times 
the proper frame energy density associated with the wind luminosity $L$, 
then $B^{\prime 2}=2\epsilon_B L/(\beta c R^2 \Gamma^2)$,
and 
\begin{equation}
E_{max} < \left({Ze\over \Gamma}\right)\sqrt{2\beta \epsilon_B L\over  c}\;,
\label{Emax17}
\end{equation}
implying
\begin{equation}
L_\gamma \gtrsim \left({3\times 10^{45}\over Z^2\beta}\right)\;\Gamma^2\;E^2_{20}
{\rm ~erg/s}\;
\label{Emax18}
\end{equation}
\cite{wax04,fg09,2010ApJ...724.1366D}, noting that the apparent $\gamma$-ray luminosity $L_\gamma < L$.

Eqs.\ (\ref{tprimesyn}) and (\ref{epsilonprimesyn})  
imply 
\begin{equation}
E_{ionsyn,~max}\cong m_ec^2 {A\over \phi Z^2}\;{\Gamma\over 1+z}\; {m_p\over m_e}\;\left({27\over 8\alpha_f}\right) \cong 0.44\;{A(\Gamma/1000)\over \phi Z^2(1+z)} \;{\rm PeV}\;. 
\label{esyngammaion}
\end{equation}
Thus proton or ion synchrotron can make a $\gamma$-ray component in jet sources provided the ion power and energy is sufficiently 
great.

\subsection{$L$-$\Gamma$ diagram}

Using the minimum Lorentz or Doppler factors implied from $\gamma$-ray opacity arguments, 
eqs.\ (\ref{eq2}) and (\ref{deltaD}), along with the apparent $\gamma$-ray luminosity 
corresponding to the time that $\Gamma_{min}$ is measured, we can 
construct an $L$-$\Gamma$ diagram.

\begin{figure}
\begin{center}
 \includegraphics[width=3.4in]{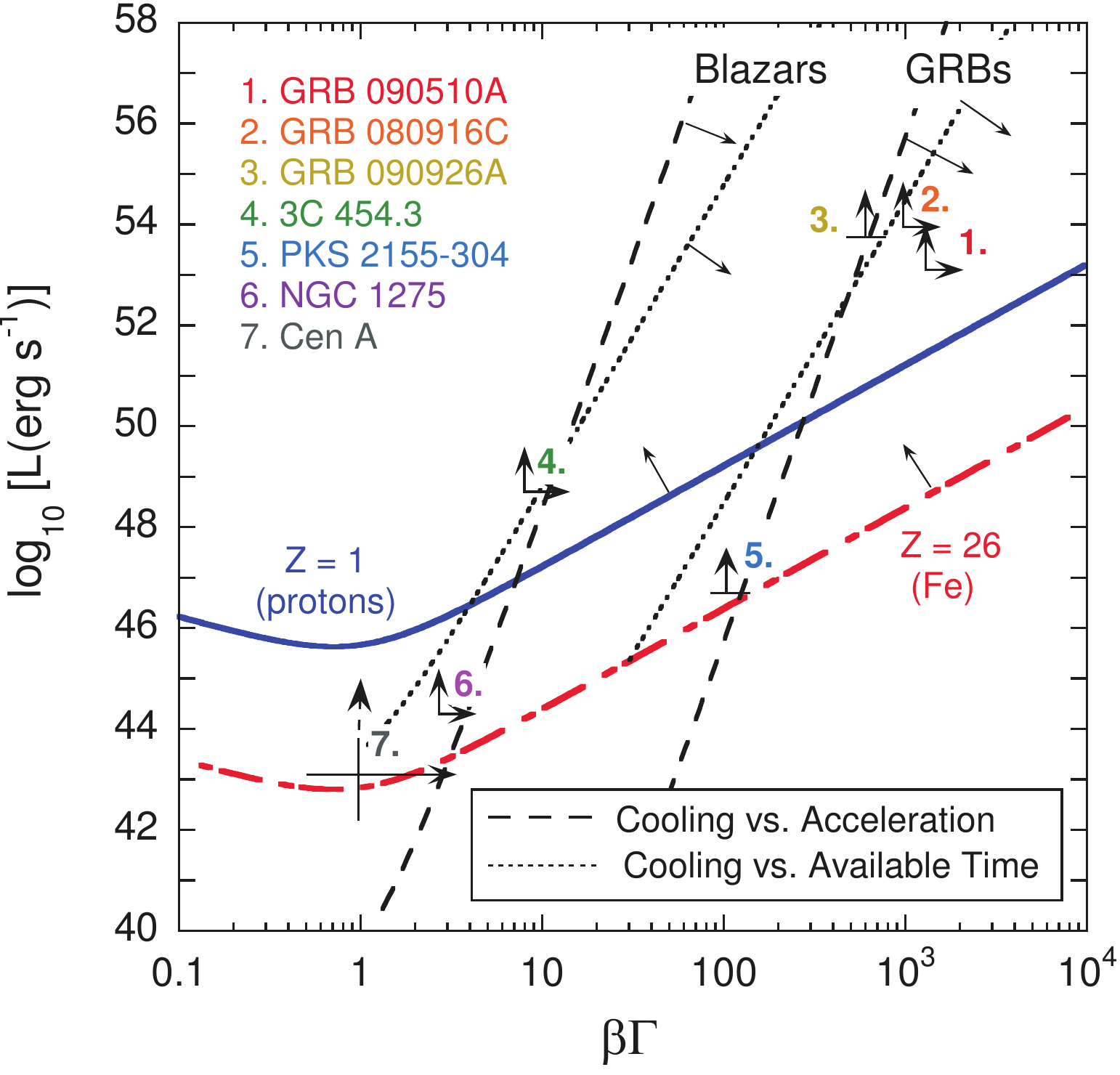}
 \caption{Data shows apparent isotropic $L$ versus $\Gamma_{min}$ for blazars, radio galaxies, and GRBs. Solid and dot-dashed curves plot the constraint given by eq.\ (\ref{Emax18}).  }
\label{geometr2y}
\end{center}
\end{figure}

Figure \ref{geometr2y} shows such a plot for blazars and GRBs \cite{2010ApJ...724.1366D}. Note that GRB 090510A is a short hard GRB, whereas  GRB 080916C and GRB 090926A are long-duration GRBs.  Except for PKS 2155-304, which uses HESS data for the giant outbursts in 2006, all the data were measured with the {Fermi} LAT.  As can be seen, the powerful Fermi GRBs have more than adequate luminosity to accelerate either protons or ions to ultra-high energies, even with Lorentz factors $\Gamma_{min}\sim 10^3$ as inferred from $\gamma\gamma$ opacity arguments. By comparison, the blazars and radio galaxies have smaller apparent luminosities and also smaller $\Gamma_{min}$. Acceleration of high-$Z$ ions like Fe to ultra-high energies is possible for these sources on the basis of eq.\ (\ref{Emax17}), but acceleration of protons appears unlikely, except possibly during flaring episodes.

\subsection{Luminosity density of extragalactic $\gamma$-ray jet sources}

Maintaining the intensity of UHECRs against photopion losses with the CMB requires a luminosity density in UHECRs of $\sim 10^{44}$ erg/Mpc$^{3}$-yr \cite{wb99}. This value can easily be obtained by dividing the energy density of UHECRs at $\approx 10^{20}$ eV, which is $\sim 10^{-21}$ erg/cm$^3$, by the photopion loss timescale $t_{p\pi}\sim 100$ Mpc/c. Because $t_{p\pi}$ increases rapidly with decreasing energy, the required luminosity density remains at this level even when fitting to the ankle of the cosmic-ray spectrum at $ E\approx 4\times 10^{18}$ eV \cite{dermer07}, though precise values of the local luminosity density depends in detail on the assumed cosmic star-formation rate factor applicable to GRBs. 

The standard argument for inferring the luminosity density of a source class is to assume that the nonthermal $\gamma$-ray luminosity represents, to an uncertain factor of order unity, the luminosity in UHECRs. This seems reasonable because some radiative losses into $\gamma$ rays are likely to occur during acceleration. If those losses are small, then the UHECR power could exceed the $\gamma$-ray luminosity, but if the particles experience severe radiative losses during acceleration, then they could not be the sources of the UHECRs  \cite{wb99}. This estimate is furthermore independent of beaming factor, because for every beamed source we detect, a proportional number of misdirected sources will point away from us. The argument suffers, however, from the likelihood that a large percentage of the $\gamma$ radiation is produced by ultra-relativistic leptons (especially in the case of blazars where correlated variability with lower energy bands can be monitored). Moreover, it is not certain that MeV $\gamma$ rays from long duration GRBs, which comprise the bulk of the energy output, is entirely nonthermal, inasmuch as a thermal/photospheric interpretation can potentially resolve the line-of-death problem that plagues nonthermal synchrotron interpretations of GRBs \cite{rp09}. Thus the nonthermal LAT emission might be a more appropriate luminosity with which to define the nonthermal luminosity density of GRBs \cite{egp10}.

\begin{figure}
\begin{center}
 \includegraphics[width=4.5in]{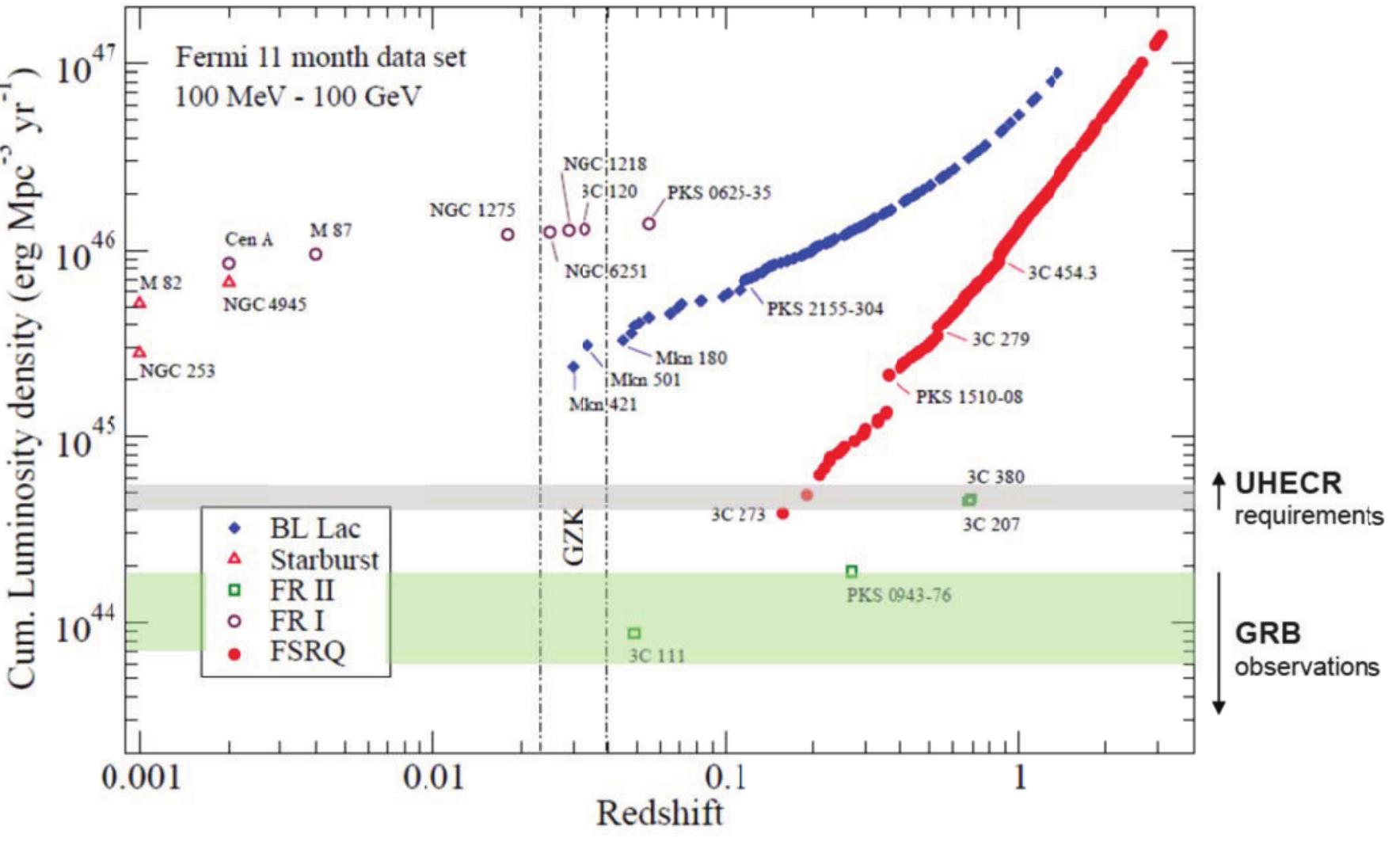}
 \caption{$\gamma$-ray luminosity density inferred from Fermi observations of various classes of $\gamma$-ray emitting AGNs \cite{2010ApJ...724.1366D}, 
compared with the local luminosity density of long-duration GRBs (Table \ref{tab:4}), and UHECR source requirements.}
\label{figlumden}
\end{center}
\end{figure}

The luminosity density of GRBs has been evaluated in detail in Section 8.2. 
The luminosity density of radio galaxies and blazars, based on the average 100 MeV -- 100 GeV flux 
measured in the 1LAC \cite{abd101LAC}, is shown in Fig.\ \ref{figlumden}.

\subsection{Origin of UHECRs}

The sources of the UHECRs are unknown. With the anisotropy of the UHECRs observed with Auger becomes
less significant that originally reported, 
leaving the cluster of UHECRs in the vicinity of Cen A as the only outstanding hotspot \cite{pao10}, even Galactic sources have been considered 
as plausible UHECR candidates \cite{ckn10}. Indeed, a wide variety of possible source classes could contribute
to the UHECRs, including magnetars and young pulsars \cite{aro03,ghi08}, particle acceleration in structure-formation shocks \cite{min08},
and new physics candidates. Criteria that the sources of UHECRs should meet are \cite{der11}
\begin{itemize}
\item extragalactic origin;
\item 	mechanism to accelerate to ultra-high energies;
\item  adequate luminosity density;
\item sources within the GZK radius; and
\item  UHECR survival during acceleration, escape, and transport.
\end{itemize}
On this basis, we \cite{dm09} suggest that the very luminous, very energetic extragalactic blazar and GRB jet sources are the 
most plausible candidates, with the BL Lac objects and FR1 radio galaxies perhaps  favored \cite{2010ApJ...724.1366D}. Electromagnetic signatures
of ultra-relativistic hadrons in GRBs and blazars are not clearcut. High-energy neutrino detections from transient or bursting 
$\gamma$-ray sources will be crucial to finally solve this puzzle. 

\subsection{Black holes, jets, and the extreme universe}


\subsubsection{Black hole physics}

The Kerr metric for an uncharged rotating black hole, written in the Boyer-Lindquist
coordinates $\{t,r,\theta,\varphi\}$ useful for asymptotic analysis in the exterior region
of the horizon of the black hole, 
 takes the form \cite{wal84}
\begin{equation}
ds^2=( \beta^2 - \alpha^2 )\;dt^2 \;+  \;2 \;\beta_\varphi \;d\varphi
\;dt
+\gamma_{rr}\; dr^2 + \;\gamma_{\theta \theta}\; d\theta^2 +
\;\gamma_{\varphi\varphi}\;d\varphi^2 \;,
\end{equation}
where the metric coefficients are given by
$$\beta^2-\alpha^2 \;= \;g_{tt} \;=\; -1 + \frac{2Mr}{\rho^2}\;,$$$$\beta_\varphi \; \equiv g_{t \varphi}\; = \;\frac{-2Mr a
\sin^2\theta}{\rho^2}\;,\;\;\;\gamma_{rr} =
\frac{\rho^2}{\Delta}\;,$$
$$
\gamma_{\theta \theta} = \rho^2, \;\; {\rm and}\;\; \gamma_{\varphi \varphi} = \frac{\Sigma^2 \sin^2\theta}{\rho^2}\;.
$$
Here, $\alpha$  is the lapse function and $\beta$ is the shift vector,
$$\rho^2 = r^2 + a^2
\cos^2\theta\;,\;\;\;\Delta = r^2 -2 M r + a^2\;,\; {\rm and}\;\;\Sigma^2 = (r^2 + a^2)^2 -\Delta \; a^2 \sin^2\theta\;.$$
 Additionally
$$
\alpha^2 = \frac{\rho^2 \Delta}{\Sigma^2}, \;\;\; \beta^2 =
\frac{\beta_\varphi^2}{\gamma_{\varphi \varphi}}\; ,\;{\rm and}\;\;\sqrt{-g}=\alpha\; \sqrt{\gamma} = \rho^2 \sin\theta\;.$$
The parameter $a$ is the angular momentum per unit
mass of the Kerr black hole, so $aM$ is the angular momentum of the 
black hole. The horizons  are located at   $ r_\pm = M \pm \sqrt{M^2 - a^2} $, 
as shown in Fig.\ \ref{horizons}.

\begin{figure}
\begin{center}
 \includegraphics[width=3.4in]{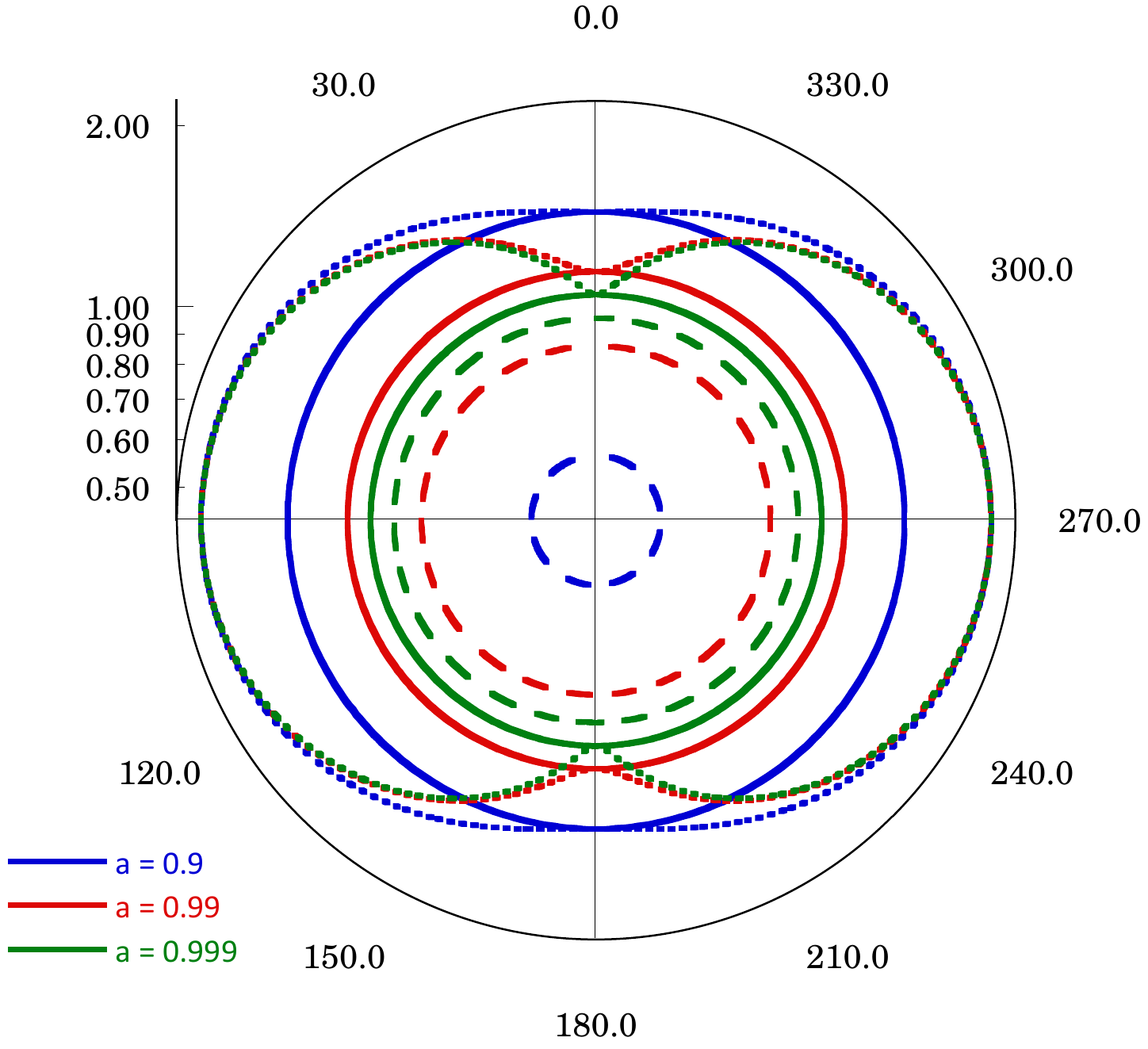}
 \caption{Sketch of locations of the horizons $r_+$ and $r_-$ and the ergosphere $r_{erg}(\theta )$ 
in  the Kerr geometry.}
\label{horizons}
\end{center}
\end{figure}

The set of points described by $g_{tt} = 0$ defines the ergosphere
\begin{equation} 
r_{erg} (\theta) = M + \sqrt{M^2 - a^2 \cos^2 \theta}\;.
\end{equation}
Note that $g_{tt} > 0$ in the region
\begin{equation} 
r_+ < r < r_{erg} (\theta)\;,
\label{ergo}
\end{equation}
and consequently $\tilde t \equiv \partial_t$ becomes spacelike in the above region.
 When $r > r_{erg}$,  $\tilde t$ is timelike and future directed. See Fig.\ \ref{horizons}.

In elementary electrodynamics \cite{jac62}, the Poynting vector in flat space far away from
a gravitating object is given by
\begin{equation}
S =  {c\over 4\pi}\;(E\times H)\;,
\label{S}
\end{equation}
and 
\begin{equation}
{d\over dt}{\cal E}_{tot} = {d\over dt}({\cal E}_{field} + {\cal E}_{mech} )\;, 
\end{equation}
where
\begin{equation}
{\cal E}_{field} = {1\over 8\pi} \int_V d^3 x\,(E^2 +B^2)\;,\;{\rm and}
\label{Efield}
\end{equation}
\begin{equation}
{d\over dt}{\cal E}_{mech} =  \int_V d^3 x\;J\cdot E 
\label{Efield1}
\end{equation}
give the field and mechanical energies, respectively. 
Thus we see that $J\cdot E$ is an emissivity, so conservation of 
energy requires 
\begin{equation}
{\partial u\over \partial t} + \nabla
\cdot S = - J\cdot E\;,
\label{pupt}
\end{equation}
where the field energy density is $u = (E\cdot D + B\cdot H)/8\pi$.

Electromagnetic energy extraction through the Blandford-Znajek process
involving Penrose processes in curved space can be expressed in the case of a  
stationary axisymmetric force-free magnetosphere as 
\begin{equation}
{d^2{\cal E}\over dA dt}|_{\rm BZ} = S^r \sqrt{\gamma_{rr}} = -H_\varphi \Omega B^r \sqrt{\gamma_{rr}}\;
\label{dEdtbz}
\end{equation}
\cite{kom04,md05}. The fields must satisfy the Znajek regularity condition \cite{zna77} 
\begin{equation}
H_\varphi \left |_{r_+} = \frac{\sin^2\theta}{\alpha}\; B^r\; (2Mr\; \Omega -a) \right |_{r_+}
\label{znaregcond}
\end{equation}
at the event horizon. We can suppose that the 
polodial function $\Omega$ defining the fields is radially independent, so 
\begin{equation}
{d{\cal E}\over dt}|_{\rm BZ} = 4\pi M r_+ \int_0^\pi d\theta \sqrt{\gamma}\;{\sin^2\theta\over \alpha}  \Omega (\Omega_H - \Omega) B_+^{r 2} \;, 
\label{dEdtbz1}
\end{equation}
where $B^r_+$ is the radial magnetic field at the event horizon and 
\begin{equation}
\Omega_H = {a\over 2Mr_+ }
\label{Omegah}
\end{equation}
is the angular velocity of the event horizon \cite{lev06}.

The form of $\Omega$ in eq.\ (\ref{dEdtbz1}) is found by solving the constraint equation, 
originally derived by Blandford and Znajek \cite{bz77}, governing fields and  currents in a 
black-hole magnetosphere under the force-free condition.  The constraint equation in the $3+1$ formalism takes 
the form  \cite{md05}
\begin{equation}
{1\over 2\Lambda } \,{dH^2_\varphi\over d\Omega} = \alpha (\rho_c\Omega\gamma_{\varphi\varphi} - J_\varphi )\;,
\label{constraint}
\end{equation}
where the charge density $\rho_c$ and current density $J$ is expressed in terms of $\Omega$ and metric coefficients.
By satisfying the Znajek regularity condition, eq.\ (\ref{znaregcond}), we \cite{md05}
 obtained  solutions
\begin{equation} 
\Omega_\pm \equiv \frac{a}{2 M r_+ \pm \rho_+ ^2}\;,
\label{Omegapm}
\end{equation}
($\rho_+^2 = r_+^2 +a^2\cos^2\theta)$ by considering the behavior of the constraint equation in the far-field
limit for radially-independent $\Omega$. Thus
\begin{equation} 
\Omega_+ \equiv \frac{a}{2 M r_+ + \rho_+^2}\;,
\label{Omegap}
\end{equation}
and
\begin{equation} 
\Omega_- \equiv \frac{1}{a\sin^2\theta}\;.
\label{Omegam}
\end{equation}
The $\Omega_+$ solution implies a rate of electromagnetic energy extraction 
of
\begin{equation}
\frac{d \cal E}{dt}|_{\rm BZ} =\frac{\pi Q_0^2}{a r_+}\left( \arctan \frac{a}{r_+}- \frac{a}{2M}\right)\;\rightarrow
\;{{8\pi\over 3} B_0^2 M^2} \times \cases{\left({a\over 2M}\right)^2\;,\; &  $a/2M\ll 1$ \cr\cr \left( {2\over 9}\right) \left( {\pi\over 4} - {1\over 2}\right)\; , \;&  $a\rightarrow M\;,$ \cr}\;\;
\label{fracdEdt}
\end{equation}
after relating $Q_0$ and $B_0$ \cite{dm09}.
The $\Omega_+$ solution generalizes the Blandford-Znajek \cite{bz77} split monopole solution  for all values of $0< a < M$.
Inspection of this solution shows that Poynting flux outflow is greatest along the equator, so the energy flux has 
a pancake geometry. The jet geometry, as found ubiquitously in nature, is not described by this solution. 

The exact $\Omega_-$ solution implies an electromagnetic energy flux peaking towards the poles of the rotating
black hole, but describes an inward energy flux, so is unphysical. The $\Omega_-$ solution has an additional
freedom in the assignment of an arbitrary poloidal field $\Lambda$ in the prescription for the poloidal magnetic field
\begin{equation}
B_{P} = {\Lambda\over \sqrt{\gamma}}(-\Omega_{,\theta}\partial_r + \Omega_{,r}\partial_\theta )\;
\label{BP}
\end{equation} 
\cite{md07}. Energy extraction is accomplished not only by Poynting flux outflow, but also by the matter currents that
sustain the black-hole magnetosphere. Analysis of the timelike geodesics consistent with the Znajek regularity condition \cite{md09}, 
and symmetry transformation of the $\Omega_-$ solution yields a dual class of exact solutions with positive energy extraction of
Poynting and matter  outflows. The radial magnetic field for this dual class of solutions takes the form
\begin{equation}
B^r = {2\over a}\Lambda {\cos\theta\over\sqrt{\gamma}\sin^3\theta}\;,
\label{Brdual}
\end{equation}
The extra freedom in the solutions with general $\theta$-dependent $\Lambda$ provides formal solutions devoid of 
strict physical content, and with arbitrary $a$ dependence.  

Performing dimensional analysis of energy extraction using 
eq.\ (\ref{dEdtbz1}) (see \cite{md05,md11,dm09} for detailed treatments) yields an estimate of 
the BZ power given by
\begin{equation}
{d{\cal E}\over dt}|_{\rm BZ} \approx 4\pi M r_+ \;{ \sqrt{\gamma}\over \alpha}  \Omega (\Omega_H - \Omega) B_+^{r 2}\approx \pi M r_+ \;{ \sqrt{\gamma}\over \alpha}  \Omega_H^2 B_+^{r 2} \;, 
\label{dEdtbz2}
\end{equation}
where $B_+^r$ is the radial component of the magnetic field threading the event horizon. Note that the expression peaks at $\Omega = \Omega_H/2$ \cite{lev06a}.  Along the equatorial direction, $\theta \cong \pi/2$, $\sqrt{\gamma}/\alpha \cong \Sigma^2/\Delta$. 
Supposing the expression is evaluated on a size scale 
somewhat larger than the event horizon, to cancel the  divergence, 
we have
\begin{equation}
{d{\cal E}\over dt}|_{\rm BZ} \approx \pi c\; \left({a\over M}\right)^2 \; r_+^2  \; B_+^{r 2}\;\approx 10^{47}\; \left({a\over M}\right)^2 \; M_9^2  \; B_4^2\;\;{\rm erg/s}\;, 
\label{dEdtbz3}
\end{equation}
where $M_9$ is the mass of the black hole expressed in units of $10^9 M_\odot$, 
and the magnetic field is given in units of $10^4$ G. 

As expressed, and often found in the literature, the value of $10^4$ G is artificial.  The magnetic field 
threading the ergosphere can be scaled to the Eddington luminosity
\begin{equation}
L_{\rm Edd} = 1.26\times 10^{47} M_9\;{\rm erg/s}
\label{Ledd}
\end{equation} 
by the expression 
\begin{equation}
{B^2\over 8\pi } = \epsilon_B \ell_{\rm Edd} \;{L_{\rm Edd}\over 4\pi r_+^2 c}\;,
\label{b28pi}
\end{equation}
where the parameter $\ell_{\rm Edd}$ reflects not only different accretion rates but different radiative efficiencies, including the reduction in radiative efficiencies in the low 
Eddington ratio, advection-dominated regime.
Thus
\begin{equation}
{{d{\cal E}\over dt}|_{\rm BZ}\over L_{\rm Edd} } \approx \epsilon_B \ell_{\rm Edd}\ll 1\;.
\label{dEdtbz4}
\end{equation}

\subsubsection{Jets and the extreme universe}

The time-averaged luminosities of FSRQs  extend to values in excess of $L_\gamma\approx 10^{49}$ erg s$^{-1}$ (Figure \ref{GammavsL}). In the extraordinary 2010 November flare of 3C 454.3 \cite{abd11}, $L_\gamma$ reached apparent isotropic luminosity of $(2.1\pm 0.2)\times 10^{50}$ erg s$^{-1}$ over a period of a few hours, making it the most luminous blazar yet observed. Black-hole mass estimates for 3C~454.3 are in the range $0.5 \lesssim M_9 \lesssim 4$ \cite{bon10}, where $10^9 M_9M_\odot$ is the mass of the black hole powering this AGN. For this range of masses, the Eddington luminosity therefore  ranges from $\approx 6\times 10^{46}$ erg s$^{-1}$ to $\approx 5\times 10^{47}$ erg s$^{-1}$. During this extreme outburst, the apparent luminosity of 3C 454.3 was more than a factor of $\approx 400$ greater than its Eddington luminosity. Even its time-averaged luminosity of $L_\gamma\approx 10^{49}$ erg s$^{-1}$ is super-Eddington by a factor of $\approx 20$. 

Assuming that the Eddington condition {\it does} limit accretion flow onto the black hole, which is likely to be the case for the long-term average luminosity if not for the flaring luminosity, then the absolute radiant luminosity is limited to a value of $L_{abs} \lesssim  5\times 10^{47}$ erg s$^{-1}$. This is consistent with the large apparent luminosities if the emission is highly beamed. For a simple top-hat jet beaming factor, a jet opening angle $\theta_{j}$ implies a beaming factor $f_b = \theta_j^2/2$ for a two-sided jet with $\theta_j \ll 1$. A mechanism for collimation is, however, required. Should this arise from the Blandford-Znajek process, then we are still restricted to values of the absolute Blandford-Znajek power, eq.\ (\ref{dEdtbz3}).
 As we have seen, by scaling the energy density of the magnetic field to the energy density of accreted matter near the event horizon shows that the Blandford-Znajek power is likewise Eddington-limited. Making the hypothesis \cite{dermerlott} that the extraction of energy through black-hole rotation collimates the jet outflow with $\cos\theta_j \approx a/M$, then $f_b \cong 1 - (a/M)$ and $\theta_j \cong \sqrt{2(1-a/M)}$, implying $a/M > 1 - (L_{\rm Edd}/L_\gamma)$. If the jet opening angle is a consequence of the bulk Lorentz factor $\Gamma$ of the outflow, then $\Gamma \gtrsim \sqrt{L_\gamma/2L_{\rm Edd}}$.

For the case of 3C 454.3 in its flaring state, when $L_\gamma/L_{\rm Edd}\gtrsim 10^3$, this hypothesis then implies that $a/M > 0.999$ and $\Gamma \gtrsim 23$, which is consistent with the value $\Gamma_{min} \approx 14$ from $\gamma\gamma$ opacity arguments \cite{abd11}. In the most conservative case with $M_9 = 4$, $a/M \cong 0.998$ and $\Gamma \gtrsim 15$. This is  marginally consistent with the limiting maximum value $a/M \cong 0.996$ suggested by Aschenbach from analyses of microquasars and the Galactic Center black hole \cite{2004A&A...425.1075A}. The smaller black-hole mass estimate,   $M_9 = 0.5$ \cite{bon10}, implies a value of $a/M$ that violates this limit by a large margin.  Much work, both numerical and theoretical, has been devoted to jet formation from the Blandford-Znajek process, and it is unclear if jet collimation can be described by the guess that  $\cos\theta_j \approx a/M$, but it is interesting to suggest this possibility, which leads to values of $\Gamma$ consistent with separate inferences regarding the outflow Lorentz factor.

\begin{figure}
\begin{center}
 \includegraphics[width=3.4in]{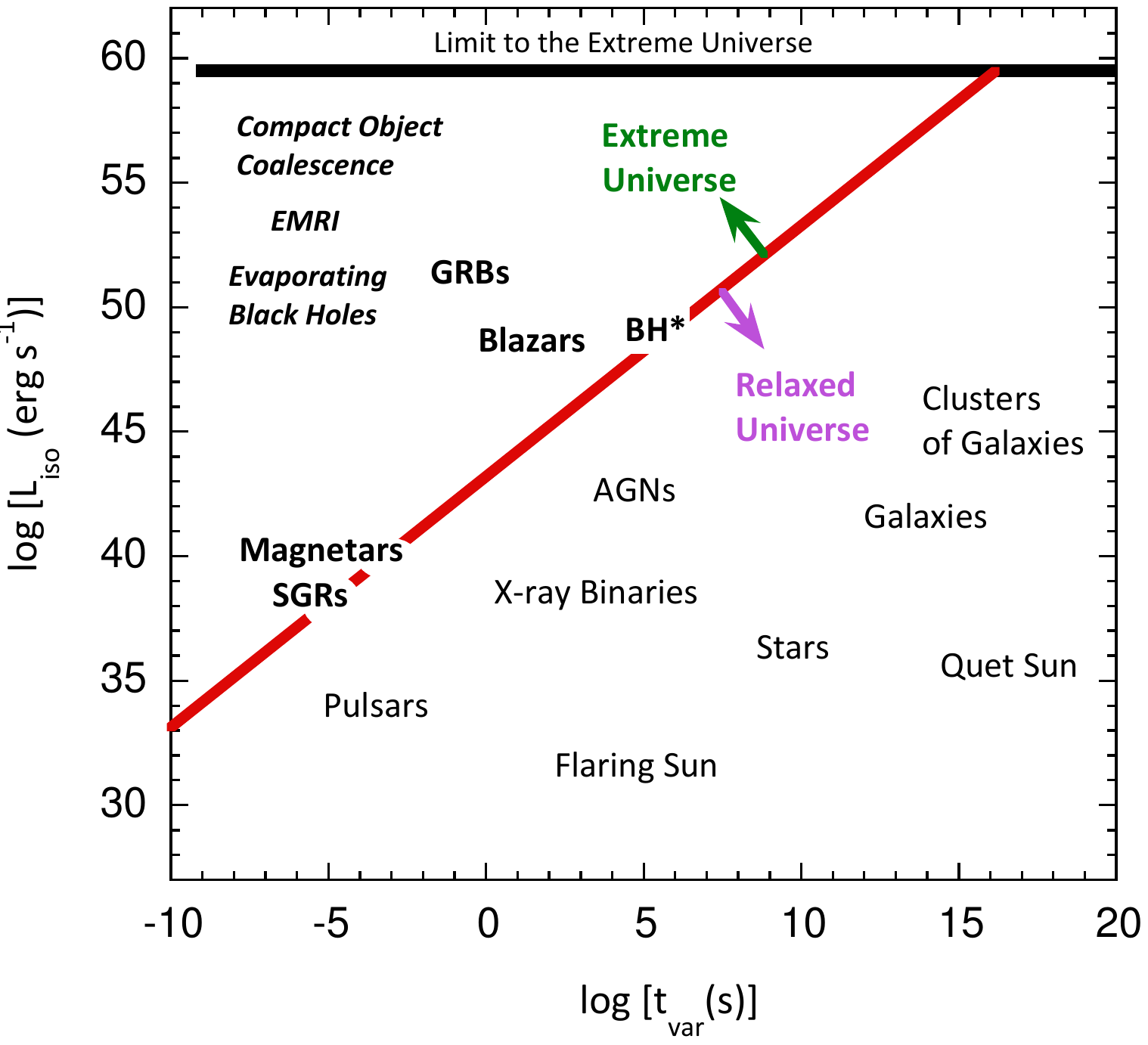}
 \caption{Luminosity versus characteristic timescales for variability or duration
 for various types of astronomical sources and cosmic events. Upper limit is from eq.\ (\ref{Lmax}) and
the horizontal line separating the extreme and moderate universe is defined by eq.\ (\ref{LEDDtS}).}
\label{extremeuniverse}
\end{center}
\end{figure}

Figure\ \ref{extremeuniverse} summarizes some various source classes 
and their residence in a diagram of L vs.\ time. The extreme universe, 
as contrasted with the moderate, or relaxed universe, is
defined by eq.\ (\ref{LEDDtS}), when 
the ratio of the apparent luminosity and variability timescale exceeds 
the Eddington luminosity divided by the light-crossing time for the Schwarzschild 
radius, $L_{\rm Edd}/t_{\rm S} = 1.26\times 10^{43}$ erg/s$^2$. 
``BH*" stands for events where normal main sequence stars are tidally captured
by a black hole, as in the 28 March 2011 event \cite{2011Sci...333..203B}, and ``EMRI"
stands for extreme mass-ratio inspiral events \cite{2009CQGra..26i4034G}. The luminosity limit of the 
extreme universe is given by eq.\ (\ref{Lmax}), namely 
$c^5/G = 3.63\times 10^{59}$ erg/s. Through observations
of blazars and GRBs, $\gamma$-ray astronomy is pushing toward this limit. A coordinated assault involving
gravitational wave observatories, the Fermi $\gamma$-ray telescope, neutrino observatories and multiwavelength 
campaigns might over the next decade  reach this limit of the extreme universe.

\vskip0.2in

{\it Acknowledgements}\\
The results presented in these lectures would not have been possible without the dedicated efforts of 
hundreds of Fermi Collaboration members. I would like to specifically 
thank the following persons for correspondence and use of presentation material:\\
1. GeV instrumentation and the GeV sky with the Fermi Gamma-ray Space Telescope \\
\indent		Thanks to  S.\ Ritz, P.\ Michelson, J. McEnery,  S.\ Digel, D. Thompson, W. Atwood\\
2. First Fermi Catalog of Gamma Ray Sources and the Fermi Pulsar Catalog\\		
\indent		Thanks to L.\ Guillemot, S.\ Digel, P.\ Ray\\
3. First Fermi AGN Catalog \\		
\indent			Thanks to B.\ Lott, P.\ Giommi\\
4. Relativistic jet physics and blazars\\
\indent			Thanks to C.C.\ Cheung, J.\ Finke\\
5. $\g$ rays from cosmic rays in the Galaxy\\
\indent			Thanks to S. Funk, A. Strong, I. Moskalenko, N. Giglietto, W. Atwood, S. Digel\\
6. $\g$ rays from star-forming galaxies and clusters of galaxies, and the diffuse extragalactic $\g$-ray background \\
\indent			Thanks to K. Bechtol, J. Kn\"odlseder, P. Martin, M. Ackermann\\
7. Microquasars, radio galaxies, and the extragalactic background light \\		
\indent			Thanks to R. Dubois, C.C. Cheung, J. Finke, S. Razzaque\\
8. Fermi Observations of Gamma Ray Bursts\\		
\indent			Thanks to A. von Kienlin, V. Connaughton, K. Hurley, C. Fryer, C. Kouveliotou, N. Omodei, S. Razzaque, D. Eichler\\
9. Fermi acceleration, ultra-high energy cosmic rays, and Fermi at 2\\
\indent			Thanks to A. Levinson, M.\ Barkov, G. Menon

I would especially like to thank Paola Grandi, Benoit Lott, David Paneque, Stefan Funk, Marco Ajello, and Aous Abdo and Damien Parent for providing, and in some cases
modifying from the original, Figs.\ \ref{GammavsL}, \ref{3c454ltcv} and \ref{3c454spct}, \ref{mrk501fig}, \ref{Fermi_LAT_SNRs},  \ref{fig:edb}, and \ref{B1259_lc}, respectively.
I would also like to thank Justin Finke for the use of Fig.\  \ref{EBL+figs}, Soeb Razzaque for the use of Fig.\ \ref{figlumden}, and Jean-Marc Casandjian for correspondence.
Detailed comments by  Tyrel Johnson and Igor Moskalenko are gratefully acknowledged. In addition, I would like to thank
Marco Ajello, Armen Atoyan, Markus B\"ottcher, and Kohta Murase for collaboration, and Roland Walter and Marc T\"urler for the opportunity to lecture in the Saas-Fee course.

This work was supported by the Office of Naval Research and NASA.
 
\vspace*{0.5 in}

\input{reference}


\printindex
\end{document}

%% file: reference.tex
%
%

%
%
